# Phonons and Anomalous Thermal Expansion Behaviour in Crystalline Solids

R. Mittal[1,2], M. K. Gupta[1] and S. L. Chaplot[1,2]
[1]Solid State Physics Division, Bhabha Atomic Research Centre, Trombay, Mumbai 400 085
[2]Homi Bhabha National Institute, Anushaktinagar, Mumbai 400094, India

**Abstract**

Anomalous thermal expansion behaviour of several open frame-work compounds has been extensively investigated using the techniques of inelastic neutron scattering and lattice dynamics. These compounds involve increasing level of structural complexity and flexibility, which leads to increased values of thermal expansion coefficients approaching colossal values. In several compounds, neutron inelastic scattering experiments have produced quantitative estimates of the anharmonicity of phonons over a range of low energies, and thereby explained the observed thermal expansion quantitatively. The anharmonicity is found to be an order of magnitude larger than that in usual materials. Lattice dynamical calculations have correctly predicted the observed anharmonicity in the neutron experiments and revealed the overall nature of phonons involved. In compounds showing negative thermal expansion, the phonons responsible have rather low energies up to 10 meV. In most compounds, the anharmonic phonons span all over the Brillouin zone, while in some cases the specific phonons are limited to certain wave-vectors. The nature of specific phonons responsible for anomalous behavior is found to be different in all these compounds. These phonons generally involve transverse vibrations, librations and internal distortions of the polyhedral units. The paper reviews recent advances in the understanding of anomalous thermal expansion behaviour.

# CONTENTS

# 1. Introduction

Compounds that contract upon heating over a certain temperature range are exceptional. Such materials find extremely useful technical applications in composites. Compensating the usual (positive) thermal expansion of ordinary materials by contracting compounds allows tailoring the thermal expansion behavior of the ensemble. The technological relevance of these compounds and their composites principally derives from their ability to withstand thermal shock without any damage. The negative thermal expansion (NTE) phenomenon typically originates [1-39] from the presence of low-energy anharmonic vibrations of atoms.

The phenomenon of negative thermal expansion (NTE) is well known in nature, most notably in water around its melting point. Amongst solids, invar, which is an alloy of Fe and Ni, has very low or negative volume thermal expansion coefficient depending on its composition. The popular semiconductor materials silicon and germanium also exhibit the NTE behavior at low temperature. However, the discovery of large negative thermal expansion (NTE) in $ZrW_2O_8$[40] over a wide range of temperatures has lead to extensive research in this area[1-25, 28, 34, 36, 40-99]. Over the last two decades anomalous thermal expansion behavior has been found in large number [12, 63, 70, 75, 79, 83, 84, 87, 100-112] of open frame work compounds. These compounds find applications in forming the composites with tailored thermal expansion coefficients useful for applications such as in fiber optics, coatings, electronics, and mirror substrates to tooth fillings.

NTE has been found in framework oxides[40, 100, 101] with M-O-M (M = metal) linkages. For example, $ZrW_2O_8$ has a NTE coefficient of $-27 \times 10^{-6}$ K$^{-1}$ at 300 K. Further, the NTE coefficients of compounds with M–CN–M linkages was found to reach a much higher value of $-51 \times 10^{-6}$ K$^{-1}$ for $Zn(CN)_2$[102] at 300 K. Apparently, the linkage between two neighboring polyhedral units by two atoms (CN) instead of a single atom (O) provides a greater framework flexibility. The concept of increasing framework flexibility has led to the discovery[103, 104] of colossal positive and negative thermal expansion in $Ag_3Co(CN)_6$ and $Ag_3Fe(CN)_6$. The crystal lattice expands remarkably in the hexagonal a-b plane ($\alpha_a \sim 120 \times 10^{-6}$ K$^{-1}$), while it contracts equally strongly along the c-axis ($\alpha_c \sim -110 \times 10^{-6}$ K$^{-1}$). The structure thus expands and contracts at rates more than ten times larger than what is found in normal materials.

A number of theoretical and experimental studies[1, 42, 57, 113, 114] have been performed to identify these kinds of materials and understand the phenomenon of anomalous thermal expansion. Contraction with increasing temperature could arise due to proximity to phase transitions[18, 76, 115-117]. Sometimes changes in electronic configuration[117, 118] and open framework nature of compounds [89, 116, 119] may cause the negative thermal expansion behavior. The contraction may be isotropic or anisotropic depending upon the structure and bonding in the compound[40, 115, 120, 121]. $ZrW_2O_8$, $ScF_3$ and $Cd(CN)_2$ are known to show isotropic negative thermal expansion behavior[36, 62]. These materials are characterized by strong atomic bonds and NTE arises from the dynamical deformation which consumes open spaces in a crystal lattice[122, 123]. However, a number of minerals and cyanides like quartz, $LiAlSiO_4$, $Ag_3Co(CN)_6$, $ZnAu_2(CN)_4$, $KMn[Ag(CN)_2]_3$ and MCN (M=Ag,Au,Cu) are known for anisotropic negative thermal expansion[1, 44, 103, 124, 125].

Various mechanisms have been proposed in the literature to understand the NTE. In certain cases, the atomic radius may change significantly with changing temperature and may be responsible for the NTE behaviour. This could happen when charge transfer occurs between constituent atoms with temperature and give rises to significant contraction in one or more participating atoms. This type of mechanism is reported in $Sm_{2.75}C_{60}$[117]. In magnetically driven NTE mechanism, the magnetic moment of atoms changes with temperature such that it causes contraction in the volume. The very famous compound invar[126, 127], which has a very small thermal expansion (~$1\times10^{-6}$ $K^{-1}$), works on magneto-volumetric driven NTE mechanism. In this review, we do not focus on the few materials that exhibit NTE via the above two mechanisms.

Geometric models of thermal contraction are obviously of much interest. One such model[128, 129] has been based on rigid librations of the linked polyhedra that would shrink the lattice with increase in the libration amplitude. This model, however, leads to the prediction[128] of negligible magnitude of shear elasticity, that is contrary to experiments in case of $ZrW_2O_8$[130, 131]. In fact, $ZrW_2O_8$ appears to be a perfectly normal elastic material with a bulk modulus of 74 GPa at 300 K, that is, with no abnormally low values of any of the elastic constants[130], and possesses Debye-like specific heat at very low temperatures[131]. Sleight [123] has provided a mechanism of NTE in open network structures in terms of large transverse thermal motion of oxygens in the M-O-M linkages. In $ScF_3$ compound, Hu et al [55] proposed the Guitar string model of NTE, i.e. if the string is pulled in transverse direction it will create tension along the string.

The dynamics of atoms in solids plays a very important role[7, 15, 132]in unusual phenomena like negative thermal expansion. Thermal expansion, positive or negative, in insulators is known to arise from anharmonic lattice vibrations. The vibrational entropy is important in this context and its value at a given temperature would increase if any phonon frequency decreases. If phonons soften on contraction of the lattice, it could lead to thermal contraction or negative thermal expansion.

Liu et al [133, 134] explained the NTE behaviour in terms of configurational entropy of metastable states, which might be viewed as the excited states of the anharmonic phonon oscillators. In NTE compounds, the mean atomic distance in the metastable state is smaller than that in the ground state configuration. Thermodynamically if entropy of the system increases with pressure, the system will exhibit NTE behaviour.

Anharmonicity of phonons has two components, one due to change in lattice volume, and another due to increase in vibrational amplitude. The change in phonon frequencies due to change in volume or pressure at a fixed temperature, is known as implicit anharmonicity. It is represented by the so called Grüneisen parameters. The other part, known as the explicit anharmonicity, has the dominant role in determining the phonon frequencies at a finite temperature, and consequently the phonon free energy. We note that in many NTE materials, the comparison of the measured temperature and pressure dependence of phonon spectrum revealed that the explicit anharmonicity, not only dominates but has an opposite sign to the quasiharmonic (implicit) anharmonicity. This shows the dominant role of the explicit anharmonicity in determining the temperature dependence of the phonon frequencies and so also the free energy.

On the other hand, the equilibrium volume at a given temperature is determined by the minimum of the free energy with respect to volume. Therefore, one has to look at the variation of the free energy as a function of volume at a fixed given temperature. This would be reflected in the variation of the phonon frequencies with respect to volume at the given temperature (mode Grüneisen parameters). At low temperature, the Grüneisen parameters may be determined essentially by the quasiharmonicity. At a given finite temperature, one should use the Grüneisen parameters at that temperature. However, the temperature dependence of the Grüneisen parameters is ignored in most calculations of thermal expansion. The hope is that, although a phonon frequency may strongly depend on temperature (both the implicit and the explicit parts), the difference in the value of the frequency at two different volumes (implicit part) at a fixed given temperature, may not as strongly depend on the temperature. This would be expected if the implicit and the explicit parts do not couple with each other. In other words, the explicit part may largely cancel out in determination of the Grüneisen parameters. The coefficient of thermal expansion as a function of temperature is determined by the mode Grüneisen parameters multiplied by the temperature-dependent specific heat due to the phonon mode (summed over all the modes). The explicit anharmonicity has a significant role in the latter specific heat term at high temperatures, as for example, revealed in the calculations by Glensk et al[135].

To summarize, the implicit anharmonicity is mainly relevant in determining the equilibrium volume at a given temperature, since the phonon frequency changes with volume determine the changes in the vibrational entropy leading to maximum in the free energy. The role of the explicit anharmonicity is significant at high temperatures [38, 39, 136]. At high temperature, the frequencies would renormalize both due to the implicit and the explicit part. However, while calculating the changes in the phonon frequencies with respect to volume at a fixed high temperature, the explicit part may largely cancel out. The implicit anharmonicity may also be termed as 'quasiharmonic' anharmonicity as it may be calculated in the quasiharmonic approximation.

The above discussion indicates that the quasiharmonic approximation should be useful in calculation of the mode Grüneisen parameters, and so the NTE, at least at low temperatures. This is actually borne out in practical calculations of NTE in a large number of compounds as reviewed in the present article. In fact, in most compounds, the quasiharmonic calculations are found to produce a good quantitative estimate of the NTE up to fairly high temperatures. On the other hand, anharmonicity can be a source of thermal expansion (positive or negative) even if quasiharmonicity (implicit anharmonicity) is zero. In some of the cases careful comparisons of the magnitude of anharmonic and quasi-harmonic effects on phonons show that the anharmonic part may be significantly high. These calculations are based on the many- body theory of anharmonicity and molecular dynamics simulations. Recently Glensk et al[135] has shown that strong anharmonic as well as many-body effects are important for understanding the accurate thermodynamic description for fcc metals (Al, Ag, Au, etc). Hence these effects are also expected to cause influence on the thermal expansion behavior. Further, In case of $Ag_2O$ and $ScF_3$, quasiharmonic lattice dynamics [2, 43] could not quantitatively reproduce the observed thermal expansion at high temperature, while the molecular dynamics simulations [62, 68] were quite successful.

In harmonic theory of lattice dynamics, we expand the crystal potential in terms of atomic displacement up to second order and ignore the higher order terms. At high temperature, the higher order terms become significant, which are responsible for phonon-phonon interaction. The phonon frequency shift due to these anharmonic terms may be treated using perturbation method. When the phonon frequency changes because of its own anharmonicity (i.e., as an isolated oscillator), it is called self-energy correction. The frequency shifts due to self anharmonicity and phonon-phonon interaction are referred as explicit anharmonicity (sometimes as true anharmonicity), whereas, the quasiharmonic approximation yields only the implicit anharmonicity. The former aspects have been extensively discussed in the literature[137-140]. As noted above, the implicit anharmonicity is most relevant to NTE since it mainly contributes to the changes in phonon entropy with pressure. However, in calculation of the phonon frequencies with temperature, both the terms are important. Since NTE is contributed by mainly phonons with large implicit anharmonicity, it may turn out that the phonons which show large explicit anharmonicity are different than those involved in NTE.

Quantitative estimates of the implicit part are best obtained by high-pressure experiments at a fixed given temperature. Temperature dependent measurements may not, however, yield good estimates of the implicit part of the anharmonicity. For example, in case of $ZrW_2O_8$, the measured phonon spectrum at low energies showed hardening with increase of temperature, unlike the softening seen in pressure dependent measurements, which clearly indicates that the implicit anharmonicity may be the less dominant one, and may work opposite to the other explicit part.

The magnitude of thermal expansion coefficient arising from any one phonon is proportional to the magnitude of the specific heat arising from that phonon. So the temperature dependence of the thermal expansion coefficient due to a phonon has the quantum mechanical behavior similar to the specific heat of an Einstein mode. It is needless to say that a purely classical description would not produce the temperature dependence at low temperatures.

The classical molecular dynamics as well as the ab-initio molecular dynamics simulations involve solving Newton's equations of motion[122, 141]. These simulations can produce anharmonic behaviour of vibrations at high temperatures. Thermal expansion has been calculated by the simulations in $Ag_2O$ [68], which produces a good agreement with experiments.

In several compounds, anharmonicity of phonons has been determined from measurements of the pressure dependence of phonon density of states using inelastic neutron scattering experiments on powder samples, and thereby explained the observed NTE quantitatively. Anharmonicity was found to be an order of magnitude larger than that in usual materials. The pressure dependent phonon spectra in the entire Brillouin zone have been calculated using ab–initio density functional theory and potential based model.It has been shown that a proper description of NTE requires consideration of both the acoustic and optic phonon modes in the entire Brillouin zone.

The anomalous thermal expansion behaviour of several polyhedral as well as low dimensional compounds has been extensively studied [1-6, 8-16, 24, 25, 142, 143] using the techniques of inelastic neutron scattering and lattice dynamics. It has been shown that the nature of phonons responsible for anomalous behavior is found to be different pertaining to the compounds described in this review. A brief description about the scientific interest and results obtained for various compounds which are reviewed in this article are given below:

Isotropic negative thermal expansion has been known [40, 100, 101] in cubic $ZrW_2O_8$, $HfW_2O_8$ and $ZrMo_2O_8$ upto 1443, 1050 and 600 K respectively. The basic structure of $MW_2O_8$ consists of corner-sharing $MO_6$ octahedra and $WO_4$ tetrahedra. In these compounds one of the oxygen atoms belonging to $WO_4/MoO_4$ tetrahedra is coordinated with only one W atom, which plays an important role in governing the thermal expansion behavior in these compounds. The high pressure inelastic neutron scattering experiments and ab-inito as well as empirical lattice dynamical calculations show [3, 5, 6, 8, 9, 142] that large softening of several low energy phonons is mainly responsible for the NTE in $ZrW_2O_8$, $HfW_2O_8$ and $ZrMo_2O_8$. The specific anharmonic phonons have been identified that are responsible for large NTE in terms of translation, rotation and distortion of polyhedron units. The density functional calculations for $ZrW_2O_8$ are used to interpret the experimental phonon spectra as a function of pressure and temperature as reported in literature. It is observed that the phonons showing anharmonicity with temperature are not necessarily the same as those showing anharmonicity with pressure although both are of similar frequencies. As discussed above, the latter phonons are mainly associated with NTE.

There has been considerable interest in the negative thermal expansion properties of $AV_2O_7$ (A=Zr,Hf). Like the $ZrW_2O_8$ family, the $AV_2O_7$(A=Zr, Hf) family [100, 101] also has a cubic crystal structure and shows a large isotropic NTE. The high-temperature phase of $AV_2O_7$ shows NTE from about 400 and 900 K. The basic structure of $AW_2O_8$ and $AV_2O_7$ consists of corner-sharing $MO_6$ octahedra and $WO_4/VO_4$ tetrahedra. In case of $AV_2O_7$ each $VO_4$ tetrahedron shares three of its four O atoms with an $MO_6$ octahedron, while the fourth is shared with another $VO_4$ tetrahedron leading to a $V_2O_7$ group. All oxygen atoms in $MV_2O_7$ have twofold coordination. The $AW_2O_8$ and $AV_2O_7$ compounds differ in terms of the oxygen coordination around the V/W atoms leading to differences in the nature of soft phonons under compression that are responsible for the NTE. Lattice dynamical calculations [11] quantitatively reproduce the negative expansion over a range of temperatures. The calculations also reveal the relationship of the soft phonons with the phase transitions observed in the $MV_2O_7$ family. Especially, the calculations show a soft-phonon mode at a wave vector of 0.31<1, 1, 0>, which is in excellent agreement with the known incommensurate modulation in ZrV$_2$O$_7$ below 375 K. In the low-T phase below 400 K, the soft phonons of the high-T phase would freeze and may no longer have the negative Grüneisen parameters. The low-T phase has positive thermal expansion coefficient.

In the case of M$_2$O (M=Ag, Cu and Au) compounds, the metal ion M acts as terminal atom to connect M$_4$O polyhedra and plays important role in governing the thermal expansion behavior[83, 84]. The calculated thermal expansion in Ag$_2$O and Cu$_2$O shows negative NTE behavior in agreement with available experimental data, while Au$_2$O shows positive thermal expansion. It is observed [2, 4, 10]that although low energy phonon modes of similar energies are present in all the M$_2$O compounds, the difference in nature of bonding as well as open space in the unit cell are important in governing the thermal expansion behavior. The ab-initio calculations are used for understanding the temperature and pressure dependence of phonon modes in M$_2$O. The calculated quartic anharmonicity of phonons is able to account for temperature dependence of phonon modes in M$_2$O.

β- eucryptite (LiAlSiO$_4$) shows one-dimensional super-ionic conductivity as well as anisotropic thermal expansion behavior[106]. The inelastic neutron scattering measurements in β-eucryptite over 300-900 K and calculation of the phonon spectrum using ab–initio density functional theory has been reported. The anisotropic stress dependence of the calculated phonon spectrum is used to obtain[16]the thermal expansion behavior along various axes. The high-energy modes involving translational motions of Li result in a positive thermal expansion in the hexagonal a-b plane and contraction along the c-axis. Such an anisotropic behavior is a result of anisotropic elasticity and anisotropic Grüneisen parameters. On the other hand, low-energy phonons involving rotational degrees of freedom as well as distortion of polyhedral are responsible for the negative thermal expansion behavior at low temperature along all axes.

ReO$_3$ is an example of simple metallic oxide. The compound has a simple perovskite-like cubic structure and its conductivity is comparable to that of Ag. The structure consists of corner-linked ReO$_6$ octahedral units with Re at the centers and linear Re-O-Re links. The energy of the M3 phonon mode is found to decreases with decreasing temperature contrary to the normal hardening behaviour. The lattice dynamical calculations are performed to investigate the temperature and pressure variation of the M3 phonon mode of ReO$_3$. The frequency of the M3 mode decreases on contraction of the lattice with increasing pressure due to the implicit anharmonicity. However, the same mode frequency increases with volume contraction with increasing temperature due to its explicit quartic anharmonicity. The implicit anharmonicity of the M3 mode leads to NTE in ReO$_3$[12, 13]. The M3 mode involves rigid anti-phase rotations of the neighbouring ReO$_6$ octahedra. Similar anti-phase rotational mode at R-point has been found to be related to the negative expansion in ScF$_3$[43].

A comparative study [143] of thermal expansion behavior of isostructural[107-110] copper halides, CuX (X=Cl, Br, I) has been performed. Certain low energy phonons are found to soften with pressure and are responsible for NTE. Based on the eigen-vector analysis of low energy mode it is suggested that transverse acoustic phonons would lead to NTE in these compounds. The difference in the NTE behavior in the three compounds is associated to difference in their volume[143].

Recently metal cyanides have gained attention. Zn(CN)$_2$ and Ni(CN)$_2$ are known for exhibiting anomalous thermal expansion [102, 111] over a wide temperature range. The volume thermal expansion coefficient for the cubic Zn(CN)$_2$ is negative ($\alpha_V = -51 \times 10^{-6}$ K$^{-1}$). However, for Ni(CN)$_2$, a tetragonal material, the thermal expansion coefficient is negative in the two dimensionally connected sheets ($\alpha_a = -7 \times 10^{-6}$ K$^{-1}$), but the overall thermal expansion coefficient is positive ($\alpha_V = 48 \times 10^{-6}$ K$^{-1}$). The pressure and temperature dependence of phonon spectra in these compounds has been reported [14, 25] using neutron scattering measurement. *Ab-initio* calculations indicate that phonon modes of energy of about 2 meV are major contributors to NTE in both the compounds. NTE has been explained in terms of rotations, translations and deformations of M(C/N)$_4$ (M=Zn, Ni) coordinated tetrahedra.

Structural flexibility plays a key role in the physical properties of framework materials. The covalent framework structure of compounds exhibiting colossal thermal expansion behavior [103, 104] in Ag$_3$M(CN)$_6$ (M=Co, Fe) consists of M–CN–Ag–CN–M linkages. These compounds feature alternating layers of octahedral [M(CN)6]$^{3-}$ and Ag$^+$ cations. Bonding is found to be very similar in both the compounds. It has been shown [24] that chemical variation at the MC$_6$ octahedral site does not change thermal expansion behavior in confirmation with the experimental observation. The thermal expansion seems to be driven by the anharmonicity of the Ag atoms that are involved in the low-energy phonon modes below 5 meV. The strong anharmonic behavior of the Ag vibrations is corroborated by the experimental pressure dependence of the phonon spectra. Very similar colossal thermal expansion has also been observed in KMnAg$_2$(CN)$_6$[75]. Raman scattering experiments and ab-initio lattice dynamics calculations have been reported[44] that reveal the anisotropic properties in this compound.

The thermal expansion behavior in low dimension MCN (M=Cu, Ag and Au) compounds shows [79] anisotropic thermal expansion behavior. The structure of cyanides is chain like and resembles a quasi one-dimension structure. These chains consist of C≡N units connected via metal ions (M-C≡N-M). The compounds show C/N disorder along the chain in terms of random flipping of C/N sequence. The differences in the phonon mode behavior is found [1]to explain the differences in the thermal expansion behavior among the three compounds. The chain-sliding modes are found to contribute maximum to the negative thermal expansion along hexagonal c-axis in the Cu- and Ag- compounds, while the same modes contribute to positive thermal expansion in the Au- compound. Several low energy transverse modes lead to positive thermal expansion in the hexagonal a-b plane in all the compounds. The calculated elastic constants and Born effective charges are correlated with the difference in nature of bonding among these metal cyanides[1].

It may be noted that an understanding of anomalous thermal expansion, as also many thermodynamic properties of solids, requiresan accurate description of the phonon spectrum. In the following Sections 2 and 3 we briefly recapitulate the experimental techniques to measure the phonon spectrum and to calculate the same using lattice dynamical model.The phonon spectrumis used to derive the thermodynamic properties. The calculation is founduseful to understand the microscopic origin of exotic material properties and their response to various thermodynamic conditions. Sections 4-14 provide a review of studies on a variety of compounds, which is followed by overall conclusions in Section 15.

## 2. Experimental Techniques

Experimental studies of lattice vibration include techniques like Raman spectroscopy, infrared absorption, inelastic neutron scattering andinelastic X-ray scattering. Unlike Raman and infrared studies which probe only the long wavelength excitations in one-phonon scattering, inelastic neutron and X-ray scattering can directly probe the phonons in the entire Brillouin zone. Here we briefly describe the various experimental techniques for measurements of phonons.

### 2.1 Raman and Infra-red Scattering

Raman scattering[144], is due to inelastic scattering of the incident photons whereby energy is transferred to or received from the sample due to changes in the vibrational modes of atoms or sample molecules in the sample, causing a change in the energy, and therefore the frequency of the scattered light. If the incident photon gives up energy to the sample it is scattered with a red shifted frequency and referred to as stokes shift. If the solid/molecule is already in an exited energy state, and gives energy to the scattered photon, the output has a blue-shifted frequency, and is referred to as anti-stokes shift. The selection rule governing Raman scattering is determined by changes in polarizibility during the vibration, which is different from another vibrational spectroscopic technique – infrared spectrometry (IR). In the case of IR spectroscopy[145, 146], the frequency of incident light has to match the energy differences between ground and excited vibrational states; and the subsequent energy loss of the incident light is detected. The molecular vibration where there is a change in dipole moment can only be observed in the IR spectroscopy. Raman scattering spectrum provides essentially the same type of information as the infrared (IR) absorption spectrum, namely, the energies of vibrational modes in solid.

**2.2 Inelastic Neutron Scattering**

Thermal neutrons have typically energy of about 1-500 meV and wavelength around 0.04 to 1 nm. While scattering from a sample, thermal neutrons can exchange a part of their energy and momentum with an excitation in the system. During the process, the sum of energy and momentum of the neutron and system will remain conserved. This can be expressed mathematically by using following equations[147, 148]:

$$E_i - E_f = \hbar\omega(\mathbf{q}, j) \qquad (1)$$

$$\hbar(\mathbf{k}_i - \mathbf{k}_f) = \hbar\mathbf{Q} = \hbar(\mathbf{G} \pm \mathbf{q}) \qquad (2)$$

where $\mathbf{k}_i$, $\mathbf{k}_f$ are incident and the scattered neutron wavevectors and $\mathbf{Q}$ is the wavevector transfer (scattering vector) associated with the scattering process. $\mathbf{q}$ is the wavevector of the excitation with energy $\hbar\omega$ and $\mathbf{G}$ is a reciprocal lattice vector of the system under study. $E_i$ and $E_f$ are the incident and scattered neutron energies and $\hbar\omega$ is the energy transfer to the system in the scattering process. The +(-) sign indicates that the excitation is annihilated (created) in the scattering process.

In the scattering process, the inelastic scattering cross-section of the process is measured and this is directly proportional to the dynamical structure factor $S(\mathbf{Q}, \omega)$ (characteristic of the system), which is the double Fourier transform of the space-time correlation function of the constituents of the system including the phonons. Peaks in $S(\mathbf{Q}, \omega)$ correspond to these elementary excitations[147, 149, 150]. The measurements on single crystals give information about the $\mathbf{q}$ dependence of phonon (phonon dispersion relation), while polycrystalline samples provide frequency distribution of the phonons (phonon density of states $g(\omega)$). In order to obtain a complete picture of the dynamics, it is useful to determine the phonon density of states.

The coherent inelastic neutron scattering data from a powder sample are usually analyzed in the incoherent approximation. In this approximation one neglects the correlations between the motions of atoms, and treats the scattering from each atom as incoherent with the scattering amplitude $b_k^{coh}$. However, this is valid only for large $\mathbf{Q}$. In practice, the data are averaged over a large range of $\mathbf{Q}$ values. The data include scattering from both the one-phonon and multi-phonon processes; however, corrections can be made for the multi-phonon scattering. In the incoherent approximation, the expression[7, 147, 150-152] for coherent one-phonon scattering ($S_{coh}^{(1)}(\mathbf{Q},\omega)$) from a powder sample is given by

$$S_{coh}^{(1)}(\mathbf{Q},\omega) = A' \sum_k \exp(-2W_k(\mathbf{Q})) \frac{(b_k^{coh})^2}{m_k} \sum_{\mathbf{G}qj} \frac{1}{3} Q^2 |\xi(\mathbf{q}j,k)|^2 \frac{\hbar}{2\omega(\mathbf{q},j)} \{n(\omega) + \frac{1}{2} \pm \frac{1}{2}\} \quad (3)$$
$$\delta(\mathbf{Q}-\mathbf{G} \pm \mathbf{q})\delta(\omega \mp \omega(\mathbf{q}j))$$

where $A'$ is the normalization constant and $b_k$ and $m_k$ are neutron scattering length and mass of the $k^{th}$ atom, respectively. $\xi$ is eigenvector of excitation, $\exp(-2W_k(\mathbf{Q}))$ is the Debye-Waller factor. $\hbar\mathbf{Q}$ and $\hbar\omega$ are the momentum and energy transfer on scattering of the neutron, respectively, while $n(\omega)$ is the phonon- population factor.

Incoherent inelastic scattering also contributes to the scattering from a powder sample in almost the same way as coherent scattering in the incoherent approximation. Thus, the so-called neutron-weighted density of states involves weighting by the total scattering cross-section of the constituent atoms. The measured scattering function [7, 147, 150-152] in the incoherent approximation is therefore given by

$$S_{inc}^{(1)}(Q,\omega) = \sum_k \frac{b_k^2}{<b^2>} e^{-2W_k(Q)} \frac{Q^2}{2m_k} \frac{g_k(\omega)}{\hbar\omega} (n(\omega) + \frac{1}{2} \pm \frac{1}{2}) \quad (4)$$

where the partial density of states $g_k(\omega)$ is given by

$$g_k(\omega) = \int \sum_j |\xi(\mathbf{q}j,k)|^2 \delta(\omega - \omega_j(\mathbf{q})) d\mathbf{q} \quad (5)$$

Thus, the scattered neutrons provide the information on the density of one-phonon states weighted by the scattering lengths and the population factor. The observed neutron-weighted phonon density of states is a sum of the partial components of the density of states due to the various atoms, weighted by their scattering length squares.

$$g^n(\omega) = B' \sum_k \{\frac{4\pi b_k^2}{m_k}\} g_k(\omega) \quad (6)$$

where $B'$ is a normalization constant. Typical weighting factors $\frac{4\pi b_k^2}{m_k}$ for the various atoms in the units of barns/amu are: Al: 0.055; Mg: 0.150; Fe: 0.201; Ca: 0.075; Si: 0.077 and O: 0.265. By comparing the experimental phonon spectra with the calculated neutron-weighted density of states obtained from a lattice-dynamical model, the dynamical contribution to frequency distribution from various atomic and molecular species can be understood.

A number of different techniques [147, 149-151, 153] have been developed at steady and pulsed neutron sources to obtain the vibrational spectra of compounds. For the polycrystalline samples, the neutron time-of-flight technique (TOF) is rather convenient for the measurements. A typical time-of-flight spectrometer, namely, IN6 at the Institut Laue Langevin (ILL), Grenoble, France is shown in Fig. 1. Similar spectrometers are available at many of the reactors and also at the spallation neutron sources at ISIS, Oak Ridge, JPARC, etc. The energy of the neutrons is fixed either before or after the scattering process. The change in energy and the scattering vector **Q** is obtained by measuring the flight time and the scattering angle of the neutrons from a beam pulsing device (chopper) to the detectors. For measurement of phonon density of states the scattered neutrons from the sample are collected over a wide range of scattering angles, typically from 10º to 110º. By choosing a suitable incident neutron energy, measurement of the scattering function S(**Q**, ω) over a wide range of momentum and energy transfers can be undertaken and the data can be averaged over a wide range of **Q**. A large number of temperature dependent measurements have been reported[4, 25, 68] using various TOF spectrometers at several reactor and spallation sources.

Several high pressure inelastic neutron scattering experiments[6, 9] have been carried out on polycrystalline samples using the time-of-flight IN6 spectrometer (Fig. 1) at the Institut Laue Langevin (ILL), France. An incident energy of 3.12 meV with an elastic resolution of 80 μeV is chosen and the measurements are performed in the energy gain mode. The polycrystalline sample of about 2 cc was compressed using argon gas in a pressure cell. The use of argon gas as a pressure transmitting medium allowed to perform measurements above its critical point (say 160 K at 2.5 kbar). The measurements are carried out at several pressures up to 2.5 kbar. The inelastic neutron scattering signal is corrected for the contributions from argon at the respective pressures, absorption from the sample, and for the empty cell.

**2.3 Inelastic X ray Scattering**

Similar to the Inelastic Neutron Scattering (INS), Inelastic X ray scattering (IXS) is used[154, 155] to measure phonon mode in the entire Brillouin zone. The energy of the X-rays is in the order of KeV, however the energy of phonons is of the order of meV. Hence one would need a very high-resolution instrument available at a synchrotron radiation source, where the incident flux is very high. The IXS has specific advantage over INS as it requires microgram samples hence suitable for high pressure experiments in diamond cell. The access to small samples follows from the very high flux and brilliance of synchrotron radiation sources, with the option to focus beams easily to ~100 microns in diameter, and, with some losses to ~10 microns.

Neutrons remain advantageous when high - energy resolution is needed, as backscattering spectrometers provide sub-meV resolution, at least for smaller energy transfers. Neutrons remain extremely competitive when large single crystals of heavier materials are available, whereas, x-rays are limited by the short penetration length into the sample.

The above methods differ fundamentally in mechanism and selection rules, and each has specific advantages and disadvantages. The major disadvantage of optical spectroscopy techniques is due to large wavelength (~5000 A) of the incident radiation. That would allow probing only a tiny region of the Brillouin zone close to zone centre. However, various phase transitions and properties of the materials are contributed from entire Brillouin zone. Further selection rule for Raman and IR techniques restrict their applicability, whereas there is no such limitation is neutron and X-ray inelastic technique that can measure any phonon mode in the entire Brillouin zone. The advantage of the Raman and IR techniques is that the radiation flux in these techniques is much higher than that from neutron sources, hence data can be collected in very short times. Moreover, these techniques can easily be set up in small laboratories, whereas neutron and X-ray scattering requires major facility of a neutron reactor or a spallation source or a synchrotron source.

**3. Lattice Dynamics and Thermodynamic Properties of Solids**

It is very important to analyze and interpret the measured data as well as understand the various physical and thermodynamical properties like elastic constant, specific heat, thermal expansion etc. The character and energy of atomic dynamics in solid as well as the individual atomic contribution to the spectra and other thermodynamical properties are difficult to obtain from measurements. Theory of lattice dynamics is a well-established technique which helps us to understand the measured spectra and their atomic origin. The simulations are very useful to interpret the experimental data. Simulation and experimental techniques complement each other and provide complete information about a system. Lattice dynamics calculations of vibrational properties may be carried out using either an ab-initio density functional theory (DFT) or an atomistic approach involving semiempirical interatomic potentials. In following section, we will describe the lattice dynamics techniques to compute the vibrational spectra and derived thermodynamical quantities.

**3.1 Theory of Lattice Dynamics**

The theory of lattice dynamics[152, 153] is based on two basic approximations namely, the Born-Openheimer approximation and the harmonic approximations. According to Born-Openheimer approximation in a system of electron and ion, we can treat the equation of motion of electron and ion separately since the electronic degree of freedom is much faster than ionic degree of freedom because of the huge mass difference. Here one can assume that electron will follow instantaneously the ionic motion. The Born-Openheimer approximation will fail if the ionic and electronic motions become comparable.

In the harmonic approximation, the potential energy is Taylor expanded in term of the atomic displacements and terms above the second order (harmonic term) are assumed to be negligible. Thus, the potential energy Φ of crystal can be Taylor expanded in terms of atomic displacement $u_\alpha \begin{pmatrix} l \\ k \end{pmatrix}$ of $k^{th}$ atom in $l^{th}$ cell along α Cartesian direction as follows.

$$\phi = \phi_0 + \sum_{\substack{ll',k,k' \\ \alpha}} \phi_\alpha \begin{pmatrix} l & l' \\ k & k' \end{pmatrix} u_\alpha \begin{pmatrix} l \\ k \end{pmatrix} + \frac{1}{2} \sum_{\substack{ll',k,k' \\ \alpha\beta}} \phi_{\alpha\beta} \begin{pmatrix} l & l' \\ k & k' \end{pmatrix} u_\alpha \begin{pmatrix} l \\ k \end{pmatrix} u_\beta \begin{pmatrix} l' \\ k' \end{pmatrix} \qquad (7)$$

Where *α, β= x, y, z* and *k'=1, n* (*n* is number of atoms in unit cell).

In equilibrium configuration, the force on every atom will be zero. Hence this will lead to the first derivative of potential energy to be precisely zero.

Under harmonic approximation the equation of motion is given by

$$m_k \ddot{u}_\alpha \begin{pmatrix} l \\ k \end{pmatrix},t = -\sum_{l',k',\beta} \phi_{\alpha\beta}\begin{pmatrix} l & l' \\ k & k' \end{pmatrix} u_\beta\begin{pmatrix} l \\ k \end{pmatrix},t \quad (8)$$

Atomic displacement of $k^{th}$ atom of $l^{th}$ unit cell at any instantaneous time $t$ can be expanded as sum of plane waves

$$u_\alpha\begin{pmatrix} l \\ k \end{pmatrix},t = \frac{U_\alpha(k,\boldsymbol{q})}{\sqrt{m_k}} \exp^{i(\boldsymbol{q}\cdot r(l,k)-\omega(q)t)} \quad (9)$$

Now using equation (9) in (8), we will get 3n simultaneous equations

$$\omega^2(\boldsymbol{q})U_\alpha(k,\boldsymbol{q}) = \sum_{k',\beta} D_{\alpha\beta}\begin{pmatrix} \boldsymbol{q} \\ kk' \end{pmatrix} U_\beta(k',\boldsymbol{q}) \quad (10)$$

$$D_{\alpha\beta}\begin{pmatrix} \boldsymbol{q} \\ kk' \end{pmatrix} = \frac{1}{\sqrt{m_k m_{k'}}} \sum_{l'} \phi_{\alpha\beta}\begin{pmatrix} l & l' \\ k & k' \end{pmatrix} \exp^{i\boldsymbol{q}\cdot(r(l',k')-r(l,k))} \quad (11)$$

The above equation can be written in matrix form

$$\Omega(\mathbf{q})U(\mathbf{q}) = D(\mathbf{q})U(\mathbf{q}) \quad (12)$$

D(q) is known as dynamical matrix, which is a Hermitian matrix. The solution of the above matrix equation will have the form

$$\left\| \left( D(\boldsymbol{q}) - \omega^2(\boldsymbol{q}) 1_{3n} \right) \right\| = 0 \quad (13)$$

Solving the equation (13), 3n eigenvalues are obtained, which are $\omega_j^2(\boldsymbol{q})$, (j=1,2,…,3n). Because, the dynamical matrix is Hermitian, the eigenvalues are real and its eigenvectors are orthogonal to each other. The eigenvalues are the square of phonon frequency at given q, hence we have 3n phonon at any given q. For a stable system all the phonon frequencies at any q must be positive. The components of the eigenvectors $\xi_j(\mathbf{q})$ determine the direction and relative amplitude of displacement of the atoms in a particular $j^{th}$ mode of vibration. The eigenvector is also known as the polarization vector. Corresponding to every direction in **q**-space, there are 3n curves $\omega=\omega_j(\mathbf{q})$, (j=1,2,…3n). Such curves are called the phonon dispersion relation. The index j, which distinguishes the various frequencies characterizes various branches of the dispersion relation.

Dispersion relation must satisfy the crystal symmetry. Though, some of these branches are degenerate because of symmetry, in general they are distinct. The form of dispersion relation depends on the crystal structure as well as on the nature of the interaction between atoms. Because of the translational invariance, three phonon frequencies are zero at q=0. These three branches are referred to as acoustic branches. The remaining (3n-3) branches have finite frequencies at *q*=0, which are labeled as optic branches.

The distribution of ω(q) is known as density of states. Mathematically it can be represented as

$$g(\omega) = \frac{1}{3nN} \sum_{j,q} \delta(\omega - \omega_j(q)) \qquad (14)$$

Here δ is the Dirac delta function.

The calculation of force constants matrix (eq. (13)) for computation of phonon frequencies may be carried out using either density functional theory (DFT) approach or involving semiempirical interatomic potentials. These are discussed below.

### 3.1.1 Semiempirical Interatomic Potentials

The compounds discussed here have several atoms at low symmetry sites in a rather complex crystal structures. An approach based on an empirical interatomic potential function is used in some of the calculations. The inter atomic potential[15] consists of long-range Coulombic interaction, short range Born-Mayer type repulsive terms, and weakly attractive van der Waals terms. The form of the interatomic potential used in model is given by the following expression:

$$V(r) = \frac{e^2}{4\pi\varepsilon_0} \frac{Z(k)Z(k')}{r} + a\, \exp(\frac{-br}{R(k)+R(k')}) - \frac{C}{r^6} \qquad (15)$$

where r is the separation between the atoms of a type k and k'. Z(k) and R(k) are effective charge and radius parameter of the atom type k. The last term in eq.(1.25) is applied only between certain atoms. a=1822eV and b=12.364 are empirical constants. We have successfully used [7, 132] this set of parameters in the lattice dynamical calculations of several complex solids.

In some cases, a more general form of interatomic potential has been used:

$$V(r) = \frac{e^2}{4\pi\varepsilon_0} \frac{Z(k)Z(k')}{r} + a_{kk'} \exp(-b_{kk'}r) - \frac{c_{kk'}}{r^6} \qquad (16)$$

where $a_{kk'}$, $b_{kk'}$ and $C_{kk'}$ are the empirical parameters of the potential.

As a further improvement, a covalent potential[37] is also included between certain atoms.

$$V(r) = -D\exp[-n(r-r_o)^2/(2r)] \qquad (17)$$

where $n$, $D$ and $r_o$ are the empirical parameters of the potential.

**3.1.2 Density Functional Theory**

The ab-initio density functional theory (DFT) method is used to calculate the total energy from first principles[156] at T=0K. The theory is based on solving the Hamiltonian of the system including electrons and ions together. The valence electrons are the fundamental entity controlling the nature of bonding, volume, charge and various other properties. The behavior of electrons in the vast environment of other electrons and ions controls the many physical, chemical and thermodynamical properties of the material. The many–body Hamiltonian operator that governs the behavior of a system of interacting electrons and nuclei in atomic units takes the form

$$\hat{H} = -\frac{1}{2}\sum_i \frac{\nabla_i^2}{m_e} - \sum_{i,I} \frac{Z_I}{|r_i - R_I|} + \frac{1}{2}\sum_{i \neq j} \frac{1}{|r_i - r_j|} - \sum_I \frac{\nabla_I^2}{m_I} + \frac{1}{2}\sum_{I \neq J} \frac{Z_I Z_J}{|R_I - R_J|} \qquad (18)$$

The summations over $i$ and $j$ correspond to the electrons, and summations over $I$ and $J$ correspond to nuclei of the system $r_i$, $R_I$ and $Z_I$ are the $i^{th}$ electrons spatial coordinate, and the position and charge on $I^{th}$ nuclei respectively.

The mass of an electron is negligible in comparison to the atomic masses in the system, hence the nuclei can be assumed to remain stationary from the point of view of an electron. We also effectively disregard the kinetic energy of the nuclei because of the heavy mass. Now the Hamiltonian becomes:

$$\hat{H} = \hat{T}_e(r) + \hat{V}_{eN}(r,R) + \hat{V}_{ee}(r) + \hat{V}_{NN}(R) = \hat{H}_e + E_{NN} \qquad (19)$$

The $\hat{H}_e$ is known as the electronic Hamiltonian, as it describes the motion of electrons in a fixed environment of atomic nuclei.

$$\hat{H}_e \psi = E\psi \qquad (20)$$

Determining the ground state wave–function for the Schrödinger Eq. (1.39) in terms of the many–body electron wave function is only really tractable for simple systems with relatively very few electrons. One approach that transforms the many–body Schrödinger wave equation into a simple one–electron equation is known as the density functional theory formalism, or DFT. A central property of DFT is that it recasts the basic variable of equations from being the ground state electronic wave function $\Psi(r_1, r_2, ...r_N)$ to that of the ground state electron density $n_0(r)$, where

$$n_0(r) = N \int |\Psi_0(r, r_2, ...r_N)|^2 dr_2....dr_N \qquad (21)$$

and hence it effectively reduces the 3N degrees of freedom to just 3 for an N–electron system. Density functional theory is able to predict the ground state energy and wavefunction of the system, all the ground state properties of systems can be determined using the theory. In principle DFT is an exact theory of the ground state of a system, however due to lack of the exact form the exchange–correlation functional, the theory works on certain assumptions and approximations to this functional (LDA and GGA)[157] are a few famous approximations).

Density functional theory is based on two pillars namely Hohenberg-Kohn theorem[158] and Kohn-Sham theorem[159]. In figure 2 the flow chart of the DFT calculation scheme is shown.

The output of the DFT calculation is total energy and force on each atom. The forces on atoms are calculated by displacing the various atoms in different symmetry directions in a supercell. The calculated forces are used to compute the force constants matrices between various pairs of atoms. The phonon frequencies at given wave-vector q is obtained by diagonalizing the dynamical matrix. The flow chart for the calculation is shown in Fig. 3.

## 3.2 Thermodynamic Properties

By knowing the phonon spectrum one can calculate the various thermodynamical quantities contributed from phonons. The theory of lattice dynamics described in previous section allows us to determine the phonon frequencies in the harmonic approximation. The calculation is also performed in the quasiharmonic approximation[15], where, the vibrations of atoms at any finite pressure and temperature are assumed to be harmonic about their mean positions appropriate to the corresponding temperature. The thermodynamic properties of a crystal are based on the energy distribution of the phonon vibrations. We do not discuss here the contribution from electron as in metals, or the magnetic excitation in a magnetic system. The phenomenon of thermal expansion is possible in the quasiharmonic approximation, which is not so strictly in the harmonic approximation. Anharmonic effects also become important at high temperatures of $k_B T >> \hbar\omega_j(\mathbf{q})$.

In a crystalline system, vibrational modes (phonon mode) are independent of each other. The partition function of the system is given by

$$Z = \exp^{-\phi(V)} \prod_{j,q} \frac{\exp^{-(\hbar\omega_j(q)/2K_B T)}}{1-\exp^{-(\hbar\omega_j(q)/K_B T)}} \qquad (22)$$

Various thermodynamic properties of the crystal derived from the partition function involve summations over the phonon frequencies in the entire Brillouin zone and can be expressed as averages over the phonon density of states. The Helmholtz free energy F and entropy S are given by

$$F = -k_B T \ln Z = \phi(V) + \int [\frac{1}{2}\hbar\omega_j(q) + k_B T \ln\{1-\exp(\frac{-\hbar\omega_j(q)}{k_B T})\}] g(\omega) d\omega \qquad (23)$$

$$S = -\frac{dF}{dT} = k_B \int \{-\ln[1-\exp(\frac{-\hbar\omega_j(q)}{k_B T})] + \frac{(\frac{\hbar\omega_j(q)}{k_B T})}{[\exp(\frac{\hbar\omega_j(q)}{k_B T})-1]}\} g(\omega) d\omega \qquad (24)$$

The energy E of the crystal is

$$E = F - T\frac{dF}{dT} = \phi(V) + E_{vib} \tag{25}$$

where $\phi(V)$ is the static lattice energy and $E_{vib}$, the vibrational energy at temperature T.

$$E_{vib} = \int \{n(\omega) + \frac{1}{2}\}\hbar\omega_j(q)g(\omega)d\omega \tag{26}$$

Where $n(\omega)$ is the population factor given by

$$n(\omega) = \frac{1}{\exp(\hbar\omega_j(q)/k_B T) - 1} \tag{27}$$

The specific heat $C_V(T)$ is given by

$$C_V(T) = \frac{dE}{dT} = k_B \int \left(\frac{\hbar\omega_j(q)}{k_B T}\right)^2 \frac{e^{\left(\frac{\hbar\omega_j(q)}{k_B T}\right)}}{(e^{\left(\frac{\hbar\omega_j(q)}{k_B T}\right)} - 1)^2} g(\omega)d\omega \tag{28}$$

The calculated phonon density of states can be used to compute the specific heat. While lattice dynamical calculations yield $C_V$, the specific heat at constant volume, experimental heat capacity data correspond to $C_P$, the specific heat at constant pressure. The difference $Cp-Cv$ is given[137] by

$$C_P(T) - C_V(T) = [\alpha_V(T)]^2 BVT \tag{29}$$

where $\alpha_V$ is the volume thermal expansion[137] and B is the bulk modulus defined as

$$B = -V\frac{dP}{dV} \tag{30}$$

$$\alpha_V(T) = \frac{1}{BV_0}\Sigma_{q,j} C_V(q,j,T) \Gamma_{q,j} \tag{31}$$

The Grüneisen parameters are often defined and used in different ways in the literature depending on the context. We are dealing with the mode Grüneisen parameter that is different for each phonon mode. Experimentally these are measured from pressure dependence of phonon frequencies i.e. isobaric process. Theoretically, these are derived from phonon frequency change with volume (isotropic system) or lattice parameters (anisotropic system). The phonon frequencies are calculated in the quasi-harmonic approximation at fixed (zero) temperature.

The linear thermal expansion coefficients along the 'a' 'b' and 'c' -axes can be calculated within the quasiharmonic approximation. The calculations require anisotropic pressure dependence of phonon energies in the entire Brillouin zone. These calculations are subsequently used to obtain the anisotropic mode Grüneisen parameters. According to Grüneisen's original work for anisotropic system[160], the Grüneisen parameter is given by a general expression as

$$\Gamma_l(E_{q,j}) = -\left(\frac{\partial \ln E_{q,j}}{\partial \ln l}\right)_{T,l'} ; \; l, l' = a, b, c \; \& \; l \neq l' \quad (32)$$

Where $E_{q,j}(=\hbar\omega_j(q))$ is the energy of $j^{th}$ phonon mode at point q in the Brillouin zone. The anisotropic Grüneisen parameters are calculated by applying a small anisotropic stress (~0.25 GPa) by changing one of the lattice constants keeping others fixed, and vice versa, where temperature is held constant at zero. The symmetry of the unit cell is kept invariant on introduction of strain, while performing the calculations. The anisotropic linear thermal expansion coefficients are given by[160]:

$$\alpha_l(T) = \frac{1}{V_0} \sum_{q,j} C_v(q,j,T) \sum_k S_{ll'} \Gamma_l \quad (33)$$

Where $S_{ll'}$ are elements of elastic compliances matrix $S=C^{-1}$, $V_0$ is volume at ambient conditions and $C_v(q,j,T)$ is the specific heat at constant volume for $j^{th}$ phonon mode at point $q$ in the Brillouin zone. The anisotropy of Grüneisen parameters and elasticity give rise to the anisotropic thermal expansion behavior of the compound. Since $C_v(T)$ is positive for all modes at all temperatures, it is clear that the NTE would result only from large negative values of the Grüneisen parameter for certain phonons; the values should be large enough to compensate for the normal positive values of all other phonons. The frequencies of such phonons should decrease on compression of the crystal rather than increase which is the usual behavior. It is very interesting to note that the quantization of the phonon energies is important to the phenomena of NTE, just as in the case of specific heat. Although thermal expansion and specific heat are highly averaged quantities, their temperature dependence involves progressively the role of low energy to higher energy phonons.

## 4. ZrW$_2$O$_8$, HfW$_2$O$_8$ and ZrMo$_2$O$_8$ compounds

Isotropic negative thermal expansion (NTE) has been known in cubic ZrW$_2$O$_8$, HfW$_2$O$_8$[40, 100, 161] (space group $P2_13$, $Z=4$) and ZrMo$_2$O$_8$[101, 162, 163] (space group $Pa\bar{3}$, $Z=4$) over a wide range of temperature upto 1443 K, 1050 K and 600 K, respectively. The structure (Fig. 4) of AX$_2$O$_8$ (A=Zr,Hf: X=W,Mo) consists of corner linked AO$_6$ octahedral and XO$_4$ tetrahedral units.

ZrW$_2$O$_8$ and HfW$_2$O$_8$ undergo an order-disorder phase transition [40, 100, 161] at about 400 K. One of the terminal oxygens of WO$_4$ tetrahedra becomes disordered over two crystallographic sites and the space group changes from $P2_13$ (ordered phase) to $Pa\bar{3}$ (disordered phase). ZrMo$_2$O$_8$ has been synthesized in the disordered phase only. The volume thermal expansion coefficient for ZrW$_2$O$_8$, HfW$_2$O$_8$ and ZrMo$_2$O$_8$ at 300 K is about $-29 \times 10^{-6}$ K$^{-1}$, $-29 \times 10^{-6}$ K$^{-1}$ and $-15 \times 10^{-6}$ K$^{-1}$ respectively. In these compounds one of the oxygen atoms belonging to WO$_4$/MoO$_4$ tetrahedra is coordinated with only one W atom, which plays an important role in governing the thermal expansion behavior in these compounds. The NTE has been associated with the presence of A-O-W transverse vibrational modes [40] and rigid unit modes [129, 164]. It has also been suggested [165] that certain translational motion of WO$_4$ tetrahedra is responsible for NTE in ZrW$_2$O$_8$.

Neutron diffraction data[166-168] show a cubic to orthorhombic (space group $P2_12_12_1$, $Z=12$) phase transition at pressures around 2.1 kbar and 6 kbar for ZrW$_2$O$_8$ and HfW$_2$O$_8$, respectively. For disordered ZrMo$_2$O$_8$ there is no such phase transition [101] below 6 kbar. An anomalous thermal expansion is also observed in the orthorhombic phase [166-168] of ZrW$_2$O$_8$ and HfW$_2$O$_8$. The compounds AX$_2$O$_8$ (A=Zr, Hf; X=W, Mo) also crystallize in orthorhombic, monoclinic and trigonal structures [163, 169-173]. Trigonal ZrMo$_2$O$_8$ transforms [88] to a monoclinic symmetry phase at about 1 GPa. Further high-pressure transformation from monoclinic to a triclinic phase occurs at about 2 GPa. X-ray diffraction measurements of trigonal ZrMo$_2$O$_8$ show highly anisotropic thermal expansion coefficients [174].

Specific heat [131, 175] measurements for cubic AW$_2$O$_8$ (A=Zr, Hf) and Raman [85, 86, 176] and inelastic neutron scattering measurements [6, 36] for cubic ZrW$_2$O$_8$ were reported. An analysis of specific heat data[175] for the two compounds suggests that the mass difference between the Zr and Hf would lead to difference in the low energy phonon modes in these compounds.

Some estimates of the Grüneisen parameters were reported[35, 36, 175] for $AW_2O_8$ (A=Zr, Hf) on the basis of the observed NTE and certain assumptions about the energy dependence of the Grüneisen parameters or the phonon density of states. These estimates indicated fairly large negative values of the Grüneisen parameters. The NTE in both the Zr and Hf compounds was also calculated [85, 86] using the pressure dependence of the Raman active phonon modes, which provided overestimates of NTE compared to the experimental values. It appears that a proper description of NTE requires consideration of both the acoustic and optic phonon modes in the entire Brillouin zone.

It has been shown [3, 5, 6, 8, 9, 142] via high pressure inelastic neutron scattering experiments and ab-inito as well as empirical lattice dynamical calculations that large softening of several low energy phonons is mainly responsible for the NTE in $ZrW_2O_8$, $HfW_2O_8$ and $ZrMo_2O_8$. The calculations helped to identify the specific anharmonic phonons that are responsible for large NTE in terms of translation, rotation and distortion of polyhedron units. The density functional calculations for $ZrW_2O_8$ are also used to interpret the experimental phonon spectra as a function of pressure and temperature. It is found that the phonons showing anharmonicty with temperature are not necessarily the same as those showing anharmonicity with pressure although both are of similar frequencies. Only the latter phonons are associated with NTE. Therefore, the cubic and/or quartic anharmonicity of phonons is not directly relevant to NTE but just the volume dependence of frequencies.

In these compounds, there are as many as 44 atoms in the cubic cell. The lattice dynamics calculations were initially performed using an empirical interatomic potential that provided an excellent description of the relevant phonons and consequently the NTE behavior[5]. Later density-functional theory calculations were carried out that provided a more accurate description of NTE[3].

**4.1. Thermal expansion from experimental pressure dependence of phonon spectra**

**4.1.1 $ZrW_2O_8$**

Pressure dependence of the zone centre Raman active phonons of energy above 5 meV were estimated [85, 86, 176] in both the Zr (Figs. 5 (a,b)) and Hf compounds. For $ZrW_2O_8$ the Grüneisen parameters of zone centre phonon modes and their weighting factors obtained from the experimental data [36] of phonon density of states (Fig. 6(a)) were used to obtain the temperature dependence of linear thermal expansion coefficient. A model [86] based on these estimates, however, yields a rather large linear thermal expansion coefficient (Fig. 6(b)) in ZrW$_2$O$_8$ (of -15 × 10$^{-6}$ K$^{-1}$ at 300 K compared to the experimental value [40, 86] of -9.5 × 10$^{-6}$ K$^{-1}$). A similar model for HfW$_2$O$_8$ reproduces thermal expansion coefficient at room temperature [85], but underestimates the same at low temperatures [142].

Here a description of NTE using pressure dependence of inelastic neutron scattering experiments is presented that include contribution from both acoustic and optic modes in the whole Brillouin zone. The experimental neutron-cross-section weighted phonon density of states $g^{(n)}(E)$ for ZrW$_2$O$_8$ at 160 K and different pressures are shown in Fig. 7. The ambient pressure results are in agreement with previous measurements [36]. The spectra at high pressures show an unusually large softening. In conformity with the predictions (Fig. 7), the phonon modes of energy below about 5 meV soften by about 0.15 meV at 1.7 kbar with respect to ambient pressure. At energies above 5 meV, the shift of the spectrum is much less than that at lower energies.

The experimental neutron-weighted phonon density of states $g^{(n)}(E)$ (Fig. 7) were converted to phonon density of states $g(E)$ using the ratio of the calculated $g^{(n)}(E)$ to $g(E)$. The spectral weight upto 10 meV is about 11.2%, which corresponds to about 15 phonon modes out of the total 132 modes in cubic ZrW$_2$O$_8$ per unit cell. The averaged Grüneisen parameter $\Gamma(E)$ for phonons of energy $E$ was obtained [Fig. 8] using the cumulative distributions for the density of states.

The weighted average value of $\Gamma$ for $E$=1.5 to 8.5 meV phonons is -10.6 which is in good agreement with a value of -14 ± 2 reported [85] from the analysis of the NTE and the ambient pressure phonon density of states. Again a negative $\Gamma$ of -32.7 as reported [35] for a representative Einstein mode at 3.3 meV, is in good agreement with the experimental data of -28 at the same energy. Grüneisen parameter of some Raman active mode at the Brillouin zone center are available [86, 176]; however, these cannot be compared with the present data which correspond to the average over the whole Brillouin zone.

The above data on $g(E)$ and $\Gamma(E)$ obtained from the phonon spectra at ambient pressure and 1.7 kbar were used to derive [130] a temperature dependent volume thermal expansion coefficient ($\alpha_V$). The $\alpha_V$ thus derived from the inelastic neutron scattering data is in good agreement (Fig. 9) with that directly observed by diffraction[35]. The analysis shows that the large negative Grüneisen parameters of modes below 10 meV are able to explain the low temperature thermal expansion coefficient and its nearly constant value [35, 40] above 70 K. In Fig. 10(b) the contribution of phonons of energy $E$ to the thermal expansion was shown as a function of $E$ at 20 K and 300 K. The maximum contribution to $\alpha_V$ is found to be from phonon modes of energy $4 \pm 1$ meV which is consistent with the previous analysis of diffraction data [35].

**4.1.2 ZrMo$_2$O$_8$**

The experimental phonon spectra for disordered ZrMo$_2$O$_8$ at various pressures are shown in Fig. 7. For disordered cubic ZrMo$_2$O$_8$, phonon softening (Fig. 7) is maximum for modes upto 3.5 meV, while above 3.5 meV softening is much less. The phonon spectra at ambient pressure and 2.5 kbar were used for determination of the averaged Grüneisen parameter as a function of energy. The $\Gamma(E)$ values of ZrMo$_2$O$_8$ for phonon of various energies (Fig. 8) are approximately half of those in ZrW$_2$O$_8$. The $\alpha_V$ for ZrMo$_2$O$_8$ (Fig. 9) was derived [9] from the phonon spectrum and $\Gamma(E)$ using the same procedure as described above for ZrW$_2$O$_8$. The large negative Grüneisen parameters of modes below 10 meV are able to reproduce the NTE behaviour in disordered ZrMo$_2$O$_8$ upto 100 K. Diffraction data [163] show that $\alpha_V$ (Inset in Fig. 9) is less negative above 100 K which may be due to positive contributions from phonon modes above 10 meV. The measured spectra above 10 meV are heavily contaminated from contribution from the pressure cell that did not allow to obtain the experimental $\Gamma(E)$ values beyond 10 meV.

**4.2 Grüneisen Parameter and Thermal Expansion using Lattice Dynamics Based on Empirical Interatomic Potentials**

Neutron diffraction measurements show that there is no significant difference between the structures [168] of ZrW$_2$O$_8$ and HfW$_2$O$_8$. The precise structure of HfW$_2$O$_8$ has not been published. For this reason, the empirical potential parameters for the lattice dynamics calculations of HfW$_2$O$_8$ and ZrW$_2$O$_8$ were taken as the same. The interatomic potential reproduces the equilibrium crystal structural parameters of the cubic phase (Table 1) and other dynamical properties quite satisfactorily as discussed below.

The cubic phase has 44 atoms in the primitive cell and thus 132 phonon modes at each wave vector which are classified as $66\Delta_1 + 66\Delta_2$ along the [100] direction. The calculated pressure dependence of phonon dispersion relation in the ordered cubic HfW$_2$O$_8$ along [100] direction is shown in Fig. 11. The phonon dispersion relation is plotted only upto 10 meV since phonons upto 8 meV are most relevant for understanding the negative thermal expansion behavior (as shown later). A large softening of phonons is observed. The softening is very large for the lowest transverse acoustic mode in the group theoretical representation $\Delta_2$ and also for a number of optic modes of energies lying between 3 and 8 meV in both the group theoretical representations. An instability was found at high pressure of about 3 kbar as revealed by softening of a phonon mode (Fig. 11) at about $q = 0.15$ along [100] direction in $\Delta_2$ representation. The eigen vector of this mode indicates that the mode is close to a transverse acoustic vibration accompanied by some internal angular distortions of the WO$_4$ tetrahedra. The calculated pressure dependence of the phonon spectrum for cubic ZrW$_2$O$_8$ also shows similar behavior. The experimental phonon spectra of cubic HfW$_2$O$_8$(Fig. 12) are in very good agreement with the lattice dynamical calculations [5]considering the inherent limitations in experiment (e.g. incoherent approximation) and theory.

The calculated one-phonon density of states $g(E)$ was used to compute the specific heat. Figs. 13(a,b) show a comparison of the calculated specific heat with the available experimental data [131, 175] in the cubic phase of ZrW$_2$O$_8$ and HfW$_2$O$_8$. The good agreement between the calculated and experimental specific heat supports the validity of the low energy phonon density of states provided by the lattice dynamical model. The sharp increase in specific heat at low temperatures is due to a low energy peak in the phonon spectrum at about 3.5 meV. The calculations show that at low temperature, HfW$_2$O$_8$ has a larger specific heat. This is in agreement with the experimental [175] observations.

The calculated Grüneisen parameters of phonon modes in the entire Brillouin zone were suitably averaged as a function of energy. The averaged energy dependence of Grüneisen parameter $\Gamma(E)$ for both the cubic ZrW$_2$O$_8$ and HfW$_2$O$_8$ is shown in Fig. 8. The averaged Grüneisen parameter has large negative values for phonons below 8 meV. It is because of the large negative $\Gamma$ value of several phonons below 8 meV that the $\alpha_V(T)$ reaches a large negative value at very low temperature. For energies above 8 meV and upto the maximum phonon energies of 140 meV, the averaged $\Gamma$ values are very small in the range of about -1 to 1. Therefore $\alpha_V(T)$ continues to be negative and maintained at a high level. The cubic HfW$_2$O$_8$ (Fig. 8) has lower values of $\Gamma(E)$ in comparison to ZrW$_2$O$_8$. This is consistent with the trend observed with the analysis of the thermal expansion and specific heat data [175].

The calculated pressure dependence of the phonon spectrum for ordered cubic AW$_2$O$_8$ (A=Zr, Hf) (Figs. 7 and 11) as well as the calculated Grüneisen parameter (Fig. 8) show that the phonons of energies below 8 meV are the most relevant to NTE, which is consistent with the analysis [119] based on the neutron results of ZrW$_2$O$_8$. The experimental results of Grüneisen parameters (Fig. 8) are in very good agreement with predictions from lattice dynamics. The calculated bulk modulus values of 88.4 GPa for both cubic ZrW$_2$O$_8$ and HfW$_2$O$_8$ are in good agreement with the experimental[166-168]values of 72.5 and 82 GPa at 300 K for ZrW$_2$O$_8$ and HfW$_2$O$_8$, respectively.

Fig. 14 shows the quasiharmonic calculation of the thermal expansion. This reveals a very interesting result, that nearly 40% of the NTE in the cubic phase arises from just the two lowest phonon branches and almost all the NTE is contributed from the modes below 8 meV. The absolute value of thermal expansion coefficient for HfW$_2$O$_8$ is smaller in comparison with ZrW$_2$O$_8$. The calculated relative thermal expansion for both the cubic ZrW$_2$O$_8$ and HfW$_2$O$_8$ is shown in Fig. 14, which indicates an excellent agreement with the experimental data[40, 100]. The latter includes a small sharp drop in volume at about 400 K associated with aorder-disorder phase transition which also reduces the volume by about 0.4%.

The calculations in the cubic phase at about 3 kbar reveal a softening of the transverse acoustic branch near the Brillouin zone center (Fig. 11). This instability of the cubic phase appears to coincide with the observed [166, 167] cubic to orthorhombic phase transition in ZrW$_2$O$_8$ at about 3 kbar. The latter phase is retained on decompression to ambient pressure. The thermal expansion of the orthorhombic phase is observed [166, 167] to be anomalous; it is negative below 300 K and positive above 300 K. Above 400 K, the orthorhombic phase transforms to cubic phase. The calculation of thermal expansion in the orthorhombic phase of ZrW$_2$O$_8$ is compared with the experiments in Fig. 15 which shows a very good agreement. Compared to the cubic phase, the negative thermal expansion contributed by the phonons below 8 meV (Fig. 14) is much reduced, which allows the high energy modes to dominate at high temperatures and yield a net positive expansion above 270 K in calculations.

In Fig. 10(a) the averaged contribution of various phonons to the thermal expansion is shown as a function of phonon energy at 10 K and 300 K for cubic ZrW$_2$O$_8$ and HfW$_2$O$_8$. The maximum negative contribution to $\alpha_V$ at both the temperatures is from the modes of energy from 3 to 5 meV. Similar behavior is calculated from the analysis of high pressure inelastic neutron scattering results [6] (section 4.5) and diffraction data [35] for cubic ZrW$_2$O$_8$. At 10 K the contribution from modes above 5 meV is negligible. At high temperature (300 K) higher energy modes are also populated and they too contribute to the thermal expansion.

In order to understand the nature of phonons responsible for NTE the partial contributions of the phonons of different energies to the mean square vibrational amplitude of the various atoms at 300 K in HfW$_2$O$_8$ has been plotted (Fig. 16). The modes upto 1 meV are largely acoustic in nature, but O4 atoms have larger amplitudes than other atoms. Above 1 meV, the O1 and O4 atoms connected to W1 have larger amplitudes in comparison with O2 and O3 connected to W2 atoms. Further, the various oxygen atoms constituting the tetrahedra have significantly different values of their vibrational amplitudes, which indicates distortions of the tetrahedra. Above 4 meV the amplitude of all the atoms is relatively small. The phonon modes of energy about 4 meV contribute (Fig. 10(a)) maximum to the negative thermal expansion. Similar behavior was observed for cubic ZrW$_2$O$_8$.

**4.3 Ab-initoCalculation of Phonon Spectra and Thermal Expansion Behavior in ZrW$_2$O$_8$**

Earlier neutron scattering data [6] as well as theoretical [5] estimates of the anharmonicty of the phonons in ZrW$_2$O$_8$ using interatomic potential model indicated that low energy phonon are largely responsible for observed NTE, while certain high energy modes also contributed based on Raman spectroscopy [6]. DFT calculations[177] at the Brillouin zone centre as well as non-zone-center modes resulting from 4 ×4 × 4 grid of q points found that the lowest energy optic modes at 45 and 46 cm$^{-1}$ have large negative Grüneisen parameters and thus contribute significantly (Fig. 17) to low temperature NTE in α-ZrW$_2$O$_8$. Other modes, between 96 and 164 cm$^{-1}$ become important to NTE near room temperature. Later ab-initio calculations were performed [3]for the phonons in the entire Brillouin zone for understanding the mechanism of NTE and to identify specific zone-boundary modes that are highly anharmonic. The calculations are able to reproduce the observed NTE as well as anomalous trends of the phonon spectra with increase in temperature and pressure.

The ab-initio calculated phonon spectrum[3] is found to be in excellent agreement with the experimental phonon spectrum [36] as shown in Fig. 18. The low energy range of phonon dispersion up to 50 meV contains large number of non-dispersive phonon branches, which give rise to several peaks in density of states. The lowest optic mode is calculated at 40 cm$^{-1}$ (~5 meV), which is in excellent agreement with the experimental value of 40 cm$^{-1}$ from Raman [176] as well as infra-red measurements [177]. The low-energy peak also leads to a sharp increase in the specific heat at low temperatures [3]. To emphasis the anharmonic nature of low energy phonons the phonon dispersion up to 10 meV is shown (Fig. 19) at 0 and 1 kbar. Several phonon branches have been found to soften with increasing pressure.

As mentioned above, the thermal expansion behavior is computed from the pressure dependence of phonon modes around ambient pressure.

The calculated $\alpha_V$ at 300 K from ab-initio calculation is $-22.5 \times 10^{-6}$ K$^{-1}$, while the experimental value [40] is about $-29 \times 10^{-6}$ K$^{-1}$. The calculated relative volume thermal expansion is shown in Fig. 20. The discontinuity in the experimental data at about 400 K is associated with an order-disorder phase transition. The contribution of phonon density of states at energy E to the thermal expansion has been determined as a function of phonon energy at 300 K. The maximum contribution to $\alpha_V$ is from phonon modes of energy $4.5 \pm 1$ meV (Fig. 20(b)), which is consistent with the previous analysis of high pressure inelastic neutron scattering measurements [6] as well as diffraction data [36]. The eigenvectors of a few of the low energy modes that contribute most to NTE have also been plotted (Fig. 5 and in [3]). The nature of these phonons can be best understood by the animations [3].

Hancock et al [30] proposed two types of modes for understanding the mechanism of NTE. In one of the mode both ZrO$_6$ as well as WO$_4$ in a chain rotate and also translate along the <111> axis. For the two lowest optic modes, it is noted that for two of the X-point modes of 3.90 meV and 4.16 meV, the motion of polyhedral units is found similar to that proposed by Hancock et al [30]. The modes show (Fig. 21) in-phase translation and rotation of WO$_4$ and ZrO$_6$ in a single chain. The motion of tetrahedral and octahedral units in two different chains is also in-phase. While the two modes seem to be of similar nature, the relative amplitudes of Zr, W atoms and O atoms are found to be different.

The M-point modes of 4.51 meV and 4.65 meV energy have negative Grüneisen parameters. The mode at 4.51 meV involves in-phase translation and bending of WO$_4$ and ZrO$_6$ network. The mode is very similar to that previously described by Cao et al [165], where a correlated motion between WO$_4$ and its nearest ZrO$_6$ is shown to lead to NTE. However, the 4.65 meV mode involves (Fig. 21) out-of-phase translation of WO$_4$ and ZrO$_6$ in two chains. In general, in most of the modes, amplitude of the free oxygens O3 and O4 are larger as compared to that of shared oxygen's O1 and O2. This means that rotation of WO$_4$ and ZrO$_6$ is accompanied by distortion of these polyhedra. In summary, the first principles calculations revealed that a large number of phonon modes contribute to the NTE in ZrW$_2$O$_8$. These include the specific modes that were earlier identified based on the analysis of data from various experimental techniques.

## 5. $Y_2Mo_3O_{12}$, $Sc_2W_3O_{12}$ and $Sc_2Mo_3O_{12}$

The orthorhombic $Y_2Mo_3O_{12}$(Fig. 22) is known to exhibit negative thermal expansion (NTE) behavior [178]. To reveal the relationship between NTE and polyhedral movements and distortions, the vibrational properties of $Y_2Mo_3O_{12}$ have been studied using first-principles calculations[178]. The lowest optical branch, which involves a translational mode of the O bridge atom in Y–O–Mo linkage, has the largest negative Grüneisen parameter and therefore contributes most to the NTE behavior. The different vibrational eigenvectors of oxygen atoms relative to Y or Mo atoms can cause the polyhedra to distort unevenly. The $YO_6$ octahedra and $MoO_4$ tetrahedra distort unevenly on the Y-O-Mo linkages, along with both polyhedra being closer which yields the overall volumetric contraction[178].

### 5.1 Phonon Spectra and Thermal Expansion

The first-principles calculation of the electronic charge density and Bader charges indicate that the bonding between Mo and O atoms is more covalent, while the Y–O bonds are mainly ionic in character. The calculated phonon density of states (Fig. 23) is in a good agreement with the experientially measured Raman spectrum[179]. The phonon modes with negative Gruneisen parameters (Fig. 24) are distributed up to high energies of 294.5 $cm^{-1}$. The lowest frequency optical branch (34.5 $cm^{-1}$) has the strongest negative Gruneisen parameter and therefore contributes most to the NTE behavior. The calculated temperature dependence of the unit cell volume and volume thermal expansion coefficient is in agreement (Fig. 25) with the experimental data.

In the absence of a direct measurement of the phonon spectrum useful information was derived from analysis of accurate heat capacity data as a function of temperature[26]. The heat capacities of $Sc_2W_3O_{12}$ and $Sc_2Mo_3O_{12}$ were measured and analysed to provide their effective phonon spectrum[26]. The spectrum of orthorhombic $Sc_2W_3O_{12}$ (Fig. 22) shows three features (Fig. 26): low-energy phonon modes with negative mode- Gruneisen parameter ($\gamma_i$) around 5 meV, high-energy phonon modes, and separation of phonon spectrum into two regions with a wide gap. The authors identified (Fig. 27) the relative contribution of $\gamma_i C_i$, where $C_i$ is heat capacity of each vibrational mode i, and observed that the low-energy phonon modes with negative $\gamma_i$ cause the NTE and that the latter two features are necessary to maintain the NTE in a wide temperature range. The monoclinic phase of $Sc_2Mo_3O_{12}$(Fig. 22) also has the low-energy mode with negative $\gamma_i$ (Figs. 28, 29) but show overall positive thermal expansion.

## 6. ZrV$_2$O$_7$ and HfV$_2$O$_7$

There has been considerable interest in the negative thermal expansion properties of AV$_2$O$_7$ (A=Zr, Hf). Like the ZrW$_2$O$_8$ family, the AV$_2$O$_7$ (A=Zr, Hf) family[87, 105] also has a cubic crystal structure and shows a large isotropic NTE. The high-temperature phase of AV$_2$O$_7$ shows NTE from about 400 and 900 K. The basic structure of AW$_2$O$_8$ and AV$_2$O$_7$ consists of corner-sharing MO$_6$ octahedra and WO$_4$/VO$_4$ tetrahedra. In case of AV$_2$O$_7$ each VO$_4$ tetrahedron shares three of its four O atoms with an MO$_6$ octahedron, while the fourth is shared with another VO$_4$ tetrahedron leading to a V$_2$O$_7$ group. All oxygen atoms in MV$_2$O$_7$ have twofold coordination. The AW$_2$O$_8$ and AV$_2$O$_7$ compounds differ in terms of the oxygen coordination around the V/W atoms leading to differences in the nature of soft phonons under compression that are responsible for the NTE. The calculations[11] quantitatively reproduce the negative expansion over a range of temperatures. The soft phonon in AV$_2$O$_7$ is found to be related with the observed phase transitions. Especially, as discussed below, the calculations show a soft-phonon mode at a wave vector of 0.31<1, 1, 0>, which is in excellent agreement with the known incommensurate modulation in ZrV$_2$O$_7$ below 375 K. In the low-T phase below 400 K, the soft phonons of the high-T phase would freeze and may no longer have the negative Grüneisen parameters. The low-T phase has positive thermal expansion coefficient.

### 6.1 Phonon Dispersion Relation and Density of States

The temperature dependent x-ray diffraction measurements[105, 180, 181] for ZrV$_2$O$_7$ and HfV$_2$O$_7$ show that the phase below 350 K corresponds to a primitive 3×3×3 superstructure phase of $Pa\bar{3}$ space group symmetry while the high temperature phase above 375 K corresponds to the normal parent structure. The intermediate incommensurate phase is characterized by a modulation wave vector of 0.314<110>[180].

The high-T cubic phase of MV$_2$O$_7$ has 40 atoms in the primitive cell and thus 120 phonon modes at each wave vector. The calculated pressure dependence of the phonon dispersion relation (up to 10 meV) in the high-T phase of MV$_2$O$_7$ in the quasi-harmonic approximation is shown in Fig.30. The significant larger mass of Hf (178.49 amu) in HfV$_2$O$_7$ in comparison to Zr (91.22 amu) in ZrV$_2$O$_7$ results in slightly lower phonon energies in HfV$_2$O$_7$. Large softening with pressure is observed for the phonon branches lying in the energy range of 2.5 to 9 meV. The transverse acoustic modes along (100) direction do not show large softening, which is in contrast to ZrW$_2$O$_8$. The maximum softening occurs for transverse acoustic mode at 0.31<110>. These modes have been calculated in the high-T phase in the quasi-harmonic approximation. The soft modes at 0.31<110> stabilize in the high-T phase due to anharmonicity and freeze in the intermediate incommensurate phase. Freezing of these soft modes at the 0.31<110> points in the reciprocal space at low temperature leads to the known incommensurate phase[180]. It may be noted that the calculated soft-mode wave vector is in excellent agreement with the observed [180] incommensurate modulation. Subsequent freezing of the soft modes at the <1/3, 1/3, 0> points in the reciprocal space at lower temperature could lead to the known 3×3×3 superstructure transition[105] . Once the <1/3, 1/3, 0> points become lattice points due to the phonon freezing, other points such as <1/3, 0, 0> and <1/3, 1/3, 1/3> would also become lattice points due to the symmetry of the low-T phase even though the modes at the latter points are not soft in the high-T phase.

The experimental Raman data [182, 183] has been obtained from polycrystalline samples in the low temperature phase. The range of phonon frequencies in the high temperature phase is expected to be same as that in the low temperature phase. The calculated range (Fig. 31) of phonon frequencies in the high temperature phase of $ZrV_2O_7$ and $HfV_2O_7$ is nearly same as that obtained from the experimental data[182, 183] in the low temperature phase. The potential reproduces the equilibrium crystal structure (Table 2) and other dynamical properties quite satisfactorily as discussed later.

The contribution of various atoms to the phonons at various energies is illustrated by the partial density of states shown in Fig. 31, in both the Hf and Zr compounds. The Zr and Hf atoms contribute in energy range of 0–50 meV, while the vanadium and oxygen atoms contribute in the entire energy range upto 130 meV. Above 105 meV the contributions are mainly due to V-O stretching modes. The Zr-O and Hf-O stretching modes occur around 30-35 and 25-30 meV respectively. The significant larger mass of Hf (178.49 amu) in $HfV_2O_7$ in comparison to Zr (91.22 amu) in $ZrV_2O_7$ give rise to the shift towards lower energies in the total density especially at low energies.

## 6.2 Grüneisen Parameters and Thermal Expansion

The calculated pressure dependence of phonon spectra is used for the calculation of the Grüneisen parameter $\Gamma(E)$, averaged for all phonons of energy E, in Fig. 32(b) for cubic $ZrV_2O_7$ and $HfV_2O_7$. Above 10 meV, the $\Gamma(E)$ values are small and lie between -1 and 1.The cubic $HfV_2O_7$ has slightly lower values of $\Gamma(E)$ in comparison to $ZrV_2O_7$. This is similar to the trend deduced from the analysis of the thermal expansion and specific heat data [100]in cubic $ZrW_2O_8$ and $HfW_2O_8$. The calculated values of $\Gamma(E)$ for $ZrW_2O_8$ are also shown in Fig. 32(b) and show a similar variation as in $ZrV_2O_7$.

The calculation of the temperature dependence of the volume thermal expansion coefficient (Fig. 33) indicates that in cubic $ZrV_2O_7$ almost all the NTE (about 95%) is contributed from the phonon modes below 9 meV, among which nearly 50% of the NTE arises from just two lowest modes. The absolute value of thermal expansion coefficient for $HfV_2O_7$ is slightly smaller in comparison with $ZrV_2O_7$. The comparison between the calculated and experimental data for cubic $ZrV_2O_7$[180] and $HfV_2O_7$[87] is shown in Fig. 34. The agreement between the calculations and experimental data is excellent in the high-T phase between 400 and 900 K. In the low-T phase below 400 K, the soft phonons of the high-T phase would freeze and may no longer have the negative Grüneisen parameters. The low-T phase has positive thermal expansion coefficient. Above 900 K, the experimental data show a sharp drop in the volume at about 900 K, which probably signifies another phase transition.

In Fig. 35 the contribution of various phonons to the thermal expansion is shown as a function of phonon energy at 500 K for $ZrV_2O_7$ and $HfV_2O_7$. The maximum negative contribution to $α_V$ at 500 K is from the modes of energy from 4 to 7 meV. The nature of the phonons may also be visualized from the calculated partial contributions of the phonons of different energies to the mean square vibrational amplitude (Fig. 36) of the various atoms. The modes up to 2 meV involve equal displacement of all the atoms, which correspond to the acoustic modes. Above 2 to 15 meV, the O1 and O2 atoms connected to $ZrO_6$ and $VO_4$ have larger amplitudes in comparison of Zr and V. Further, the various oxygen atoms constituting the tetrahedra have nearly same values of their vibrational amplitudes, which indicate translation and rotation of the $ZrO_6$ octahedral and $VO_4$ tetrahedral units. This is quite different from cubic $ZrW_2O_8$ where the important modes around 4 meV involved the oxygen atoms constituting the $WO_4$ and $ZrO_6$, and these oxygen atoms have significantly different values of their vibrational amplitudes, indicating distortions of $ZrO_6$ octahedral and $WO_4$ tetrahedral units.

## 7. $M_2O$ (M=Au, Ag, Cu)

The compounds $M_2O$ (M=Ag, Cu and Au) [83, 84] crystallize in a simple cubic lattice (space group Pn-3m) (Fig. 37). The M atoms are linearly coordinated by two oxygen atoms, while oxygen is tetrahedrally coordinated by four M atoms. $Ag_2O$ shows a large isotropic NTE over its entire temperature range of stability, i.e. up to ~ 500 K, while $Cu_2O$ only shows a small NTE below room temperature. At the moment, no experimental data are found in the literature for thermal expansion behaviour of $Au_2O$. EXAFS measurements on $Ag_2O$ and $Cu_2O$ [184-187] indicate that the mechanism at the origin of their NTE involves deformations of the $M_4O$ tetrahedral units (M = Ag, Cu), rather than simple rigid units vibrations. Also it has been suggested[187] that the large difference between the NTE coefficient of $Ag_2O$ and $Cu_2O$ not only originates from a mass effect but also from differences between the chemical interaction.

The ab-initio lattice dynamics calculations show that $Ag_2O$ and $Cu_2O$ have a NTE while $Au_2O$ has a large positive expansion with increasing temperature[2, 4]. The origins for these large differences in the amplitude and sign of the thermal expansion coefficients between these isostructural compounds are discussed [2]. It turns out that $Au_2O$ has a less ionic and more covalent character than $Ag_2O$ and $Cu_2O$ which has consequences for Au-O bonds on the lattice expansion and anharmonicity of the crystal.

**7.1 Phonon Spectra**

The DFT calculations for $Cu_2O$[188]agree well with the experimental dispersion curves, and with the measured phonon spectra (Fig. 38) [4]. Similarly the calculated and measured phonon spectra in $Ag_2O$ (Fig. 38) agree very well. The comparison validates the approach using ab-initio calculations, and gives confidence into the properties derived from them, especially those extended to $Au_2O$. As expected for isostructural compounds, the dispersion curves (Fig. 39) below 10 meV in $Ag_2O$ and $Au_2O$ are found to be similar, while in $Cu_2O$ these modes are shifted to higher energies, which can be understood considering the smaller mass of Cu.

The different spectral range for phonons reflect the different M-O bond lengths and difference in nature of M-O bonding which is discussed latter. The smallest Cu-O bond (1.866 Å) results in shifting of energies up to the highest spectral range of 75 meV. However, Ag-O and Au-O bond lengths 2.082 Å and 2.078 Å respectively in $Ag_2O$ and $Au_2O$ are similar but the stretch mode of Au is at larger frequencies. This suggests that the Au-O bond may have a more covalent nature as compared to the Ag-O bond for which an ionic nature dominates.

The temperature dependence of the inelastic neutron scattering measurements (Fig. 40)of the phonon spectrum on $Ag_2O$ and Fourier transform far-infrared spectra have been reported that showed unusually large energy shifts with temperature, and large line width broadenings[68]. First principles molecular dynamics calculations are in excellent agreement (Fig. 40)with the neutron scattering data[68]. The lifetime broadenings of $Ag_2O$ were explained by anharmonic perturbation theory, which showed rich interactions between the Ag-dominated modes and the O-dominated modes in both up- and down-conversion processes.

## 7.2 Thermal Expansion Behavior

For $Ag_2O$ and $Au_2O$, the energy range for negative $\Gamma(E)$ extends (Fig. 41) up to ~3.5 meV. However, the magnitude is much larger for the former compound, reaching a value of -40 for the lowest modes, while for $Au_2O$ the maximum negative $\Gamma(E)$ reaches -10. For $Cu_2O$ the phonons below 6.5 meV have negative $\Gamma(E)$ with a maximum negative value of -4.5.

The comparison between the available experimental data of volume thermal expansion along with the calculations is shown in Fig 42. Negative thermal expansion obtained (Fig 42, 43) in the calculations in $Ag_2O$ over its temperature range of stability of about 500 K, while $Cu_2O$ and $Au_2O$ show negative $\alpha_V(T)$ below 300 K and below 16 K, respectively. The most negative $\alpha_V(T)$ values for $Ag_2O$ (-44 ×$10^{-6}$ $K^{-1}$) and $Cu_2O$ (-8 × $10^{-6}$ $K^{-1}$) are respectively obtained at 40 K and 75K. The maximum negative value of $\alpha_V(T)$ for $Au_2O$ is much reduced compared to the other two compounds and reaches ~ -2 ×$10^{-6}$ $K^{-1}$ at T ~ 8 K. As mentioned above, one may understand the absence of NTE in the $Au_2O$ lattice as resulting from the combination of two effects: 1) reduced absolute values of negative $\Gamma(E)$ (compared to $Ag_2O$) and 2) reduced energy range for the phonon modes with negative $\Gamma(E)$ (compared to $Cu_2O$).

The maximum negative contribution to volume thermal expansion coefficient arises(Fig 44) from the modes of energy around 4 to 5 meV. The nature of the low energy phonon modes contributing to the NTE can be visualized through animations[189, 190]. The eigenvectors of a selection of them have also been plotted on Fig. 45. The lowest $\Gamma$-point optical mode corresponds to the rotation of $M_4O$ tetrahedral and the lowest X and M point modes have negative Grüneisen parameter in all the three compounds. X-point mode involves bending of M-O-M chains. The M atoms connected to various $M_4O$ have different displacements indicating significant distortion of $M_4O$ tetrahedra. This mode seems to contribute maximum to NTE in $Ag_2O$. The M-point mode involves rotation and translation as well as distortion of the $M_4O$ tetrahedra, while for R-point the amplitude of all the atoms is similar and it indicates translational motion of $M_4O$ as a rigid unit.

Based on other ab- initio lattice dynamics calculations[188]it was reported that NTE in $Cu_2O$ is dominated by low frequency vibrations of rigid O-Cu-O rods along with some smaller contribution of higher frequency motion. The authors also predict that the primary NTE mode drives a proper ferroelastic phase transition at high pressure.

The first principles MD calculations have also been reported (Fig 46) that reproduce accurately accounted for the NTE and local dynamics of $Ag_2O$, such as the contraction of the Ag-Ag shell and the large distortion of the $Ag_4O$ tetrahedral [68]. The NTE at temperatures above 250 K is due largely to the anharmonicity of phonon-phonon interactions and is not predicted with volume dependent quasiharmonicity. In particular, the strong interactions of O-dominated modes with Ag-dominated modes cause the second stage of NTE at temperatures above 250 K.

**7.3 Bonding in $M_2O$ (M=Ag, Au, Cu)**

It is clear that the large difference in the calculated thermal expansion for the three $M_2O$ compounds reflects a difference in the bonding from one compound to the other. In addition, the presence of large voids in the unit cell renders the structure even more sensitive to subtle differences in bond strength. The calculated charge density for the three compounds (Fig. 47) reveals that the bonding character of the Ag-O bond is more ionic than that of the Cu-O bond. The Au-O bond is found to be highly directional with the charge density elongated towards the O atom *i.e.* indicating a covalent nature[191]. The change of bonding from an ionic to a covalent character is due to different intra-atomic hybridization between the d, s and p states[189, 190]. The computed Born effective charges for oxygen atoms in $Ag_2O$, $Cu_2O$ and $Au_2O$ are -1.28, -1.18 and -0.54 respectively. The latter values also reflect the larger ionic character for the Ag-O bond compared to Au-O.

The compounds $Ag_2O$ and $Au_2O$ have very similar lattice parameters ($Ag_2O$ =4.81 Å and $Au_2O$ =4.80 Å) and also similar Ag/Au-O bond lengths. However, the covalent and directional Au-O bond rigidifies the $Au_4O$ tetrahedra, making them less susceptible to distortion, bending or rotation than their $Ag_4O$ counterpart. This is suggestive of the microscopic origin of the large NTE in $Ag_2O$. As discussed above, it is found that $Ag_2O$ shows a large softening of its transverse acoustic modes along the Γ-X-M line with increasing pressure while in $Au_2O$ this softening is not observed. Also in case of $Ag_2O$, high energy optical modes also show softening in contrast to that in $Au_2O$, where these modes become hard with pressure. The $Cu_2O$ and $Ag_2O$ compounds have similar the nature of bonding and both exhibit negative thermal expansion. However, there is a large difference in the magnitude of the thermal expansion coefficient. The Cu-O (1.87 Å) bond length is much smaller than the Ag-O (2.08 Å) bond. The $Cu_4O$ tetrahedral units are therefore much more compact than $Ag_4O$, rendering distortion less favorable in $Cu_4O$ as compared to $Ag_4O$. In addition, the difference in the open space in the unit cell between the two compounds leads to differences in the magnitude of the distortions and hence difference in the NTE coefficient. Here the open structure nature of the lattice regulates the extent of the NTE.

## 8. β- Eucryptite (LiAlSiO$_4$)

The β- eucryptite (LiAlSiO$_4$) shows anisotropic thermal expansion and very low volume thermal expansion[192, 193]. These properties make it useful in domestic cookware, high-precision machines, optical devices and as a good thermal shock resistance material. At room temperature, β-eucryptite crystallizes[124, 125] in the space group P6$_4$22 (Fig. 48). It has 84 atoms in a unit cell, and is basically a stuffed derivative of β-quartz[194]. β- eucryptite contains double helices of SiO$_4$ and AlO$_4$ tetrahedra. This produces alternating layers of Al and Si normal to the hexagonal c-axis[125, 195]. Li is present in one-dimensional channels parallel to the c-axis[196]. There are three secondary (S) and one primary (A) channel in a single unit cell[195]. Each channel has six available tetrahedral sites; however, at ambient temperature, only three of the available sites are occupied[124] in an alternating sequence (Fig.48). The Li atom in the secondary channels is translated by c/6 with respect to that in the primary channel[195].

The compound β-eucryptite shows a strong anisotropic thermal expansion behavior[124, 193, 197]($\alpha_c \approx -2\alpha_a$) with linear thermal expansion coefficients of about $\alpha_a=8.6\times10^{-6}K^{-1}$ and $\alpha_c=-18.4\times10^{-6}K^{-1}$, which results in a very low volume thermal expansion over a wide temperature range of 300-1400 K. Earlier lattice dynamics calculations[197] of β-eucryptite provided a fair description of the thermal expansion behavior. Perhaps due to limited computational resources at that time and structural complexity, the accuracy of the calculation was limited. Later an improved calculation of the phonon spectrum in the entire Brillouin zone was reported[16] that accurately reproduce the observed anomalous thermal expansion in β-eucryptite. It also enabled to identify the specific phonon modes to understand the microscopic mechanism of the anomalous thermal expansion.

The experimental phonon density of state for β- eucryptite is obtained from inelastic neutron scattering for the first time[16]. The inelastic neutron spectra are analyzed using ab-initio lattice dynamics calculations using the density functional theory. The pressure dependence of phonon spectrum in the entire Brillouin zone is calculated and used it to obtain the anisotropic Gruensien parameters. The eigenvectors of phonon modes are analysed to understand the microscopic mechanism of anomalous thermal expansion in β- eucryptite. The anisotropic elasticity has an important role in anisotropic thermal expansion. The specific phonon mode at the Brillouin zone centre may also contribute to the high pressure phase transition[16].

## 8.1 Phonon Spectrum

The inelastic neutron scattering measurements are performed at room temperature. The comparisons between the measured phonon spectrum and the ab-initio calculated spectrum is shown in Fig. 49. The general characteristics of the experimental features are very well reproduced by the calculations. A previous infrared study of β-eucryptite showed absorption peaks in the energy range of 100 - 1200 cm$^{-1}$ (12 to 150 meV)[32], which is consistent with the neutron data. The calculated as well as the experimental phonon spectra show that the phonon modes belong to two distinct regions, and exhibit a significant band gap. The first region extends up to 100 meV, and has contributions from the vibrations of all the atoms in the unit cell. On the other hand, the high-energy part of the spectrum, between 110-150 meV, largely corresponds to the internal vibrations of the SiO$_4$ and AlO$_4$ tetrahedra. It may be noted that the peaks in the calculated spectrum show a tiny downward shift in energy as compared to the measured inelastic spectrum. This shift is due to the overestimation of the unit cell volume in GGA approximation.

The partial density of states (PDOS) are calculated by projecting the phonon mode eigenvectors on individual atoms as shown in Fig. 50. The neutron weighted PDOS (Fig. 49) obtained by taking care of neutron scattering cross section of individual atom can be used to identify the atomic contributions in the neutron inelastic spectrum. It can be seen that the major contributions from Li atoms lie below 70 meV, while Si, O and Al atoms contributed in the whole spectral range, up to 140 meV. The partial density of states imply that the double peaks in the experimental spectra, around 10-30 meV, are due to the O atoms. The eigenvector analysis of these modes implies vibrations of the T-O-T (T=Si, Al) bonds, giving rise to a rotational motion of the tetrahedral units as it was also observed in a previous study by Zhang et al[32]. The most intense peak in the spectrum is around 40 meV. It shows maximum contributions from the Li and O atom (Fig.49). This peak is associated with the LiO$_4$ tetrahedral stretching modes, showing isotope shifts of 8.7 cm$^{-1}$(1.08 meV) to 10.6 cm$^{-1}$(1.31 meV), respectively, through $^6$Li/$^7$Li isotope substitution[198]. The eigenvectors of these modes confirmed the vibrational motion of Li atoms along the hexagonal c-axis. The next peak in the spectrum around 50-60 meV originates mainly from the O atoms. This frequency domain is associated with the stretching and bending of Si-O-Al linkages[32], in corner shared AlO$_4$ and SiO$_4$ tetrahedra as the direct Si-O-Si linkage is not favored in β-eucryptite due to the Al avoidance principle[199]. The humps in the experimental spectrum around 65 - 75 meV are due to the Li and O atoms contribution. These phonon modes correspond to the Si-O-Li bridge vibrations as reported in previous studies performed in the range 500-600 cm$^{-1}$ (62-74.5meV)[198].

The maximum contribution from Al and Si is seen from around 70 meV and onward, which ultimately forms a double peak in the experimental spectrum at around 75-95 meV. These types of vibrations have also contributions from the O atoms. The related eigenvectors indicate that the modes originate from vibrations of Al and Si tetrahedra. The high-energy spectrum at 110-140 meV is largely dominated by Si-O and Al-O stretching modes. However, the contribution from Al is relatively small (Fig.49), as compared to that of Si. This indicates that the highest energy peak is mainly governed by $SiO_4$ intra-tetrahedral vibrational modes with a small component from $AlO_4$ tetrahedral modes, in accordance with previous spectroscopic studies[32]. Dispersive nature of the stretching mode spectrum is attributed to anisotropic bonding in $SiO_4$ and $AlO_4$ polyhedra.

The partial density of states of various atoms (Fig. 50) is used for the calculation of mean squared displacements of various atoms, $<(u^2)>$, (Fig. 51). As shown in Fig. 51(a) the equal amplitude of all the atoms up to 6 meV indicates that the modes are acoustic. For energy range 6-20 meV, oxygen atoms have larger amplitudes in comparison to Al and Si atoms. This indicates that these modes involve a rotation of the $AlO_4$ and $SiO_4$ tetrahedra. Further, above 20 meV, the amplitude of Li atoms is found to be very large as compared to other atoms.

## 8.2 Anomalous Thermal Expansion Behavior

The anisotropy in the elastic constants and the mode Grüneisen parameters are important in deriving the anisotropic thermal expansion. The calculated elastic constants and bulk modulus (Table 3) are in good agreement with the reported experimental[200] data as well as other estimations from the literature[201]. The calculated anisotropic mode Grüneisen parameters as a function of phonon energy averaged over the Brillouin zone are shown in Fig. 52 (a). The Grüneisen parameters has large negative value for the low-energy phonon modes around 10 meV and has positive value for the high-energy modes in the range 30 to 70 meV. The calculated linear thermal expansion coefficients and the lattice parameters as a function of temperature are shown in Fig.52(b) and 52(c) respectively. The calculated temperature dependence of the lattice parameters is in excellent agreement (Fig. 52(c)) with the experimental data[193, 197]. The calculated anisotropic linear thermal expansion coefficients at 300 K are $\alpha_a= 6.0\times10^{-6}$ $K^{-1}$ and $\alpha_c=-13.7\times10^{-6}$ $K^{-1}$. They compare very well with available average experimental values[25][193] of $\alpha_a= 8.6\times10^{-6}$ $K^{-1}$ and $\alpha_c=-18.4\times10^{-6}$ $K^{-1}$ in the temperature range from 300 to 1400 K. The thermal expansion coefficient in the a-b plane ($\alpha_a(T)$), upto 200 K, has a small negative value (Fig. 52(b)). It evolves to a positive value at higher temperatures.

The calculated lattice parameters in the temperature range up to 600 K are in a very good agreement with the available experimental data. However, above 600 K, the calculated parameters deviate slightly (Fig. 52 (c)) from the experiments. This might be due to the increasing role of the explicit anharmonicity of phonons at high temperature. This effect has not been considered in the thermal expansion calculation, which includes only the volume dependence of phonon energies (implicit contribution). It can be seen that the compound shows negative volume thermal expansion behaviour below 300 K, and a positive expansion at higher temperatures. The compound has a very small volume thermal expansion coefficient ($\alpha_V$=-1.7 ×$10^{-6}$ $K^{-1}$) at 300 K, which makes it suitable for application as a high thermal shock resistant material.

**8.3 Phonon Eigenvectors and Thermal Expansion**

The phonon dispersion relation (Fig.53) in the RT-phase has been calculated along the high symmetry directions of the hexagonal cell at ambient, as well as at 0.5 GPa. It can be seen that some of the zone centre modes are highly anharmonic. The eigenvectors of some of these anharmonic modes are analyzed (Fig. 54) to understand the mechanism of anisotropy in the thermal expansion behavior of the compound.

The Γ-point mode of 7.0 meV (Fig. 54) has Grüneisen parameters values of $\Gamma_a$=-31.0 and $\Gamma_c$=-30.8. This mode shows an anti-phase rotation of adjacent polyhedral units ($AlO_4$ or $SiO_4$) about all the three directions. The central atom (Al or Si) of the polyhedral units has a small displacement in comparison with oxygen atoms. The next Γ-point mode of 9.6 meV ($\Gamma_a$=-13.6, $\Gamma_c$=-15.3) also involves anti-phase rotation of $AlO_4$ and $SiO_4$ but only about b-axis. Here the amplitude of rotation for $SiO_4$ is large in comparison to $AlO_4$ polyhedral units. The Li atoms in both these modes vibrate in the ab-plane.

The high energy Γ-point, modes of energy 31.7 meV ($\Gamma_a$=2.3, $\Gamma_c$=1.0) and 55.2 meV ($\Gamma_a$=3.6, $\Gamma_c$=1.5) (Fig. 54) have significant positive Grüneisen parameter value. These modes get populated as temperature is increased and contribute to the expansion in the a-b plane at higher temperature. These modes involve large Li motion with very small component of Al and Si tetrahedral rotation and distortions. The mode at 31.7 meV involves the translational motion of Li along the hexagonal channel and might be correlated with the Li diffusion along c-axis, while the other mode at 55.2 meV, involves the Li translation motion in a-b plane.

The M-Point mode (Fig. 54) of 8.7 meV ($\Gamma_a$=-3.9, $\Gamma_c$=-3.3) reflects a sliding motion of SiO$_4$ tetrahedral units in the a-b plane. The adjacent SiO$_4$ layers move in opposite directions along the b-axis. This gives rise to a rotation of the involved AlO$_4$ polyhedra. The Li atoms in adjacent layers move along b- and c-axis. This type of dynamics induces folding of spirals containing (Al,Si)O$_4$ polyhedra producing contraction in the system. The A-Point mode at 8.3 meV ($\Gamma_a$=-18.0, $\Gamma_c$=-18.9) shows (Fig. 54) AlO$_4$ polyhedral rotation about a-axis while SiO$_4$ polyhedral rotation about b-axis. Here, oxygen atoms connected to the tetrahedral units have different amplitude and may produce polyhedral distortion. The lithium motion lies in the a-b plane.

At low temperature below 200 K, low-energy modes specifically those below 20 meV contribute significantly to the thermal expansion behavior. Although the above mentioned low-energy modes have almost similar values of $\Gamma_a$ and $\Gamma_c$ ($\Gamma_a \approx \Gamma_c$), yet the anisotropy in thermal expansion behavior is there (Fig 52(b-d)). This anisotropy can be understood in terms of the anisotropy in the elastic properties of β-eucryptite.

$$\alpha_a \propto [s_{11} + s_{12}]\Gamma_a + s_{13}\Gamma_c = 0.008\Gamma_a - 0.005\Gamma_c \qquad (36)$$

and

$$\alpha_c \propto [s_{31} + s_{32}]\Gamma_a + s_{33}\Gamma_c = -0.010\Gamma_a + 0.014\Gamma_c \qquad (37)$$

where $[s_{11} + s_{12}] = 0.008$, $[s_{31} + s_{32}] = -0.010$, $s_{13}$= -0.005 and $s_{33}$= 0.014 in GPa$^{-1}$ units (Table 4).

It may be noted that for low-energy modes $\Gamma_a \approx \Gamma_c < 0$, which imply a negative thermal expansion both along a- and c-axis at low temperatures [Fig.52(b-c)]. However, at high temperatures the entire spectra will contribute to the thermal expansion behaviour and larger positive values of $\Gamma_a$ than those of $\Gamma_c$ result in a positive thermal expansion in *a-b* plane and negative thermal expansion along c-axis.

## 9. ReO$_3$

Transition metal oxides have been the subject of renewed interest ever since high temperature superconductivity was discovered in copper oxide materials. The transition metal oxides with narrow d bands form strongly correlated Mott-Hubbard system for which conventional band theory is no longer valid. The number of transition metal oxides where band theory alone can provide adequate account of their properties is quite limited, and is confined to compounds of the 4d and 5d series. $ReO_3$ is an example of such a simple metallic oxide. $ReO_3$ crystallizes [12] in the cubic space group (Pm3m). The structure consists of corner-linked $ReO_6$ octahedra with Re at the centres and linear Re-O-Re links. Among the numerous perovskite like compounds $ReO_3$ belongs to a small family of undistorted cubic structures which is stable at ambient pressure and at all temperatures up to its melting point. Also the $ReO_3$ structure has a completely vacant A cation site of the $ABO_3$ perovskite structure. This empty structure is therefore expected to allow rigid rotation of the $ReO_6$ octahedral.

High pressure X-ray and neutron diffraction measurements[202-204] established that $ReO_3$ undergoes a pressure-induced second order phase transition at 5.2 kbar at room temperature to a tetragonal (P4/mbm) intermediate phase with a very narrow stability range in pressure and then a further transition to a cubic (Im3) phase. The softening of the M3 phonon mode involving rigid rotation of the $ReO_6$ octahedra provides the mechanism of the phase transition. X-ray-absorption fine structure[205] suggests that even at ambient pressure Re-O-Re bond angle is not 180° but is about 170°. $ReO_3$ only appears cubic in a time-averaged and space-averaged structure obtained by the conventional diffraction analysis of the Bragg intensities. The local nano-scale structure of $ReO_3$ seems to be distorted.

The negative thermal expansion has been observed by neutron scattering in $ReO_3$ at low temperatures below 200 K[12]. Lattice dynamical calculations show that the negative thermal expansion of $ReO_3$ is due to the unusually large anharmonicity of the soft M3 mode, which consists of rigid anti-phase rotations of the neighboring $ReO_6$ octahedra. The lattice dynamics calculations are also found useful to understand the apparently anomalous variation of the soft phonon frequency with change in pressure or temperature. The frequency is found to increase with increasing temperature and decrease with increasing pressure. Both these observations are opposed to the usual variation of most frequencies in most materials. Moreover, the different behavior is observed with increasing pressure or temperature when the volume decreases in both cases. As explained below, the calculations[12] are able to reproduce the anomalous behavior quite satisfactorily, and bring out the underlying features of the interatomic potential.

**9.1 Phonon Dispersion**

Axe et al.[204] determined the phonon dispersions of the transverse and longitudinal acoustic phonon modes of $ReO_3$ by inelastic neutron scattering. Both the transverse T1($\zeta$00) and T2($\zeta$ $\zeta$ 0) modes have anomalous low frequencies extending to the zone boundaries. But the most remarkable feature is the pronounced reduction in frequency of the $T_2$ mode near the M-point zone boundary (1/2, 1/2, 0). Fig.55 shows the model calculations[12] of the phonon dispersion of $ReO_3$ along with the experimental data obtained from inelastic neutron scattering. The agreement appears to be satisfactory. In particular, the acoustic modes slopes have been very well reproduce and the soft mode at M- point has also been calculated qualitatively.

## 9.2 Thermal Expansion

The calculated value of the bulk modulus for $ReO_3$ is 210 GPa, which is in very good agreement with the reported experimental value of 211 GPa[206]. The Grüneisen parameter $\Gamma(E)$ averaged for all phonons of energy E has been calculated using pressure dependence of phonon spectra and is shown in Fig. 56 (Left). The calculations show that $\Gamma(E)$ has small positive values for phonons of energy up to 15 meV. For energies around $20 \pm 5$ meV phonons have average negative $\Gamma(E)$ values in the range of $-3$ and $-1$. Negative expansion is calculated (Fig. 56 (Right)) at low temperature similar to that determined by the neutron diffraction experiments. The calculations yield positive $\Gamma(E)$ values for modes above 50 meV, which results in a positive volume thermal expansion coefficient above 350 K. This result is in fair agreement with the experiments where negative to positive expansion cross-over occurs at 200 K.

Fig.57(Left) shows a comparison between the calculated and experimental thermal expansion behaviour in $ReO_3$. In Fig. 57(Right) the average contribution of various phonons of certain energy E to the thermal expansion is shown as function of E at 100 K. The maximum negative contribution to $\alpha_V$ arises from the modes of energy around 14 meV. The energy of about 14 meV corresponds to the M3 mode at the zone boundary along [110]. This mode is plotted (Fig. 58) using the calculated eigenvector, which shows librational motion of $ReO_6$ octahedron arising from the transverse vibrations of the O atoms in the a-b plane. Axe et al.[204] have also measured the pressure dependence of this phonon and found it to decrease substantially with increasing pressure,

The phonon spectrum in $ReO_3$ has also been calculated[207] using DFT and shows good agreement (Fig. 59, 60) with the inelastic neutron-scattering experiments at ambient pressure. Large negative values of the Grüneisen constants are calculated for the acoustic modes at the M and also at the R-point to a lesser extent. The calculated thermal expansion behavior is in fair agreement with the experimental data.

## 9.3 Temperature and Pressure Dependence of Phonon Energy

Temperature and pressure variation of the phonon frequency is known to occur due to anharmonicity of the interatomic potential. There are two effects; one the so-called 'implicit" effect due to the change in volume with temperature or pressure, and another one called the "explicit" effect due to the increase in vibration amplitude with temperature. The implicit effect can be calculated in the quasiharmonic approximation, where the vibrations are assumed to be harmonic around the equilibrium positions of the atoms in a lattice of a given volume. The calculation of the explicit effect is rather complex which in general involves phonon-phonon interactions besides self-anharmonicity. However, the latter gives a very good indication of the anharmonicity. If the anharmonicity is small, the explicit effect may be estimated by treating the anharmonic part of the phonon potential as a perturbation.

The lattice dynamics calculations are based on the shell model for $ReO_3$[13]. For calculations at high pressures the crystal structures are calculated at each pressure by minimization of the free energy. The calculated structures at various pressures are then used for the calculation of phonon frequencies at those pressures in the quasiharmonic approximation. The calculated pressure dependence of the quasiharmonic phonon frequencies for M3 mode is in very good agreement (Fig. 61) with the experimental results[13].

In case of temperature dependence both the implicit and explicit effects are present. The implicit effect is calculated from the known negative thermal expansion and the calculated volume dependence. Fig.61 allows estimating the implicit change in the frequency with change in volume. Since the volume decreases with temperature, the implicit change with temperature is estimated to be negative from Fig. 61.

For the calculation of explicit effect, the potential-well of M3 mode is calculated[13] as shown in Fig. 62. The potential-well is highly anharmonic with a large positive quartic anharmonicity, which leads to an increase of the frequency with increase in amplitude of vibration. The calculated explicit effect should be taken as qualitative. The comparison between the calculated and experimental temperature dependence of M3 mode is shown in Fig. 63, which included both the implicit and explicit effects[13].

The calculation is able to reproduce the observed anomalous trends, namely, the decrease of the soft mode frequency with pressure and its increase with temperature while both the changes involve a compression of the lattice. The former is consistent with the negative mode Grüneisen parameters associated with the known negative thermal expansion behavior. The increase of the frequency with temperature essentially results from the positive quartic anharmonic part of the phonon potential, which is satisfactorily brought out by the calculations.

**10. Scandium Trifluoride, $ScF_3$**

Large isotropic negative volume thermal expansion ($\sim -40 \times 10^{-6}$ K$^{-1}$) has been found in Cubic scandium trifluoride ($ScF_3$) [208]. Earlier such a large NTE over a large range of temperature (10 K-1100 K) was observed in $ZrW_2O_8$[40]. $ScF_3$ shows NTE up to 1100 K and beyond that it exhibits positive volume thermal expansion coefficient [208]. The compound crystallizes in cubic structure (space group *Pm-3m, Z=2*) at ambient pressure. It transform to hexagonal structure (R-3c) above 1.4 GPa[209, 210]. $ReO_3$ has similar structure to $ScF_3$. However it shows NTE only below 220 K [12]. The NTE behavior in $ReO_3$ is due to the softening of M3 phonon mode. The simple structure of these compounds makes it interesting to understand the NTE phenomenon as well as the phase transition [43, 62, 211]. The understanding of the mechanism of NTE involves the identification of anharmonic phonons and their behavior on compression[43, 62, 211].

**10.1 Phonon Spectra and Thermal Expansion**

Inelastic neutron scattering experiments [43] have been reported from 7 to 750 K that show (Fig. 64) a significant stiffening of phonon modes around 25 meV. First-principles phonon calculations showed that some of the phonon modes with motions of F atoms transverse to their bond direction account for a significant part of the negative thermal expansion[43] (Fig. 65). The same phonon also showed a quartic phonon potential (Fig. 66) that account for their stiffening with temperature.

The thermal behavior of $ScF_3$ has been studied[43] by density-functional molecular dynamics simulations (Fig. 67) that reproduce the experimentally observed trends of NTE up to 1000 K and revealed positive expansion at higher temperatures. The simulations revealed that the observed phenomena arise from uncorrelated dynamics of $ScF_6$ octahedra. The cubic-to-rhombohedral transformation under small pressure is found to be related [212] to the same atomic dynamics in $ScF_3$. The bulk moduli are derived (Fig. 68) from static total-energy versus volume calculations as well as from the fluctuations of the volume in isothermal-isobaric MD simulations, which shows [43] that the symmetry-breaking deformations and thermal behavior of $ScF_3$ are mutually related.

Inelastic x-ray scattering is used to identify (Fig. 69) a zone boundary soft mode branch that is associated with an incipient structural transition in $ScF_3$, together with a resolution-limited central peak[211].

**11. Cuprous halides, CuX (X=Cl, Br, I)**

The experimental studies on thermal expansion behaviour [107-110], phonon dispersion relation[213-215]and specific heat[216, 217] of cuprous halides, CuX (X=Cl, Br, I) have been reported. The temperature and pressure dependence of Raman spectrum of CuCl and CuI have been reported[218, 219]. Another work on pressure and temperature dependent elastic, dielectric, and piezoelectric constants of CuCl indicatesthat NTE in CuCl at low temperature is due to the softening of shear acoustic modes with pressure[220].

A lattice dynamical shell model[221] is found to reproduce the experimental dispersion relation of CuCl. So also the full-potential linear-muffin-tin-orbital method (LMTO-FP) calculations[222] of the zone centre phonons and elastic constants of copper halides are found to be in close agreement with the measurements. Another calculation[33, 223]based on the density functional perturbation theory found pressure-induced soft transverse-acoustic-phonon modes that may initiate phase transition at high pressures.

These compounds are intermediate between ionic and covalent bonded materials[224]. In cuprous halides, the filled and interacting d-orbital's provide surplus number of electrons while bonding. Cuprous halides form zinc blend type fcc crystal lattice shown in Fig. 70 (space group: F-43m). The size of anion is around twice of that of cation in cuprous halides and the cation is positioned at the centre of tetrahedral voids formed by anions. The phonon spectrum in these halides has been investigated by inelastic neutron scattering and theoretically by DFT-based calculations. These studies enable to bring out the specific features of the anharmonic atomic vibrations responsible for the anomalous thermal expansion. The specific features of the vibrations are explained in terms of the phonon eigen-vectors in Section 11.2.

**11.1 Phonon Spectra**

The measured inelastic neutron scattering spectra for CuX (X= Cl, Br and I) at 373 K, 473 K and 573 K is shown in Fig 71. The lowest energy peak in all the three halides is at about 4-5 meV. This peak may be due to transverse acoustic modes. The highest energy peak which corresponds to the optic modes of Cu-X, occur at different energies of at about 27 meV, 19 meV and 16 meV for CuCl, CuBr and CuI respectively. The peak around 4-5 meV shows significant temperature dependence. In particular, in case of CuCl, this peak reduces in intensity and broadens, which indicates significant anharmonic behaviour. The effect is less pronounced in case of CuI.

The neutron spectra of the CuX have been interpreted by calculating the phonon spectra as well as partial contribution of various atoms to the total phonon spectra. The partial contributions to the various atoms have been estimated by atomic projections of the one-phonon eigenvectors. The calculated partial density of states is shown in Fig 72. The contributions due to both the Cu and X atoms extend over whole spectral range. In case of CuCl, the Cl (35.45 amu) has smaller mass in comparison to the Cu atom (63.55 amu), while in CuBr, the mass of Br (79.90 amu) is comparable to that of Cu, and I (126.90 amu) atoms are relatively heavier in comparison to Cu in CuI. It is observed (Fig. 72) that in all three compounds the lowest peak in the phonon spectra at about 4-5 meV has contribution from both the Cu and X (=Cl, Br and I) atoms..

The calculated partial density of states shows that the contribution from the Cu atom is very much different in various halides. At low energy below 10 meV, the vibrational contribution of Cu atoms to the total density of states in CuCl is largest while it is least in CuI. The energy range of the Cu-X optic modes in these compounds simply follows the reduce mass and bond length considerations. The smallest Cu-X mass and the bond length in CuCl results in the spectral range of the optic modes to extend up to 30 meV while the range of optic modes in CuBr and CuI is up to 20 meV and 18 meV respectively.

The comparison between measured (373 K) and calculated neutron-weighted phonon spectra is shown in Fig 73. The experimental phonon spectrum in CuCl shows peaks centered at 4, 8, 12 and 28 meV. However, the calculated one-phonon spectrum shows three peak structures. The additional peak in the measured spectrum at 8 meV arises due to multiphonon contribution. The computed specific heat using the calculated one phonon spectra in all three compounds shows excellent agreement (Fig. 74) with experimental data[216, 217], which further validates the calculations.

Fig. 75provides a comparison between the ab-initio calculated and experimental dispersion relation[216, 217] of all the three CuX compounds. It can be seen that at low energies up to 15 meV there is an excellent agreement between calculations and measurement. In both the Cl and I compound the calculated longitudinal optic branch near the zone-centre is found to be underestimated, while the agreement for the same branch is found to be good in CuBr.

**11.2 Thermodynamic Behavior**

The energy dependence of Grüneisen parameter, $\Gamma$, is calculated from the average volume dependence of phonon spectra in the entire Brillouin zone. The calculated dispersion relation at ambient and 0.5 GPa pressure along various high symmetry directions in the Brillouin zone for all three compounds is shown in Fig.76. The computed phonon dispersion relation at 0.5 GPa shows (Fig. 76) that transverse acoustic modes show large softening in CuCl and least in CuI. The optic modes in all the compounds are found to harden with pressure.

The maximum negative value (Fig.77) of $\Gamma$ is -3.5, -1.7 and -0.90 for CuCl, CuBr and CuI respectively. It can be seen that low energy modes of energy 4-5 meV have maximum negative value of $\Gamma$. The calculated phonon dispersion relation shows that these modes are mainly transverse acoustic (TA) modes. The phonons of energy up to about 6-7 meV show negative $\Gamma$. The volume thermal expansion coefficient, $\alpha_v$, is calculated for all three compounds and shown in Fig.78. The maximum negative value of $\alpha_v$ in CuCl is $-12.5 \times 10^{-6}$ K$^{-1}$ at 34 K, while in CuBr the value of maximum $\alpha_v$ is $-4.5 \times 10^{-6}$ K$^{-1}$ at 24 K. CuI showed lowest negative values of $-1.0 \times 10^{-6}$ K$^{-1}$ at 10 K.

The calculated mode $\Gamma$ in all three compounds shows that the transverse acoustic modes have negative $\Gamma$. However, for the optic modes $\Gamma$ is positive. The magnitude of negative $\Gamma$ is largest in CuCl and least in CuI. The competition between the contribution from the low energy and high energy modes and magnitude of their $\Gamma$ values would result in net expansion behaviour. In case of CuCl the $\Gamma$ has large negative values and the contribution from the high energy optic modes with positive $\Gamma$(~ 4 to 5) would be at high temperature. This leads to negative thermal expansion behaviour below 100 K. In other two halides, the low energy modes have less negative $\Gamma$ in comparison to CuCl, whereas the high energy optic modes also start contributing at lower temperatures. The overall NTE coefficient in CuBr and CuI has small negative values below 20 K and 10 K respectively.

The calculated fractional change in volume of the CuX is compared with the available experimental data[225] in Fig 79.The agreement between the calculations and experiments is found to be excellent at low temperature, however at higher temperature there is small deviation between the computed and experimental data especially in case of CuCl. This might be due to the explicit anharmonic contributions, which are not included in the calculations. The calculated contribution of phonons of energy E to the volume thermal expansion coefficient at 300 K is shown in Fig 80. The maximum contribution to NTE behavior is from phonons of energy 4-5 meV. The other high energy phonons contribute to positive thermal expansion behavior of the compound. The energy range of the 4-5 meV peak matches very well with the energy of zone boundary TA modes at X and L-points in CuX compounds. The eigenvector of the TA phonons at X-point and L-point are plotted in Fig. 81. These mode involve transverse vibrations of the two sub-lattices of Cu and X-atoms.

The nature of bonds is an important factor to understand the thermal expansion behaviour. The results on $M_2O$[2] show that open structure and ionic nature of bonding results in large variation in thermal expansion behavior of the compounds. The Born effective charges have been calculated (Table 5) to understand the nature of bonding. CuBr has somewhat larger magnitude of charge (1.22) that CuCl (1.12) and CuI (1.12). Overall, the Born effective charges seem to be nearly same in all the three compounds, which indicates similar nature of bonding. Further the dielectric constant values (Table 5) in all the three compounds are also found to be nearly same.

**11.3 Negative Thermal Expansion in CuBr and CuI at High Pressures**

The calculated cubic cell length in CuCl is 5.414 Å while that in CuBr and CuI is 5.685 Å and 6.044 Å respectively. The Cu-X bond length in these compounds follow Cu-Cl (2.34 Å) < Cu-Br (2.46 Å) < Cu-I (2.62 Å). The difference in bond lengths at ambient pressure may be responsible for the difference in thermal expansion behavior (Fig. 78) among CuX compounds. While at ambient condition CuBr and CuI show positive thermal expansion behavior, at lower volume (CuBr, a=5.58 Å and P=4.2 GPa; CuI, a=5.76 Å and P=9.8 GPa) the compounds exhibit negative thermal expansion behavior. The calculated dispersion relation at lower volume in CuBr and CuI (Fig 76) shows that the transverse acoustic branch softens significantly, while other branches shift towards high energy. Mainly the transverse acoustic modes in CuBr and CuI shift to lower energies (Fig. 76) in comparison to that at ambient pressure. The large softening of transverse phonon modes in turn may lead to negative thermal expansion in CuBr and CuI at lower volume. The calculations show that at lower volume the low energy phonons below 8 meV have large negative $\Gamma$ values (Fig. 77) comparable to CuCl and are responsible for NTE behavior (Fig. 78). These analyses suggest that the magnitude of negative thermal expansion in these compounds behavior is largely dependent on volume.

## 12. Metal cyanides: Zn(CN)₂ and Ni(CN)₂

Besides oxide-based materials[40, 101], anomalous thermal expansion behavior has been observed in molecular framework materials containing linear diatomic bridges such as the cyanide anions [102, 103, 111, 226]. In another section, we discuss $Ag_3Co(CN)_6$ and $Ag_3Fe(CN)_6$, which exhibit exceptionally large ("colossal") positive thermal expansion (PTE) along the *a* direction ($\alpha_a = +140 \times 10^{-6}$ K$^{-1}$) and negative thermal expansion (NTE) along the *c* direction ($\alpha_c = -125 \times 10^{-6}$ K$^{-1}$). These thermal expansion coefficients are an order of magnitude larger than those observed in any other material. Even simple cyanides such as $Zn(CN)_2$ are reported[102] to have an isotropic NTE coefficient ($\alpha_V = -51 \times 10^{-6}$ K$^{-1}$), which is twice as large as that of $ZrW_2O_8$[40]. However, when Zn is substituted by Ni, a layered compound, $Ni(CN)_2$, is produced [226] which has NTE in two dimensions ($\alpha_a = -6.5 \times 10^{-6}$ K$^{-1}$) combined with a very large positive (PTE) coefficient ($\alpha_c = 62 \times 10^{-6}$ K$^{-1}$) in the third dimension perpendicular to the layers, to yield a large overall volume thermal expansion ($\alpha_V = 49 \times 10^{-6}$ K$^{-1}$).

It has been suggested from pair distribution function (PDF) analysis of the X-ray diffraction data that NTE in $Zn(CN)_2$ is induced by an average increase of the transverse thermal amplitude of the motion of bridging C/N atoms, away from the body diagonal [227]. Further, investigation using *ab-initio* calculations [27] of the geometry and electronic structure of $Zn(CN)_2$ shows that the naturally stiff C≡N bond is paired with weak Zn–C/N bonds. This type of bonding allows large transverse thermally excited motions of the bridging C/N atoms to occur in (M-CN-M) bridges within metal-cyanide frameworks. Structural studies[228] of $Zn(CN)_2$ show that two different models having cubic symmetry with space group *Pn3m* (disordered model) and *P43m* (ordered model), give equally good account of the diffraction data. The ordered structure (Fig. 82) consists of a $ZnC_4$ tetrahedron (at the centre of the cell) linked to four neighboring $ZnN_4$ tetrahedra (at the corners of the cell) with CN groups along four of the body-diagonals.

$Ni(CN)_2$ is fundamentally different from $Zn(CN)_2$, in that it forms a layered structure with average tetragonal symmetry. Nickel cyanide has a long-range ordered structure in two dimensions (*a-b* plane) (Fig. 83) but a high degree of stacking disorder in the third dimension. The relationship between neighbouring layers is defined, but there is a random element to the relationship between next nearest neighbours. A crystallographic model in *P4₂/mmc*[226] reproduces the structure well and the disorder in the stacking is dealt comprehensively by Goodwin *et al.*[111]. The grid like layers (*a-b* plane) (Fig. 83) consists of $NiC_4$ square-planar units (as shown at the centre of the figure), which are linked by vertex sharing to four neighboring $NiN_4$ square-planar units (shown at the corners of the figure) with the resultant

CN groups along the diagonals in the *a-b* plane. The covalent bonding within the layers is much stronger than the van der Waals' bonding between the layers.

Raman spectroscopy at high pressure [227] has been used to experimentally determine Grüneisen parameters values of $Zn(CN)_2$ for modes above 200 cm$^{-1}$ (25 meV). The low-energy part of the phonon spectra has been measured[29] by neutron scattering that indicate the existence of dispersion less modes at about 2 meV(~16 cm$^{-1}$). The anharmonicity of phonons in $Zn(CN)_2$[14] is investigated by high-pressure inelastic neutron scattering. The temperature dependence of the spectra in $Zn(CN)_2$ and $Ni(CN)_2$ have also been measured. The analysis of the experiments is performed with the help of state-of-the-art *ab-initio* lattice dynamical calculations. In this way, one obtains a clear and detailed insight into the phonon mechanisms responsible for thermal expansion in $Zn(CN)_2$ and $Ni(CN)_2$.

For lattice dynamical calculations of $Zn(CN)_2$, the available structure having the cubic space group (*P43m* (215), $\left[T_d^1\right]$) is considered [9]. For $Ni(CN)_2$, a periodic model system is used (Table 6) to generate the layers within the tetragonal space group (*P4* (75) $\left[C_4^1\right]$) [111]. This model for $Ni(CN)_2$ is an approximation of the real situation as the interlayer spacing used is double that found in the actual material. Such a model results in no interaction between the layers. By separating the sheets, one is able to employ *P4* symmetry and achieve a great saving in computational resource. Structures with the sheets at the correct separation and alignment can only be described in *P1*. The use of *P1* symmetry would be prohibitively expensive in computing time in *ab-initio* calculations However, the *P43m* model can be used to reproduce most features of the Raman spectrum, the phonon density of states (DOS) and, in addition, can be used to investigate the in-plane negative thermal expansion.

**12.1 Phonon spectrum**

The measured spectrum [14] from $Zn(CN)_2$ as a function of temperature at various pressures is shown in Fig. 84 for energy transfer up to 30 meV. The ambient pressure measurements are in agreement with the previous measurements [29]. Ab-initio phonon calculation [228] shows that the spectral weight up to 30 meV is 46.67 %, which corresponds to 14 phonon modes out of a total of 30 modes in cubic $Zn(CN)_2$ per unit cell. The striking feature in the spectrum is the very strong low energy peak at 2 meV. There is a continuous spectrum of excitations in the measured energy transfer range, with maxima at 8 meV and 18 meV. The low energy peak at about 2 meV in the phonon spectra of $Zn(CN)_2$ (Fig. 84) appears to be from a flat transverse acoustic mode[228]. The band around 8 meV is more likely due to hybridization of acoustic and optic modes.

The measured temperature dependence of the phonon spectra for $Zn(CN)_2$ and $Ni(CN)_2$ are shown in Figs. 85 and 86, respectively. The phonon spectra for $Zn(CN)_2$ (Fig. 85) have been measured at 180, 240, 270 and 320 K. These measurements show a visible change in the phonon spectra around 50 meV with change of temperature from 240 K to 270 K that matches the temperature of the anomaly in the calorimetric measurements[229]. All the observed features for $Zn(CN)_2$ are well reproduced computationally (Fig. 85), especially the low-energy peak at ~ 2 meV. The stiffening of the phonons with temperature is indeed due to the dominant explicit anharmonicity. The measurements of temperature dependence of phonon spectra [25]along with the pressure dependence of the phonon spectra (implicit anharmonicity) have been used to separate [14]the temperature effect at constant volume (explicit anharmonicity)[14].

The experimental phonon spectrum of $Ni(CN)_2$ (Fig. 86) shows several well-pronounced vibration bands that do not show any significant temperature dependence. The calculated positions of these bands are in good agreement with the experimental data while there are slight differences in the intensities. This is probably due to the fact that the interlayer interactions in $Ni(CN)_2$ have not been included. As interlayer coupling has been neglected, the modes along the stacking axis have very low energies. All these modes are included in the calculated density of states as shown in Fig. 86. This explains the extra weight in the calculated density of states at low energies.

The comparison of the phonon spectra of $Zn(CN)_2$ and $Ni(CN)_2$ shows (Fig. 87) that there are pronounced differences. The cut-off energy for the external modes in $Zn(CN)_2$ and $Ni(CN)_2$ is at about 65 meV and 90 meV, respectively. The calculated partial density of states shows that the contributions from Zn and Ni in $Zn(CN)_2$ and $Ni(CN)_2$(Fig. 88) extend up to 60 and 75 meV, respectively. These differences can obviously not be explained by a simple mass renormalization of the modes involving Zn (65.38 amu) and Ni (58.69 amu) atoms. They thus imply that the strength and may be the character of bonding is different in both systems. The inelastic scattering data show that the first low-energy band in the Zn compound is at 2 meV, while in the Ni compound this band is at 10 meV.

The comparison between the experimental and calculated zone centre modes for $Zn(CN)_2$[80] and $Ni(CN)_2$ is given in Tables 7 and 8, respectively. The agreement is very close in each case. Ref. [111] reported five dispersion-less phonon modes below 1.0 THz (4.136 meV) arising from motions of a single

Ni(CN)$_2$ layer as obtained from Reverse Monte Carlo fitting of total neutron diffraction data. However, calculations for Ni(CN)$_2$ produces the lowest optic mode at 99 cm$^{-1}$ (~3 THz, ~12.3 meV) (Fig. 89) in agreement with the Raman data (Table 8).

The calculated phonon dispersion curves for Zn(CN)$_2$ are shown in Fig. 89. This shows a remarkable flat phonon dispersion sheet of the two lowest energy acoustic modes at about 2 meV. These flat modes give rise to the observed first peak in the density of states at about 2 meV. Further flat phonon dispersion sheets are found at relatively high energies of about 25, 30, 40 and 60 meV (Fig. 89) providing the other well isolated bands in the phonon density of states (Fig. 85). The calculated dispersion relation for Ni(CN)$_2$ (Fig. 89) is quite different as compared to Zn(CN)$_2$. As explained above, the calculations for Ni(CN)$_2$ are carried out with the layers separated, which is different from the real situation. Further, at zone boundary the acoustic modes extend up to 10 meV. The flattening of acoustic modes around 10 meV gives rise to the first peak in the density of states (Figs. 86 and 87) of Ni(CN)$_2$. The large difference in the energies of acoustic modes between the compounds indicates that bonding is quite different in both compounds. Further, the flat phonon dispersion sheets at about 18, 30, 45, 60 and 75 meV give rise to the isolated peaks in the density of states of Ni(CN)$_2$. The Bose factor corrected S(Q,E) plots for Zn(CN)$_2$ and Ni(CN)$_2$ at 180 K and 160 K respectively are shown in Fig. 90. The figure clearly shows the presence of flat acoustic modes at 2 meV in the S(Q,E) plot of Zn(CN)$_2$, while for Ni(CN)$_2$ the acoustic modes extend up to about 10 meV.

In case of Ni(CN)$_2$ the acoustic dispersion within the sheets for the transverse branches possesses in the calculation an anomalous dispersion. The curves turn upwards instead of downwards with respect to increasing q. Naturally the anomalous dispersion could become normal by including the interplanar coupling. On the other hand, the measured density of states (Fig. 86) seems to be linear. This would be compatible with an anomalous dispersion. The fact that any soft mode has not been observed in S(Q,E) (Fig. 90), which demonstrates that the interplanar coupling is certainly not negligible. If this was the case, then soft modes along the stacking direction would be inevitable. Therefore, the contraction of the plane certainly should influence the physics along the stacking axis.

**12.2 Grüneisen Parameters and Thermal Expansion**

The $\frac{\partial \ln E_i}{\partial P}\left(=\frac{\Gamma_i}{B}\right)$ values for phonons of energy $E_i$ have been obtained at 165 K and 225 K [Fig.

91] using the cumulative distributions of the density of states. The estimated $\frac{\Gamma_i}{B}$ values at 165 K and 225 K are within the experimental error bars. The $\frac{\Gamma_i}{B}$ (Fig. 91) and experimental phonon spectra, $g^{(n)}(E)$ (Fig. 84) at 165 K have been used for the calculation of $\alpha_V$ (Fig.92(a)). The comparison of the volume thermal expansion derived from the phonon data and diffraction data [102] is shown in Fig. 92(b), which shows a good agreement between them. Thus the anharmonicities of low phonon modes are sufficient to account for the negative thermal expansion coefficient of $Zn(CN)_2$.

The ab-initio calculated $\frac{\Gamma_i}{B}$ for $Zn(CN)_2$ is shown in Figs. 93. The modes up to 15 meV show negative $\frac{\Gamma_i}{B}$, with the low-frequency modes around 2 meV for $Zn(CN)_2$ showing the largest negative $\frac{\Gamma_i}{B}$. The calculations $\frac{\Gamma_i}{B}$ for $Zn(CN)_2$ (Fig. 93) are in very good agreement with the values obtained from the high-pressure inelastic neutron scattering measurements with the *ab-initio* calculations done by Zwanziger[28]. The calculated temperature dependence of the volume thermal expansion coefficient derived compares very well with that derived from the phonon data (Fig. 92(a)). The calculated volume expansion is in good agreement with the corresponding value obtained from diffraction data [102](Fig. 92(b)). The *ab-initio* calculations by Zwanziger[28] give a thermal expansion coefficient of $-12 \times 10^{-6}$ $K^{-1}$ at 5 K, in agreement with the experimental data. However, Zwanziger[28] did not report Grüneisen parameters of modes below 3 meV, or a detailed temperature dependence of $\alpha_V$. Note that the value of $\alpha_V$ changes to $-51 \times 10^{-6}$ at 300 K.

Certain Raman active optical phonons (Fig. 94) have been identified [80]that are responsible for NTE in $Zn(CN)_2$ from high pressure Raman spectroscopic studies. The Grüneisen parameters derived from the neutron experiment cannot be compared with those derived from the Raman data[227] since these measurements report observation of phonon modes in $Zn(CN)_2$ only above 200 $cm^{-1}$ (~25 meV). First principles ab-initio phonon calculations [28]show that nearly dispersion-less transverse acoustic modes do appear in the energy range of 2-4 meV and have negative Grüneisen parameters of about -7. The experimental data show that for these modes $\Gamma_i$ values lie in between -14 and -9.

The pair distribution function (PDF) analysis[227] of high-energy X-ray scattering data indicate an increase in the average transverse vibrational amplitude of C/N bridging atoms with increasing temperature, which may provide a mechanism for the NTE in Zn(CN)$_2$. The experimental $\frac{\Gamma_i}{B}$ values (Fig. 93) at 165 K have been used for the estimation of the contribution of various phonons to the thermal expansion (Fig. 95) as a function of phonon energy at 165 K. The analysis of phonon data shows (Fig. 95) that the maximum negative contribution to $\alpha_V$ is from the low-energy transverse acoustic modes of energy of about 2.5 meV, which is consistent with the PDF analysis [227] of diffraction data.

Ni(CN)$_2$ shows two-dimensional NTE [226] in the *a-b* plane with $\alpha_a = -6.5 \times 10^{-6}$ K$^{-1}$. The large positive expansion [3] along *c* ($\alpha_c = 61.8 \times 10^{-6}$ K$^{-1}$) results in an increase in volume with temperature ($\alpha_V = 49 \times 10^{-6}$ K$^{-1}$). Unfortunately, modeling produces no quantitative information on the third-dimension due to the effectively isolated sheets in modeling. The two-dimensional model gives negative $\frac{\Gamma_i}{B}$ (Fig. 96(a)) for Ni(CN)$_2$. The calculated (Fig. 96(b)) linear $\alpha_L = -5.5 \times 10^{-6}$ K$^{-1}$ in the 100-300 K compares excellently with the $\alpha_a$ value of $-6.5 \times 10^{-6}$ K$^{-1}$ from diffraction experiments [102]. A qualitative explanation of the overall positive expansion of this system is that as these layers contract in the *a-b* plane they expand into the third dimension pushing the layers apart as suggested in Hibble *et al.* [226]. The weak interactions between the layers mean that expansion in this direction is easy and explains that the overall PTE in this system.

The contribution of the various phonons to the thermal expansion as a function of phonon energy in Zn(CN)$_2$ at 165 K is shown in Fig. 95. The maximum negative contribution to $\alpha_V$ stems from the low-energy modes around 2.5 meV. Similarly, for Ni(CN)$_2$, (Fig. 95) maximum contribution to NTE is also found from phonon modes of around energy 2.5 meV. This maximum strength appears well below the first maximum in the density of states (Fig. 86), which is found at 4 meV. The low-energy modes of about 8 meV and 13 meV also contribute significantly (Fig. 95) to the NTE in Zn(CN)$_2$ and Ni(CN)$_2$ respectively.

Another density functional theory calculation of the phonon spectrum(Fig. 97) and Grüneisen parameters (Fig. 98) have been reported for Zn(CN)$_2$ and Cd(CN)$_2$. The calculations are in broad agreement with neutron scattering experiments. They confirm that the modes around 2-4 meV are the primary contributors to the negative thermal expansion in these materials. These transverse acoustic modes arise from the translation of the zinc and cyanide anions together in one direction coupled with cyanide translation in the other two directions. The energies and anharmonicities of these modes are sufficient to account for the negative thermal expansion of this material, particularly at low temperature.

The eigenvectors corresponding to low energy zone centre and zone boundary modes are shown in Figs. 99 and 100. The mode assignments, phonon energies and Grüneisen parameters are given in the figures. Examination of the calculated eigenvectors for the lowest energy optic mode (symmetry $T_1$) at 7.3 meV in $Zn(CN)_2$ (Table 7) shows that this mode arises from librational motions of $ZnC_4$ and $ZnN_4$ tetrahedra involving transverse motions of C and N atoms. The $T_1$ mode has a Γ value of -9.8. As shown above, modes with energies of about 8 meV contribute significantly to NTE in in $Zn(CN)_2$ (Fig. 14). However, Fig. 14 shows that modes of energy about 2 meV make the most significant contributions to NTE in $Zn(CN)_2$. The zone boundary modes of about 2 meV along (100), (110) and (111) are plotted in Fig. 16. At the zone boundary, eigenvectors have both real and imaginary components. The actual atomic motion is a combination of both these components. The real part of the 1.94 meV mode along (100) shows librational motion (Fig. 99). All other modes show translational and bending motions. The calculations show that zone boundary modes along [100] and [110] have very large negative Γ values of -48.4 and -85.8 respectively, while the mode at the zone boundary along [111] has a very low negative Γ value of -0.7.

$Ni(CN)_2$ has three low energy optic modes at about 12.32 meV (*E* mode), 12.35 meV (*A* mode) and 12.77 meV (*A* mode) (Table 8 and Fig. 89). The *A* type mode of energy 12.77 meV only shows librational motion (Fig. 17). The other two zone centre modes (12.32 meV and 12.35 meV) as well as the zone boundary modes (4.26 to 7.3 meV) involve (Fig. 100) translations of atoms along the *c*-axis. The negative Γ values of all the three optic modes lie between -4.5 and -6.0, while zone boundary modes along [100] have large negative Γ values (~-13.0) as compared to those along [110] (~-5.3). The modes at about 2 meV make the most significant contribution to NTE. As shown in Fig. 89, the 2 meV modes are largely acoustic in nature; which include the transverse motions to produce NTE in agreement with Chapman *et al*. [29].

**13. Colossal Thermal Expansion Compounds $Ag_3M(CN)_6$ (M=Co, Fe) and $KMnAg_3(CN)_6$**

Structural flexibility plays a key role in the physical properties of framework materials. NTE has been found[40, 101] in framework oxides (-27 × $10^{-6}$ $K^{-1}$ for $ZrW_2O_8$ at 300 K) with M-O-M (M = metal) linkages. The NTE coefficients of compounds with M–CN–M linkages between two neighboring tetrahedral units, instead of a single atom, was found to reach a much higher value of -51× $10^{-6}$ $K^{-1}$ for $Zn(CN)_2$[102] at 300 K. This appears to be due to some additional flexibility in the structure with the M-CN-M linkage, which may be understood as follows. The strong C-N triple bond acts as a stick. When two polyhedra are connected by a single shared atom, their rotational motions are correlated, and therefore restricted. However, when the polyhedra are connected to the two ends of a stick, their rotations are more flexible. This may explain why NTE appears to be inherently more common amongst cyanides than amongst oxide-containing frameworks, and also why the magnitude of the NTE coefficients can be much larger.

The concept of increasing framework flexibility has led to the discovery[103, 104] of colossal positive and negative thermal expansion in $Ag_3Co(CN)_6$ and $Ag_3Fe(CN)_6$. The crystal lattice expands remarkably along the a and b crystal axes ($\alpha_a \sim 120 \times 10^{-6}$ K$^{-1}$), while it contracts equally strongly ($\alpha_c \sim -110 \times 10^{-6}$ K$^{-1}$) along the c axis. The structure thus expands and contracts at rates more than ten times larger than what is found in normal materials. The covalent framework structure of this material consists of Co–CN–Ag–CN–Co linkages. It features alternating layers of octahedral $[Co(CN)6]^{3-}$ and $Ag^+$ cations (Fig. 101). At ambient pressure $Ag_3Co(CN)_6$ is reported to decompose [103] at 500 K. High pressure diffraction measurements[230] show a transition from a trigonal to a monoclinic structure at very low pressures of 0.19 GPa at 300 K. The high-pressure phase is known to be 16% more dense as compared to the ambient pressure phase.

The anomalous thermal expansion is related to anomalous compressibility. The large colossal negative thermal expansion [103]and negative linear compressibility (NLC)[230]are both found along the trigonal (c-axis)lattice of $Ag_3[Co(CN)_6]$. The measurements of the temperature-dependence of unit cell parameters (Fig. 1) for $KMnAg_3(CN)_6$ indicate [75, 231] that thermal expansion behavior is strongly anisotropic ($\alpha_a \sim 61 \times 10^{-6}$ K$^{-1}$, $\alpha_c \sim -60 \times 10^{-6}$ K$^{-1}$) along the trigonal cell axes. The compound also exhibits[75, 231] very strong NLC effect. $KNiAu_3(CN)_6$ alsoexhibits [231]strong anisotropic ($\alpha_a \sim 60 \times 10^{-6}$ K$^{-1}$, $\alpha_c \sim -42 \times 10^{-6}$ K$^{-1}$) thermal expansion behavior along the trigonal cell axes. The compound does not show NLC effect. The comparisonof two compounds shows that there is large difference in the thermal expansion behavior along the c-axes.

### 13.1. Phonon Spectra of $Ag_3Co(CN)_6$ and $Ag_3Fe(CN)_6$

The phonon spectra as measured by inelastic neutron scattering from $Ag_3Co(CN)_6$ and $Ag_3Fe(CN)_6$ are shown in Figs. 102 and 103 at several temperatures over 160 -300 K. The spectra show several vibrational bands centered on 3, 13, 17, 22, 45, 55 and 70 meV. The ab-initio calculations (Figs.

2 and 3) reproduce very well the positions of the bands while there are slight differences with respect to the intensities.

The temperature dependence of the overall spectra is rather weak for the interval of 160-300 K. For thermal expansion the low-frequency part, as shown in Fig. 104 for 160 K and 300 K, is particularly interesting. The low-energy modes are found to harden anomalously with increasing temperature. Below 5 meV, the total anharmonicity, sum of implicit and explicit contributions, $\left( \frac{1}{E_i} \frac{dE_i}{dT} \bigg|_P \right)$, of the low-energy modes is thus found (Fig. 104) to be positive. The implicit anharmonicity, i.e. the volume dependence of the phonon spectra, should result in a decrease of phonon frequencies for compounds with colossal thermal expansion behavior. The hardening of modes with increase of temperature gives us, therefore, evidence for the large explicit anharmonicity, i.e. changes in phonon frequencies due to large thermal amplitude of atoms, reflecting the nature of phonons in these compounds.

The phonon spectra of both the compounds are very similar (Fig. 105) which is not surprising in view of the similarity in masses and nature of bonding. In particular, the cut-off energy for their external modes is at about 80 meV. The main difference resides in a more smeared out character for the modes in the Fe compound above 40 meV. The calculation of partial density of states (Fig. 105) indicates that peaks in the phonon spectra above 40 meV (Fig. 105) are from C and N atoms. The smeared-out character for the modes above 40 meV in the Fe compound might be due to likely presence of disorder in the Fe compound. However the structural studies [104] on $Ag_3[Fe(CN)_6]$ have not reported any disorder. The structural studies [28] on $Zn(CN)_2$ show disorder in C and N atomic positions. The calculated partial density of states (Fig. 106) shows that Ag atoms mainly contribute in the low-energy modes up to 10 meV. The vibrations due to C and N span the entire energy spectrum up to 280 meV. Contributions from Co (58.93 amu) and Fe (55.84 amu) are predicted to be in the same energy range of up to 75 meV.

The temperature-dependence of the Bose factor corrected scattering function S(Q,E) is plotted in Figure 107. The figure clearly shows the presence of dispersion surfaces at 2.5 and 4.5 meV. These features produce the two-peak structure in the spectra (Figure 104). The temperature dependence of S(Q,E) plots also indicate that energies of the dispersion surfaces do not change significantly as a function of temperature.

The calculated phonon dispersion curves for $Ag_3Co(CN)_6$ and $Ag_3Fe(CN)_6$ are shown in Fig. 108. The flat phonon dispersion sheet of the two lowest energy acoustic modes near the zone boundary at about 3 meV give rise to the observed first peak in the density of states. The second low-frequency peak at about 4.5 meV corresponds to flat optic modes. It may be noticed that flat phonon dispersion sheets in the entire Brillouin zone at relatively high energies of about 40, 52, 58 and 75 meV (Fig. 108) providing the other well isolated bands in the phonon density of states (Fig. 103). The dispersion relations are found to be quite similar in both the compounds.

$Ag_3Co(CN)_6$ is known to undergo a structural phase transition [230] at very low pressure of 1.9 kbar at 300 K. The phonon spectra (Fig. 109) of $Ag_3Co(CN)_6$ have been measured at ambient pressure, 0.3, 1.9 and 2.8 kbar, at 200 K. The large absorption from the Ag and Co atoms as well as from the high-pressure cell has enabled to measure phonon spectra only up to 10 meV. At 1.9 kbar the low-energy part of the phonon spectra up to 5 meV is most significantly affected by the application of pressure. The height of the peaks in the phonon spectra is found to be reduced and the modes around 2.5 and 4.5 meV shift to higher energies. As discussed above the low-energy part of the phonon spectra features mainly contributions from Ag atoms and these modes are strongly anharmonic that contribute to the colossal thermal expansion discussed below. The measurements provide indirect evidence that atomic motions in the Ag sub-lattice may be associated with the observed phase transition in $Ag_3Co(CN)_6$ at very low pressure. This is in agreement with the diffraction measurements [230] where a weak argentophilic Ag–Ag interaction is found to drive the trigonal to monoclinic phase transition.

The temperature dependence of the average value of phonon energies of Ag, Co, and CN species in $Ag_3Co(CN)_6$ have been obtained in Ref [103] from the reciprocal-space analysis of diffraction data. The analysis of diffraction data [103] indicates (Fig. 110) a decrease of the average phonon energies of Ag with increasing temperature, whereas those for Co and CN sub-lattice are predicted to change little. However, the phonon calculations show that the average energies of all the atoms remain nearly constant. The extraction of the phonon spectra by reciprocal-space analysis of diffraction data is limited by the quality of diffraction data. The method has been shown [31] to work in MgO. However careful comparison of the inelastic scattering (INS) derived phonon dispersion relation and reciprocal-space analysis derived diffraction data indicate significant differences. The most noticeable of these differences is the absence of large LO- TO splitting of the zone centre phonon modes. The range of phonon spectra was also significantly different as compared to the data obtained from INS experiments. This demonstrates that in the case of very complex systems special care has to be taken when extracting information on phonon spectra in an indirect way from structural (diffraction) data.

## 13.2 Thermal Expansion Behaviour of $Ag_3Co(CN)_6$ and $Ag_3Fe(CN)_6$

The calculated Grüneisen parameter, $\Gamma(E)$ as shown in Fig. 111 is averaged over all phonons of energy E in the Brillouin zone[24].The low-energy modes around 2 meV have very large positive $\Gamma(E)$. The ab-inito calculations seem to underestimate thermal expansion behavior (Fig. 112) at low temperatures in the present calculations. While the frequencies of the low-energy phonons are well predicted by the calculations, the magnitude of the anharmonicity, which is a higher order effect, is not equally well predicted. However, qualitatively the calculations (Fig. 112) are useful for understanding [24]the colossal thermal expansion behavior of $Ag_3M(CN)_6$ (M=Co, Fe).

The calculated contribution of phonons of energy $E$ to the volume thermal expansion coefficient ($\alpha_V$) as a function of energy at 300 K (Fig. 113) shows that the maximum contribution to $\alpha_V$ is from modes between 2 and 5 meV. The calculated partial density of states identifies these modes as mainly involving Ag atom vibrations. It is the anharmonicity of Ag vibrations that drives the thermal expansion of these flexible lattices. The difference in the phonon spectra (Fig. 106) of Fe and Co compounds above 25 meV would not much affect the thermal expansion behavior of the two compounds.

The crystal structure of $Ag_3Co(CN)_6$ consists of Co-CN-Ag-NC-Co linkage. The trigonal unit cell has $CoC_6$ octahedral units. The C atoms are covalently connected to the N atoms. Further the two alternating layers of $[Co(CN)_6]^{3-}$ ions are connected by Ag atoms. In order to understand the nature of phonons responsible for anomalous thermal expansion mean squared displacements (Fig. 114) of various atoms, $<(u^2)>$, has been calculated arising from all phonons of energy E in the Brillouin zone. The calculated partial density of states (Fig. 106) has been used for this calculation. Equal amplitude for all the atoms up to about 1.5 meV identifies the acoustic region. The larger amplitude of C atoms as compared to Co/Fe atoms in the 2 to 5 meV energy indicates librational motion of $CoC_6$ octahedra. The large amplitude of Ag and N atoms indicates their translational motion.

As mentioned above the maximum contribution to thermal expansion coefficient is from modes between 2 and 5 meV. Figure 115 shows the mode assignments of the eigenvectors of the low energy zone centre modes along with the related energies and Grüneisen parameters, $\Gamma(E)$. The calculated $\Gamma(E)$ of modes in Fe compound are higher comparing to those in the Co compound, which is in agreement with the results shown in Fig. 111. The lowest energy $A_u$ mode at 5.0 (4.4) meV in Co (Fe) compounds have very large $\Gamma(E)$ of about 44. It can be seen that for all the three modes Ag and N atoms have very large displacements as compared to the Co and C atoms. All the three modes ($B_u$ and $A_u$) in both the compounds indicate the presence of mainly translational and stretching motions.

It has been argued [232] that the origin of the colossal values of the coefficients of thermal expansion arise from an extremely shallow energy surface that allows a flexing of the structure with small energy cost. This leads to a small value of the bulk modulus and consequently a large thermal expansion with almost normal value of the Grüneisen parameter. This work has pointed to the important role of Ag. . .Ag metallophilic interactions. $Au_3[Co(CN)_6]$ is predicted [232] to show similar behaviour to $Ag_3[Co(CN)_6]$.

Ref[48] has reported ab-initio calculation of the zone centre phonons and high pressure infrared spectroscopy experiments (Fig. 116). The pressure dependence of intense modes is shown in Fig. 117. The phonon modes below 100 cm$^{-1}$(~12 meV) are found to give the main contributions to NTE. They have been assigned to translational motion of the Co−CN−Ag−NC−Co linkages. The calculated values of the linear thermal expansion coefficients are found consistent with the experimental values.

The thermal expansion behavior for isostructural $D_3[Co(CN)_6]$ has been reported [104]. While there is colossal thermal expansion behavior in $Ag_3M(CN)_6$ (M=Co, Fe), the thermal expansion coefficient in $D_3[Co(CN)_6]$ is an order of magnitude less than that in $Ag_3M(CN)_6$. It is clear that the chemical variation at the octahedral transition metal (M=Co, Fe) site does not change the thermal expansion behavior, while the substitution of Ag by D has a large impact. The large mass difference of Ag (107.8 amu) and D (2 amu) atoms would shift the modes of D atoms to higher energies in $D_3[Co(CN)_6]$ as compared to modes of Ag in $Ag_3M(CN)_6$, which in turn lowers the entropy and thermal expansion coefficient[104].

## 13.3 Phonon Spectra and Thermal Expansion behavior of KMnAg$_3$(CN)$_6$

A study of the anisotropic compressibility and thermal expansion properties of KMnAg$_3$(CN)$_6$ using Raman spectroscopy and DFT calculations has been reported[44]. Pressure and temperature dependence of phonon modes (Fig. 117) have been measured[44]. The calculations corroborate the reported measured values. Linear thermal expansion coefficients ($\alpha_a = 56 \times 10^{-6}$ K$^{-1}$ and $\alpha_c = -47 \times 10^{-6}$ K$^{-1}$) obtained from calculated directional Gruneisen parameters ($\Gamma_a = 0.663$ and $\Gamma_c = -0.288$) are in good agreement with reported measured values ($\alpha_a = 61 \times 10^{-6}$ K$^{-1}$ and $\alpha_c = -60 \times 10^{-6}$ K$^{-1}$). Strong anisotropy is found in the compressibility, elasticity and the Gruneisen parameters[44]. These properties show that anomalous thermal expansion in KMnAg$_3$(CN)$_6$ are driven by elastic and Grüneisen anisotropies combined with anharmonic lattice vibrations of Ag atoms[44].

**14. MCN (M=Ag, Au and Cu)**

The thermal expansion behavior in low dimension MCN (M=Cu, Ag and Au) compounds shows anisotropic thermal expansion behavior. The structure of these cyanides is chain like and resembles a quasi one dimension structure. These chains consist of C≡N units connected via metal ions (M-C≡N-M). The structure seems to be simple, however the compounds shows C/N disorder along the chain in terms of random flipping of C/N sequence. This is not unusual as the C/N disorder is observed in many cyanides.

The CuCN crystallizes in two different structures named as the low-temperature and high-temperature phase that depends on the method of synthesis. The low-temperature phase (LT-CuCN) is a modulated structure of the high-temperature phase (HT-CuCN).The AgCN and HT-CuCN crystallize in the hexagonal R3m structure. However, AuCN crystallizes in P6mm structure. All the three compounds have three atoms in their primitive unit cell. The crystal structures as shown in Fig. 119 indicates that in AuCN all chains (M-C≡N-M) are parallel along the c-axis while in HT-CuCN and AgCN the adjacent parallel chains are shifted by an amount of c/3 along c-axis.

Earlier reverse Monte Carlo simulations of the diffraction data have been performed to understand the local structure of these cyanides [79]. The buckling in the M-C≡N-M chains is found [79] to increase with temperature. The magnitude of buckling is governed by nature of bonding between metal ions and C≡N unit. The analysis suggests that HT-CuCN have large distortion perpendicular to the chain direction. Similar behavior is also observed in AgCN and AuCN, however, the magnitude of such distortion is very small in AuCN. The magnitude of distortion in all three compounds increases with temperature and found to be correlated[79] with the thermal expansion coefficient along the chain direction ($\alpha_c$). The thermal expansion coefficient is positive in the a-b plane; however, large NTE is found along the chain.

**14.1 Phonon Spectra and Elastic Constants**

The phonon calculations for HT-CuCN, AgCN and AuCN are performed considering the periodic lattice model using the experimental structure parameters. The model is an approximation where the C/N disorder is neglected. The comparison between the experimental and calculated phonon spectra (with lattice parameter T= 10 K) is shown in Fig. 120. The calculated spectra are able to reproduce all the major features of the observed spectra. The structural disorder could lead to a variation of the M-C, M-N and C-N bond lengths, which would in turn broadens the peaks as observed in the experimental spectra. This might be one of the reasons for difference in the calculated and experimental spectra of MCN.

The calculated elastic constants for all three compounds are given in Table 9. The elastic constants $C_{11}$ and $C_{33}$ show large difference in all the compounds. This indicates large difference in the nature of bonding in a-b plane and along c-axis. This is in agreement with the analysis of experimental diffraction data which also shows strong one-dimensional nature of these compounds. The values of $C_{33}$ for HT-CuCN, AgCN and AuCN are 536 GPa, 387 GPa and 755 GPa respectively. Large value of $C_{33}$ in AuCN (~755 GPa) in comparison to the other two compounds indicates that bonding between the atoms of -Au-CN-Au- chains is much stronger in comparison to that in Ag and Cu compounds.

The $C_{66}$ elastic constant in all the three compounds is very small (Table 9). All these suggest that CuCN and AgCN are close to instability against shear strain. However, the AuCN shows significant stability against the shear strain. On increasing temperature, the magnitude of strain arising due to the vibrational amplitude of atoms perpendicular to chain will depend on the bond strength of -M-CN-M-. The calculated elastic constants indicate that the nature of bonding in AuCN is strongest among all the three cyanides. This is consistence with the reverse Monte Carlo analysis of the diffraction data, which indicates that AuCN does not show any shear distortion even up to 450 K, however significant distortion is observed in HT-CuCN and AgCN.

## 14.2 Thermal Expansion Behavior

The lattice parameter as a function of temperature has been reported from neutron diffraction measurements at temperature ranging from 90 K to 450K[79]. The measurements show that the *c* lattice parameter decreases with increase in temperature, however, lattice parameter *a* (*=b*) shows positive expansion behaviour. The overall volume thermal expansion is found to be positive in all three cyanides and has similar magnitude. The negative thermal expansion behavior along c axis is largest in CuCN and least in AuCN. So also, the positive expansion along a- and b-axis is largest in CuCN and least in AuCN. It can be seen in Fig. 121 that the calculated $\Gamma_a$ values for AuCN are significantly different than those calculated for HT-CuCN and AgCN. Among all the modes, the number of unstable modes in HT-CuCN and AgCN and AuCN are less than 1%.

Fig. 122 shows the comparison between the calculated and experimental anisotropic thermal expansion behavior.  It can be seen that the calculated negative thermal expansion behavior along the c-axis in all the three compounds is in good agreement with the available experimental data. As mentioned above, the ab-initio calculations performed with the ordered structures exhibit the highest number of unstable modes for HT-CuCN, while AuCN show the least number of unstable modes. The calculated Born effective charges show that AuCN has covalent nature of bonding, which may results in the least transverse distortion as well as the least number of unstable modes and also the coefficient of NTE along the c-axis in AuCN to be smallest[79] among the three compounds.

The calculated positive thermal expansion along the a-axis is underestimated as compared to experiments in HT-CuCN and AgCN, while it is in good agreement with the observations in AuCN (Fig. 122).  The underestimation in the calculated positive thermal expansion behavior along the a-axis is responsible for considerable discrepancies in the volume thermal expansion behavior. The underestimation in the calculations is the maximum in HT-CuCN and least in AuCN. The underestimate might be related to the magnitude of transverse distortion in the M-CN-M chains in these compounds. The total neutron diffraction data have been analyzed using the reverse Monte Carlo technique[79] which shows that the distortion in the M-CN-M chain is least in AuCN. This distortion is found be disordered.

The experimental[79, 233] value of the coefficient of negative thermal expansion (NTE) along the chain direction ($\alpha_c$) for HT-CuCN, AgCN and AuCN is -27.9×10$^{-6}$ K$^{-1}$, -14.8×10$^{-6}$ K$^{-1}$ and -6.9×10$^{-6}$ K$^{-1}$ respectively, while positive thermal expansion (PTE) in the a-b plane ($\alpha_a$) is  74.8×10$^{-6}$ K$^{-1}$,  65.7×10$^{-6}$ K$^{-1}$ and 57.4 ×10$^{-6}$ K$^{-1}$ respectively. As noted above, among the three compounds HT-CuCN  has the highest distortion in the M-CN-M chains and it has also the highest positive as well as negative thermal expansion coefficients.  AuCN has the least M-CN-M distortion and has the smallest values of NTE and PTE coefficients. It seems the distortion stabilizes the structure and contributes towards positive thermal expansion behavior.

As noted above, the linear thermal expansion coefficient along a- and c- axis are found to be positive and negative respectively. It is of interest to find the modes which have large negative and positive Grüneisen parameters and contribute towards thermal expansion behavior. The displacement pattern of a few zone boundary phonon modes has been plotted (Fig. 123). The mode assignments, phonon energies and Grüneisen parameters are given in the figures. As mentioned above, HT-CuCN and AgCN crystallize in the same space group (R3m), hence the eigen-vector pattern for symmetrically equivalent phonon modes would be similar. The investigation of the displacement pattern of the eigenvectors shows that the phonon modes have mainly two kinds of dynamics. One kind involves atomic vibrations along the chain and the other in which atoms vibrate perpendicular to the chain.

For HT-CuCN and AgCN, the lowest zone-boundary modes at LD-point $(-2/3,1/3,1/3)_R$ indicate that the q vector in the Brillouin zone is unstable. For the LD point mode (Fig. 123), within a chain, the M and C≡N move with equal displacements. The movement of atoms in the adjacent chains is found to be out-of-phase with each other. Such type of modes would contribute maximum to the NTE along c-axis. However in case of AuCN, the K-point $(1/3,1/3,0)_H$ mode (Fig. 123) also shows sliding of -M-C≡N-M- chains out-of-phase with each other. The mode is found to have small positive $\Gamma$ of 1.1. It seems that the chain sliding modes mainly contribute to negative $\alpha_c$ in HT-CuCN and AgCN compounds and this contribution is not seem in AuCN.

The vibrational amplitude along the chain would depend on the nature of bonding between metal and cyanide (-C≡N-) as well as on the atomic mass of metal ion. As mentioned above, this bonding in HT-CuCN and AgCN seems to be similar. The smaller mass of Cu (63.54 amu) would lead to large amplitude of thermal vibration along the chain in comparison to Ag (107.87 amu) compound, which indicates that the contraction along the -M-C≡N-M- chain would be more in the HT-CuCN in comparison to the AgCN, which is qualitatively in agreement[79] with the observed NTE behavior in these compounds. Several modes in which the atoms move perpendicular to the chain have positive Grüneisen parameters and would be responsible for positive thermal expansion behavior.

## 15. Conclusions

Here we have given an overview of anomalous or negative thermal expansion behavior and its relation with phonons in variety of compounds. The review provides the reader familiarity with some of the recent advances in the modeling ofanomalous thermal expansion of compounds using the techniques of scattering experiments and computer simulations. We have shown that the lattice dynamical calculations have succeeded in generally reproducing the anomalous thermal expansion in various compounds. The nature of specific phonons responsible for the anomalous thermal behavior is found to be different in all these compounds. The low energy librational and bending modes over a large part of the Brillouin zone are responsible for large negative thermal expansion (NTE) in $AW_2O_8$ (A=Zr, Hf)), whereas for $ZrV_2O_7$ the soft phonon occurs only at an incommensurate wave vector. The large difference in thermal expansion behaviour of $M_2O$ (M=Cu, Ag, Au) compounds is due to difference in nature of bonding as well as open space in the unit cell. The soft phonon modes involving rotations and distortion of the $M_4O$ (M=Cu, Ag, Au) tetrahedral units are related to NTE in $M_2O$. Similar types of modes are responsible for NTE behavior in β- eucryptite ($LiAlSiO_4$). In case of $ReO_3$, the zone boundary M3 mode softens and leads to anomalous thermal behavior. The softening of the transverse-acoustic modes lead to NTE in CuX (X=Cl, Br, I) compounds. The transverse motions of the bridging C/N atoms as well as low energy vibrations of translational and bending motions of tetrahedral or octahedral units are responsible for anomalous behavior in CN molecular framework compounds. The chain-sliding modes are found to govern the thermal expansion behavior in low dimension MCN (M=Cu, Ag and Au) compounds. We find that open and less covalently bonded structures are more likely to show negative thermal expansion behavior.

**Acknowledgements**

S. L. Chaplot would like to thank the Department of Atomic Energy, India for the award of Raja Ramanna Fellowship.

Table 1. Comparison of the experimental structural parameters [40] at 293 K in the cubic phase of ZrW$_2$O$_8$ with the calculations. The units used are Å for the lattice constant $a$ and Å$^2$ for mean squared amplitudes $u^2$[3, 5].

|    |       | Experimental | Interatomic potential | ab-initio |
|----|-------|--------------|----------------------|-----------|
|    | $a$   | 9.15993      | 9.2188               | 9.3200    |
| Zr | $x$   | 0.0003       | 0.0020               | 0.0015    |
|    | $u^2$ | 0.010        | 0.012                | 0.012     |
| W1 | $x$   | 0.3412       | 0.3547               | 0.3428    |
|    | $u^2$ | 0.012        | 0.011                | 0.010     |
| W2 | $x$   | 0.6008       | 0.6018               | 0.6005    |
|    | $u^2$ | 0.010        | 0.011                | 0.008     |
| O1 | $x$   | 0.2071       | 0.2071               | 0.2061    |
|    | $y$   | 0.4378       | 0.4382               | 0.4387    |
|    | $z$   | 0.4470       | 0.4475               | 0.4471    |
|    | $u^2$ | 0.022        | 0.026                | 0.020     |
| O2 | $x$   | 0.7876       | 0.7863               | 0.7867    |
|    | $y$   | 0.5694       | 0.5674               | 0.5675    |
|    | $z$   | 0.5565       | 0.5599               | 0.5564    |
|    | $u^2$ | 0.020        | 0.018                | 0.018     |
| O3 | $x$   | 0.4916       | 0.4905               | 0.4917    |
|    | $u^2$ | 0.023        | 0.021                | 0.022     |
| O4 | $x$   | 0.2336       | 0.2429               | 0.2355    |
|    | $u^2$ | 0.037        | 0.045                | 0.034     |

Table 2. Comparison of the calculated (at 0 K) and experimental structural parameters in the cubic phase of ZrV$_2$O$_7$ (at 387 K) [105] and HfV$_2$O$_7$ (at 480 K) [87]. For the space group $Pa\bar{3}$, the Zr/Hf, V, O1 and O2 atoms are located at (0, 0, 0), (x, x, x), (x, y, z) and (0.5, 0.5, 0.5), respectively, and their symmetry equivalent positions. Since the ionic radius of Hf and Zr are nearly the same, we have used the same potentials for both the Zr and Hf compounds. Therefore the calculated structures for both ZrV$_2$O$_7$ and HfV$_2$O$_7$ are the same[11].

|       |   | Expt. ZrV$_2$O$_7$ | Expt. HfV$_2$O$_7$ | Calc.  |
|-------|---|--------------------|--------------------|--------|
|       | a | 8.8194             | 8.7862             | 8.914  |
| Zr/Hf | x | 0.0                | 0.0                | 0.0    |
| V     | x | 0.3865             | 0.381              | 0.390  |
| O1    | x | 0.4366             | 0.437              | 0.439  |
|       | y | 0.2052             | 0.205              | 0.213  |
|       | z | 0.4072             | 0.408              | 0.419  |
| O2    | x | 0.5                | 0.5                | 0.5    |

Table 3. Calculated[16, 201, 234] and experimental[200] elastic constants in GPa units in β-eucryptite.

|  | $C_{11}$ | $C_{12}$ | $C_{13}$ | $C_{33}$ | $C_{44}$ | $C_{66}$ | Bulk Modulus |
|---|---|---|---|---|---|---|---|
| Calculations[16] | 171.6 | 79.0 | 92.9 | 140.0 | 60.0 | 46.3 | 113.1 |
| Experimental[200] | 176.3 | 68.5 | 89.8 | 139.9 | 61.2 | 53.9 | 109.9 |
| Calculation[234] | 178.9 | 102.8 | 118.3 | 181.3 | 47.4 | 38.1 | 134.9 |
| Calculation[201] | 165.6 | 71.0 | 78.6 | 132.8 | 58.7 | 47.3 | 101.5 |

Table 4. Calculated elastic compliance matrix components ($s_{ij}$) in GPa$^{-1}$ units for β-eucryptite[16].

| $S_{11}$ | $S_{12}$ | $S_{13}$ | $S_{33}$ | $S_{44}$ | $S_{66}$ |
|---|---|---|---|---|---|
| 0.00932 | -0.00148 | -0.00520 | 0.01405 | 0.01668 | 0.02163 |

Table 5. The computed dielectric constants and Born effective charge of various atom in CuX (X=Cl, Br, I)[143].

| Compound | Dielectric constant (ε) | Born effective charges | |
|---|---|---|---|
|  |  | Cu | X |
| CuCl | 7.3 | 1.12 | -1.12 |
| CuBr | 7.4 | 1.22 | -1.22 |
| CuI | 6.9 | 1.12 | -1.12 |

Table 6. Fractional atomic coordinates used to generate the Ni(CN)$_2$ layers within the tetragonal space group *P4* for Ni(CN)$_2$. $a = b = 6.869$ Å and $c = 6.405$ Å[25].

| Ni | 0.0 | 0.0 | 0.0 |
|---|---|---|---|
| Ni | 0.5 | 0.5 | 0.0 |
| C | 0.1909 | 0.1909 | 0. |
| N | 0.3091 | 0.3091 | 0.0 |

Table 7. *Ab-initio* calculated (Calc)[25] and observed (Exp) [80] Raman and IR frequencies (cm$^{-1}$) for Zn(CN)$_2$. Irrep, Type and M stand for irreducible representation, type of the mode and multiplicity, respectively. R, RI and S indicate if the mode is Raman active or is both Raman and IR active or optically inactive, respectively. The point group symmetry is $T_d^1$. The experimental Raman and IR frequencies are taken from Table I of Ref. [80].

| Calc | 59 | 173 | 178 | 204 | 240 | 330 | 336 | 476 | 481 | 2251 | 2261 |
|---|---|---|---|---|---|---|---|---|---|---|---|
| Exp |    |    | 178 | 216 |    | 343 | 339 | 461 |    | 2218 | 2221 |
| Irrep | T$_1$ | E | T$_2$ | T$_2$ | T1 | T$_2$ | E | A$_1$ | T$_2$ | T$_2$ | A$_1$ |
| Type | S | R | RI | RI | S | RI | R | RI | R | RI | R |
| M | 3 | 2 | 3 | 3 | 3 | 3 | 2 | 3 | 1 | 3 | 1 |

Table 8. *Ab-initio* calculated (Calc) and observed (Exp) Raman and IR frequencies (cm$^{-1}$) for Ni(CN)$_2$. Irrep, Type and M stand for irreducible representation, type of the mode and multiplicity, respectively. R and RI indicate if the mode is Raman active or is both Raman and IR active, respectively. The point group symmetry is $C_4^1$, thus all IR are Raman active. A and E Irreps (polar modes) are also IR, with polarizations lying along the *z*-axis and in the *xy*-plane, respectively. The B modes are Raman active only[25].

| Calc | 99 | 100 | 103 | 210 | 303 | 328 | 333 | 334 | 335 | 337 |
|---|---|---|---|---|---|---|---|---|---|---|
| Exp |    |    |    | 200 |    |    |    | 334 (broad) |    |    |
| Irrep | E | A | A | B | E | E | A | A | B | E |
| Type | RI | RI | RI | R | RI | RI | RI | RI | R | RI |
| M | 2 | 1 | 1 | 1 | 2 | 2 | 1 | 1 | 1 | 2 |
| Calc | 397 | 461 | 489 | 490 | 566 | 583 | 606 | 2196 | 2205 | 2238 |
| Exp |     |     |     | 508 | 561 | 604 | 607 | 2202 | 2206 | 2215 |
| Irrep | B | A | E | B | B | A | E | E | B | A |
| Type | R | RI | RI | R | R | RI | RI | RI | R | RI |
| M | 1 | 1 | 2 | 1 | 1 | 1 | 2 | 2 | 1 | 1 |

Table 9. The various elastic constants and Bulk Modulus of metal cyanides MCN (M=Cu, Ag and Au) in unit of GPa at T=0 K[1].

|         | C$_{11}$ | C$_{33}$ | C$_{44}$ | C$_{66}$ | C$_{12}$ | C$_{13}$ | B |
|---|---|---|---|---|---|---|---|
| HT-CuCN | 14.0 | 536.0 | 4.0 | 0.4 | 6.1 | 11.0 | 10.1 |
| AgCN    | 18.5 | 387.0 | 5.2 | -0.3 | 8.1 | 16.0 | 13.3 |
| AuCN    | 28.4 | 755.3 | 6.5 | 2.2 | 15.3 | 14.0 | 21.8 |

Fig. 1. Schematic diagram of the IN6 spectrometer at ILL, Grenoble, France (after www.ill.fr).

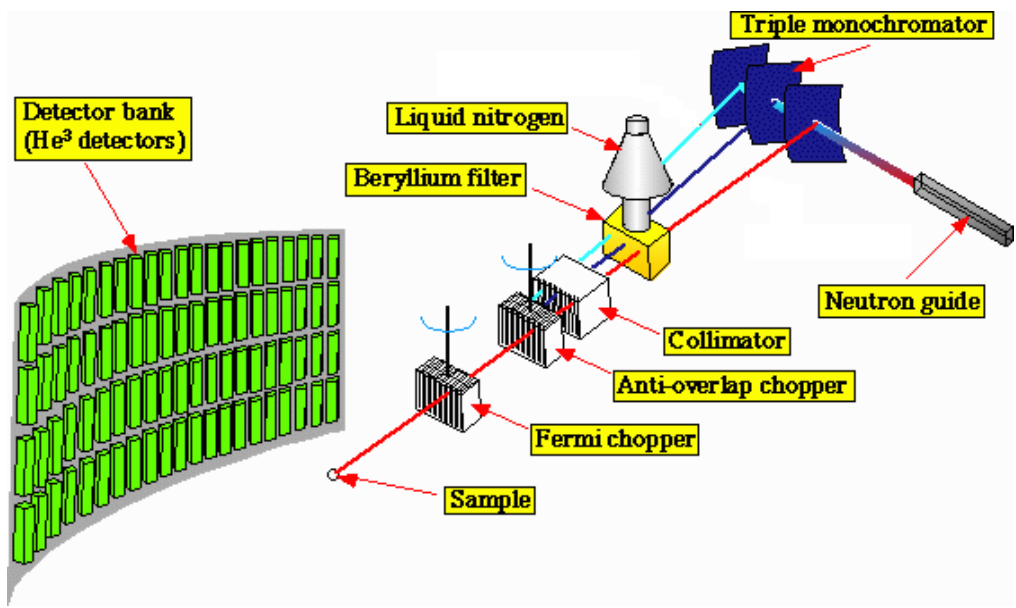

Fig. 2. Flow chart of density functional theory (DFT) calculation scheme.

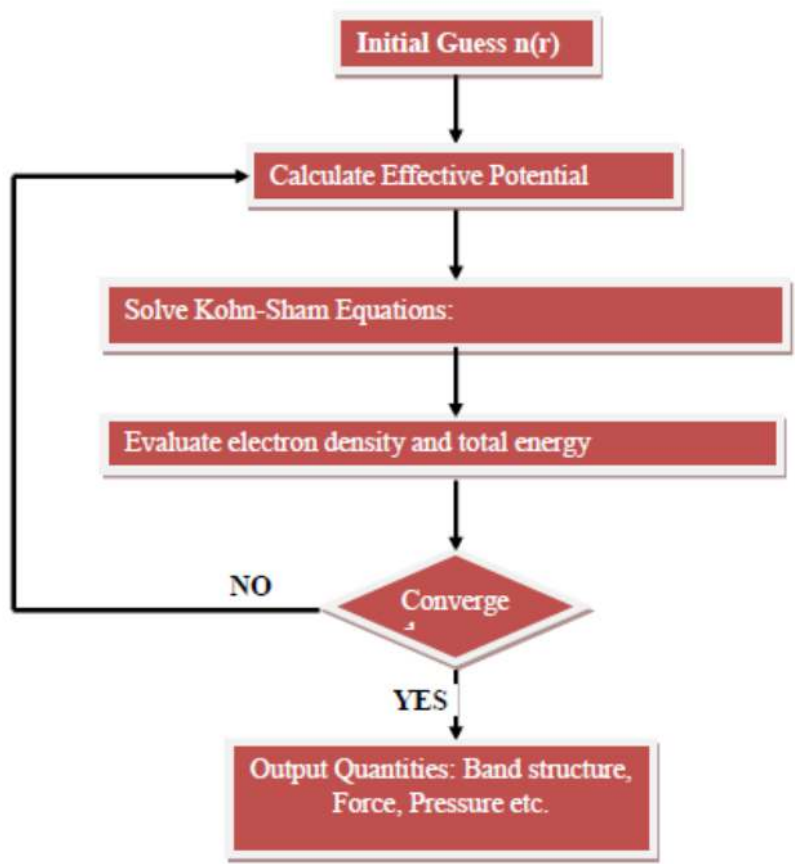

Fig. 3. Flow chart of PHONON calculation using density functional theory (DFT) methods.

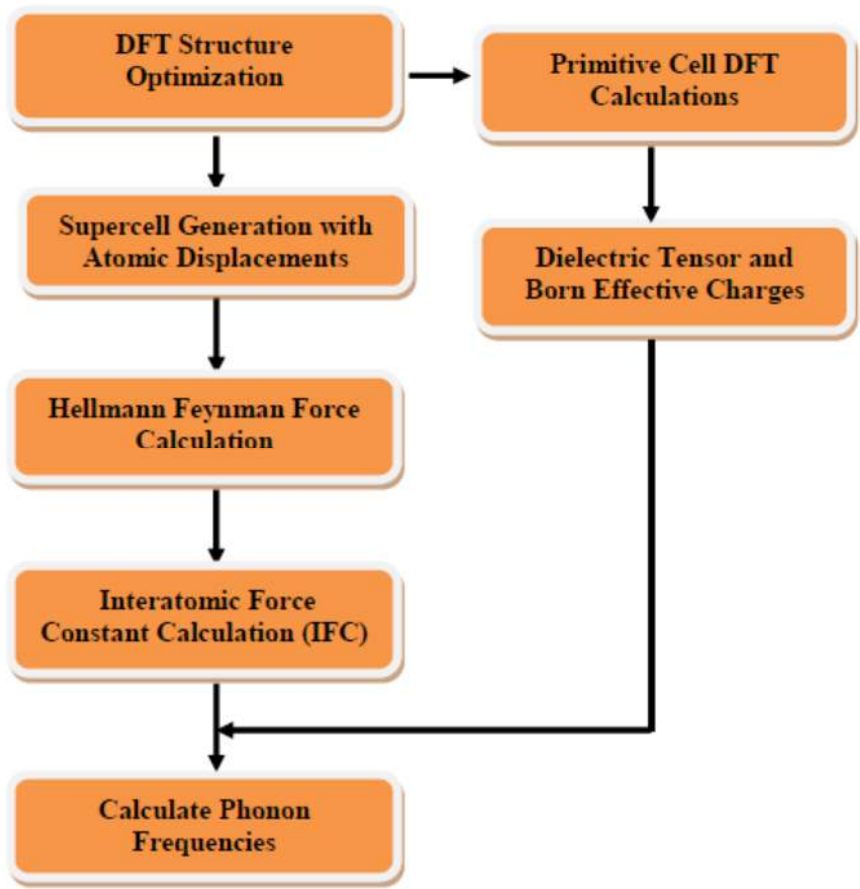

Fig. 4. The crystal structure of $AW_2O_8$ (A=Zr, Hf). The tetrahedral and octahedral units are formed around W and A atoms respectively.

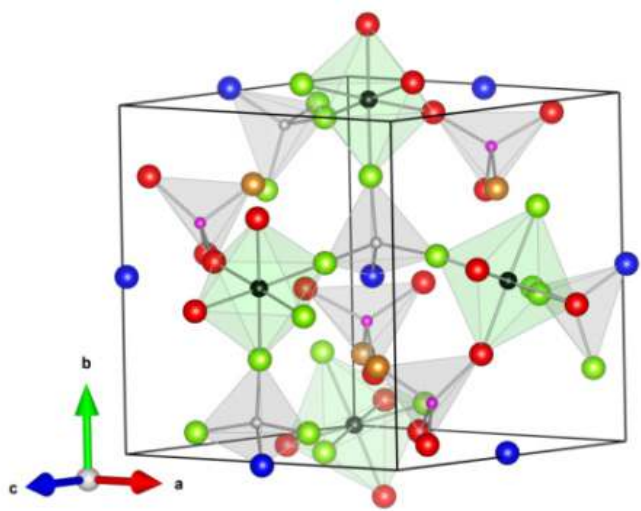

Fig. 5. Pressure dependence of the zone centre modes in ZrW$_2$O$_8$. **(a)** High frequency region; **(b)** low frequency region. The discontinuous change in the mode frequencies take place across the cubic to orthorhombic phase transition (1 cm$^{-1}$= 0.124 meV)[86].

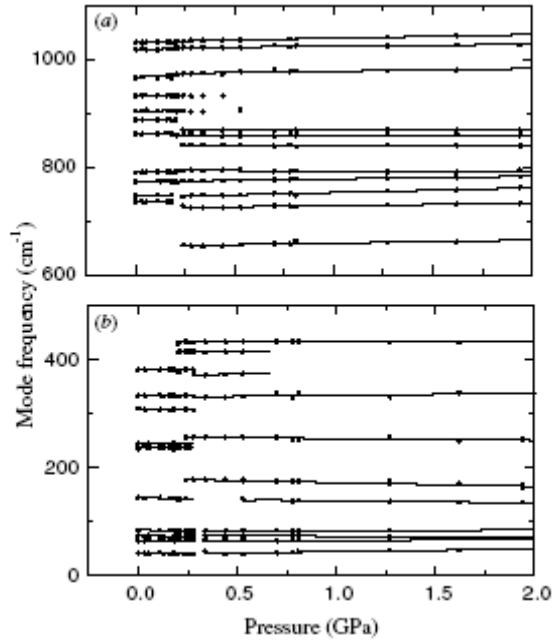

Fig. 6.(a) Comparison of reported phonon density of states [36] and the observed Raman mode energies[86]. Vertical lines represent the positions of the Raman modes. (b) Temperature dependence of the linear thermal expansion coefficient calculated [86] from the Grüneisen parameters of zone centre phonon modes and their weighting factors obtained from the experimental data of phonon density of states [36] shown in (b). The symbols (solid circles) represent the values of linear thermal expansion coefficient extracted[35] from the lattice parameter data[86].

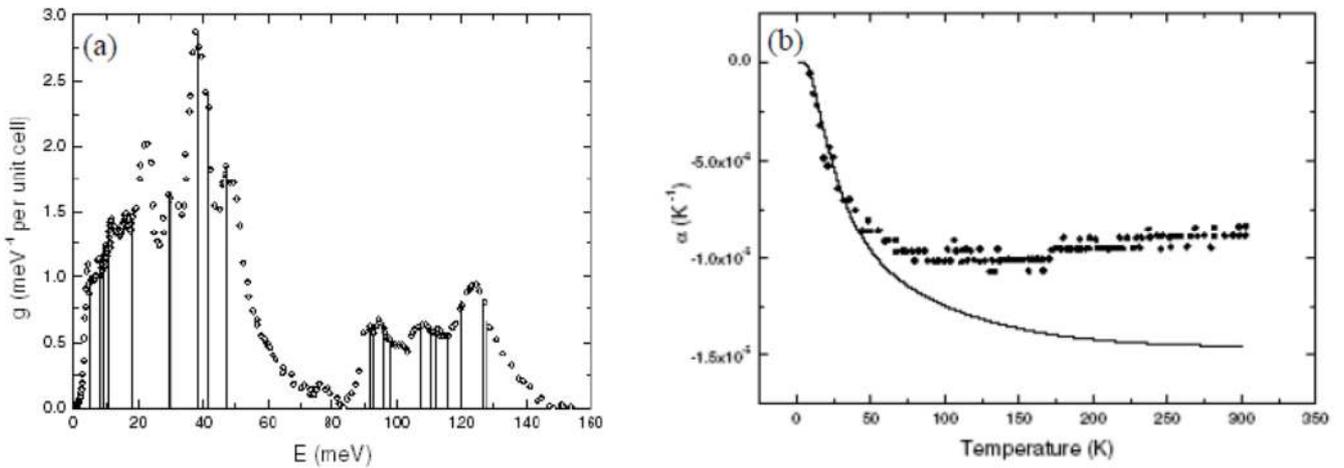

Fig. 7. The experimental (160 K) and calculated (0 K) [5] phonon spectra for cubic ZrW$_2$O$_8$[6] and ZrMo$_2$O$_8$ [9] as a function of pressure. Inset (upper) shows the spectra for ZrW$_2$O$_8$ around 4 meV on an expanded scale[5, 6, 9].

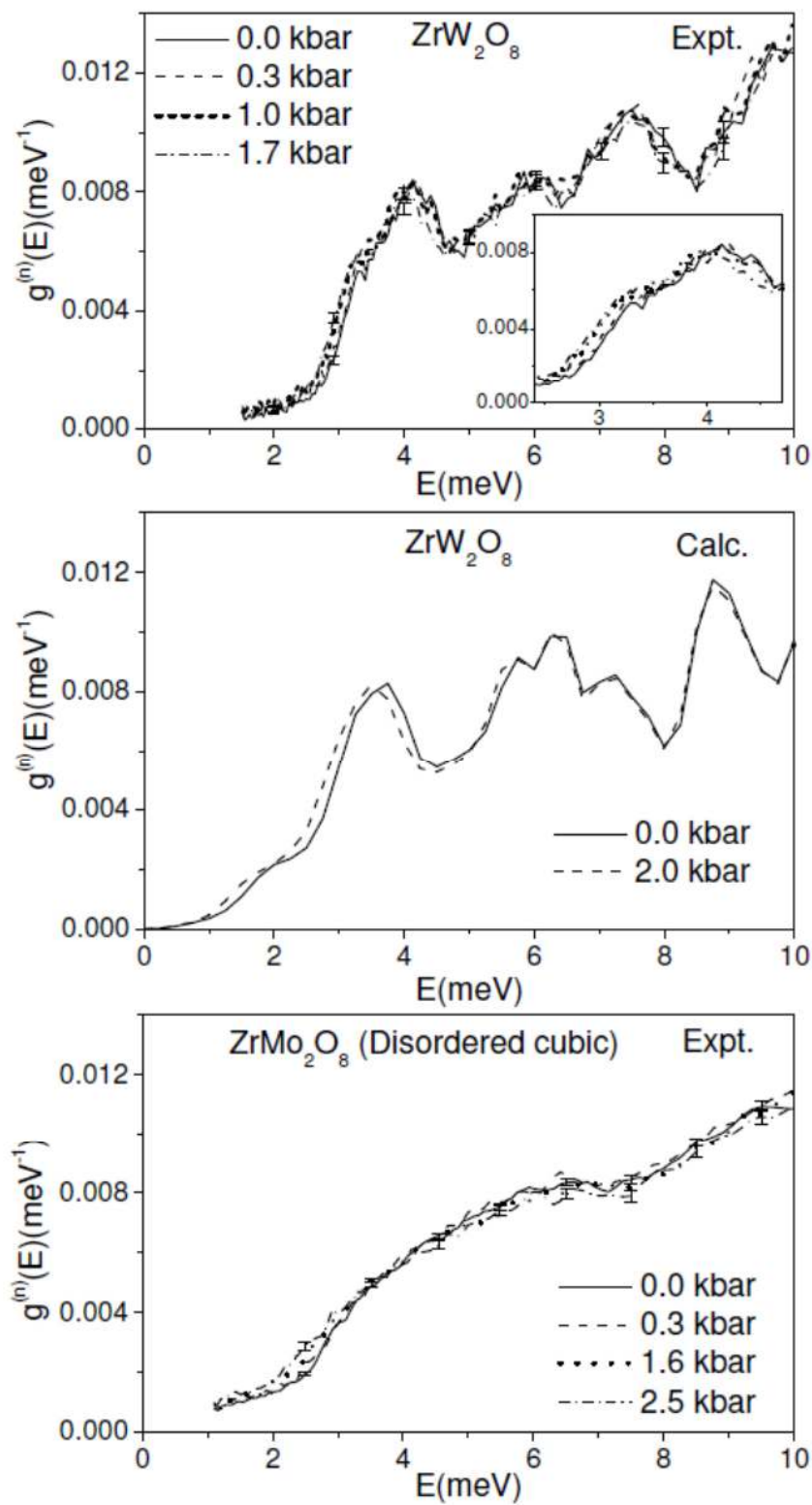

Fig. 8. The Grüneisen parameter ($\Gamma(E)$)[5, 6, 8, 9, 142] averaged over phonons of energy $E$. The experimental [6, 9] values were determined using the neutron data of phonon density of states as a function of pressure[9].

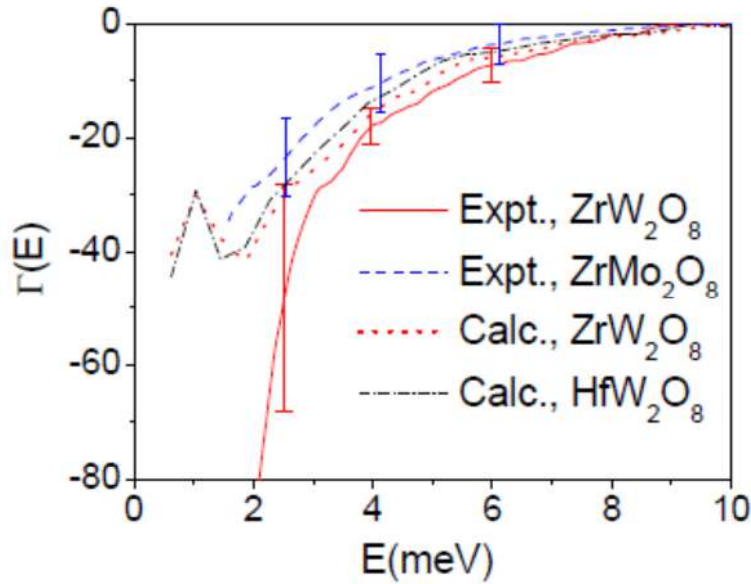

Fig. 9. The comparison between the volume thermal expansion ($\alpha_V$) for cubic $ZrW_2O_8$[6] and $ZrMo_2O_8$ [9] derived from the high pressure inelastic neutron scattering experiment (solid line) and that obtained using neutron diffraction [35, 163] (circles). Inset shows the comparison for $ZrMo_2O_8$ upto 450 K[6, 9].

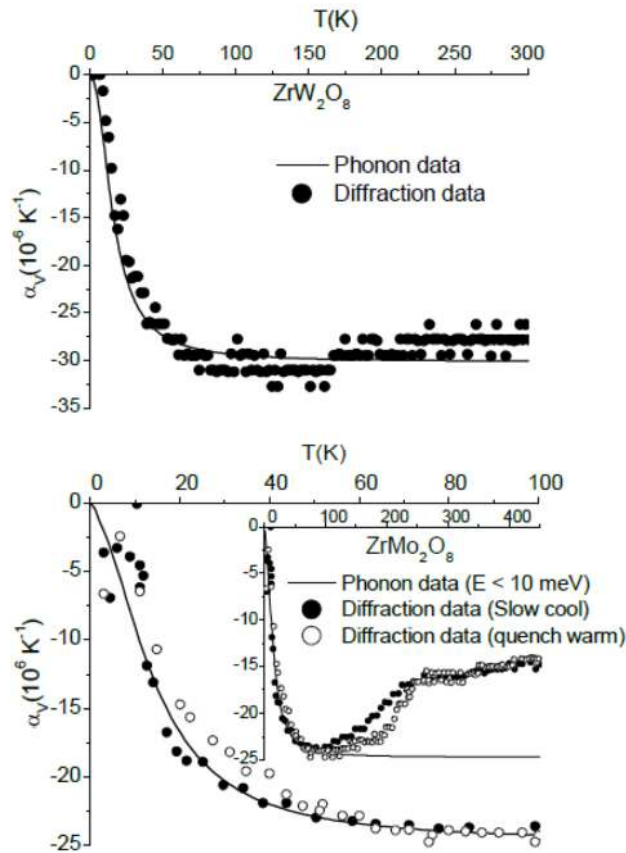

Fig. 10. The contribution of phonons of energy $E$ to the volume thermal expansion as a function of $E$: **(a)** Calculated [34] at 10 K and 300 K for cubic $ZrW_2O_8$ and $HfW_2O_8$ from lattice dynamical calculations **(b)** Calculated [6] at 20 K (dash line) and 300 K (solid line) from the phonon data of cubic $ZrW_2O_8$. A constant $\Gamma(E)$ value of -80 was assumed in the calculation from the phonon data for phonons of energy below 2 meV. We estimated from lattice dynamical calculations [5, 34] that modes upto 10 meV contribute to about 90% of the total thermal expansion.

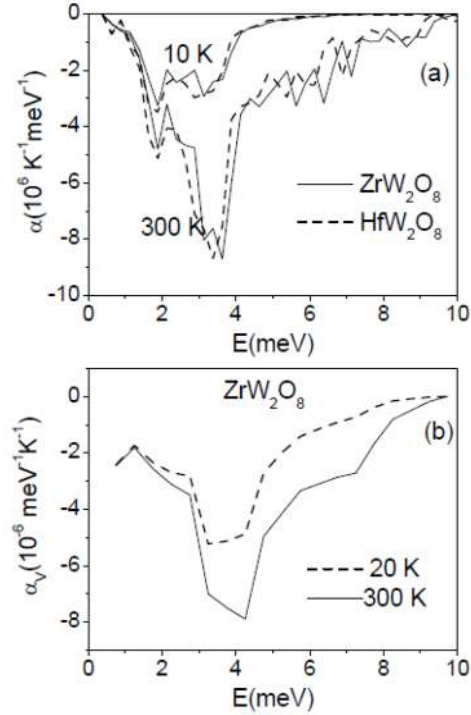

Fig. 11. The calculated phonon dispersion relation upto 10 meV for cubic $HfW_2O_8$ along [100] direction. The solid and dash lines correspond to ambient and 2.6 kbar[8].

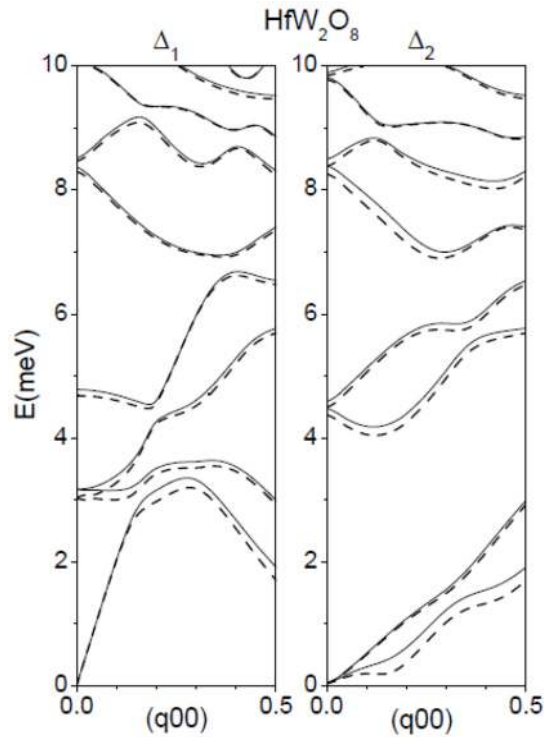

Fig. 12. The experimental (at 7 K) and calculated neutron-weighted phonon density of states of cubic $HfW_2O_8$. In order to account for the experimental resolution broadening in cubic $HfW_2O_8$ the calculated spectrum was convoluted with a Gaussian function of a full width of half maximum (FWHM) of 7 meV. The multiphonon contribution (dash line) was subtracted from the experimental data of cubic $HfW_2O_8$ to obtain the experimental one-phonon density of states[8].

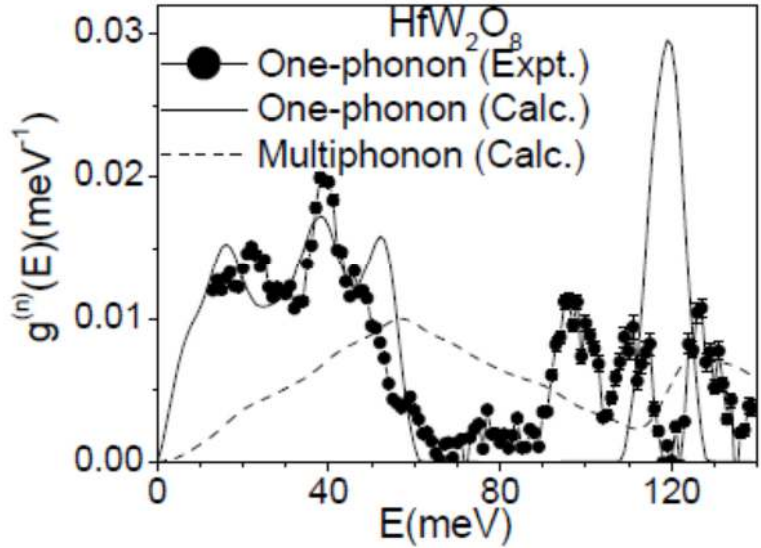

Fig. 13. The comparison between the calculated [5, 8] and experimental [131, 175] specific heat for the cubic **(a)** $ZrW_2O_8$ **(b)**$HfW_2O_8$. For $ZrW_2O_8$ two different experimental data[131, 175] are available. Our calculations (solid line) for $ZrW_2O_8$ are in agreement with the experimental data[131, 175] (open circle)[5, 8].

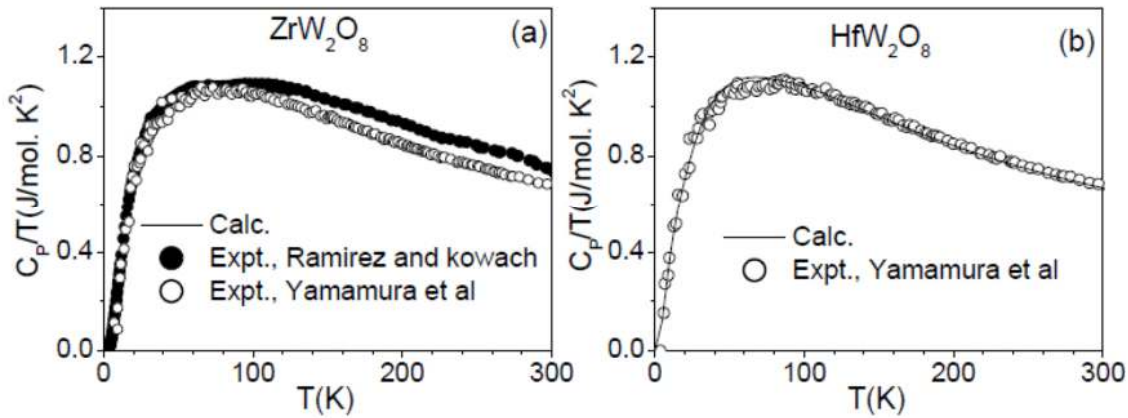

Fig. 14.The calculated volume thermal expansion (solid line) in the cubic and orthorhombic $ZrW_2O_8$ along with separate contributions from the two lowest phonon branches (dotted line) and all the phonons below 8 meV (dash line)[5].

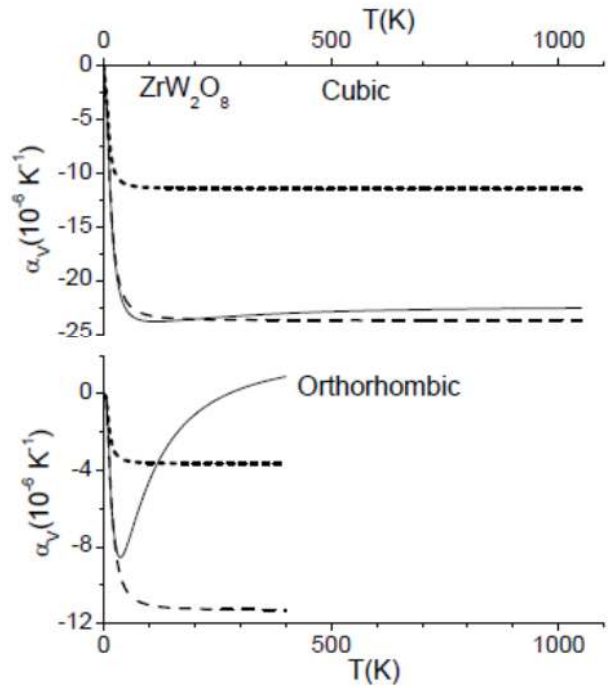

Fig. 15. The experimental and calculated [5, 8] relative volume thermal expansion for cubic $ZrW_2O_8$, $HfW_2O_8$ and orthorhombic $ZrW_2O_8$, $(V_T/V_{300}-1) \times 100\%$, $V_T$ and $V_{300}$ being the cell volumes at temperature $T$ and 300 K respectively. There is a small sharp drop in volume for cubic phase at about 400 K associated with an order disorder phase transition [40, 100, 161].It is noted in Ref.[40, 100, 161]. that the NTE in $HfW_2O_8$ continues till 1050 K; however, specific data were not given. Experimental data; cubic $ZrW_2O_8$ [40, 161], cubic$HfW_2O_8$ [100], orthorhombic $ZrW_2O_8$[166, 167].

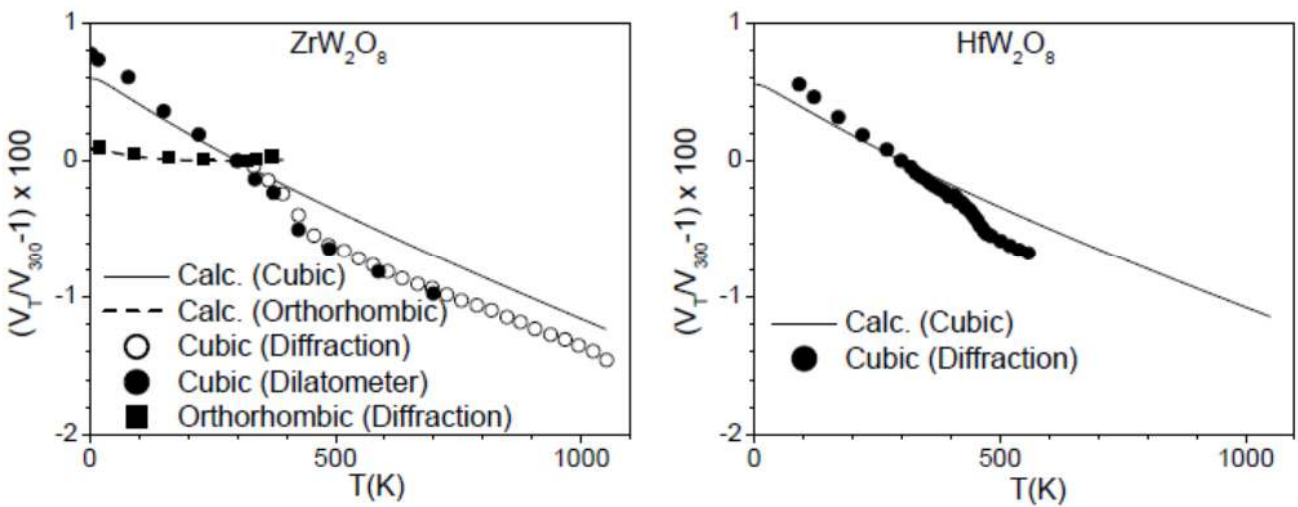

Fig. 16. The calculated [8] contribution to the mean squared amplitude of various atoms arising from phonons of energy $E$ at $T=300$ K in the cubic $HfW_2O_8$ at ambient pressure. The atoms are labeled as indicated in Refs. [40, 161].

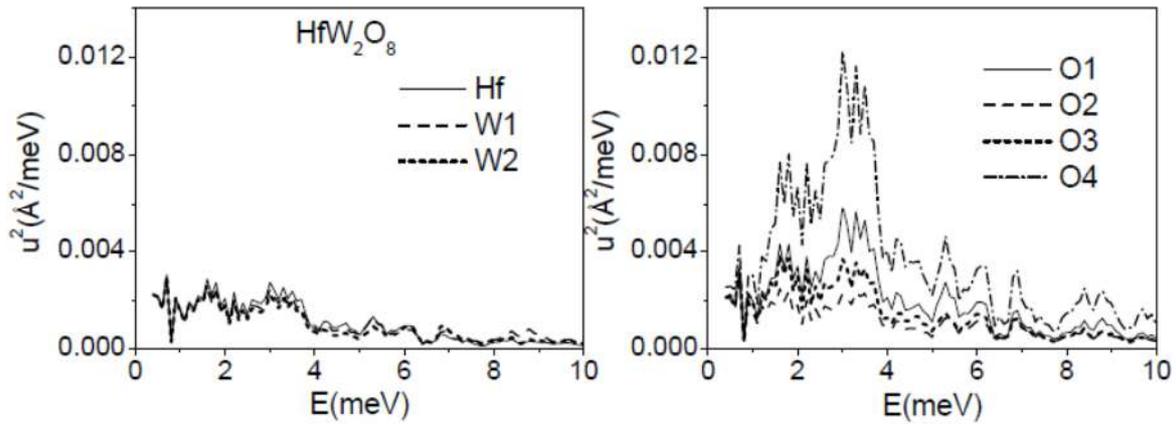

Fig. 17. Experimental and calculated (a) lattice parameter and (b) linear coefficient of thermal expansion for cubic $ZrW_2O_8$ as a function of temperature. Gaussian–zone-center and plane wave–zone-center refer to results obtained within the Debye-Einstein model of the QHA from DFT calculations using atom-centered and plane wave basis sets, respectively[177].

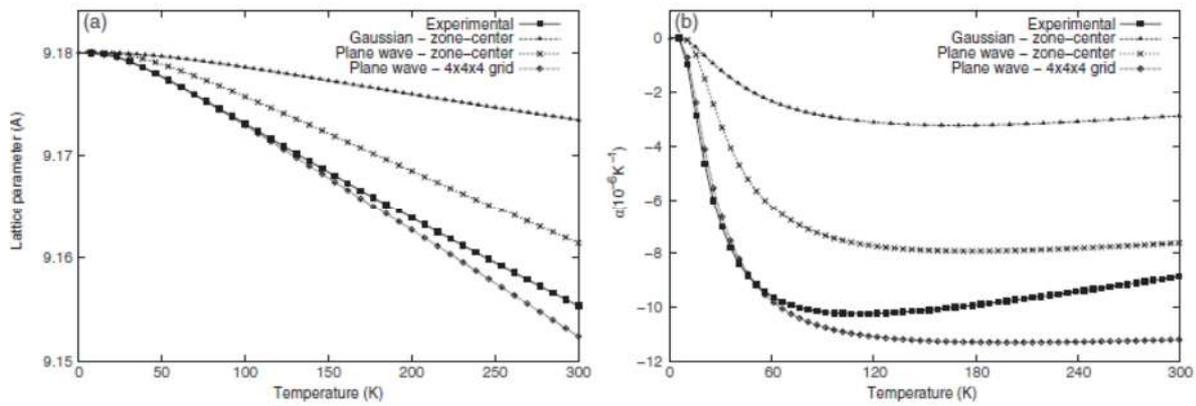

Fig. 18.The calculated (0 K) and experimental (300 K) [36] neutron-weighted phonon spectra in ZrW$_2$O$_8$. The phonon spectra are normalized to unity. For better visibility the experimental phonon spectra [36] is shifted along the y-axis by 0.03 meV$^{-1}$.The calculated zone-centre optic modes, A, E, F(TO) and F(LO) are also shown[3].

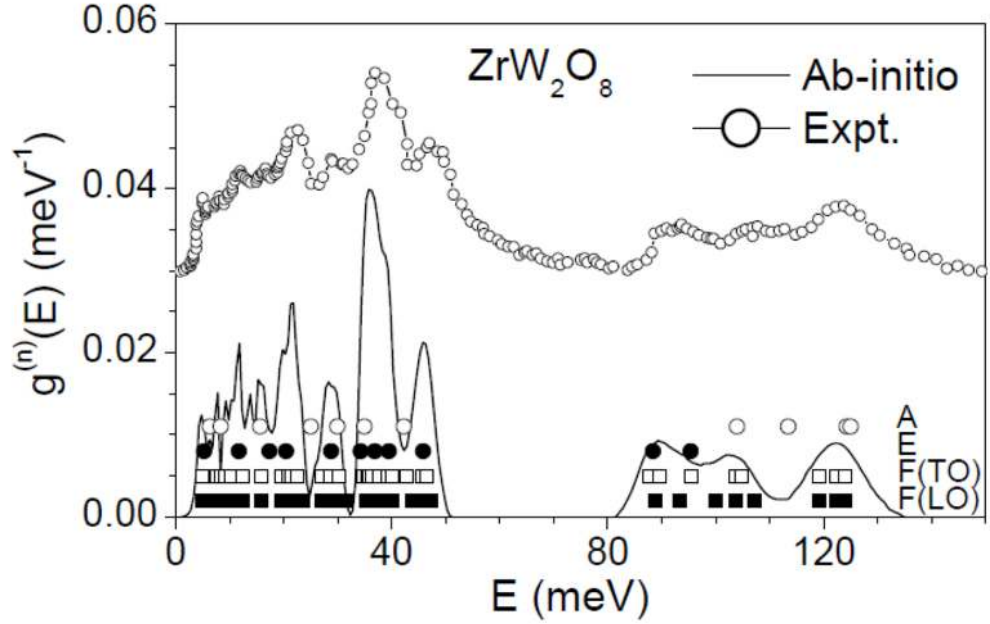

Fig. 19. Calculated low-energy part of the pressure dependent dispersion relation for ZrW$_2$O$_8$. The solid and dashed lines correspond to the calculations at ambient pressure and 1 kbar. Γ=(0,0,0); X=(1/2,0,0); M=(1/2,1/2,0) and R=(1/2,1/2,1/2)[3].

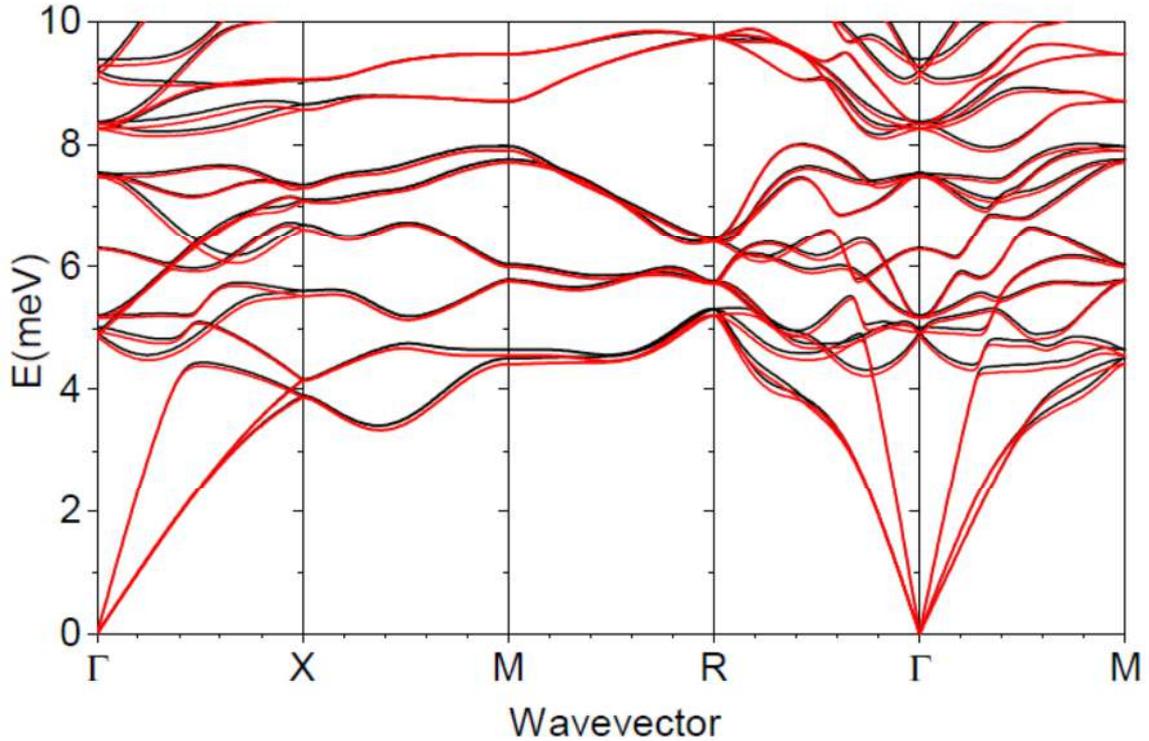

Fig. 20. The calculated[3] and experimental [40] relative volume thermal expansion for $ZrW_2O_8$, $(V_T/V_{300}-1) \times 100\%$, $V_T$ and $V_{300}$ being the cell volumes at temperature T and 300 K respectively. There is a small drop in volume at about 400 K associated with an order-disorder phase transition. (b) The contribution of phonons of energy E to the volume thermal expansion as a function of E at 300 K from the ab-initio calculation as well as phonon data [3].

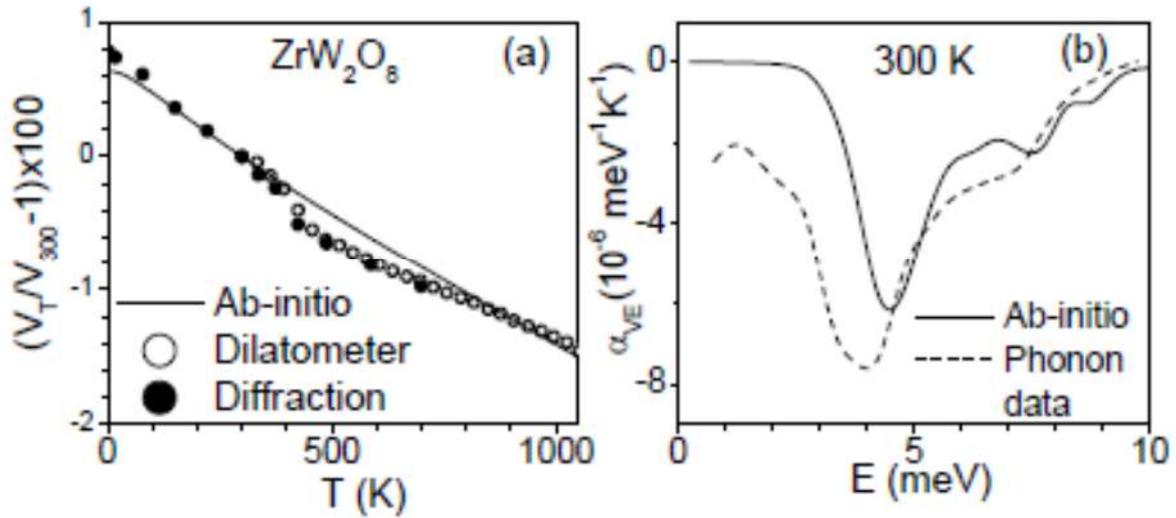

Fig. 21. Polarization vectors of selected phonon modes in $ZrW_2O_8$. The numbers after the wave vector (X and M) give the phonon energies and Grüneisen parameters respectively. The lengths of arrows are proportional to the vibrational amplitudes of the atoms[3].

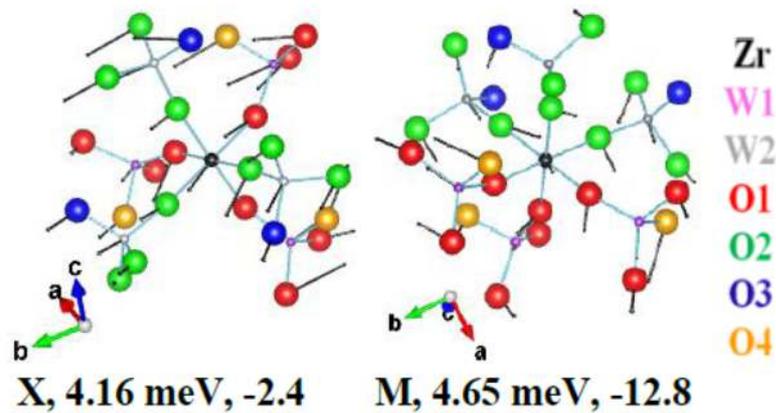

Fig. 22. The crystal structure of orthorhombic $Y_2Mo_3O_{12}$, $Sc_2W_3O_{12}$ and monoclinic $Sc_2Mo_3O_{12}$. The tetrahedral and octahedral units are formed around W/Mo and Y/Sc atoms respectively.

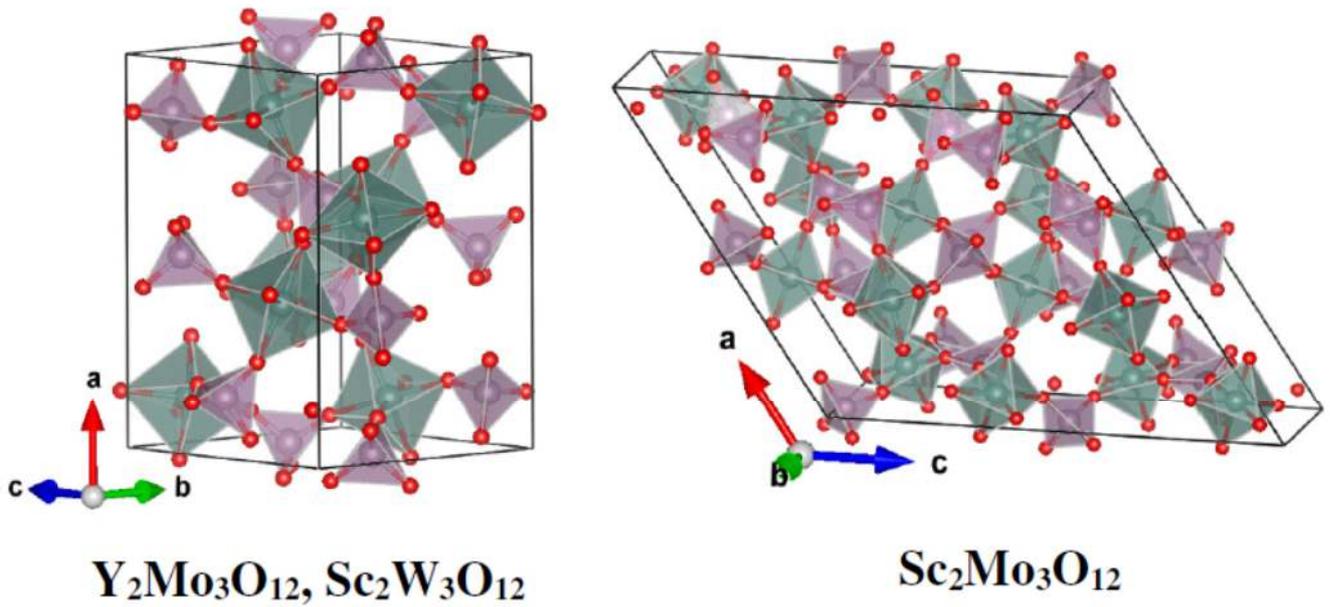

Fig. 23. (a) The comparison of the calculated total phonon DOS (black line) and measured Raman intensity (red circles). (b) The calculated partial phonon DOS of $Y_2Mo_3O_{12}$[178].

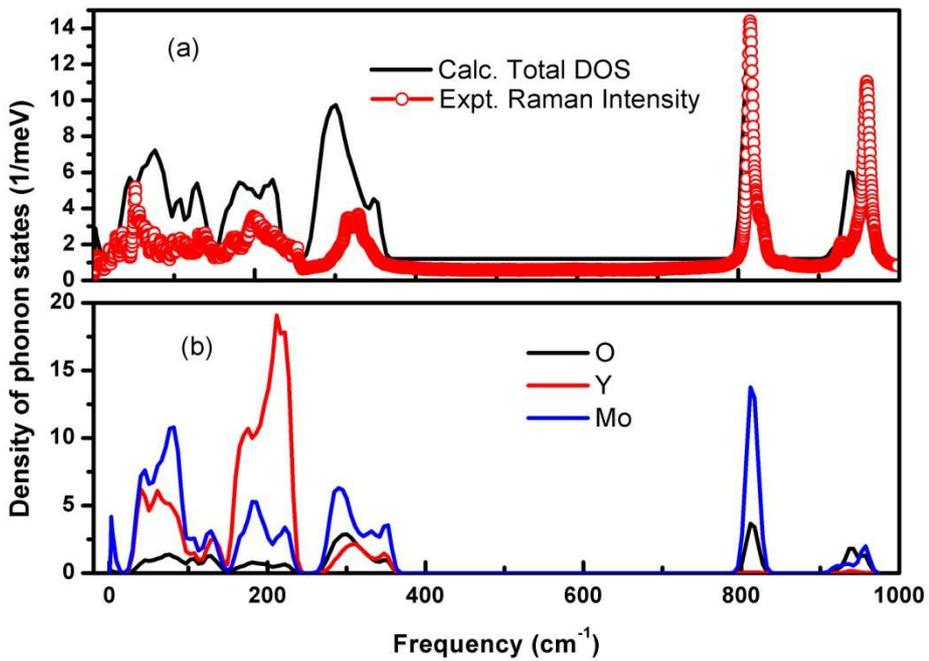

Fig. 24. The negative Gruneisen parameters for orthorhombic $Y_2Mo_3O_{12}$ denoted by semi-solid red circle. (a) Optical phonon from No. 1 to No. 57. (b) Optical phonon from No. 94 to No. 111[178].

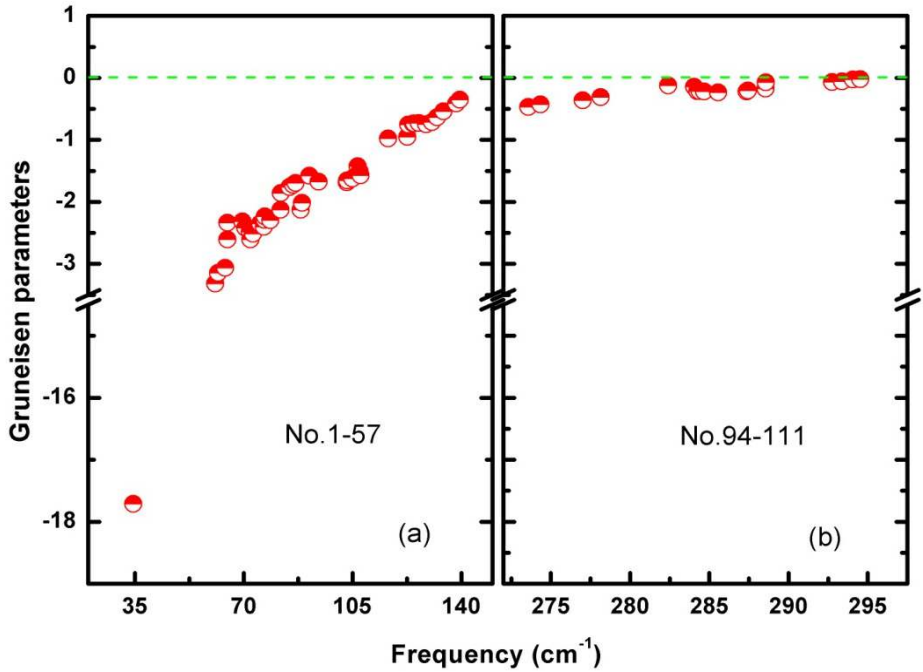

Fig. 25. The calculated temperature dependence of the unit cell volume (left panel) and volume thermal expansion coefficient (right panel) of $Y_2Mo_3O_{12}$. The experimental temperature–volume curve from Ref. [235] is shown as the inset[178].

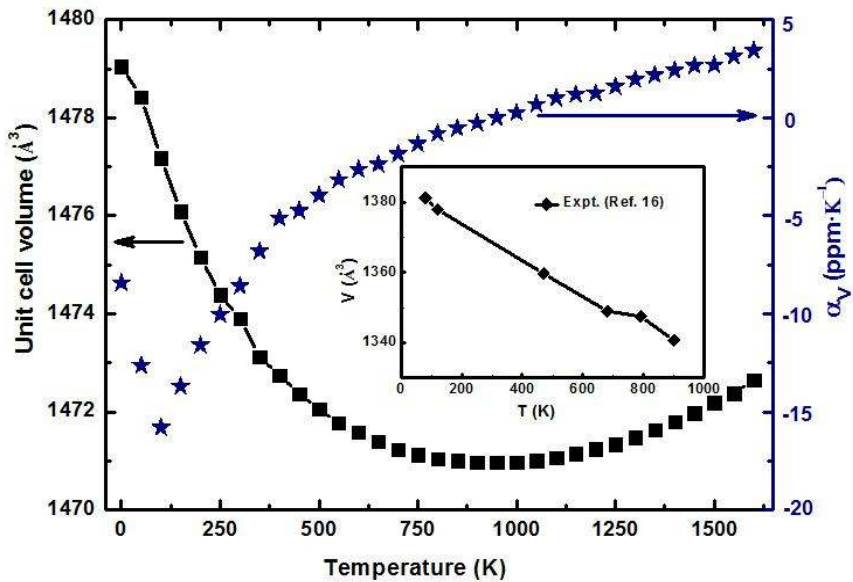

Fig. 26. Effective phonon density of states (DOS) of $Sc_2W_3O_{12}$[26].

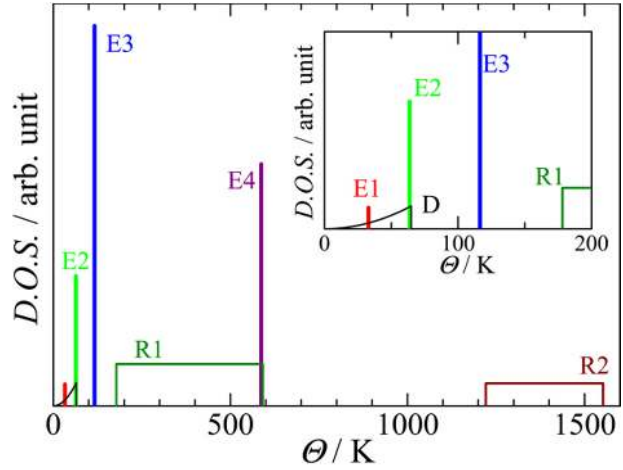

Fig. 27. Temperature dependences of $\gamma_i C_i$ of $Sc_2W_3O_{12}$, where $\gamma_i$ are mode Grüneisen parameters[26].

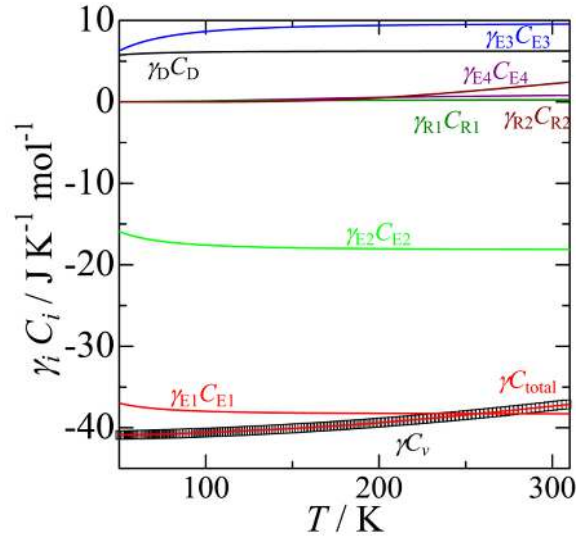

Fig. 28. Effective phonon density of states (DOS) of $Sc_2Mo_3O_{12}$[26].

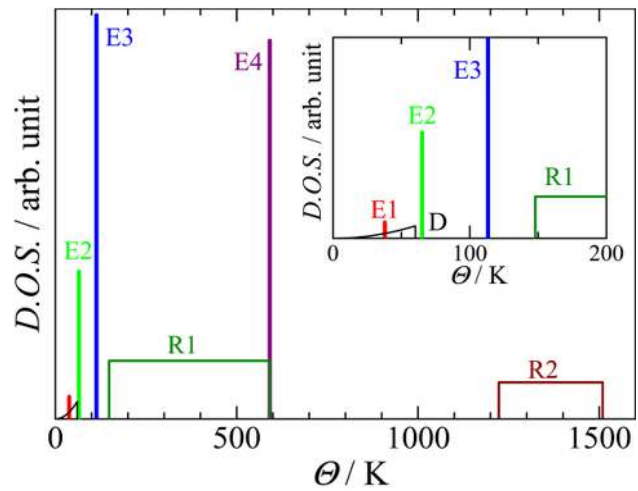

Fig. 29. Temperature dependences of γiCi of Sc$_2$Mo$_3$O$_{12}$, where $\gamma_i$ are mode Grüneisen parameters[26].

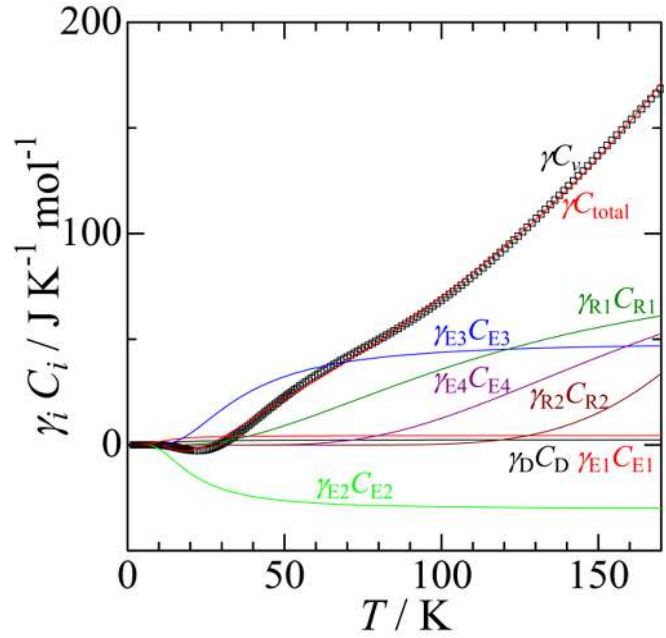

Fig. 30. The calculated phonon dispersion relation up to 10 meV for cubic ZrV$_2$O$_7$ and HfV$_2$O$_7$ along the (100), (110) and (111) directions. The solid and dashed lines correspond to ambient pressure and 3 kbar respectively[11].

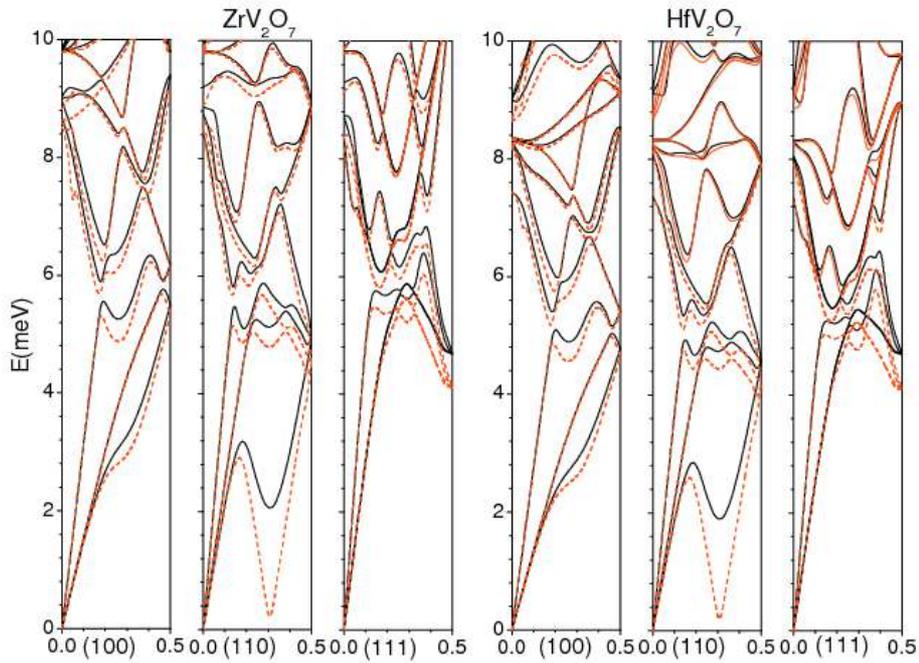

Fig. 31. Calculated partial density of states of various atoms in $ZrV_2O_7$ and $HfV_2O_7$[11].

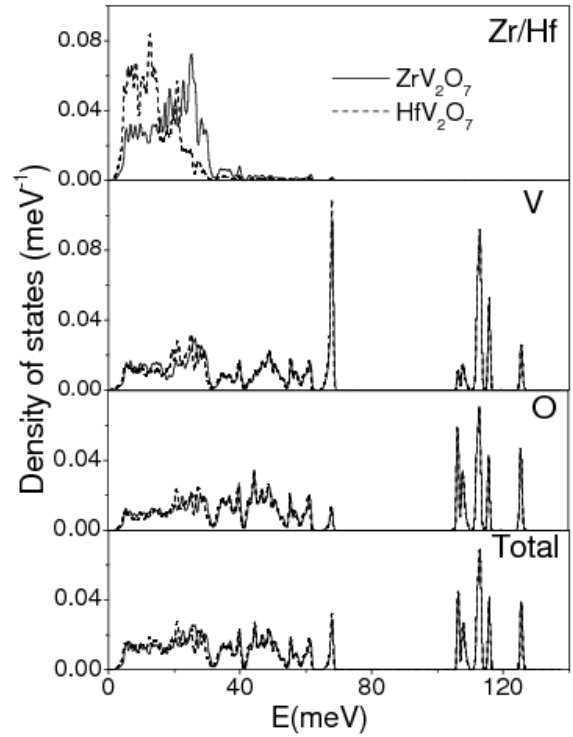

Fig. 32. (a) Calculated pressure variation of phonon density of states up to 10 meV in $ZrV_2O_7$ and $HfV_2O_7$. (b) Calculated Grüneisen parameter $\Gamma(E)$ averaged over phonons of energy E. For comparison, the calculated values of $\Gamma(E)$ for $ZrW_2O_8$[5] are also shown[11].

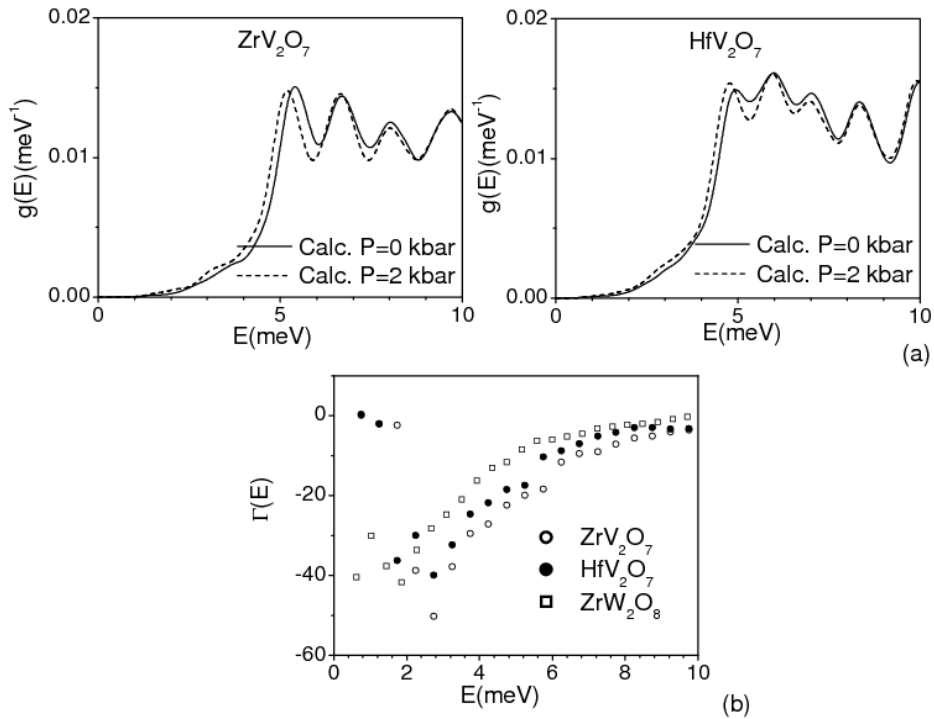

Fig. 33. The calculated volume thermal expansion (solid line) in the cubic $ZrV_2O_7$ along with separate contributions from the two lowest phonon branches (dotted line) and all the phonons below 9 meV (dashed line)[11].

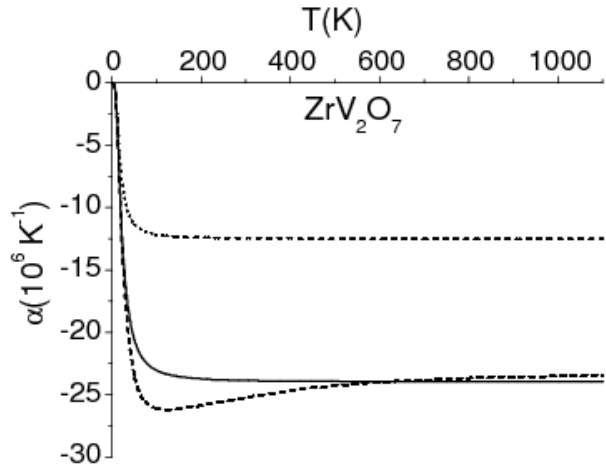

Fig. 34. Comparison between the calculated and experimental thermal expansion behavior of $ZrV_2O_7$[180] and $HfV_2O_7$[87]. The low temperature phase of $MV_2O_7$ (M=Zr, Hf) below about 400 K has positive thermal expansion coefficient. The calculations have been carried out in high temperature phase[11].

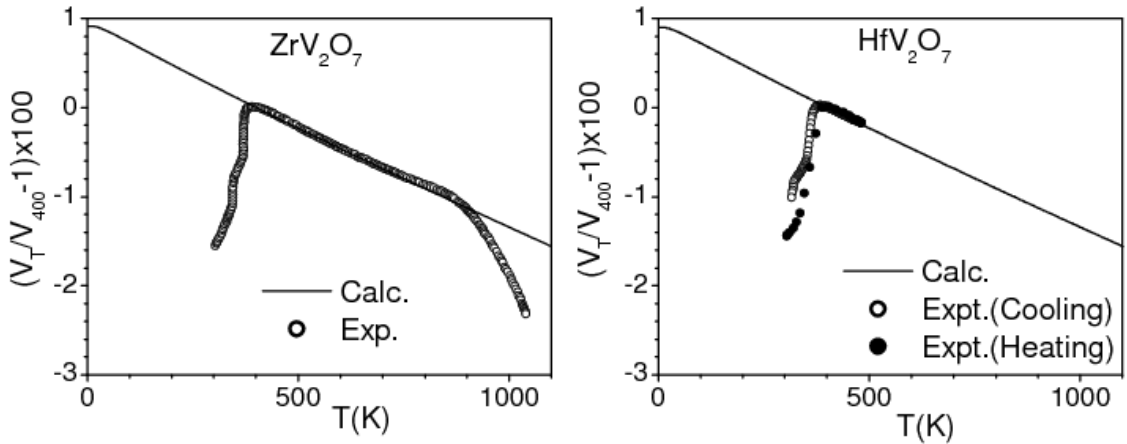

Fig. 35. Contribution of phonons of energy E to the volume thermal expansion as a function of E at 500 K[11].

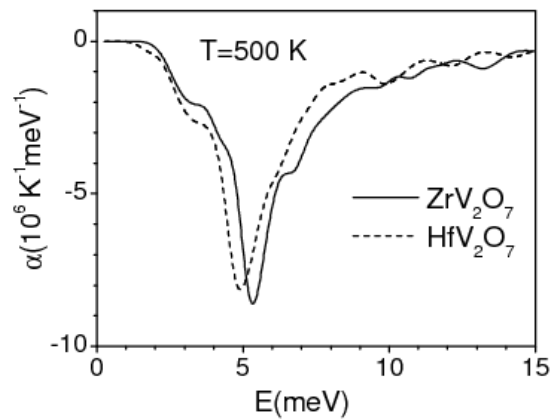

Fig. 36. The calculated contribution to the mean squared amplitude of various atoms arising from phonons of energy E at T=500 K in $ZrV_2O_7$. The atoms are labeled as indicated in Table 1[11].

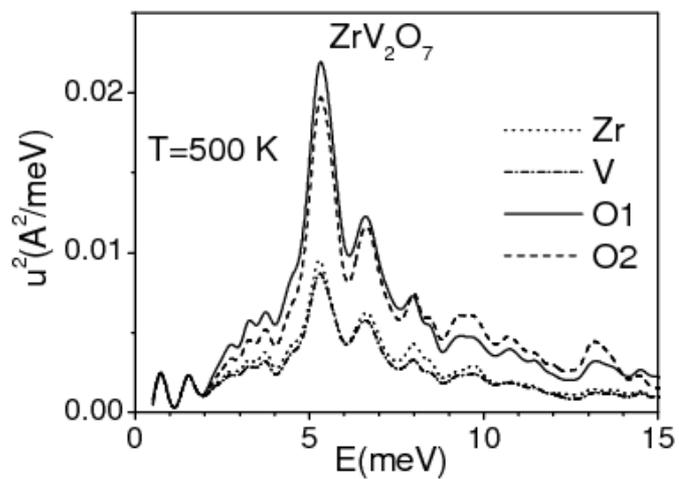

Fig. 37. The crystal structure of $M_2O$ (M=Ag, Cu and Au). Key: M, red sphere; O, blue sphere[2].

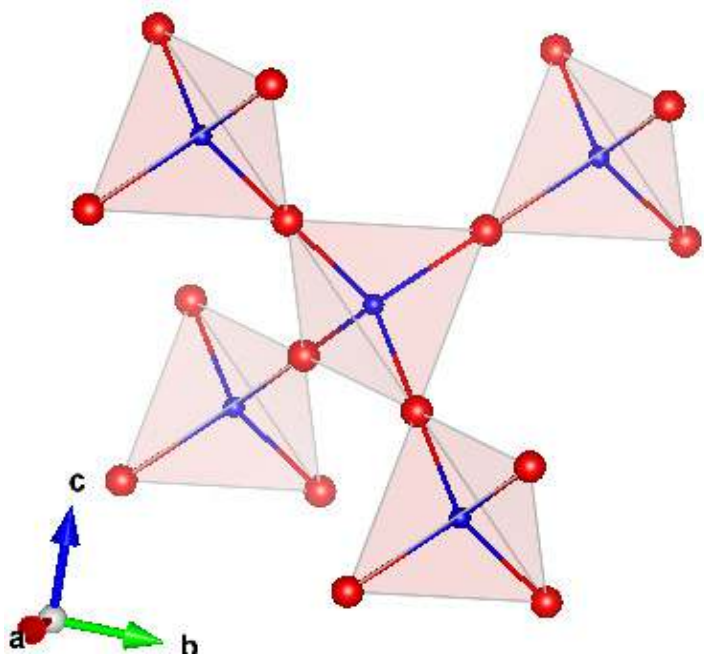

Fig. 38. Experimental (symbols plus line) and calculated (solid line) neutron-weighted phonon density of state of M$_2$O (M=Ag, Au and Cu) compounds. The experimental phonon spectra as well as ab-initio calculation of Ag$_2$O are already published [4] and shown here for comparison with Au$_2$O and Cu$_2$O. The calculated spectra have been convoluted with a Gaussian of FWHM of 15% of the energy transfer in order to describe the effect of energy resolution in the experiment[2].

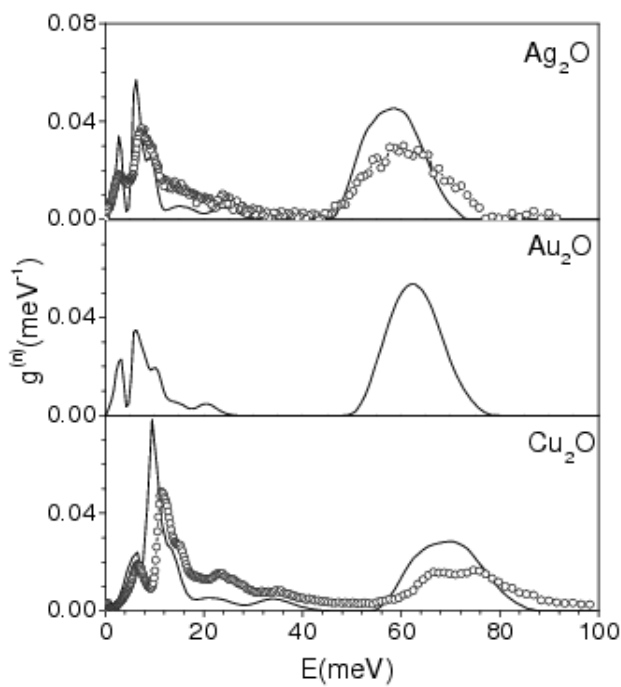

Fig. 39. The calculated low energy part of the phonon dispersion relation of $M_2O$ (M=Ag, Au and Cu). The Bradley-Cracknell notation is used for the high-symmetry points along which the dispersion relations are obtained. $\Gamma=(0,0,0)$; $X=(1/2,0,0)$; $M=(1/2,1/2,0)$ and $R=(1/2,1/2,1/2)$ [2].

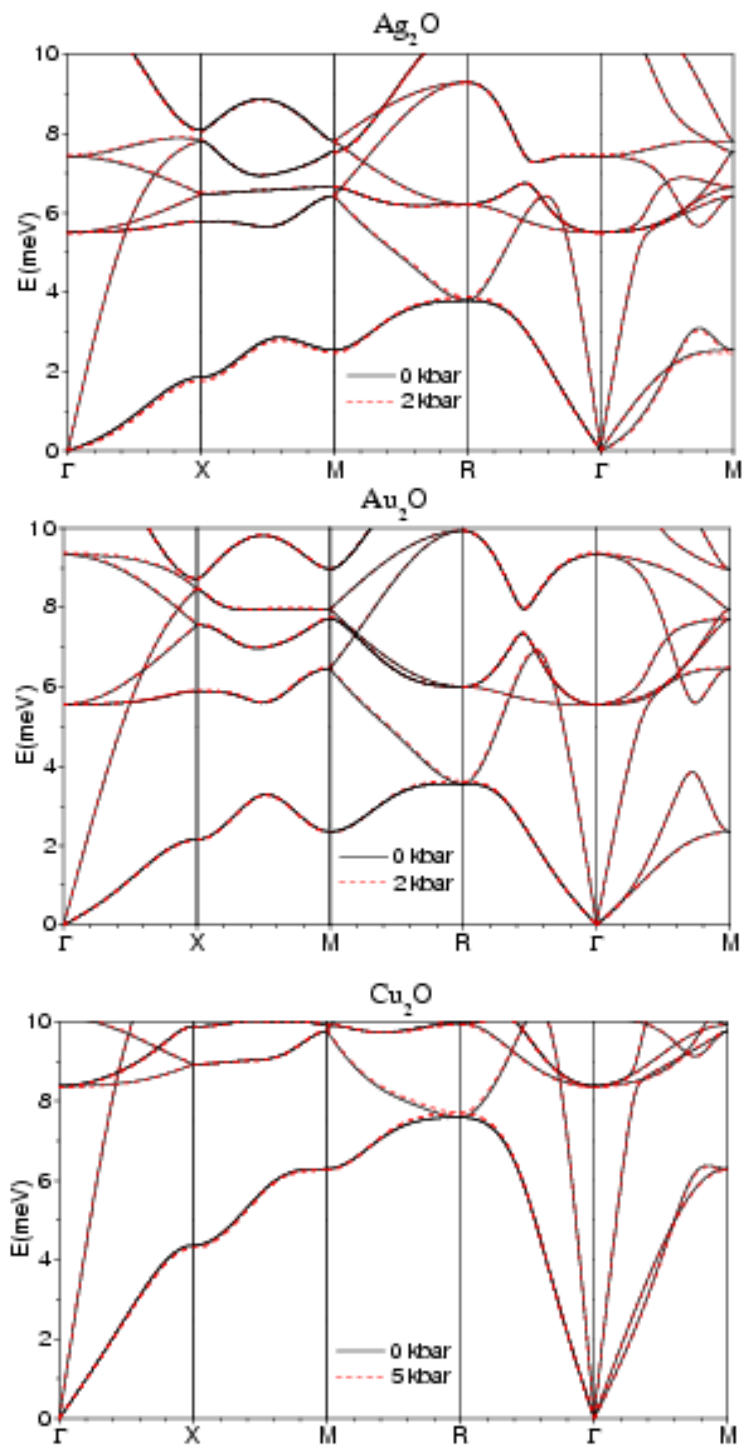

Fig. 40.  Neutron weighted phonon DOS of Ag$_2$O from experiments (black dots) and MD simulations (red curves) from 40 to 400 K. The dashed spectrum corresponds to the 40 K experimental result[68].

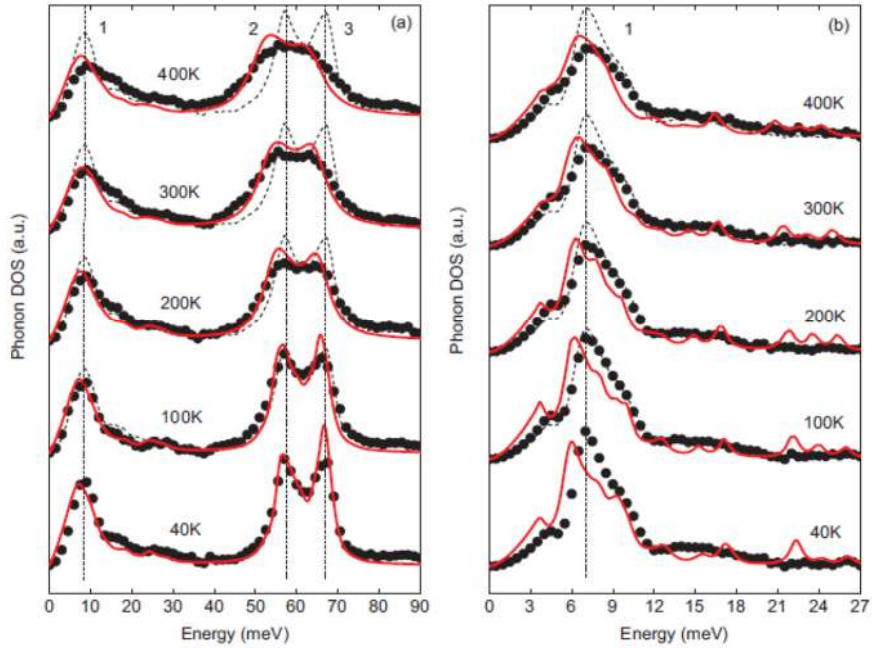

Fig. 41.  The calculated Grüneisen Parameter of M$_2$O (M=Ag, Au and Cu). The calculations for Ag$_2$O [4] are shown here for comparison with Au$_2$O and Cu$_2$O[2].

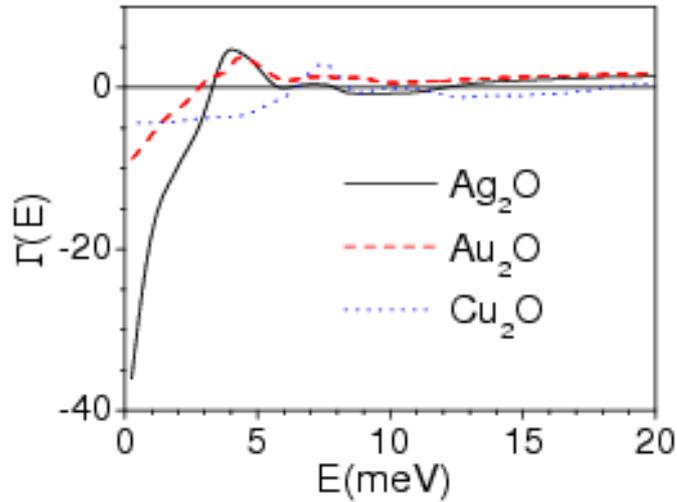

Fig. 42. The calculated and experimental[83, 84] volume thermal expansion of $M_2O$ (M=Ag, Au and Cu). The calculations for $Ag_2O$ [4] are shown here for comparison with $Au_2O$ and $Cu_2O$[2].

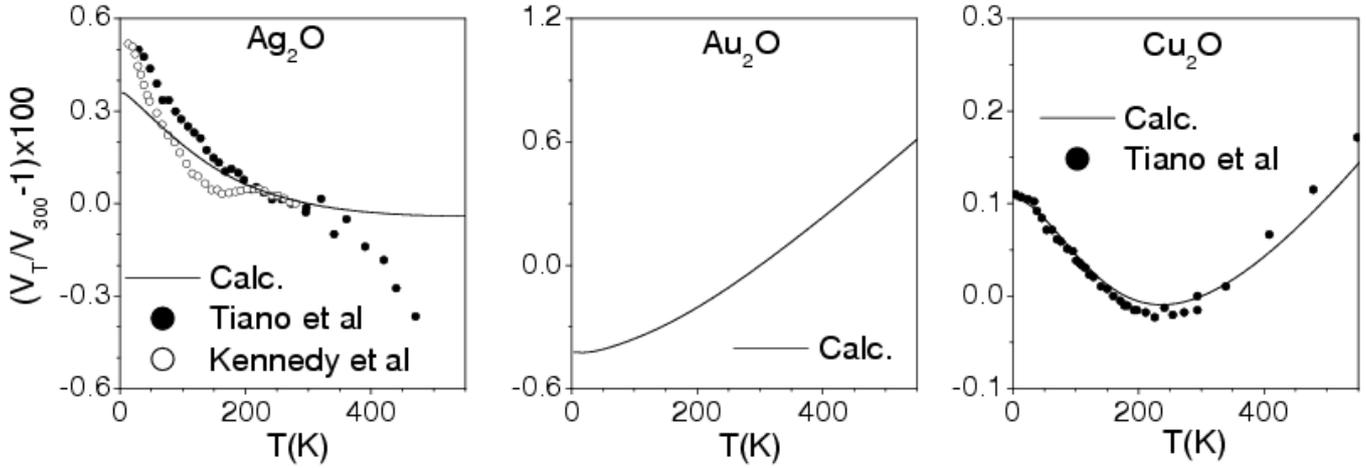

Fig. 43. Volume thermal expansion ($\alpha_V$) coefficient as a function of temperature in $M_2O$ (M=Ag, Au and Cu). The calculations for $Ag_2O$ [4] are shown here for comparison with $Au_2O$ and $Cu_2O$[2].

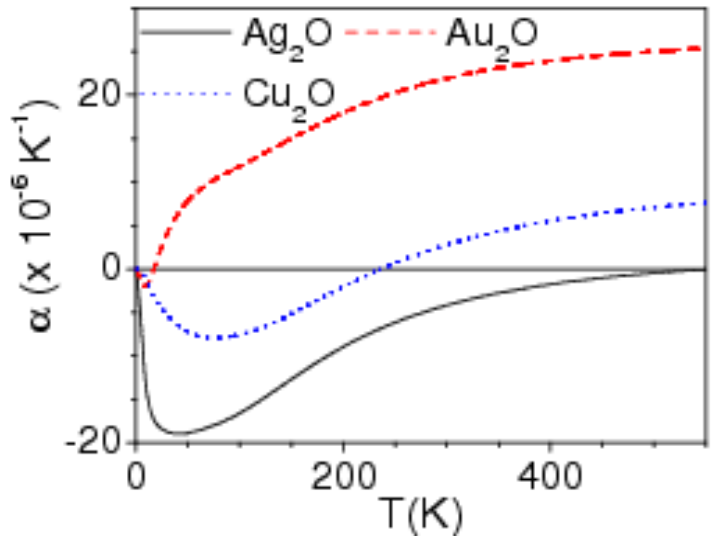

Fig. 44. Volume thermal expansion ($\alpha$) coefficient contributed from phonons of energy E[2].

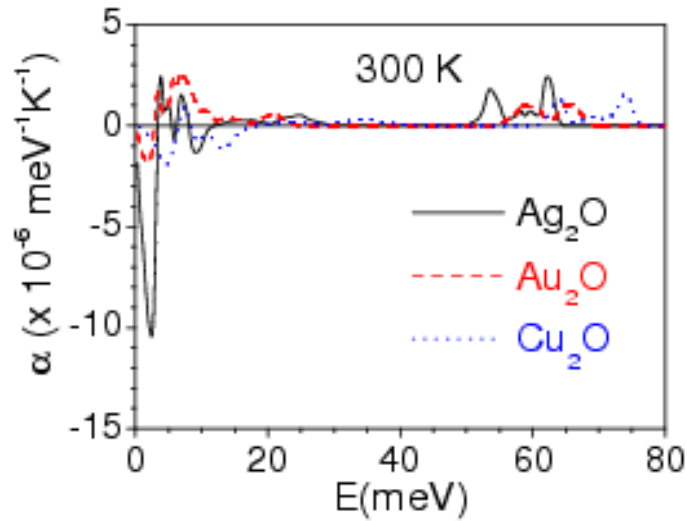

Fig. 45. Polarization vectors of selected phonon modes in M$_2$O (M=Ag, Au and Cu). The numbers after the wave vector (Γ, X, M and R) gives the Grüneisen parameters of Ag$_2$O, Au$_2$O and Cu$_2$O respectively. The lengths of arrows are related to the displacements of the atoms. Key: M, grey spheres; O, brown spheres[2].

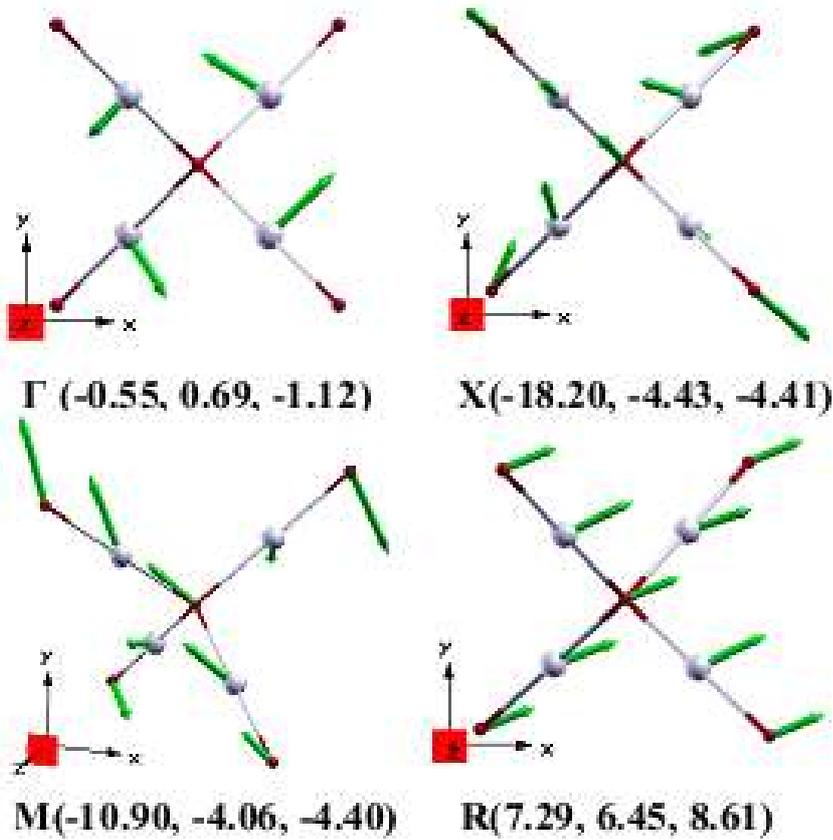

Fig. 46. Temperature dependence of lattice parameter from experimental data in Ref. [84], quasiharmonic calculations and MD calculations, expressed as the relative changes with respect to their 40 K values[68].

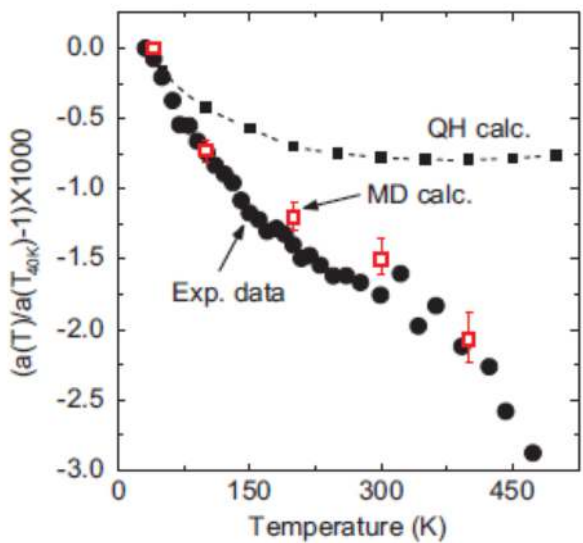

Fig. 47. The calculated charge density for $Ag_2O$, $Cu_2O$ and $Cu_2O$ in (011) plane. M (Ag, Au, Cu) are the central atoms while rest are oxygen's. The b and c-axes are in the horizontal and vertical directions respectively, while a-axis is out of plane[2].

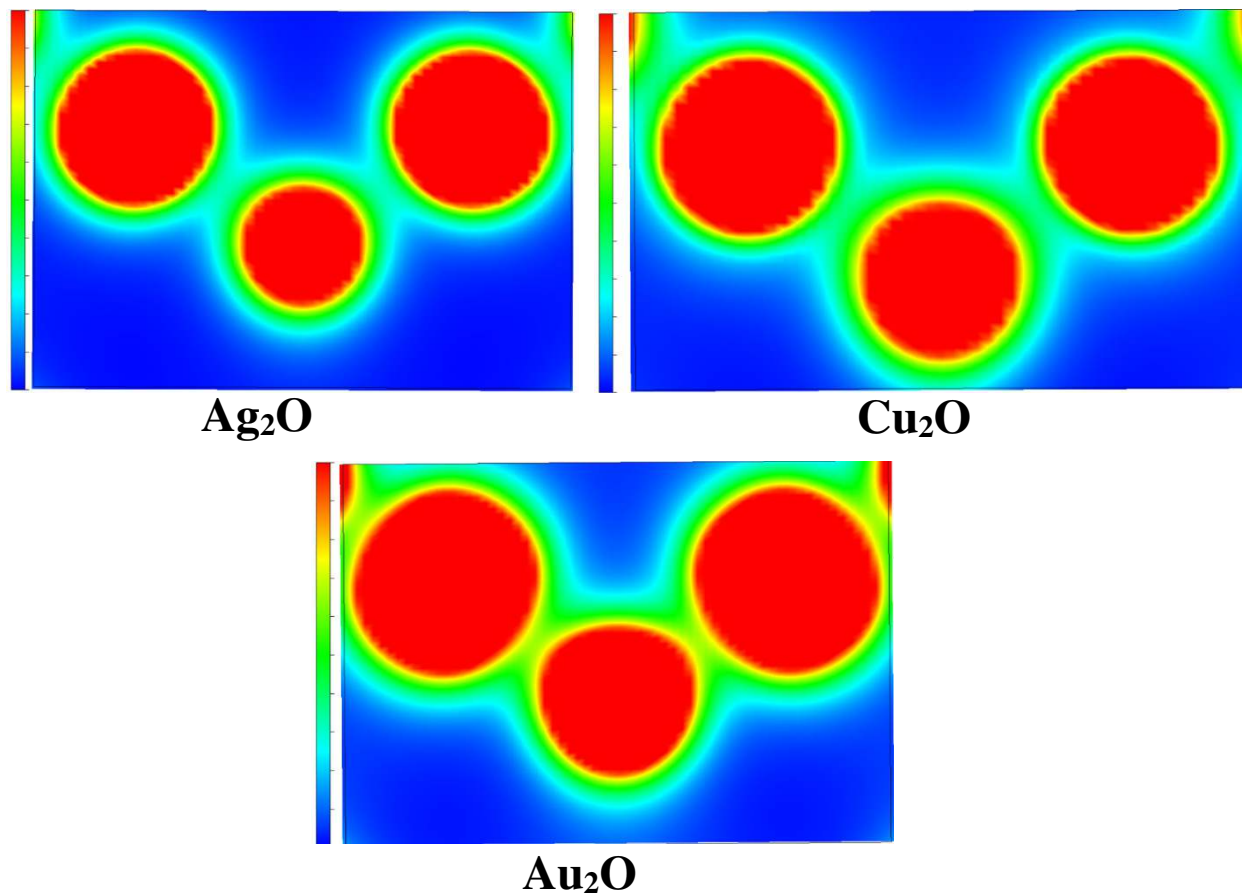

Fig. 48. Schematic representation of the unit cell structure of β-eucryptite, Li1(Red) occupies A-type channel while Li2 & Li3 ( Blue& Green) occupy S-type channels, projected along different axes. $AlO_4$ and $SiO_4$ tetrahedra are in blue and red color respectively[16].

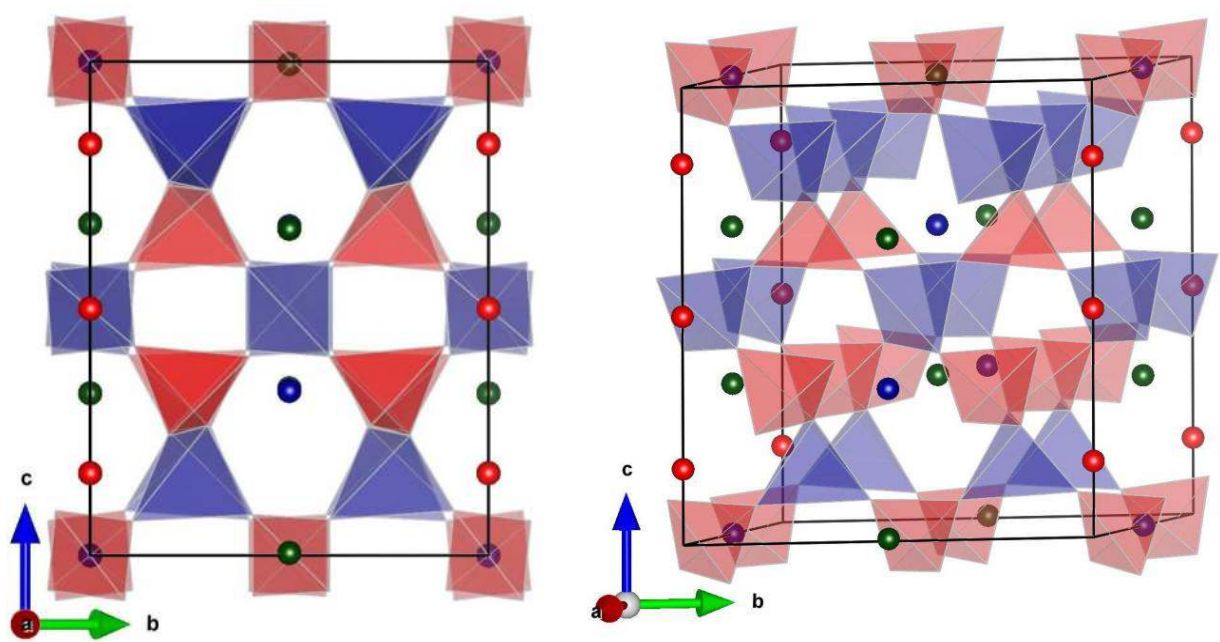

Fig. 49. Experimental and computed neutron weighted phonon density of states of β-eucryptite. The calculated neutron weighted partial phonon density of states of various atoms is also shown[16].

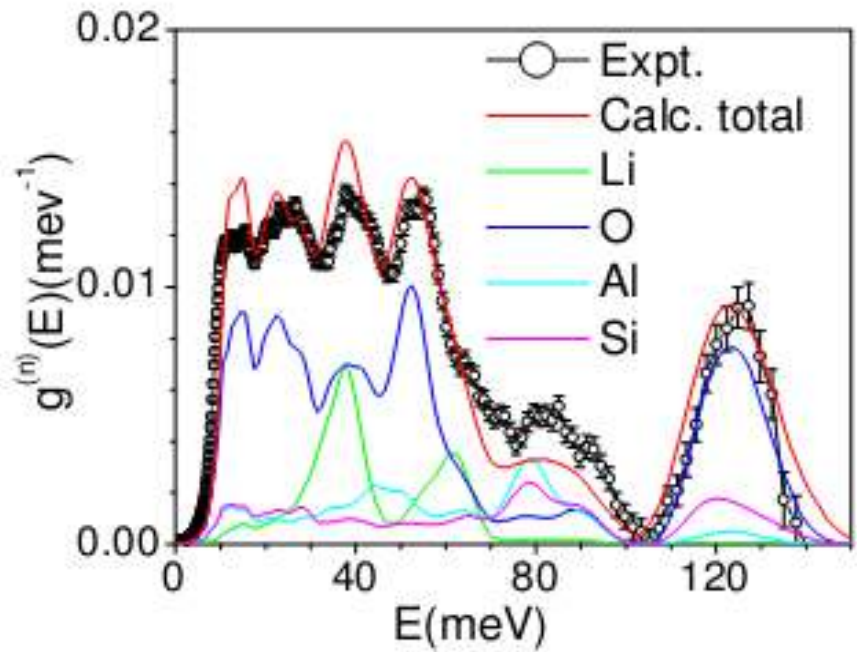

Fig. 50. Calculated phonon density of states of various atoms of β-eucryptite from ab-initio calculations[16].

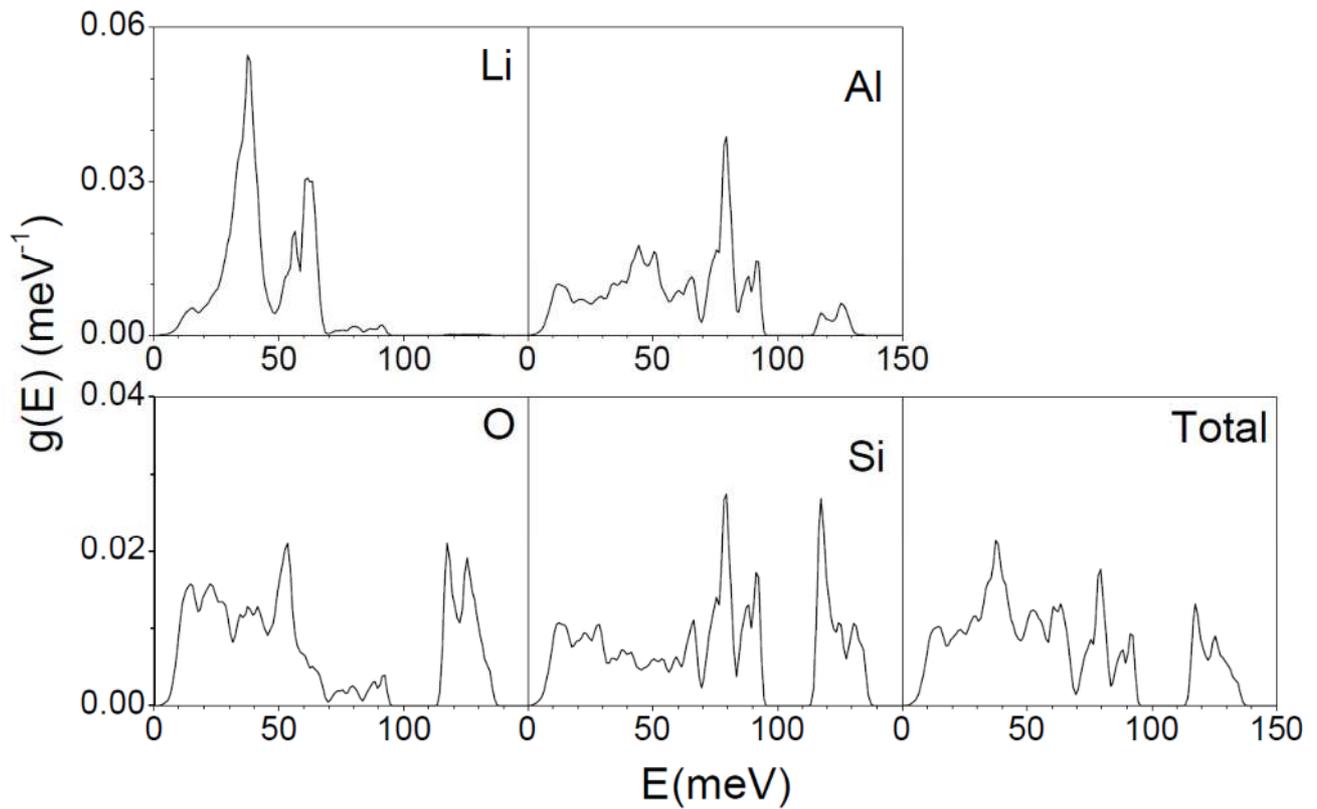

Fig. 51. The calculated contribution to the mean squared amplitude of the various atoms of β-eucryptitearising from phonons of energy E at $T$=300 K[16].

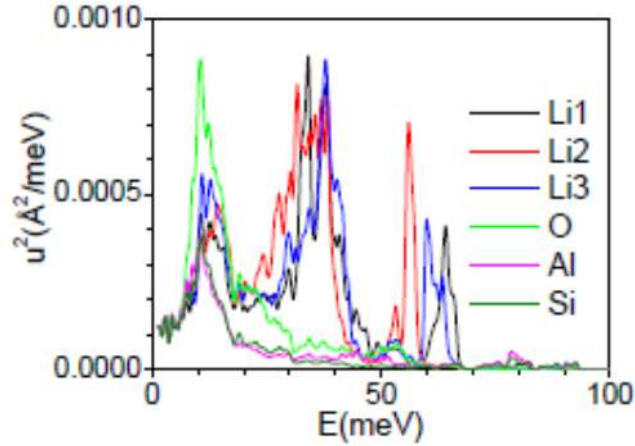

Fig. 52.(a) The calculated Grüneisen parameters ($\Gamma_l$, $l$=a, c) as obtained from anisotropic stress along '$a$' and '$c$' axes. (b) Linear thermal expansion coefficients as a function of temperature along the '$a$' and '$c$'-axes. (c) The variation of the lattice parameters ($l$= $a$ or $c$) with temperature from ab-inito calculations and experiments[197]. (d) The contribution of phonons of energy $E$ to the linear thermal expansion coefficients ($\alpha_a$ and $\alpha_c$) as a function of $E$, at 300 K, in the room temperature phase of β-eucryptite[16].

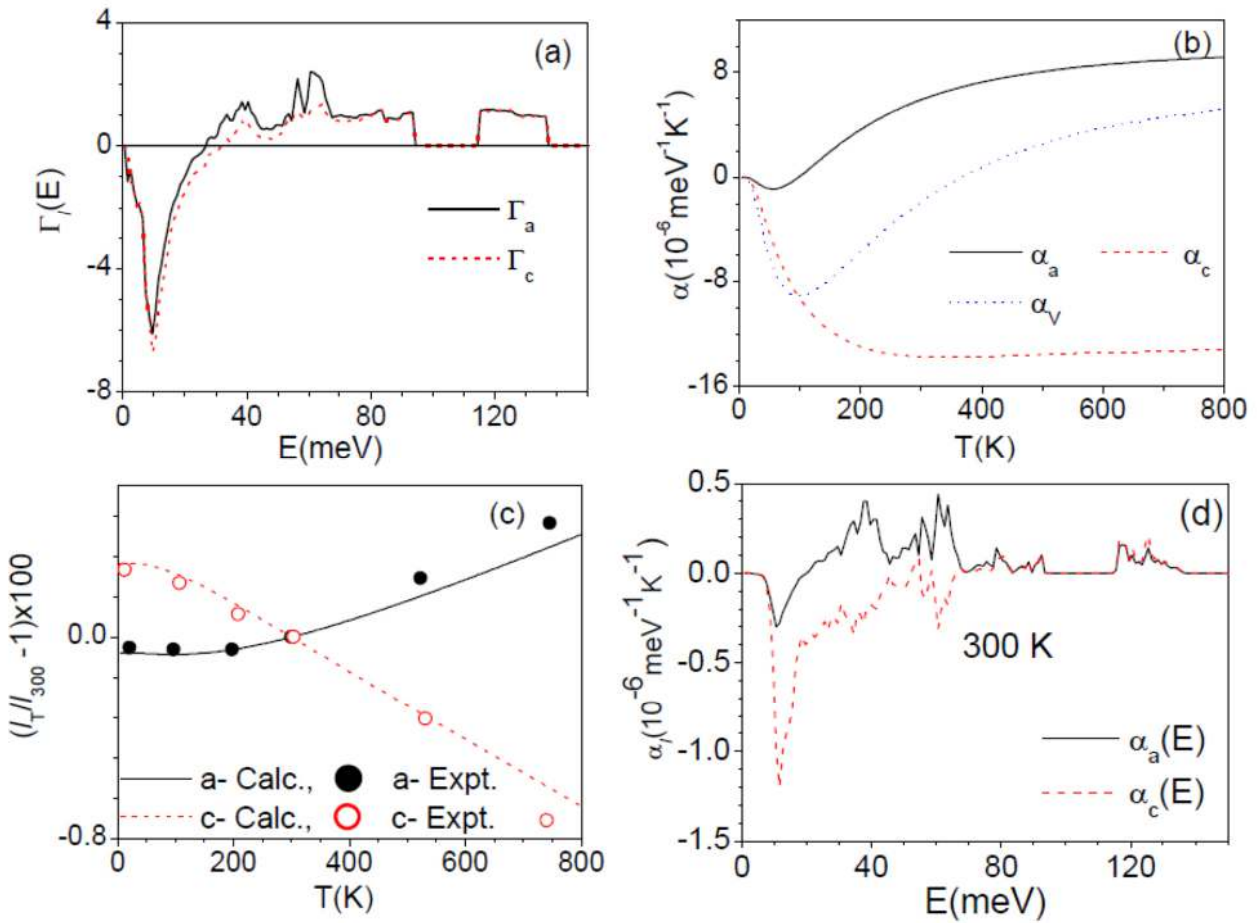

Fig. 53. Phonon dispersion curves of β-eucryptite, at ambient (solid line) and 0.5 GPa (dash line) pressures, along the high symmetry directions in the Brillion zone of hexagonal unit cell. The Bradley-Cracknell notation is used for the high-symmetry points:Γ (0,0,0), M(1/2,0,0), K(1/3,1/3,0), A (0, 0, 1/2), H (1/3,1/3,1/2) and L (1/2,0,1/2)[16].

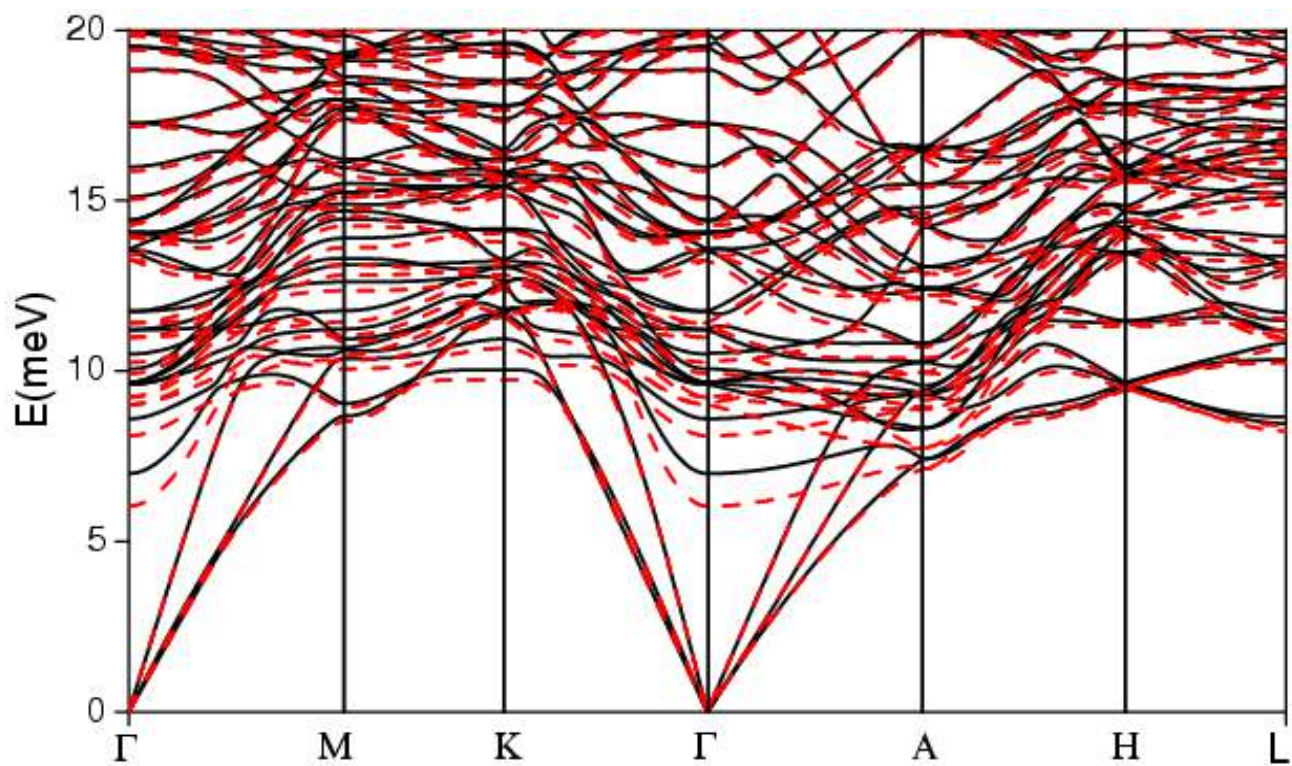

Fig. 54. Schematic representation of the polarization vectors of selected zone centre optic phonon modes in β- eucryptite. The first, and second and third values specified below each plot indicate the phonon energy and Grüneisen parameters, respectively. The tetrahedral units around Al and Si are color-coded by blue and red respectively, while the Li atoms are represented by a green color[16].

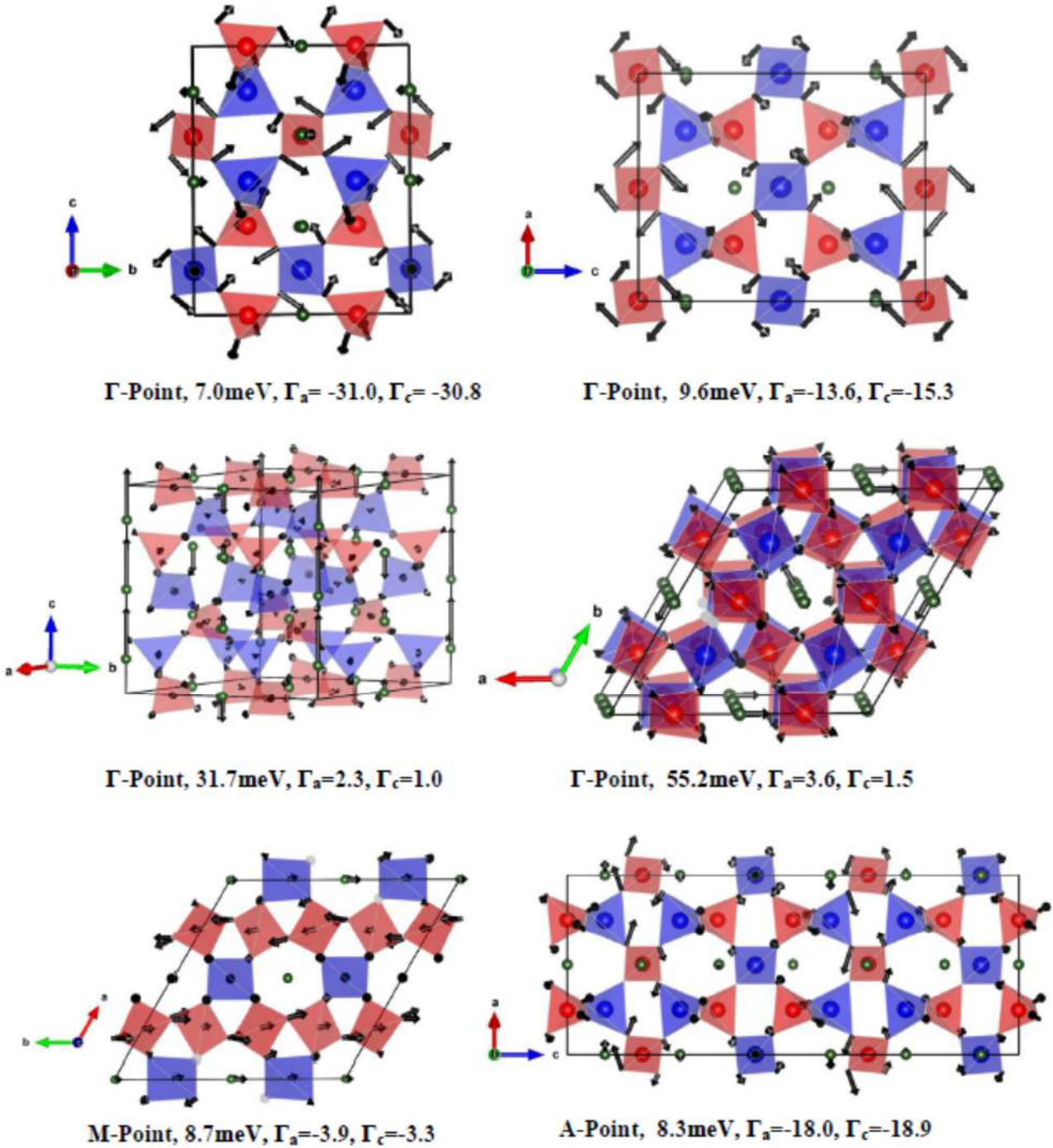

Fig. 55. Comparison between the calculated (lines) and experimental data (symbols)[204] of phonon dispersion relation for $ReO_3$[12].

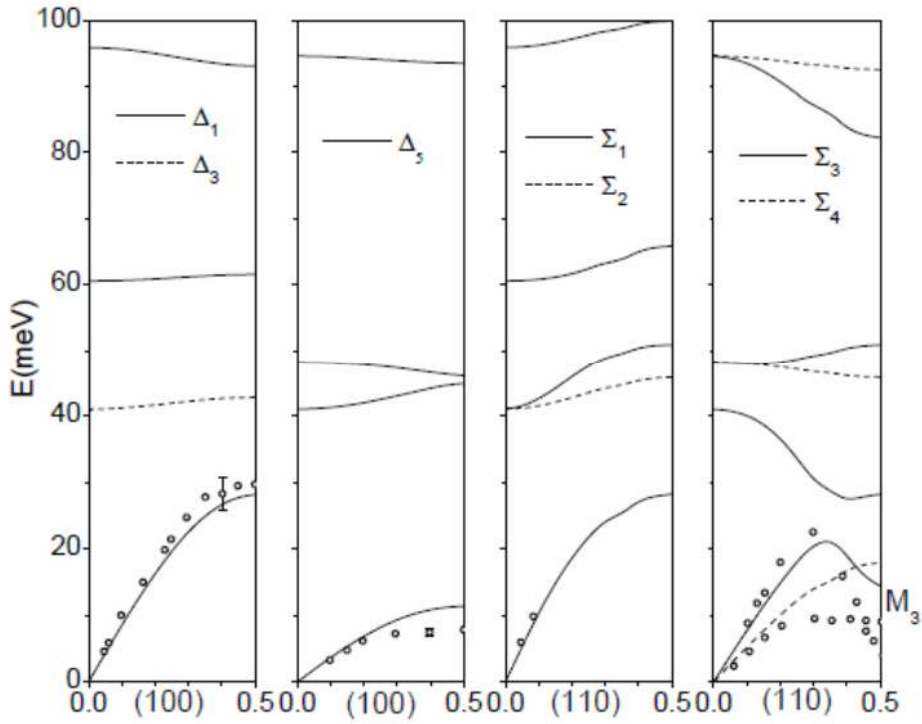

Fig. 56. Calculated (Left) Grüneisen parameter $\Gamma(E)$ averaged over phonons of energy E and (Right) volume thermal expansion coefficient $\alpha_V$ in $ReO_3$[12].

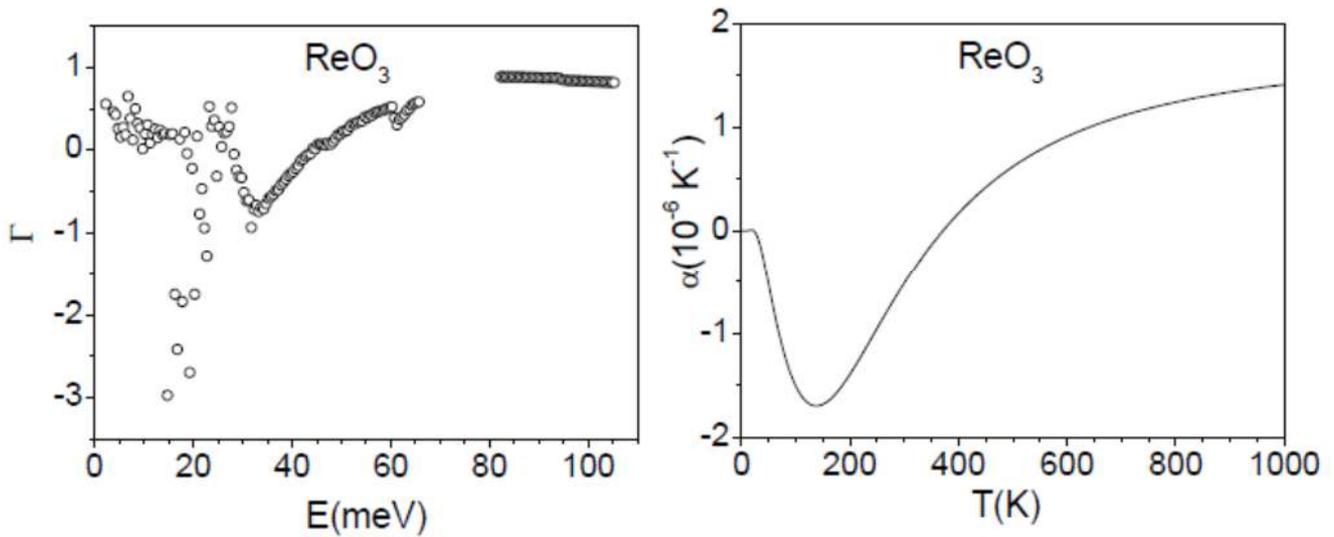

Fig. 57. (Left) Comparison between the calculated and experimental thermal expansion behavior of ReO$_3$. (Right)Contribution of phonons of energy E to the volume thermal expansion as a function of E at 100 K in ReO$_3$[12].

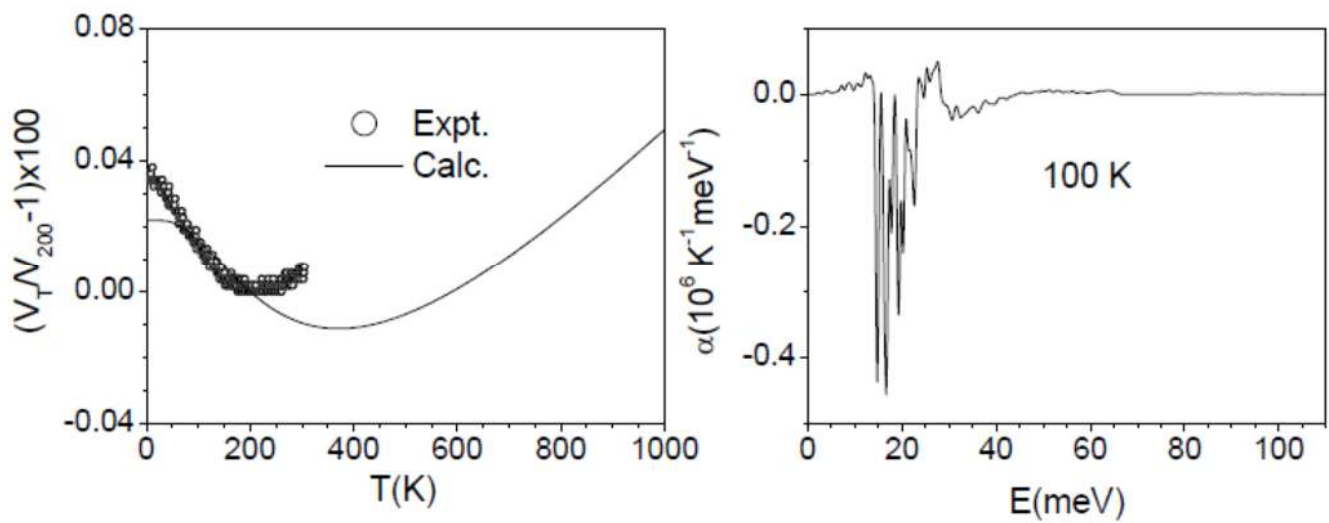

Fig. 58.Polarization vector of the M3 mode involving rotation of ReO$_6$ octahedra in the a – b plane [12].

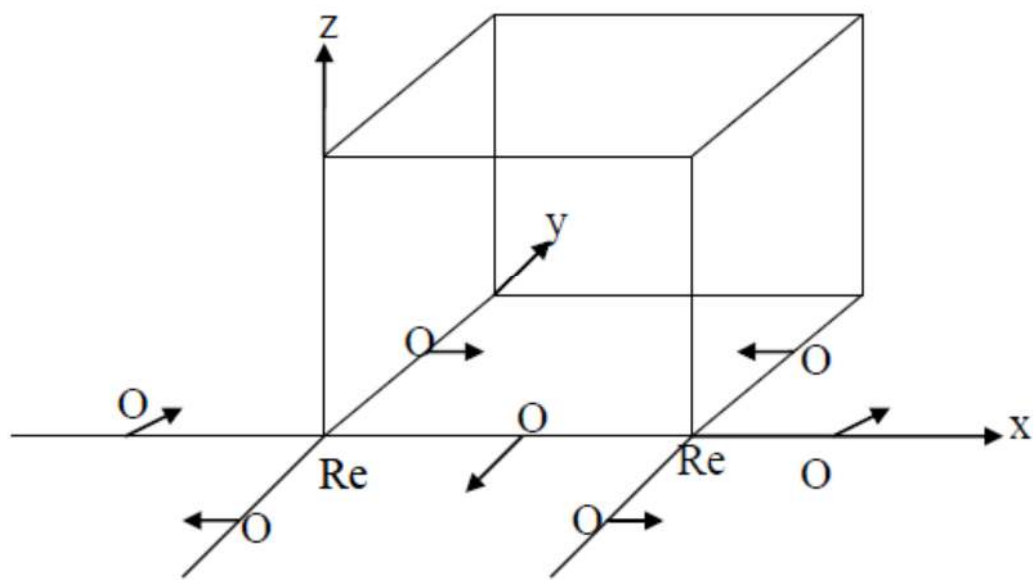

Fig. 59. Comparison between measured (solid symbols) and calculated (shaded area) GDOSs for ReO$_3$ sample at 310 K. Contributions from Re and O vibrations to GDOS are shown by solid and dashed curves, respectively[207].

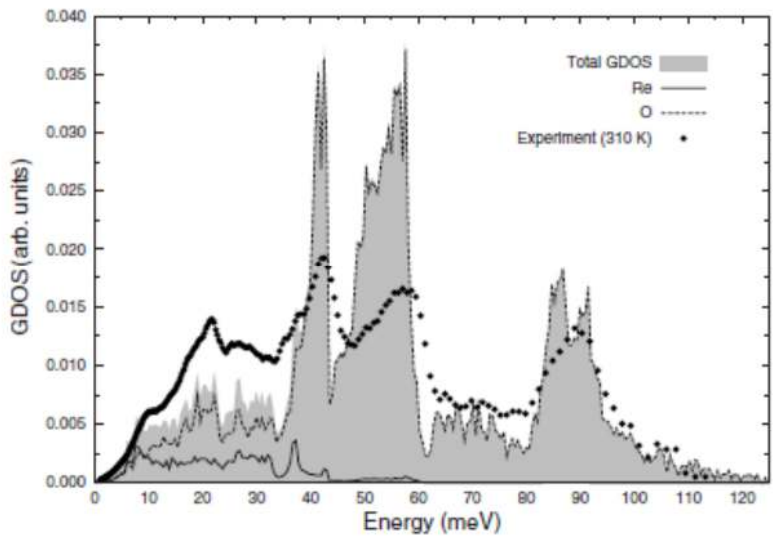

Fig. 60. Ab-initio calculated phonon-dispersion curves in ReO$_3$ at ambient pressure. Experimental data from inelastic neutron scattering [204] measured at room temperature are shown as open symbols[207].

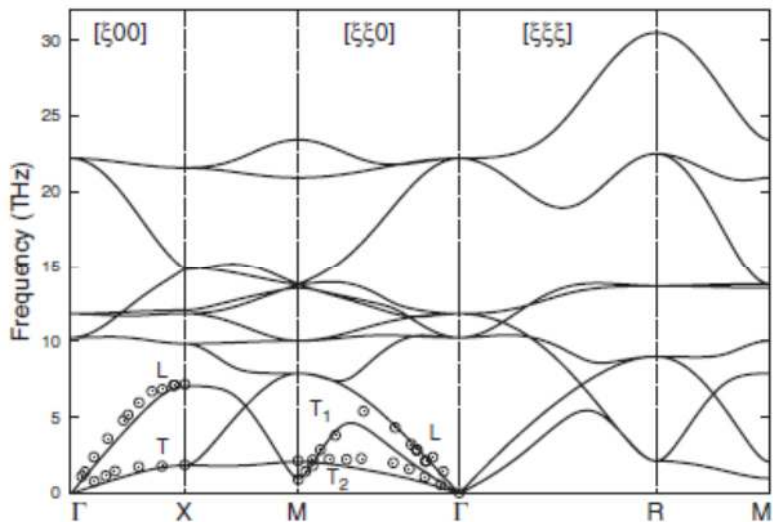

Fig. 61.Comparison of the calculated and experimental pressure dependence of M3 mode in ReO$_3$[13].

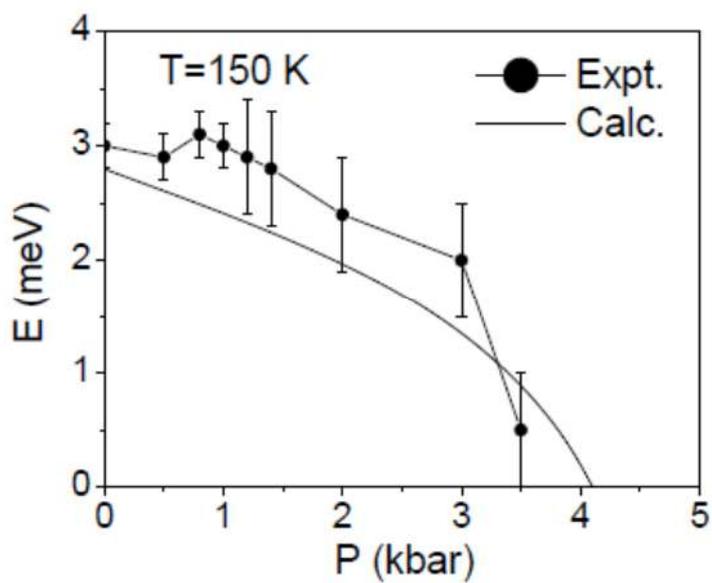

Fig. 62. The calculated potential of M3 mode in ReO$_3$[13].

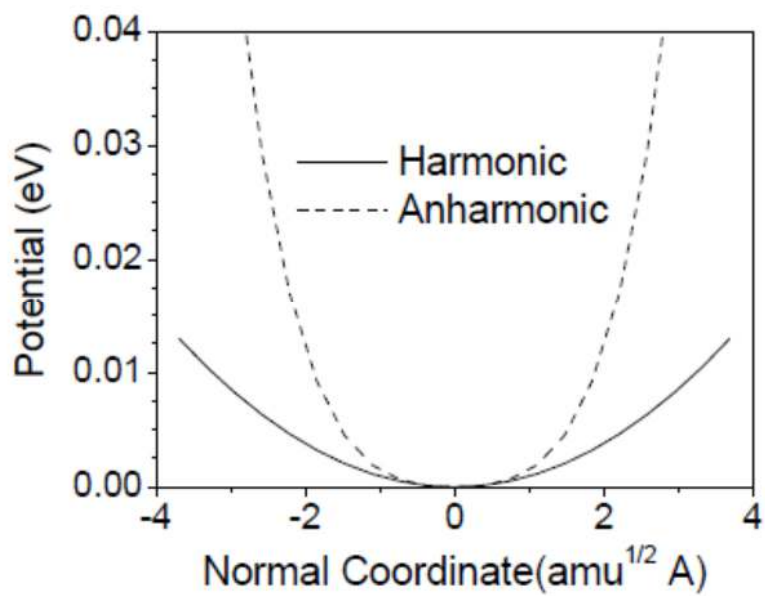

Fig. 63. Comparison of the calculated and experimental temperature dependence of M3 mode in $ReO_3$. At T=0 K the harmonic value of M3 mode is 2.8 meV[13].

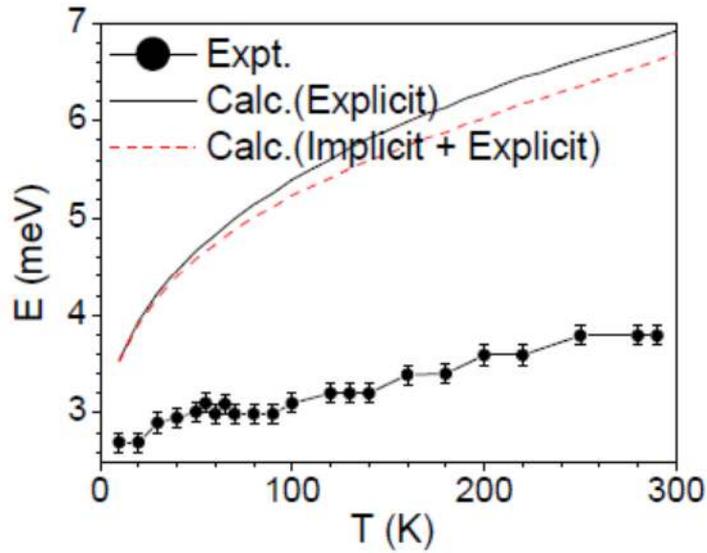

Fig. 64. Neutron-weighted phonon DOS for $ScF_3$ from incident energies of 118.7 (black lines), 79.5 (green dashed lines), and 30.0 meV (red or grey lines), scaled to conserve spectral areas and offset for clarity. Five vertical lines are aligned to peak centers at 7 K and labeled by numbers. Errors at the top are from counting statistics, and are similar at all temperatures[43].

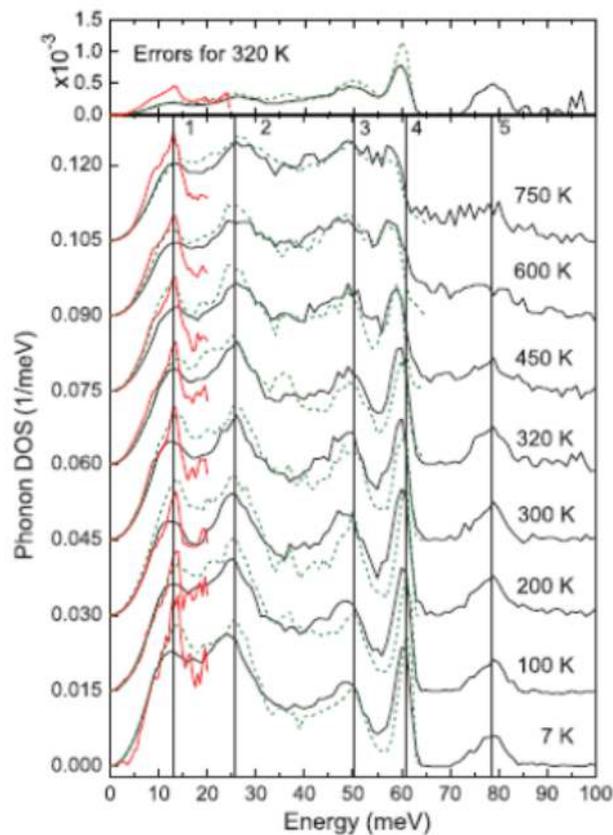

Fig. 65. Experimental [208] and calculated linear thermal expansion coefficients[43].

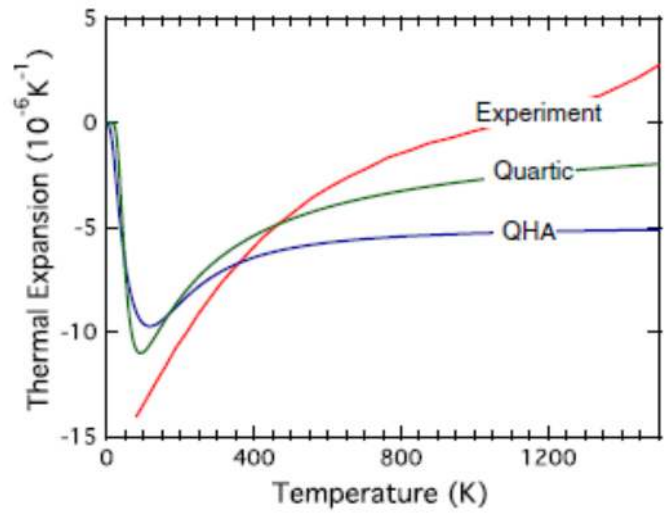

Fig. 66. Phonon mode R4 + , its frozen phonon potential, and the quadratic (harmonic) and quartic fits to the frozen phonon potential. The range of the quadratic fit is from -0.1 to 0.1 Å for the transverse displacements of F atoms[43].

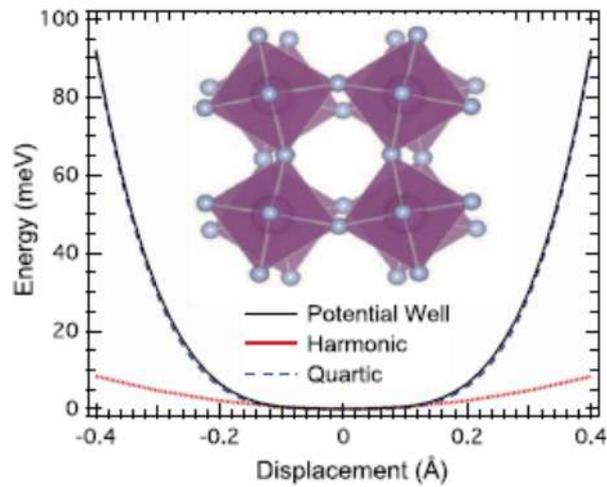

Fig. 67. Molecular dynamics calculated[62] temperature dependence of the reduced lattice constant of ScF$_3$ at pressures of $p = 0, \pm 1$ GPa and the experimental zero-pressure results [208].

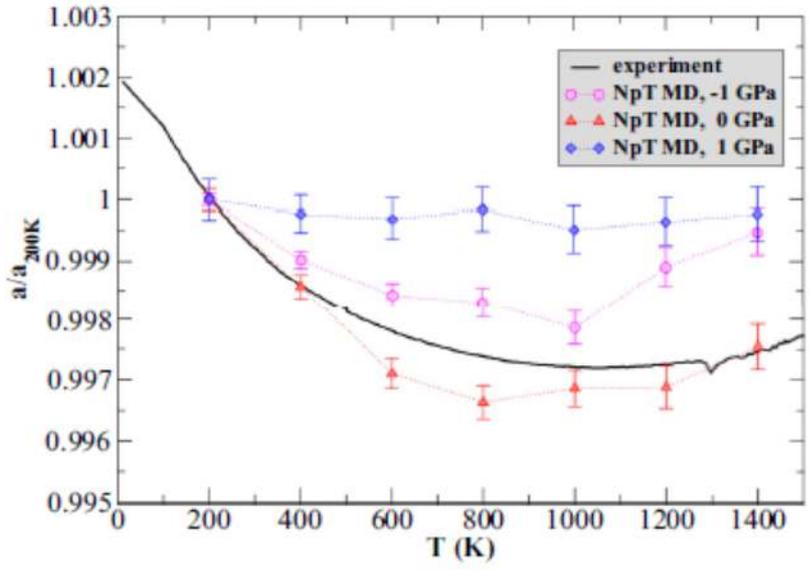

Fig. 68. Temperature dependence of the isothermal bulk modulus $K_T$ of ScF$_3$ as calculated from the MD simulations (black asterisks) and from static lattice relaxations (full symbols)[62]. The red horizontal line gives the experimental values [236].

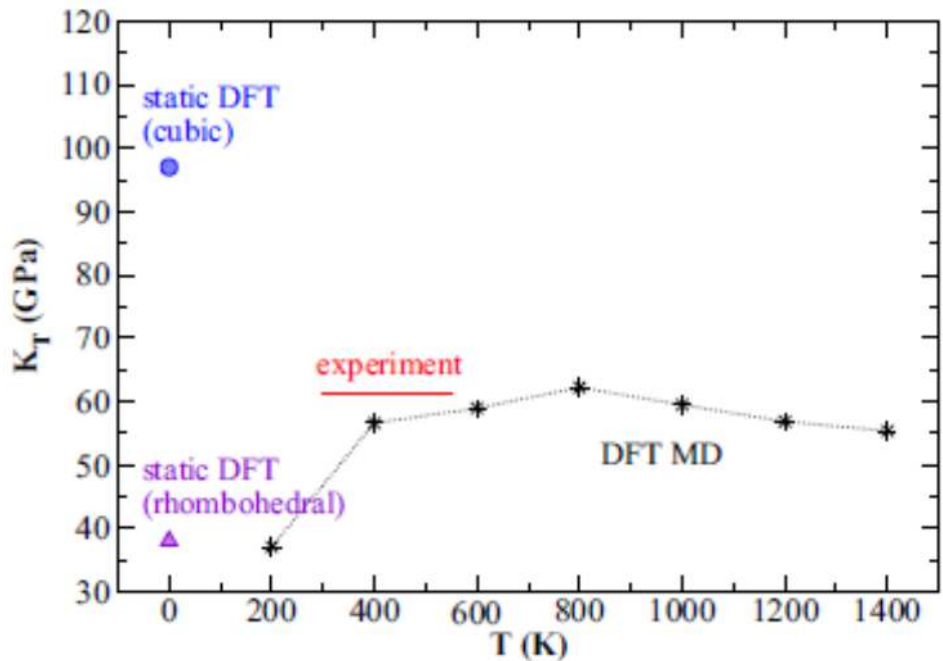

Fig. 69. (a) Lattice volume as a function of temperature determined on single-crystalline $ScF_3$. Inset: The lattice volume as determined on powder samples [208]. (b) The phonon mode dispersion with strong branch softening along the *M-R* cut. Black solid lines are a guide to the eye[211].

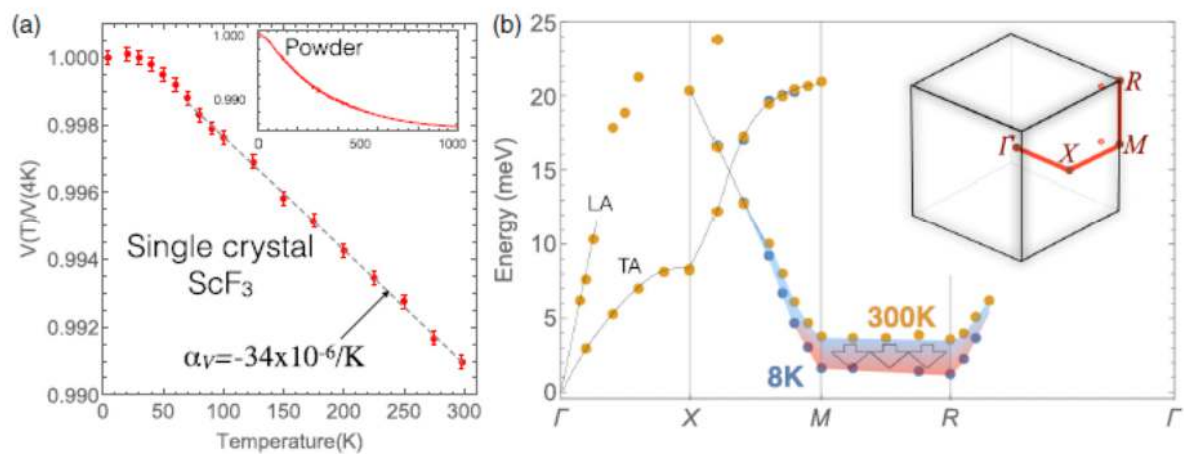

Fig. 70. The crystal structure of CuX (X=Cl, Br and I). Key: Cu, blue sphere; X, green sphere[143].

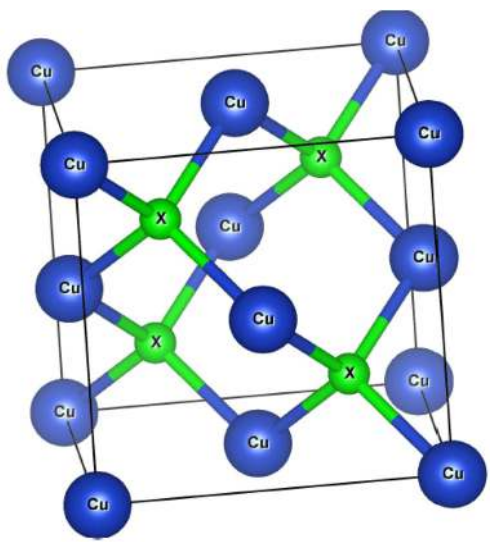

Fig. 71. The measured neutron inelastic spectra for CuX (Cl, Br and I) at 373 K, 473 K and 573 K[143].

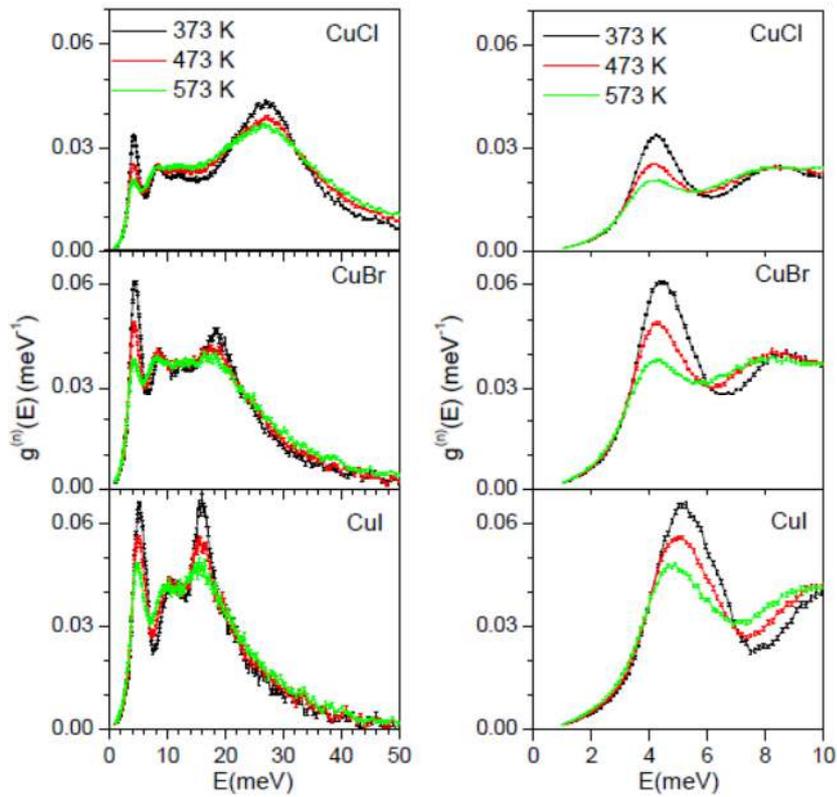

Fig. 72. The calculated partial density of states of various atom in CuX (Cl, Br and I)[143].

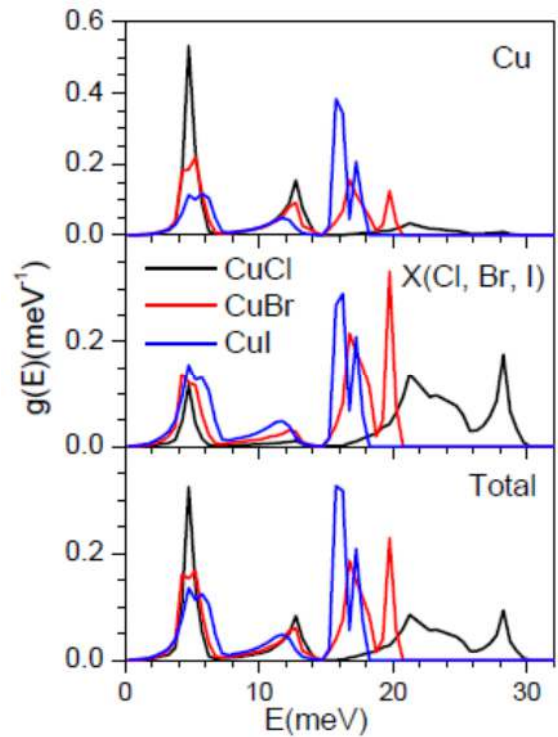

Fig. 73. The comparison between the experimental (373 K) and calculated phonon spectra of CuX (Cl, Br and I) including the multi-phonon contributions[143].

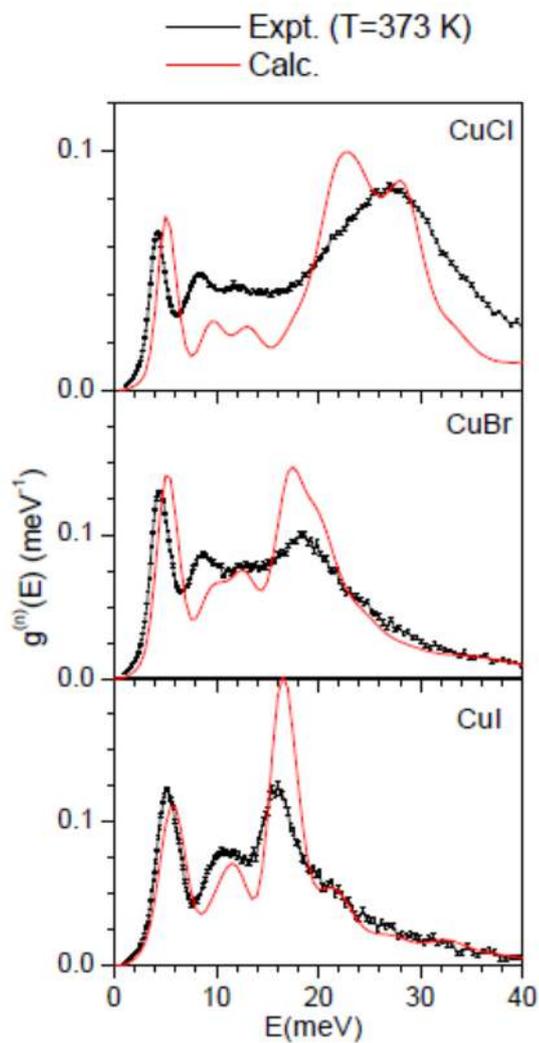

Fig. 74. The calculated (line) and experimental[216, 217](symbols) specific heat in CuX (Cl, Br and I)[143].

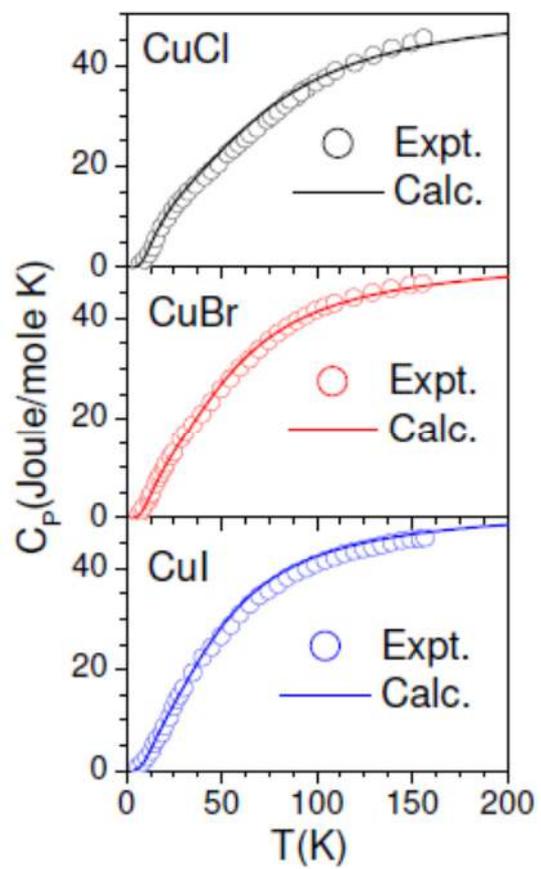

Fig. 75. The comparison between calculated (lines) phonon dispersion and measured phonon dispersion relation[213-215] (symbols) of CuX (X=Cl, Br and I)[143].

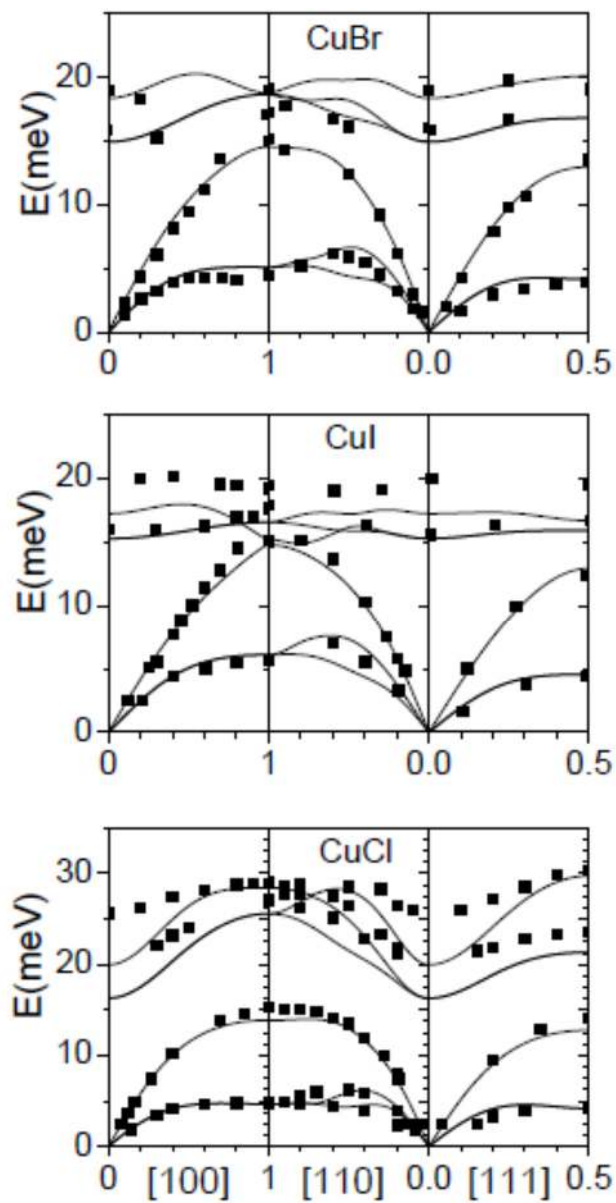

Fig. 76. The calculated dispersion relation along various high symmetry directions of CuX (Cl, Br and I) at ambient (black) and high pressures (red). The Bradley-Cracknell notation is used for the high-symmetry points. Γ(0 0 0), X (1 0 0), L( 0.5 0.5 0.5) and W(0.5 1 0) in the notation of cubic unit cell[143].

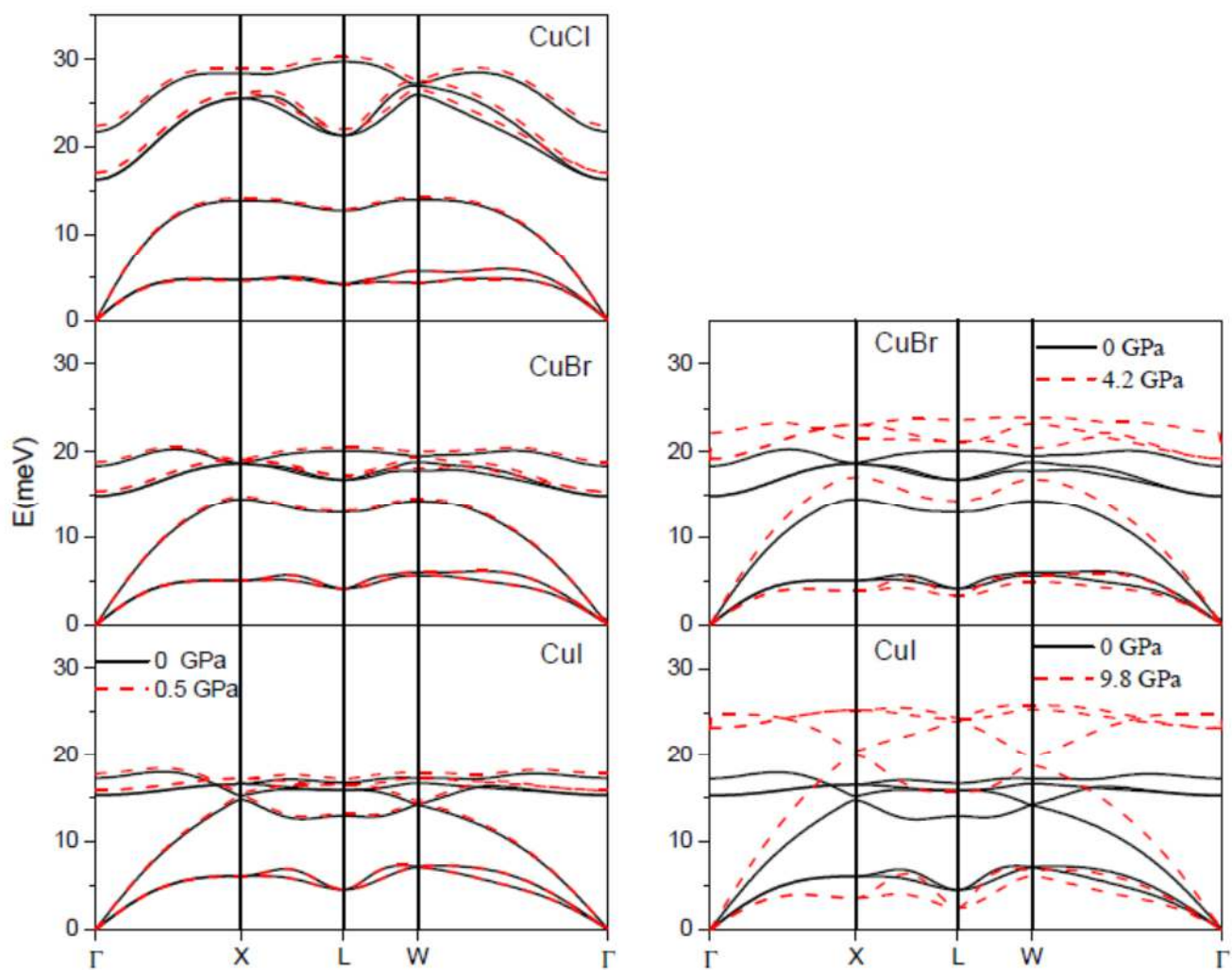

Fig. 77. The calculated Grüneisen parameters of CuX (Cl, Br and I) as obtained from pressure dependence of phonon frequency[143].

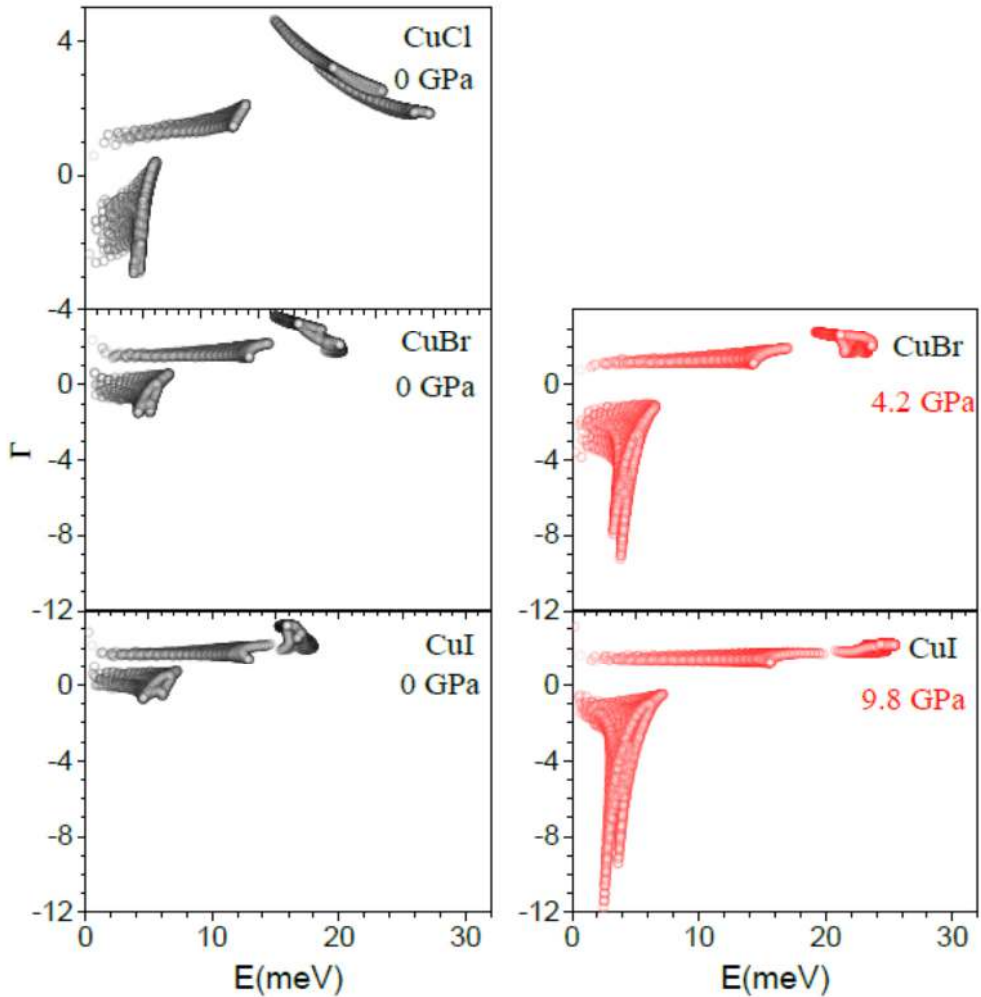

Fig. 78. The calculated volume thermal expansion coefficient as a function of temperature of CuX (Cl, Br, I) at (a) ambient pressure and (b) high pressures[143].

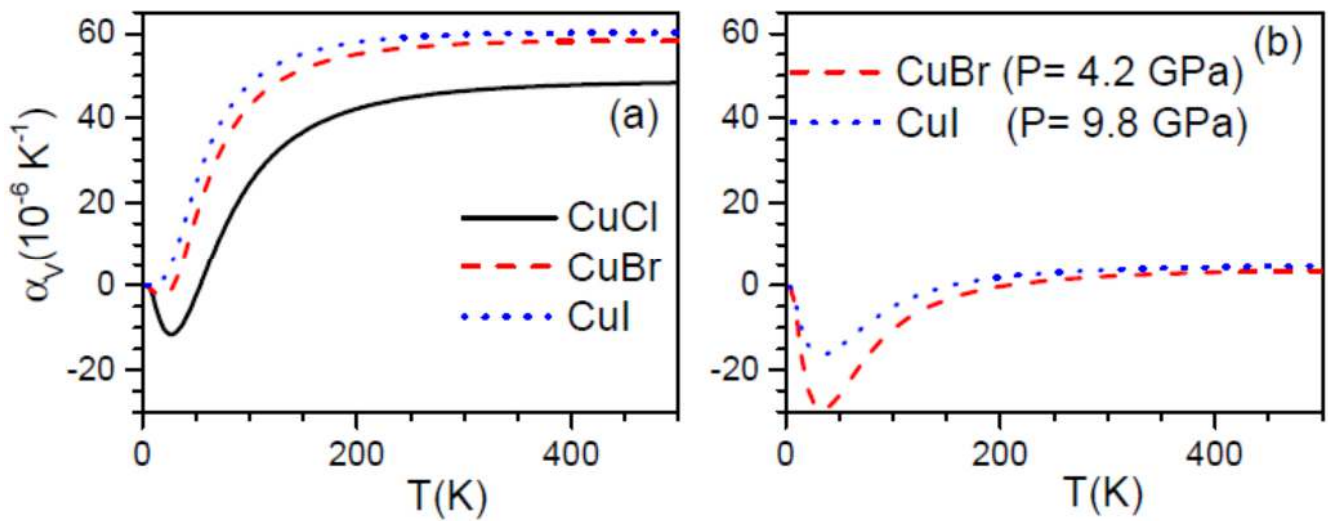

Fig. 79. The calculated and experimental[225] thermal expansion behavior of CuX (Cl, Br, I)[143].

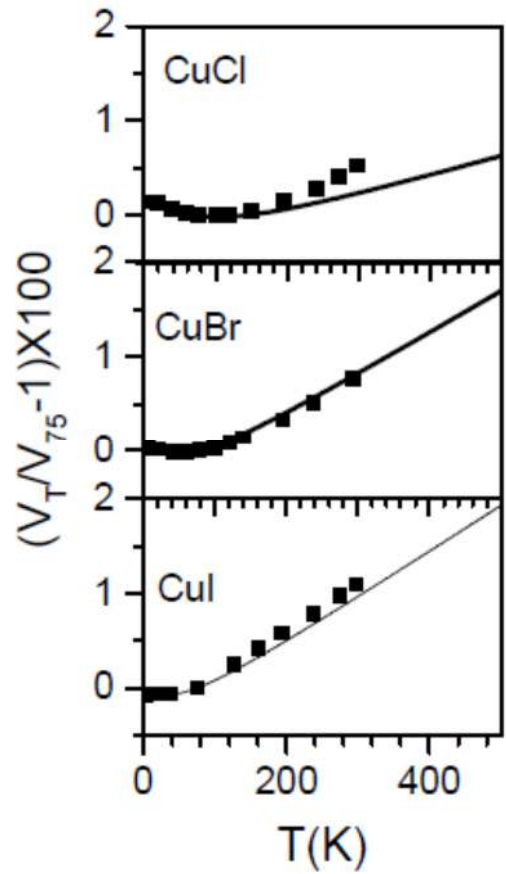

Fig. 80. The contribution of phonon of energy E to the volume thermal expansion at T=300 K in CuX (X= Cl, Br, I)[143].

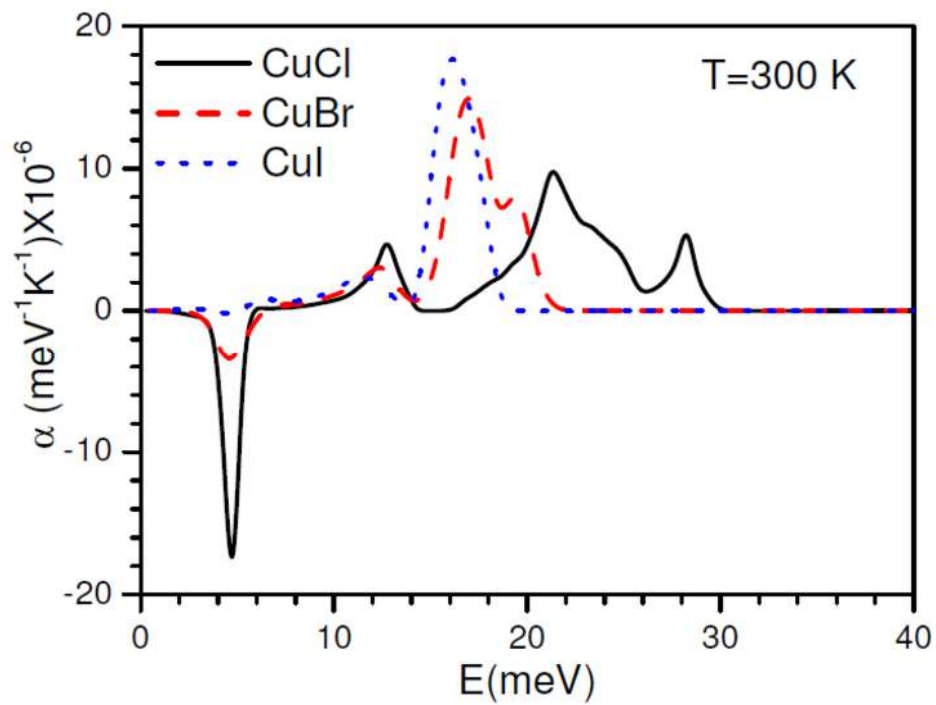

Fig. 81. The eigen vector of various phonon modes in CuX (X=Cl, Br and I). The relative displacements of various atoms are as follows:
X-point mode: (Cu:0.117, Cl:0.060 in CuCl), (Cu:0.106, Br:0.061 in CuBr) and (Cu:0.092, I:0.061 in CuI)
L-point mode: (Cu:0.118, Cl:0.058 in CuCl), (Cu:0.107 Br:0.059 in CuBr) and (Cu:0.093, I:0.060 in CuI)
Key: Cu, blue sphere; X, green sphere[143]

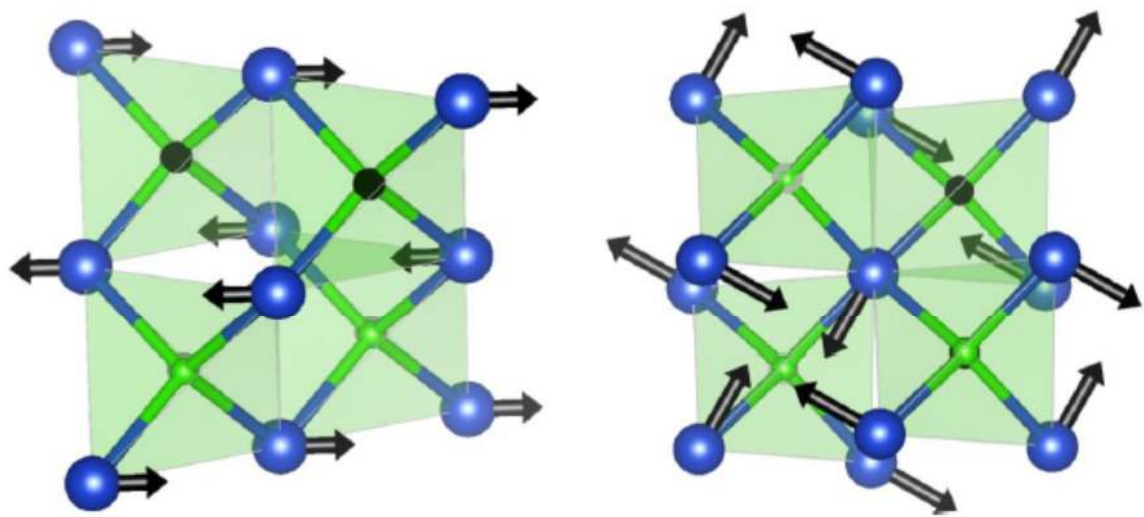

TA Phonon mode at X (1 0 0)    TA Phonon mode at L (1/2 1/2 1/2)

Fig. 82. The structure of Zn(CN)$_2$ in *P43m*. Key: Zn, grey spheres; C, green spheres; N, blue spheres[25].

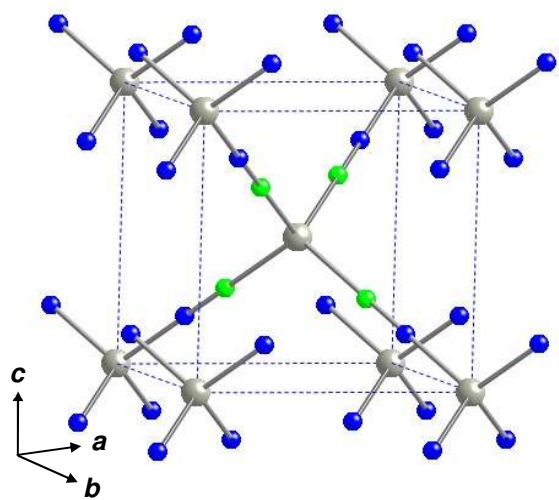

Fig. 83. The structure of part of one layer of Ni(CN)$_2$ with $D_{4h}$ symmetry. Key: Ni, grey spheres; C, green spheres; N, blue spheres[25].

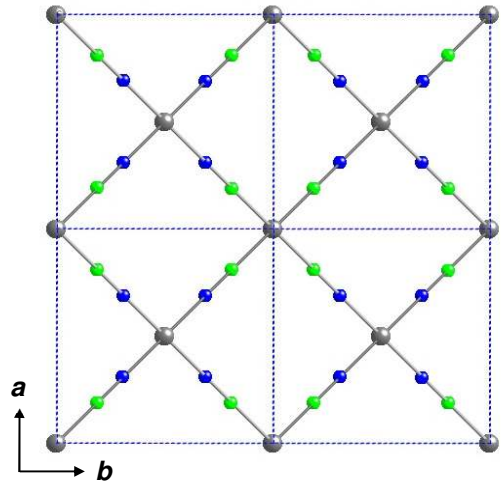

Fig. 84. The experimental phonon spectra, $g^{(n)}(E)$ for Zn(CN)$_2$ as a function of pressure at fixed temperatures of 165 K and 225 K: ambient pressure (full line), 0.3 kbar (dotted line), 1.9 kbar (dashed line), and 2.8 kbar (dash-dotted line)[14].

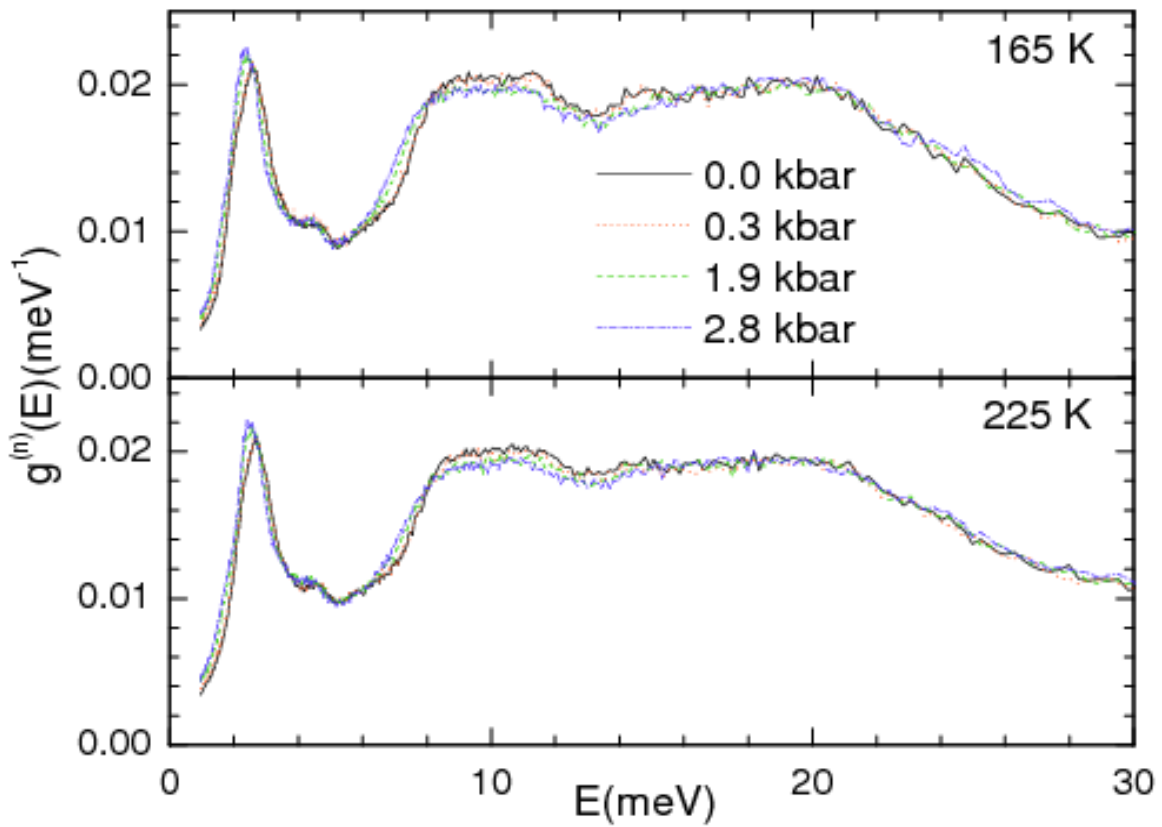

Fig. 85. The temperature dependence of the phonon spectra for Zn(CN)$_2$. The phonon spectra are measured with an incident neutron wavelength of 5.12 Å using the IN6 spectrometer at ILL. The calculated phonon spectra from *ab-initio* calculations are also shown. The calculated spectra have been convoluted with a Gaussian of FWHM of 10% of the energy transfer in order to describe the effect of energy resolution in the experiment carried out using the IN6 spectrometer[25].

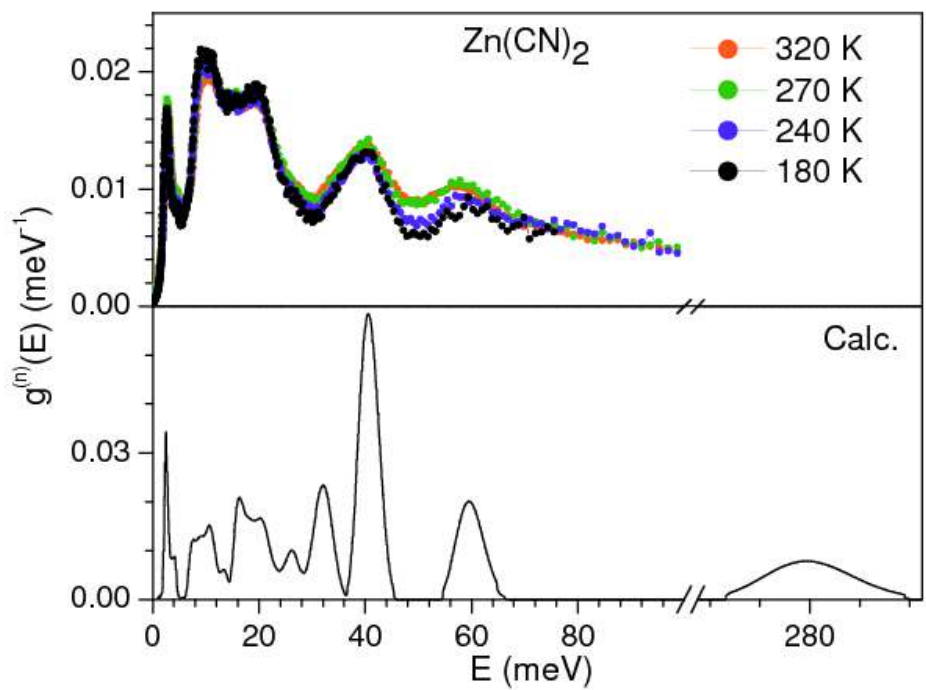

Fig. 86. The temperature dependence of the phonon spectra for Ni(CN)$_2$. The phonon spectra are measured with an incident neutron wavelength of 4.14 Å using the IN6 spectrometer at ILL. The calculated phonon spectra from *ab-initio* calculations are also shown. The calculated spectra have been convoluted with a Gaussian of FWHM of 10% of the energy transfer in order to describe the effect of energy resolution in the experiment carried out using the IN6 spectrometer[25].

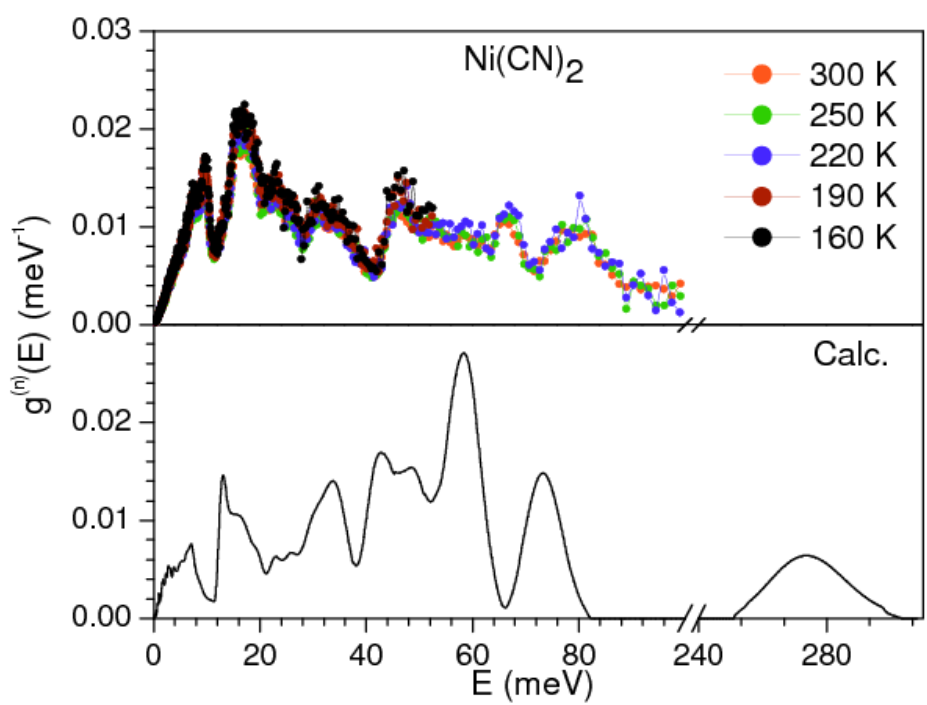

Fig. 87. Comparison of the experimental phonon spectra for Zn(CN)$_2$ and Ni(CN)$_2$[25].

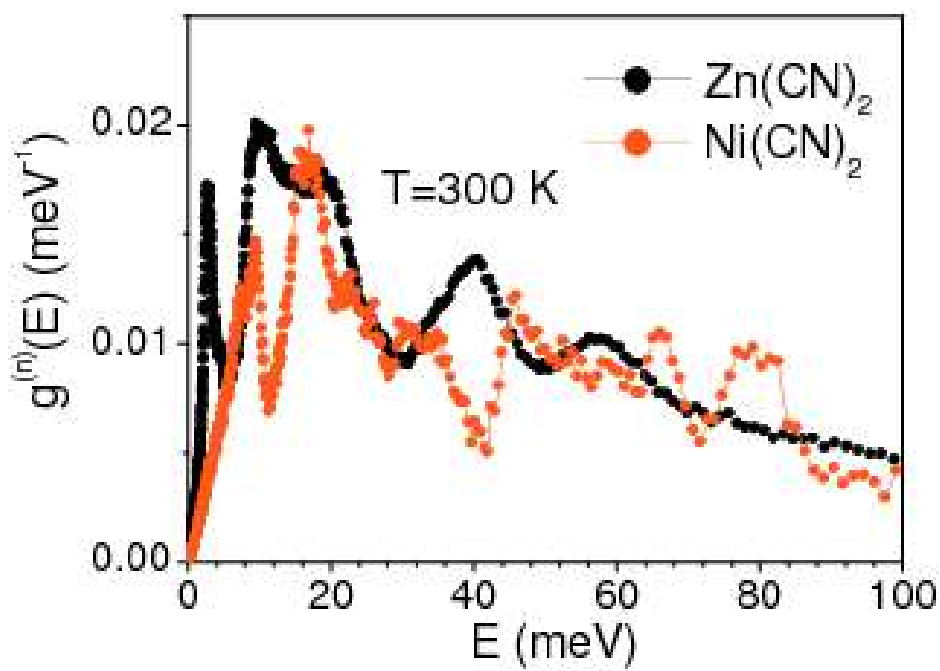

Fig. 88. The calculated partial density of states for the various atoms in Zn(CN)$_2$ and Ni(CN)$_2$[25].

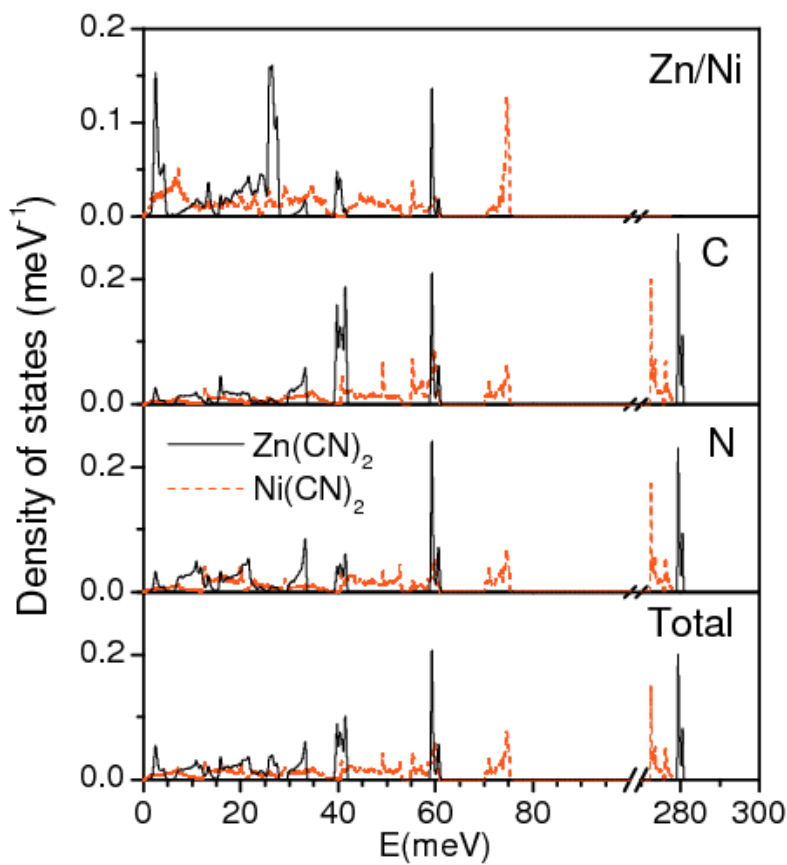

Fig. 89. The calculated phonon dispersion curves for Zn(CN)$_2$ and Ni(CN)$_2$. The Bradley-Cracknell notation is used for the high-symmetry points along which the dispersion relations are obtained. Zn(CN)$_2$: $\Gamma=(0,0,0)$; $X=(1/2,0,0)$; $M=(1/2,1/2,0)$ and $R=(1/2,1/2,1/2)$. Ni(CN)$_2$: $\Gamma=(0,0,0)$; $X(1/2,0,0)$ and $M(1/2,1/2,0)$. In order to expand the *y*-scale, the sets of four and three dispersion-less modes, respectively, in Zn(CN)$_2$ and Ni(CN)$_2$ due to the cyanide stretch at about 280 meV are not shown[25].

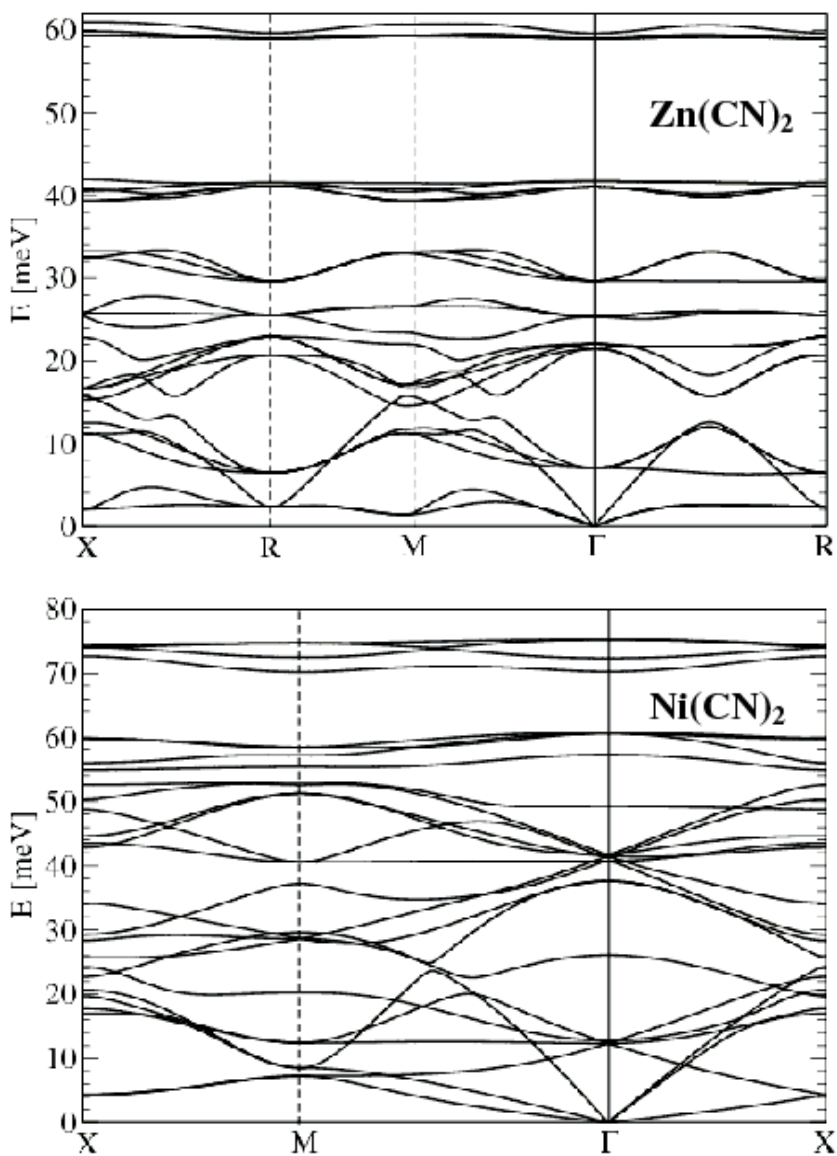

Fig. 90. The experimental Bose-factor corrected S(Q,E) plots for Zn(CN)$_2$ and Ni(CN)$_2$ at 180 K and 160 K, respectively. The values of S(Q,E) are normalized to the mass of the sample in the beam. For clarity, a logarithmic representation is used for the intensities. The measurements for Zn(CN)$_2$ and Ni(CN)$_2$ were performed with an incident neutron wavelength of 5.12 Å (3.12 meV) and 4.14 Å (4.77 meV), respectively[25].

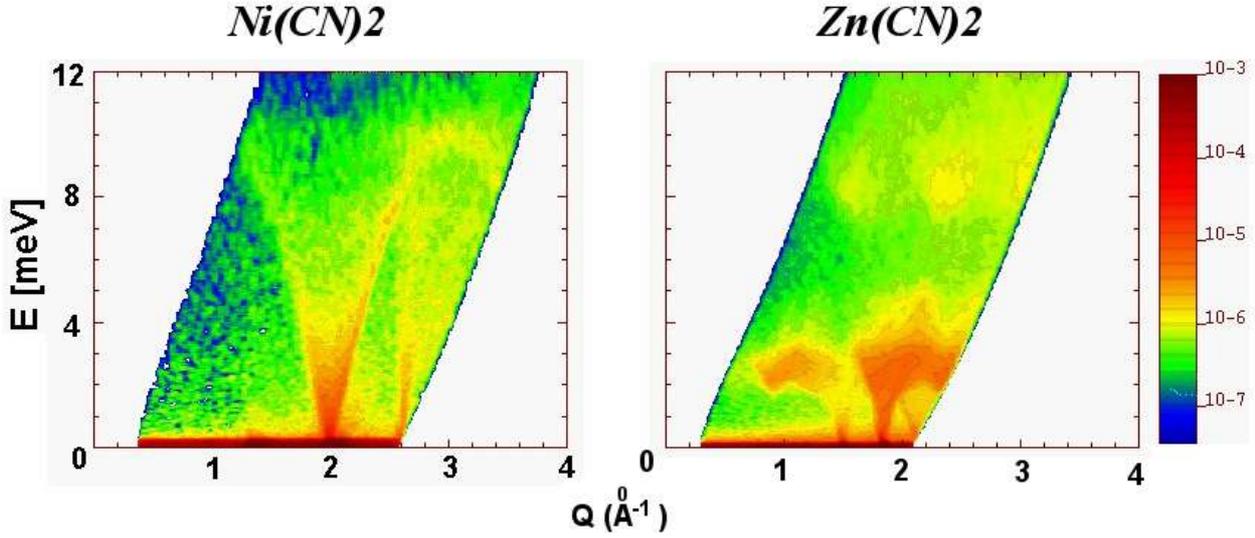

Fig. 91. The experimental $\frac{\Gamma_i}{B}$ as a function of phonon energy $E$ (averaged over the whole Brillouin zone). The $\frac{\Gamma_i}{B}$ values at 165 K, and 225 K has been determined using the density of states at $P = 0$ and 2.8 kbar (full line) which represents the average over the whole range in this study. The $\frac{\Gamma_i}{B}$ values derived from ab-initio calculations [28] are shown by open circles[14].

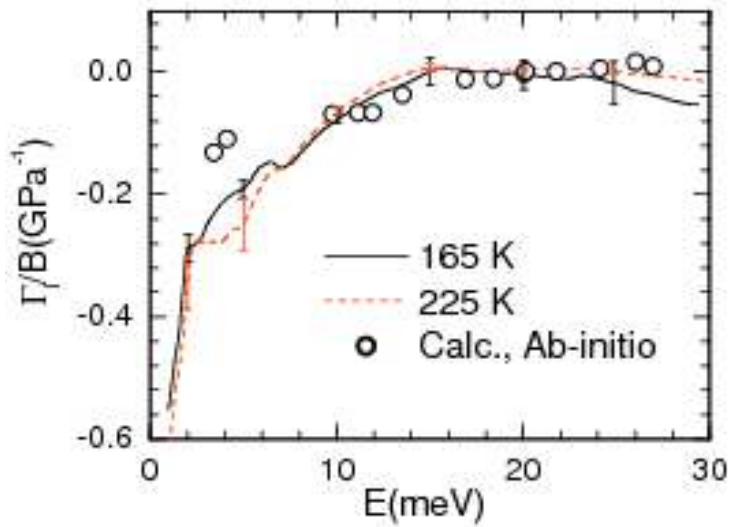

Fig. 92. (a) The comparison between the volume thermal expansion coefficient ($\alpha_V$) derived from the *ab-initio* calculations and experimental $\frac{\Gamma_i}{B}$ values [17] at 165 K. (b) The comparison between the volume thermal expansion derived from the *ab-initio* calculations (solid line), high-pressure inelastic neutron scattering experiment [14] (dashed line) and that obtained using X-ray diffraction [102] (open circles)[25].

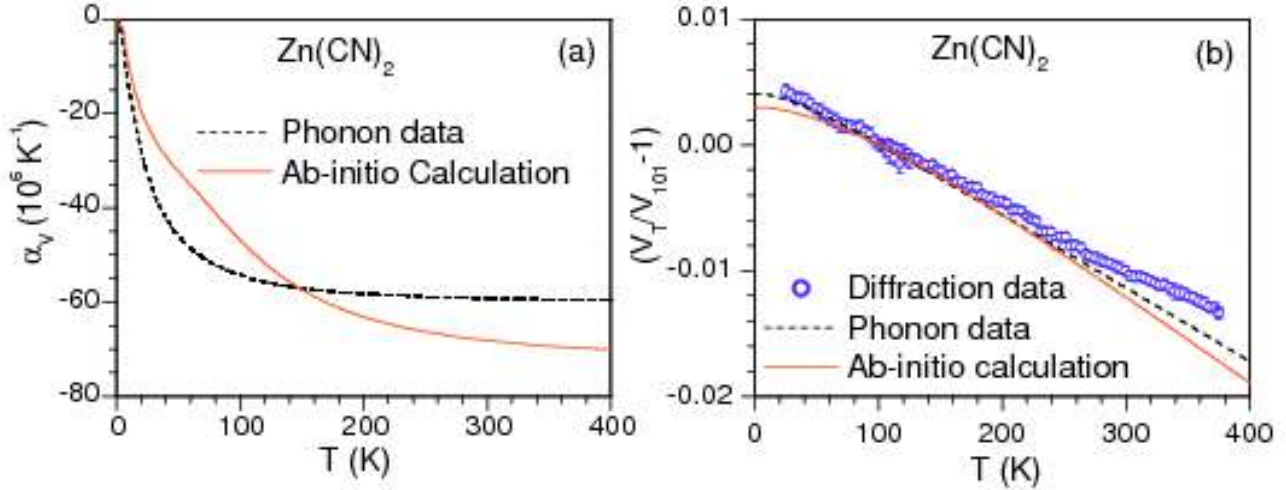

Fig. 93. The comparison between the experimental (at 165 K) and calculated $\frac{\Gamma_i}{B}$ plotted as a function of phonon energy $E$. The $\frac{\Gamma_i}{B}$ values derived from *ab-initio* calculations from Ref. [28] are shown by closed circles. The $\frac{\Gamma_i}{B}$ values represent the average over the whole Brillouin zone[25].

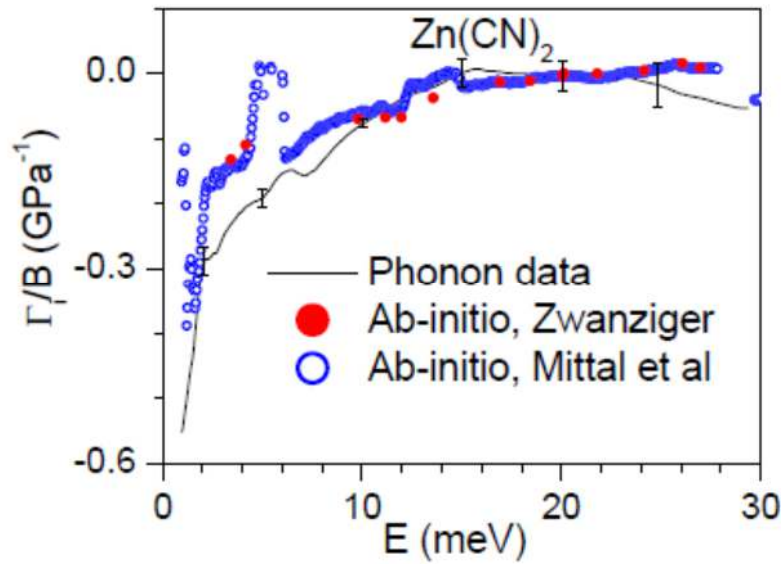

Fig. 94. Mode frequency vs pressure for the observed Raman modes in $Zn(CN)_2$. Open symbols: Results without pressure medium. The inset shows $\omega$ vs $P$ for all the 11 modes obtained from the phonon calculations[80].

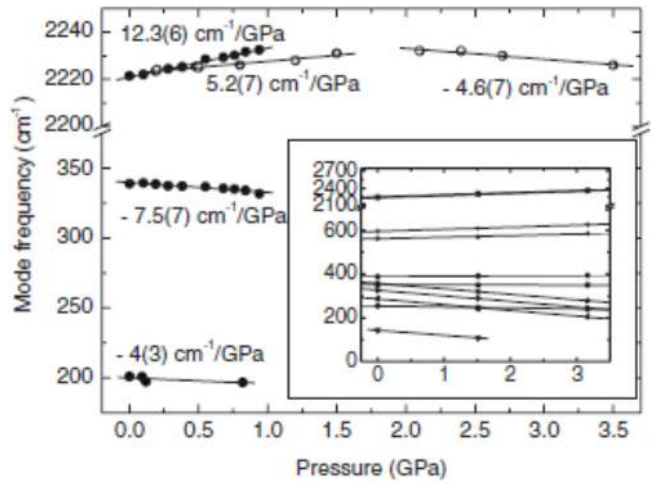

Fig. 95. The contribution of phonons of energy $E$ to the volume thermal expansion coefficient ($\alpha_V$) as a function of $E$ at 165 K in $Zn(CN)_2$ and $Ni(CN)_2$[25]. The contribution from phonons for $Zn(CN)_2$ is obtained from experimental $\dfrac{\Gamma_i}{B}$ as indicated in Ref. [14].

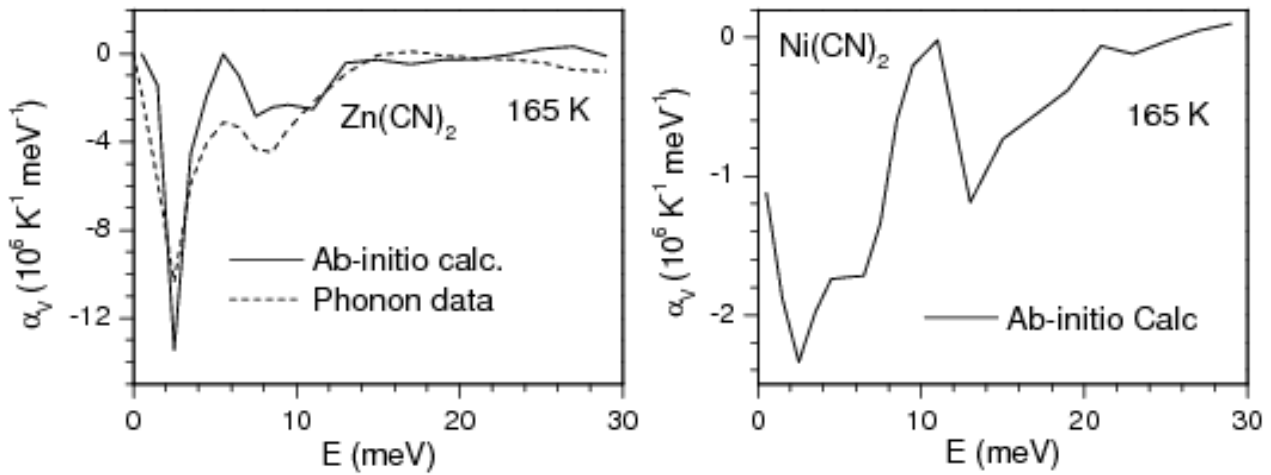

Fig. 96. (a) The calculated $\frac{\Gamma_i}{B}$ and (b) volume thermal expansion coefficient ($\alpha_V$) derived for Ni(CN)$_2$ from *ab-initio* calculations[25].

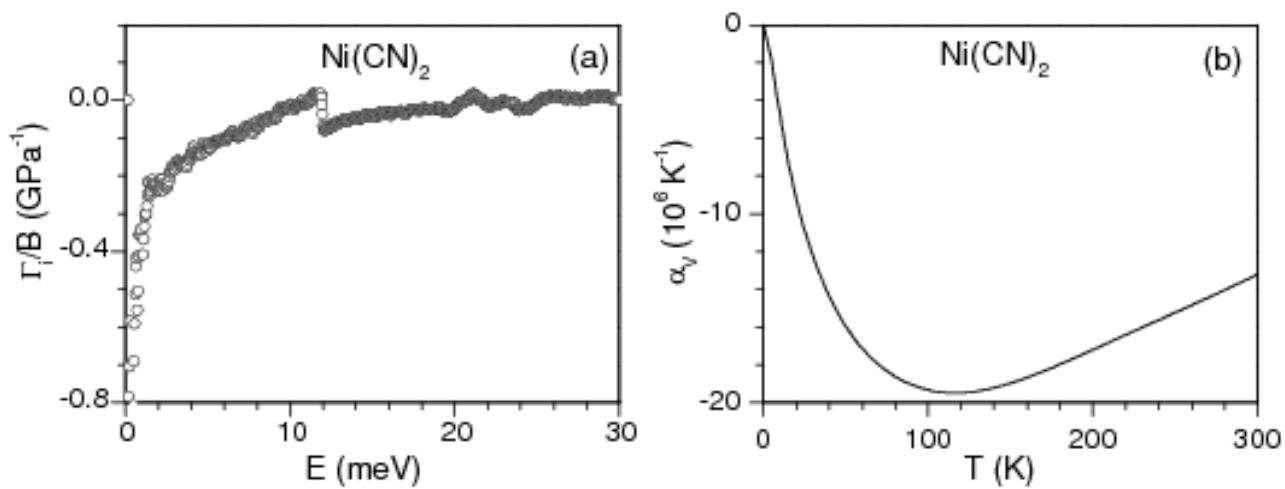

Fig. 97. Phonon dispersion curves for (a) Zn(CN)$_2$ and (b) Cd(CN)$_2$. The cyanide anion stretch modes at about 273 meV are not shown[28].

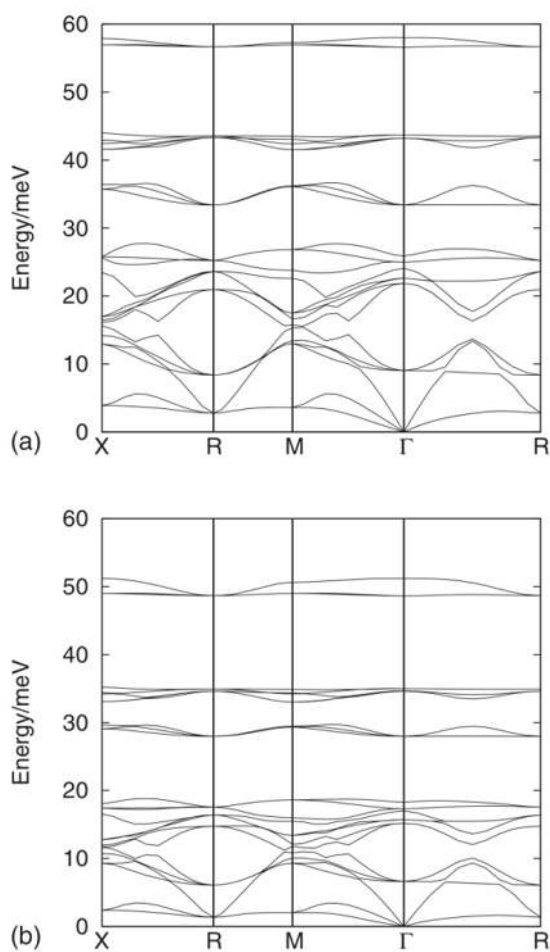

Fig. 98. The calculated Grüneisen parameters (γ) as a function of mode energy (ω). The calculations for Zn(CN)$_2$ and Cd(CN)$_2$ are plotted with squares and circles respectively. The cyanide stretch modes at 273 meV are not shown. They have Grüneisen parameters of 0.25[28].

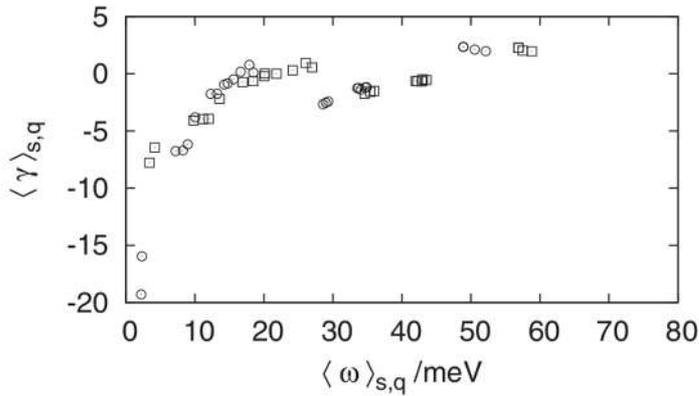

Fig. 99. Polarization vectors of the zone centre and zone boundary modes in Zn(CN)$_2$. For each mode the energy and Grüneisen parameter is indicated. The lengths of arrows are related to the displacements of the atoms. The absence of an arrow on an atom indicates that the atom is at rest. At the zone centre, where the phase factor is zero, the eigenvectors have only real components. The numbers after the mode assignments give the phonon energies and Grüneisen parameters respectively. Zone boundary eigenvectors have both real and imaginary components. The actual displacements of atoms are a combination of both these components. Key: Zn, grey spheres; C, red spheres; N, blue spheres[25].

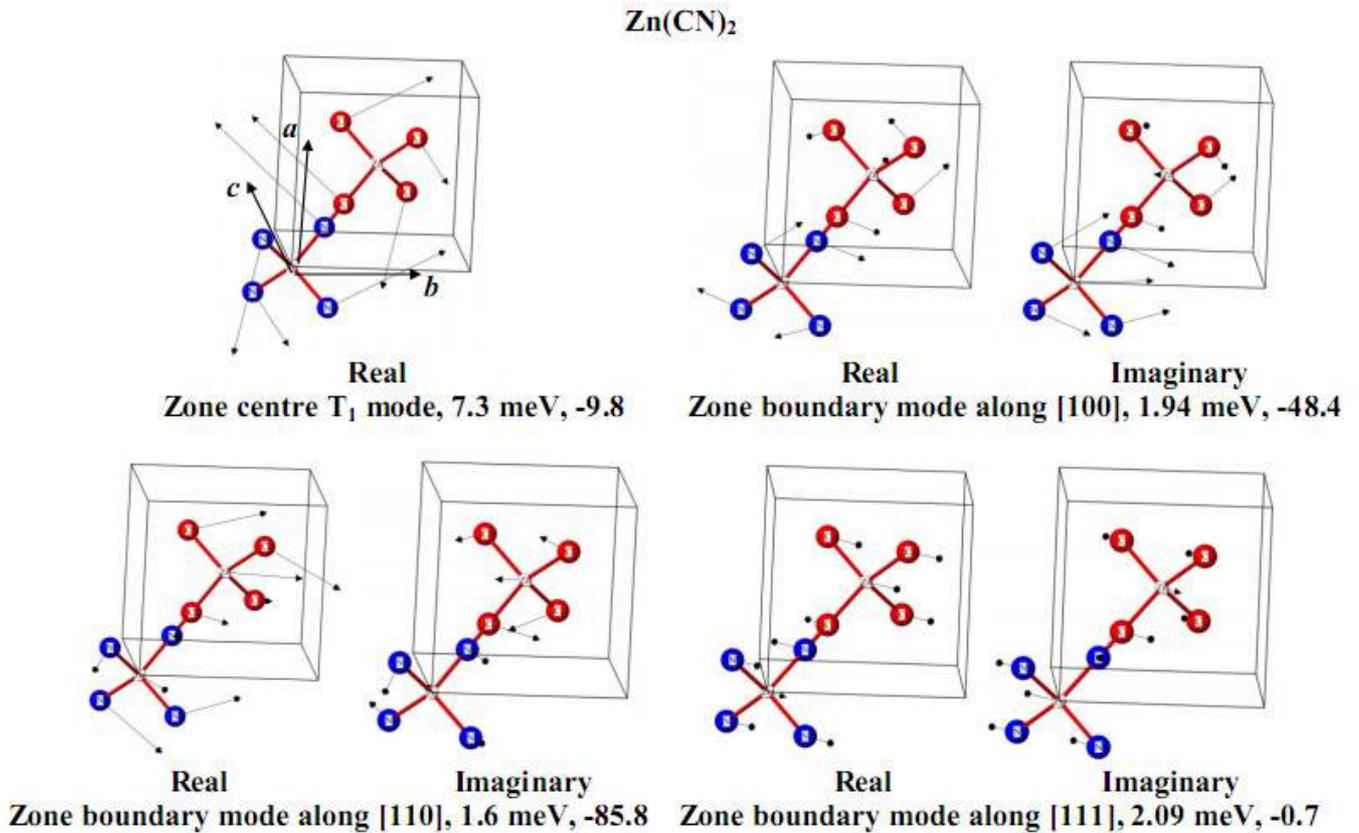

Fig. 100. Polarization vector of the zone centre and zone boundary modes in Ni(CN)$_2$. For each mode the energy and Grüneisen parameter is indicated. The lengths of arrows are related to the displacement of the atoms. The absence of an arrow on an atom indicates that the atom is at rest. At the zone centre, where the phase factor is zero, the eigenvectors have only real components. The numbers after the mode assignments give the phonon energies and Grüneisen parameters respectively. Zone boundary eigenvectors have both the real and imaginary components. The actual displacements of atoms are a combination of both these components. Key: Ni, grey spheres; C, red spheres; N, blue spheres[25].

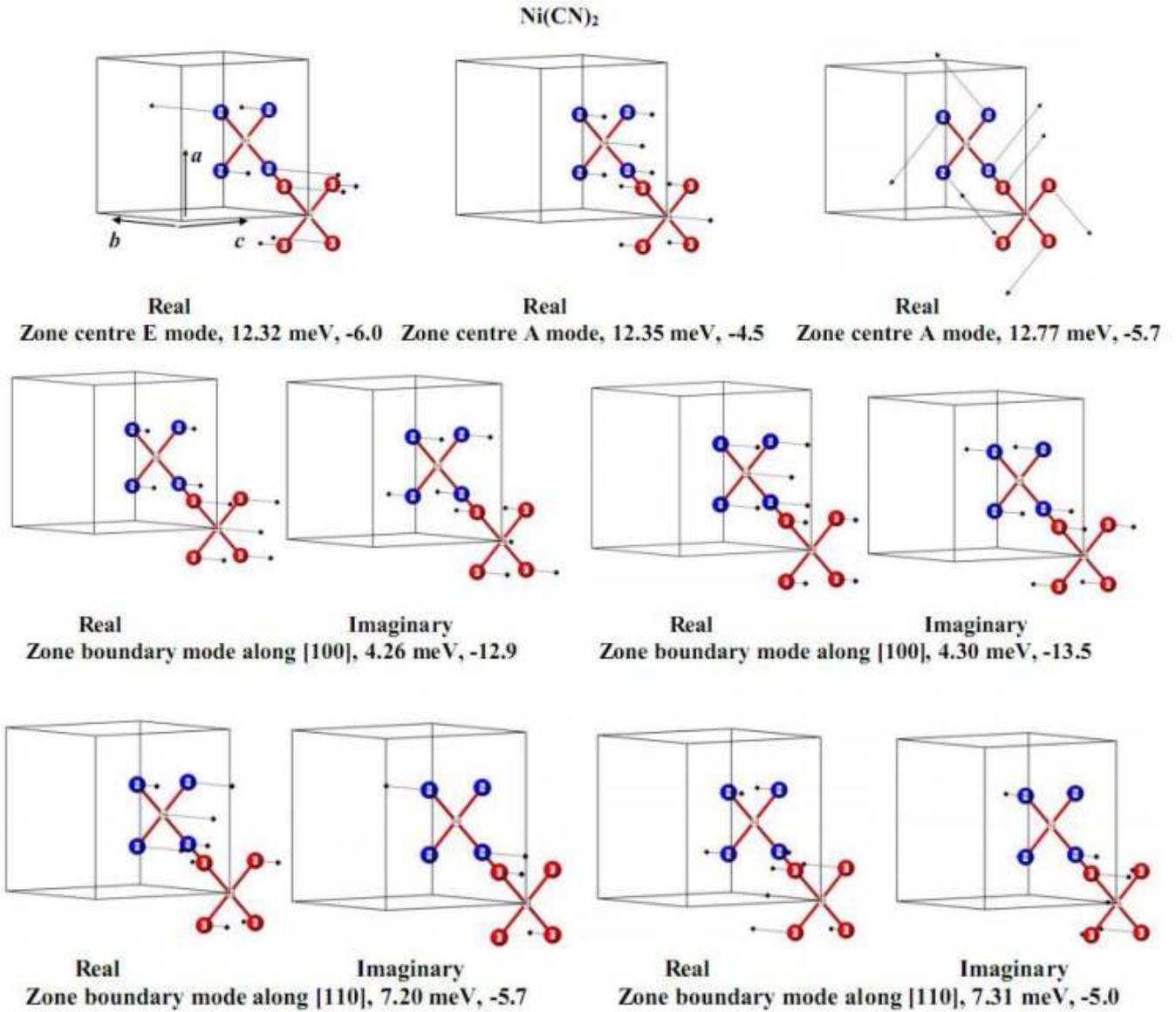

Fig. 101. The crystal structure of Ag$_3$M(CN)$_6$ (M=Co, Fe)[24].

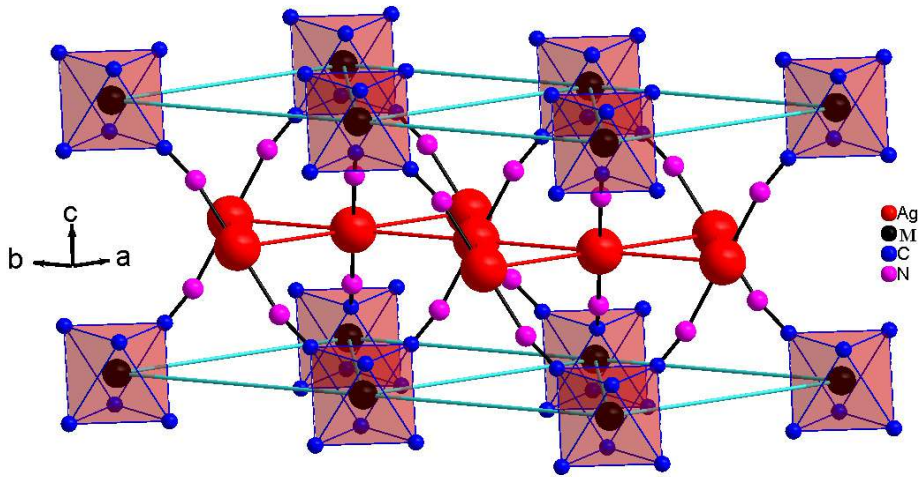

Fig. 102. The temperature dependence of the phonon spectra for Ag$_3$Co(CN)$_6$. The phonon spectra are measured with an incident neutron wavelength of 4.14 Å using the IN6 spectrometer at the ILL. The calculated phonon spectra from *ab-initio* calculations are also shown. The calculated spectra have been convoluted with a Gaussian of FWHM of 10% of the energy transfer in order to describe the effect of energy resolution in the experiment[24].

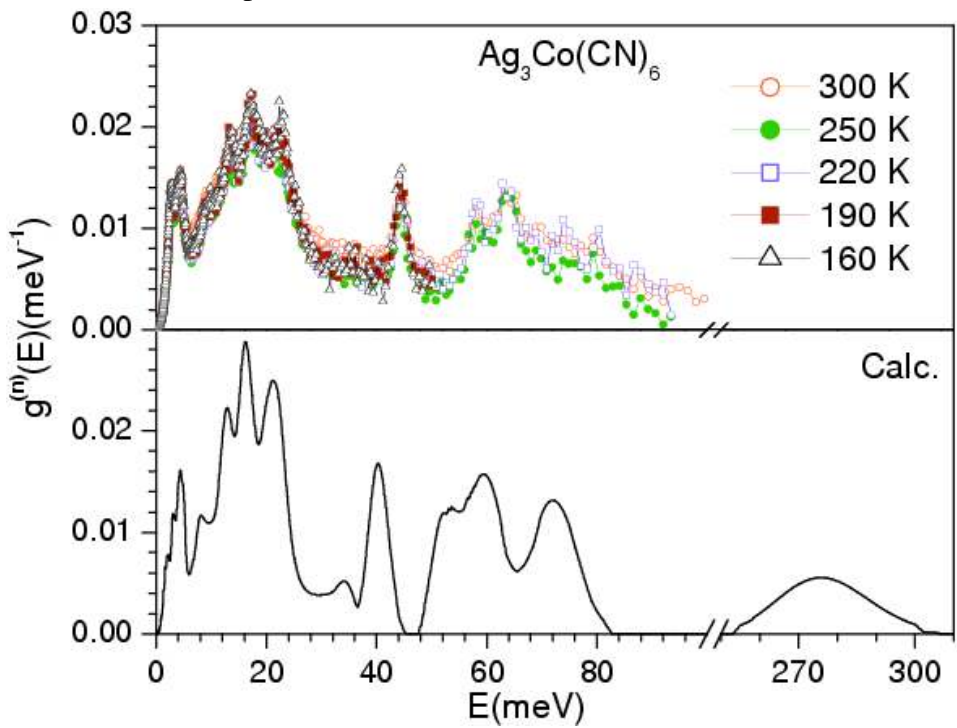

Fig. 103. The temperature dependence of the phonon spectra for $Ag_3Fe(CN)_6$. The phonon spectra are measured with an incident neutron wavelength of 4.14 Å using the IN6 spectrometer at the ILL. The calculated phonon spectra from *ab-initio* calculations are also shown. The calculated spectra have been convoluted with a Gaussian of FWHM of 10% of the energy transfer in order to describe the effect of energy resolution in the experiment[24].

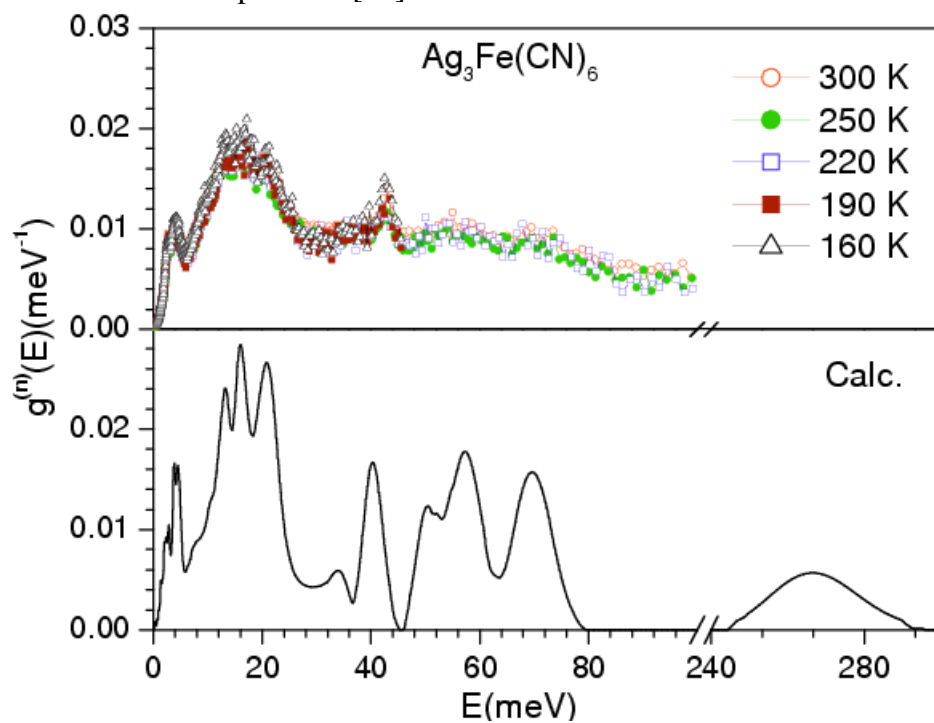

Fig. 104. The low-energy part of the temperature dependence of the phonon spectra for $Ag_3Co(CN)_6$ and $Ag_3Fe(CN)_6$, extracted from Figures 102 and 103[24].

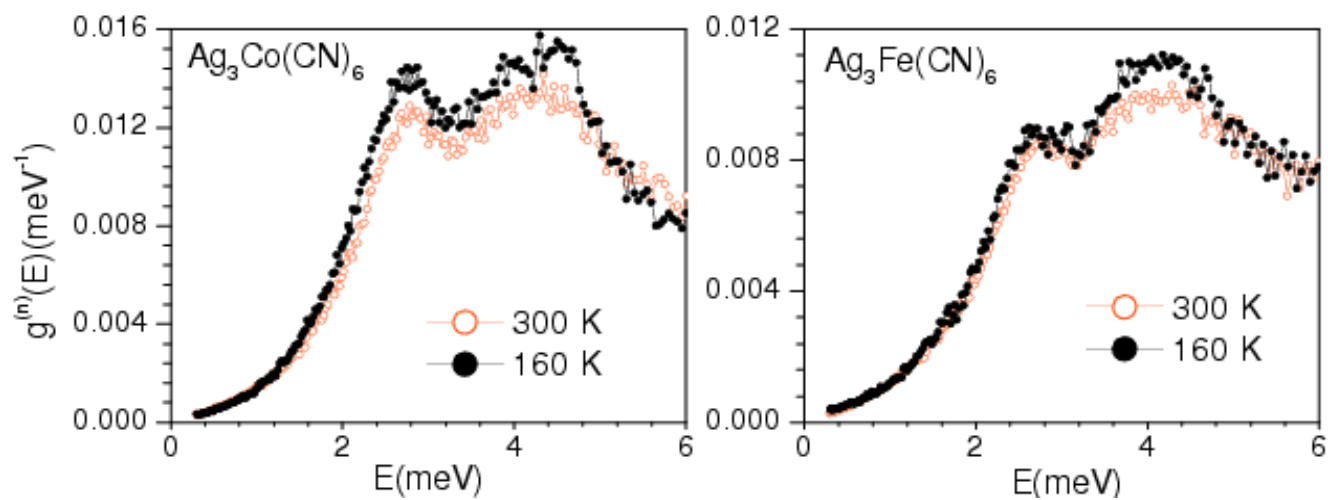

Fig. 105. Comparison of the experimental phonon spectra for $Ag_3Co(CN)_6$ and $Ag_3Fe(CN)_6$ at 300 K[24].

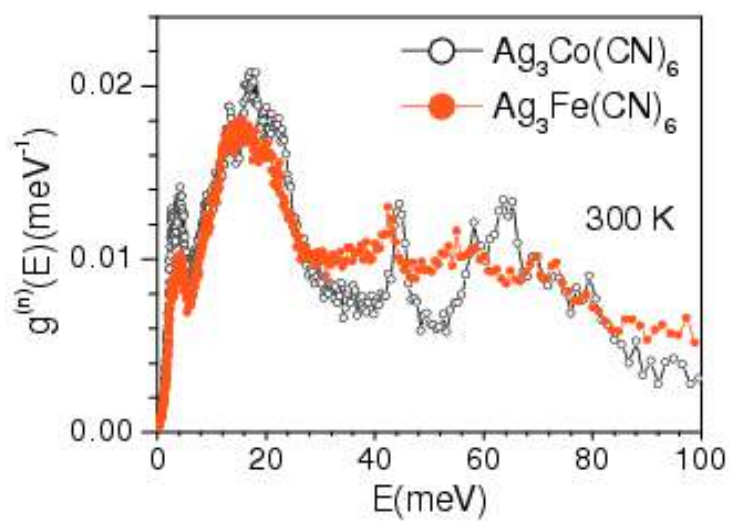

Fig. 106. The *ab-initio* calculated partial density of states for the various atoms in $Ag_3Co(CN)_6$ and $Ag_3Fe(CN)_6$[24].

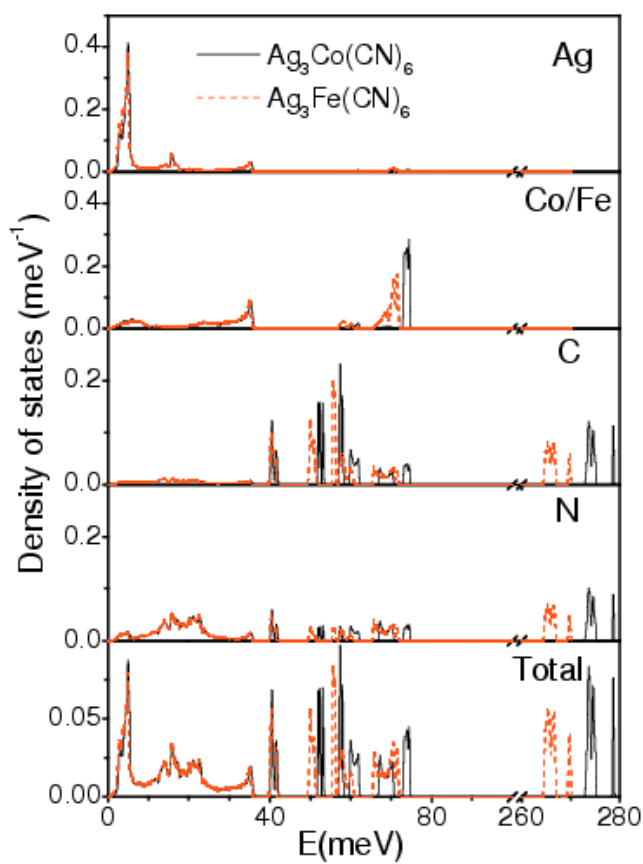

Fig. 107. The temperature dependence of the experimental Bose-factor corrected scattering function $S(Q,E)$ for $Ag_3Co(CN)_6$ (top panel) and $Ag_3Fe(CN)_6$ (bottom panel). For clarity, a logarithmic representation is used for the intensities. The measurements were performed with an incident neutron wavelength of 4.14 Å (4.77 meV)[24].

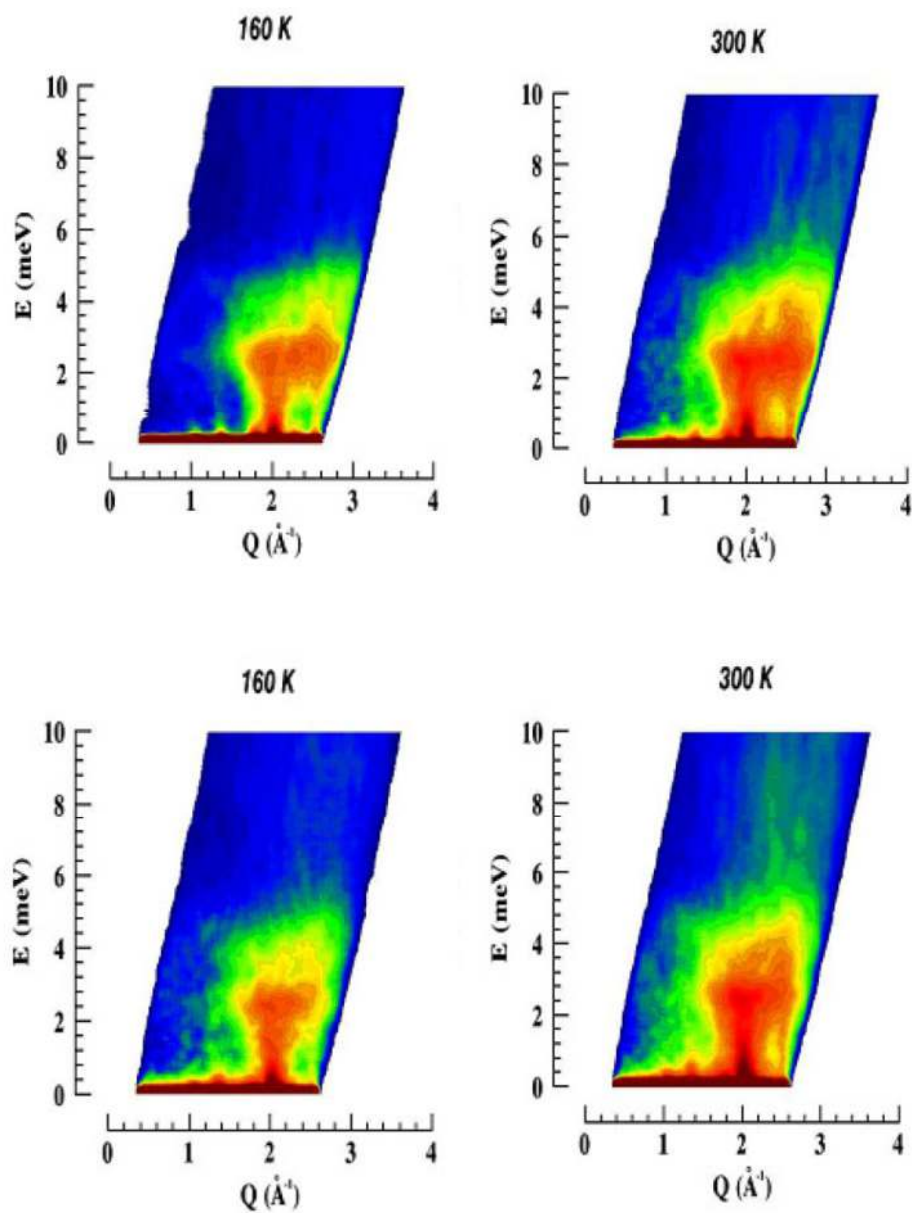

Fig. 108. The *ab-initio* calculated phonon dispersion curves for Ag$_3$Co(CN)$_6$ and Ag$_3$Fe(CN)$_6$. The Bradley-Cracknell notation is used for the high-symmetry points along which the dispersion relations: Γ=(0,0,0), M(1/2,0,0), A(0,0,1/2), and L(0,1/2,1/2). In order to expand the *y*-scale, the six dispersion-less modes due to the cyanide stretch at about 265 to 280 meV are not shown[24].

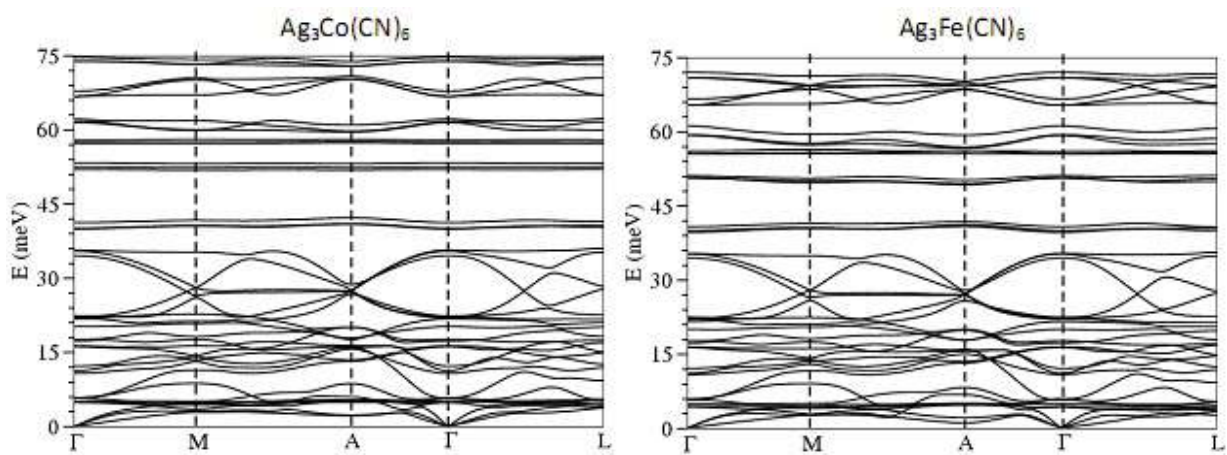

Fig. 109. The experimental phonon spectra for Ag$_3$Co(CN)$_6$ as a function of pressure at 200 K: ambient pressure (full line), 0.3 kbar (dotted line), 1.9 kbar (dashed line), and 2.8 kbar (dash-dotted line). The measurements were done at 200 K using the IN6 spectrometer at the ILL with an incident neutron wavelength of 5.12 Å[24].

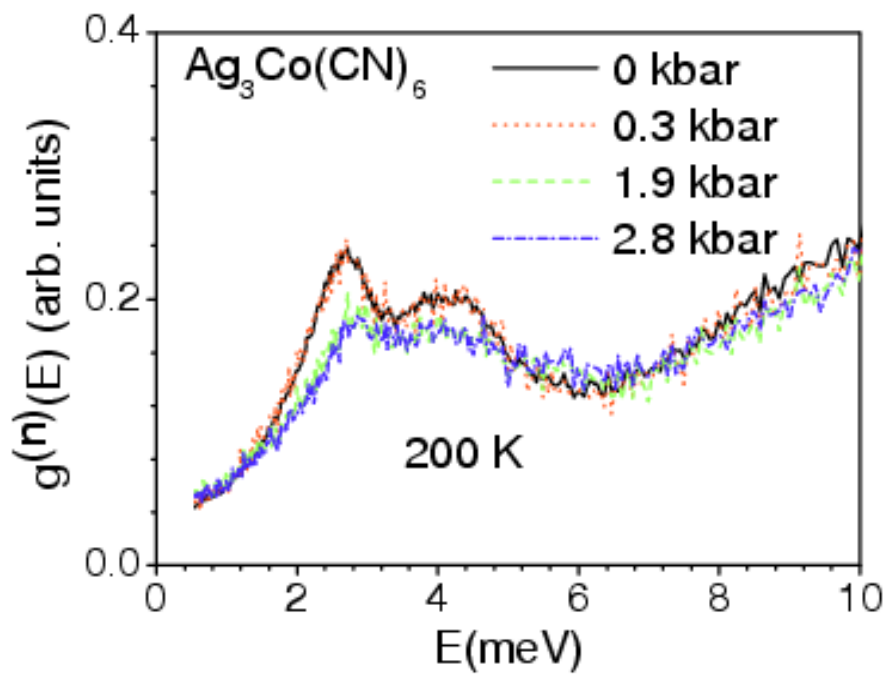

Fig. 110. The mean phonon frequencies of various atoms as obtained from ab-inito calculations in Ag$_3$Co(CN)$_6$ (full lines) and Ag$_3$Fe(CN)$_6$ ( dashed lines) and from reciprocal-space analysis of neutron diffraction data [103] in Ag$_3$Co(CN)$_6$ (symbols plus lines)[24].

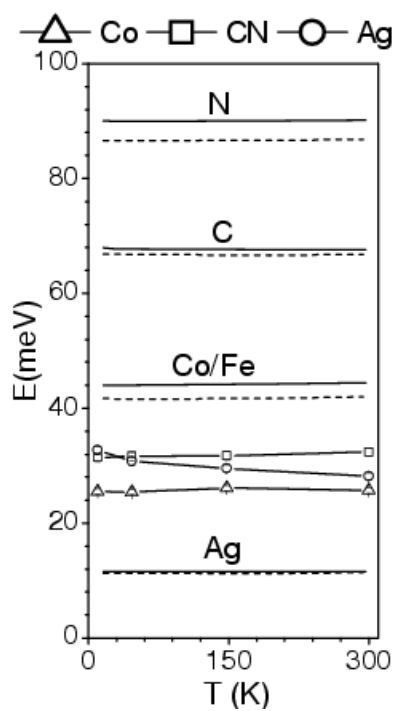

Fig. 111. The Grüneisen parameters, $\Gamma(E)$ for Ag$_3$Co(CN)$_6$ and Ag$_3$Fe(CN)$_6$ as a function of phonon energy $E$, from *ab-initio* calculations[24].

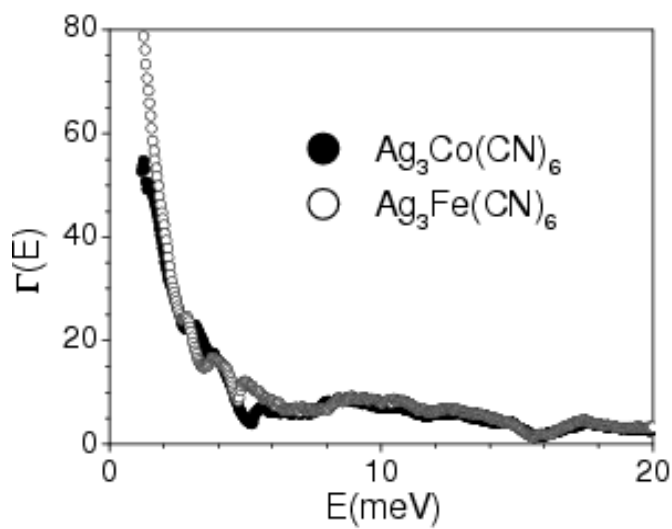

Fig. 112. The comparison between the volume thermal expansion derived from the present *ab-initio* calculations (solid line) and that obtained using X-ray diffraction [103, 104](open circles)[24].

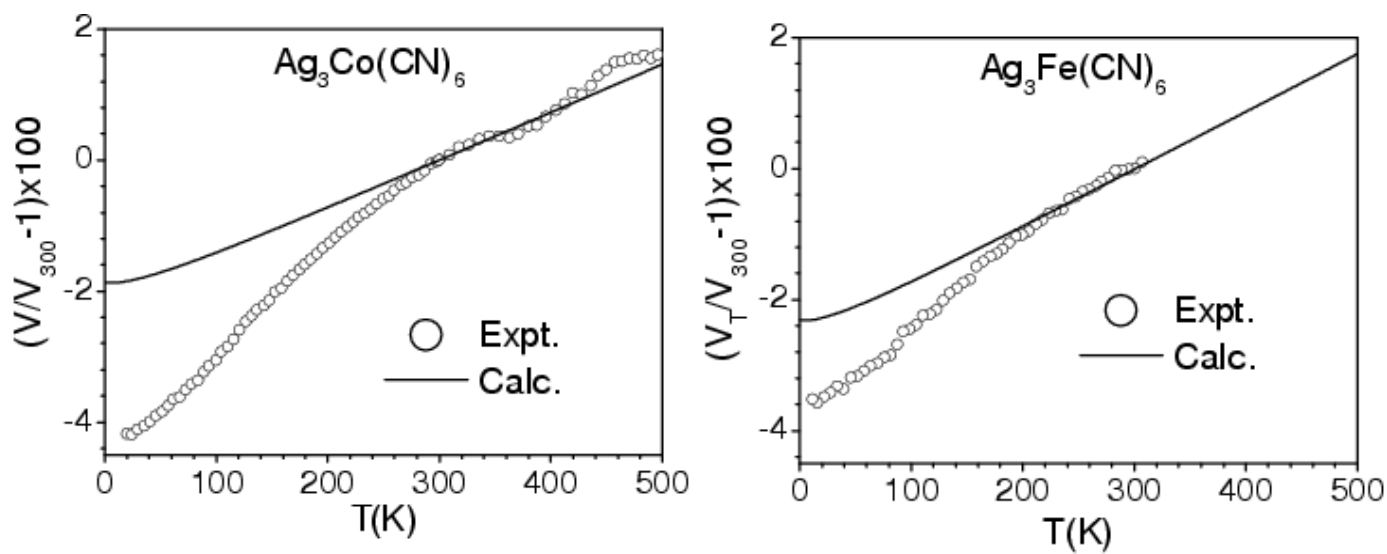

Fig. 113. The contribution of phonons of energy $E$ to the volume thermal expansion coefficient ($\alpha_V$) as a function of $E$ at 300 K in $Ag_3Co(CN)_6$ and $Ag_3Fe(CN)_6$, from *ab-initio* calculations[24].

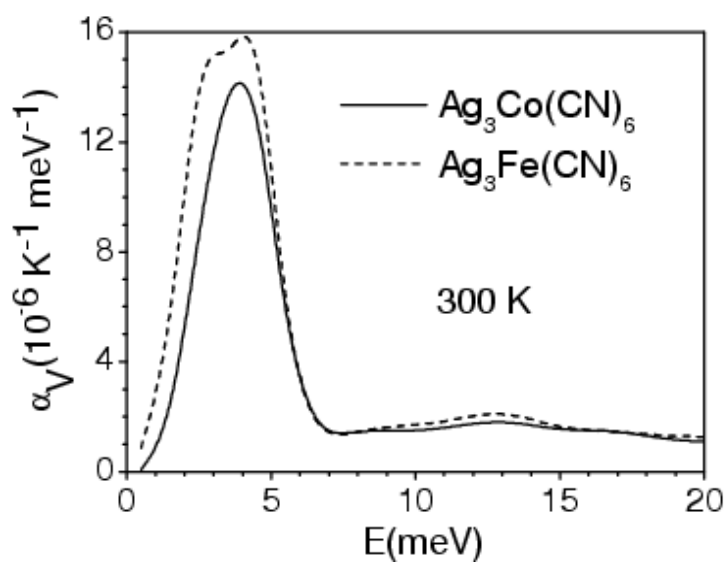

Fig. 114. The calculated contribution to the mean squared amplitude of the various atoms arising from phonons of energy E at $T$=300 K in $Ag_3[Co(CN)_6]$ and $Ag_3[Fe(CN)_6]$, from *ab-initio* calculations[24].

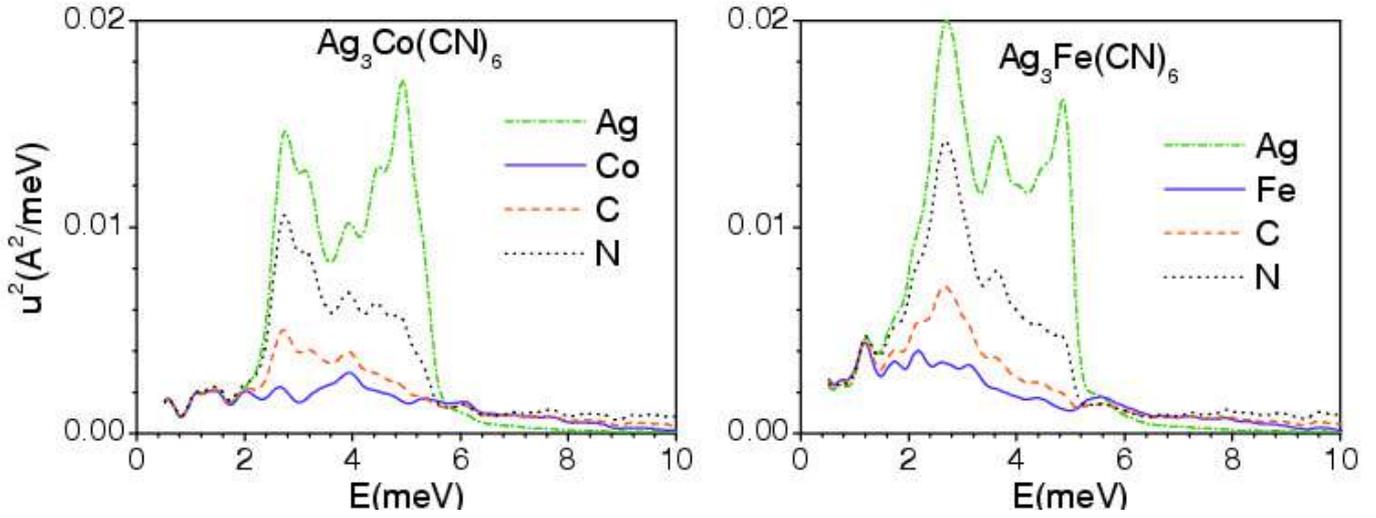

Fig. 115. Polarization vector of the zone centre modes in $Ag_3[Co(CN)_6]$ and $Ag_3[Fe(CN)_6]$. For each mode the energy and Grüneisen parameter are indicated. The lengths of the arrows are related to the displacement of the atoms. The absence of an arrow on an atom indicates that the atom is at rest. The numbers after the mode assignments refer to the phonon energies and Grüneisen parameters, respectively[24].

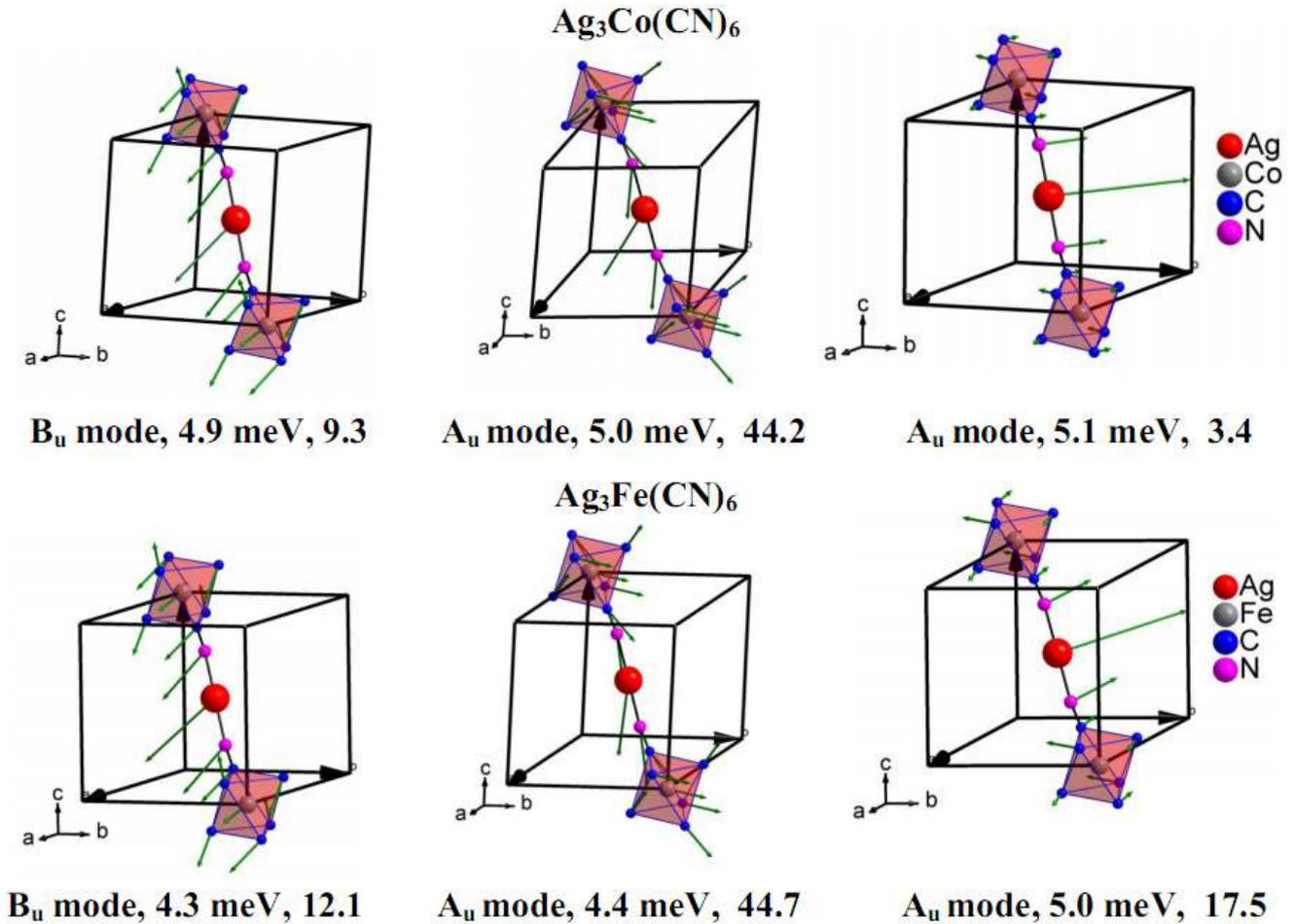

Fig. 116. The pressure dependence of Infrared spectra of Ag3[Co(CN)6] measured at room temperature. Two bands assigned to the cyanide stretching modes between 2200 and 2250 cm$^{-1}$ are not shown[48].

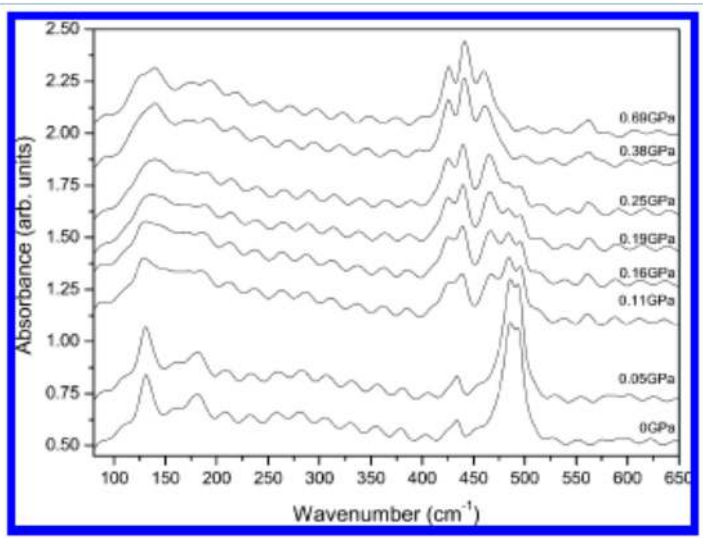

Fig. 117. Pressure dependence of the most intense infrared modes of Ag3[Co(CN)6][48].

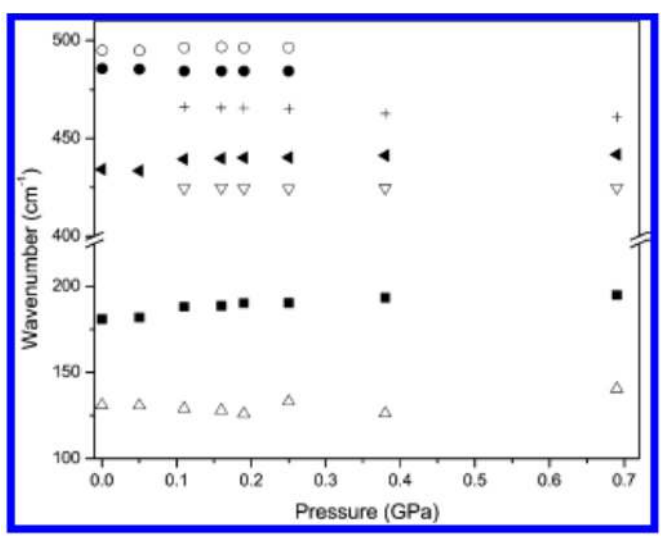

Fig. 118. Temperature and pressure variation of phonon frequencies of KMnAg$_3$(CN)$_6$ showing monotonic decrease of all mode frequencies with increase in temperature. All modes, except the CN stretch band, disappear above 13 GPa[44].

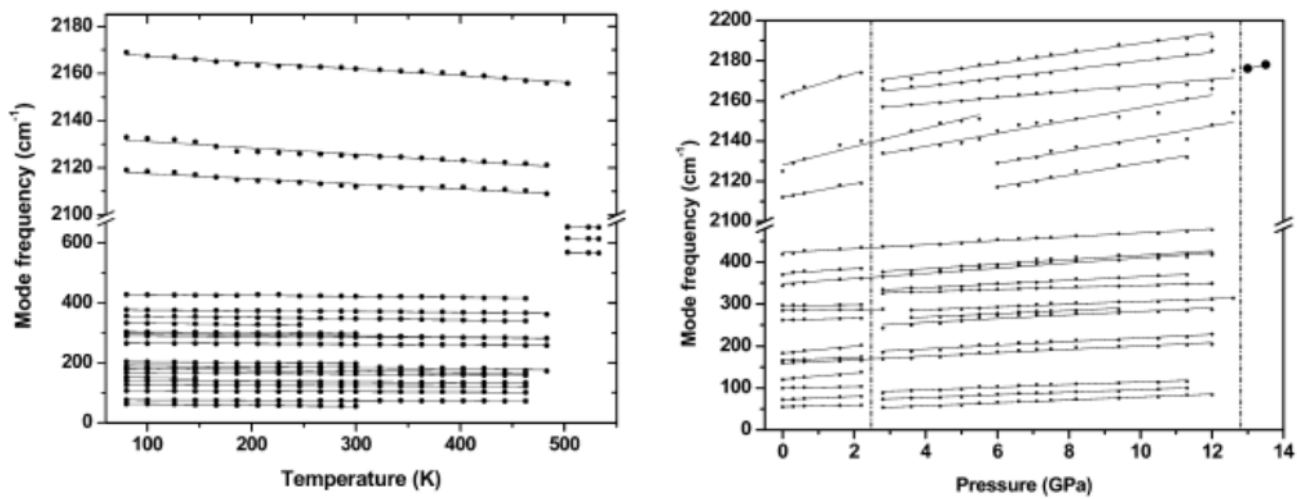

Fig. 119. The structure of AuCN and HT-CuCN/AgCN as used in the ab-intio calculations. Key: C, red sphere; N, blue sphere; Cu/Ag/Au green sphere[1]

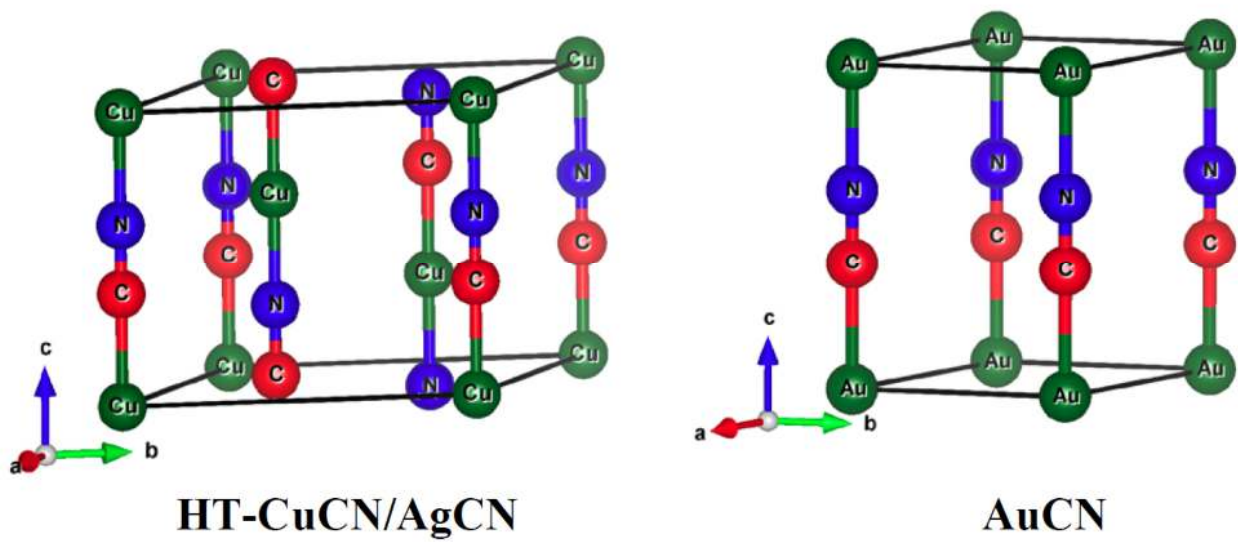

Fig. 120. The comparison between the measured (310 K) and calculated phonon spectra of MCN (M=Cu, Ag and Au)[1].

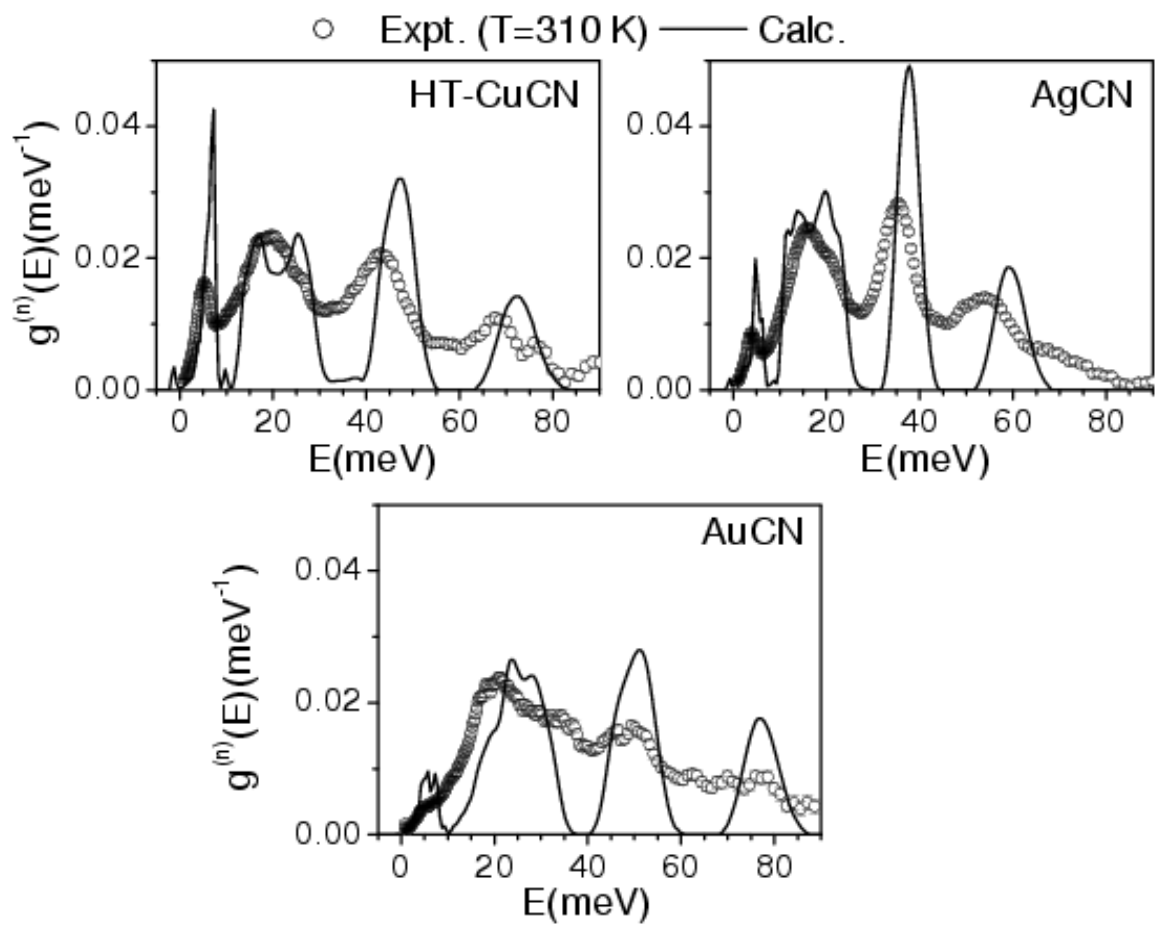

Fig. 121. (a) The calculated Grüneisen parameters ($\Gamma_l$, $l$=a, c) as obtained from anisotropic stress along '$a$' and '$c$' axes. (b) The contribution of phonons of energy $E$ to the linear thermal expansion coefficients ($\alpha_a$ and $\alpha_c$) as a function of $E$ at 300 K[1].

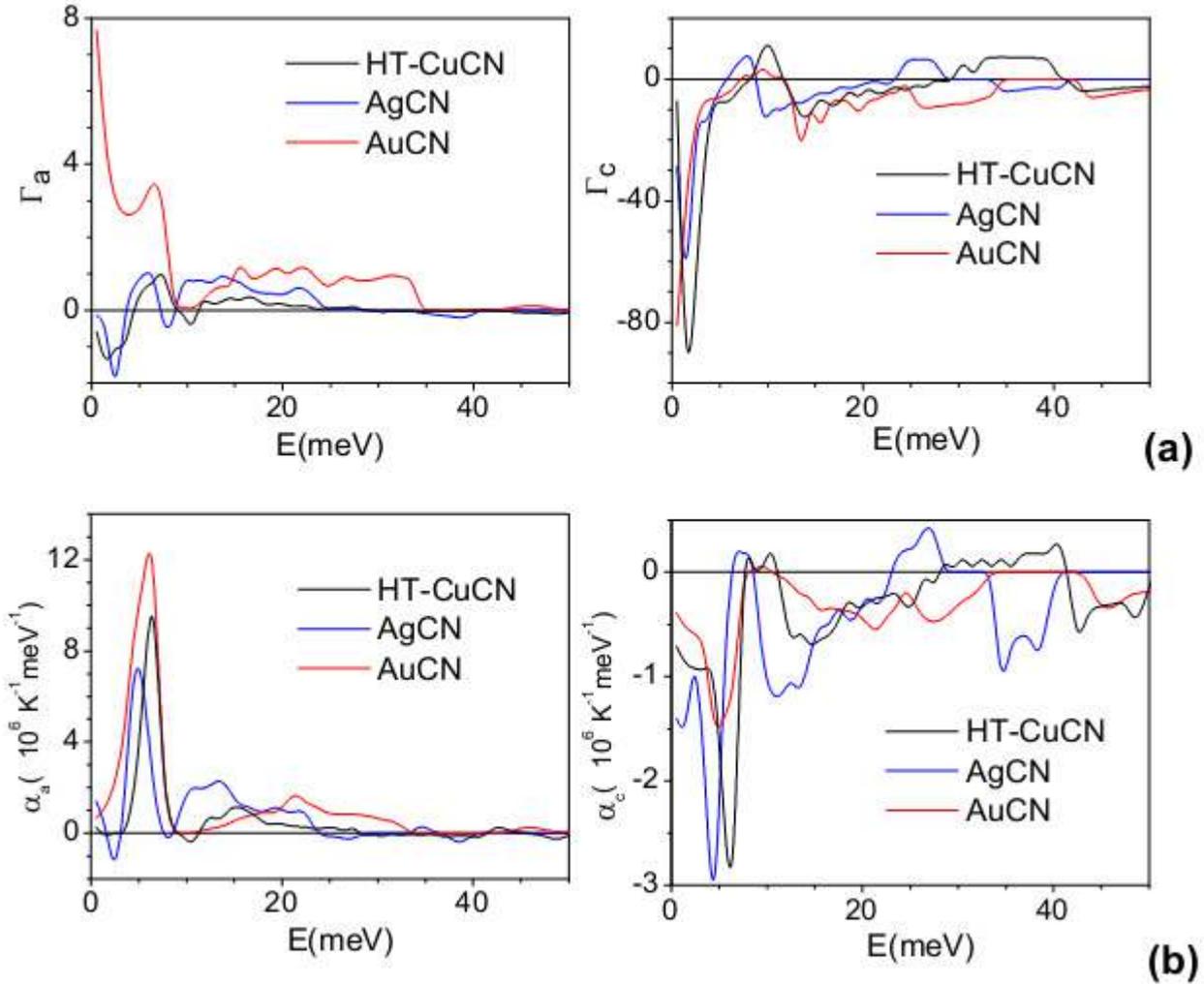

Fig. 122. The calculated and experimental thermal expansion[79] behavior of MCN (M=Cu, Ag and Au)[1].

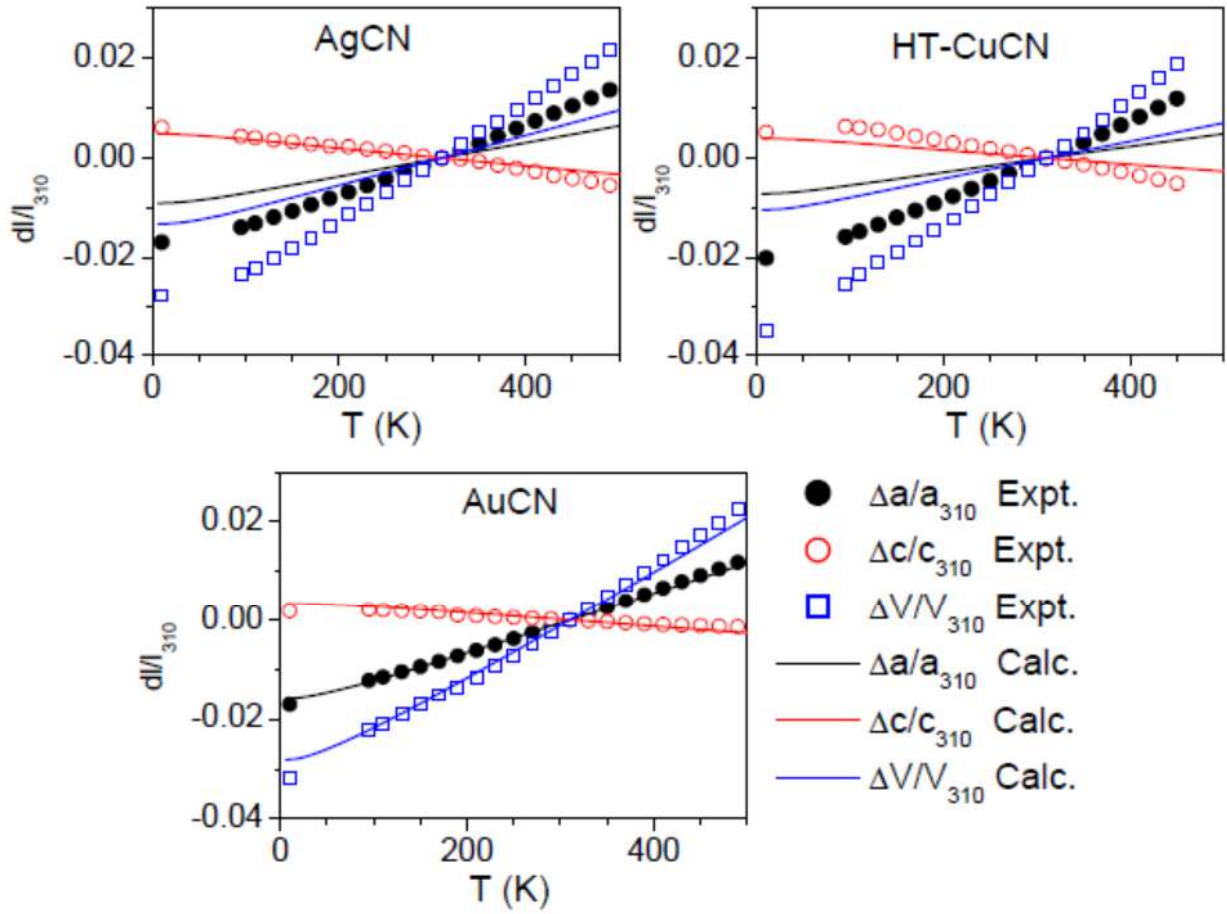

Fig. 123. The calculated displacement pattern of various phonon modes in AuCN and HT-CuCN and corresponding Grüneisen parameters. The first line below each figure represents the size of the supercell. The second line below the figure give the high symmetry point, phonon energies and Grüneisen parameters, respectively. In the bottom panel (HT-CuCN and AgCN) the second and third line below the figure corresponds to HT-CuCN and AgCN respectively. The Bradley-Cracknell notation is used for the high-symmetry points. AuCN: A=(0 0 1/2)$_H$, K(1/3,1/3,0)$_H$ and M(1/2,1/2,0)$_H$; HT-CuCN/AgCN: L(0,1/2,0)$_R$ ≡ (-1/2,1/2,1/2)$_H$, T1, T2 (1/2,1/2,-1/2)$_R$≡ (0, 1, 1/2)$_H$, F(1/2,1/2,0)$_R$≡(0,0.5,1)$_H$, LD(-2/3,1/3,1/3)$_R$≡ (1,0,0)$_H$. Subscript R and H correspond to rhombohedral and hexagonal notation respectively. The c-axis is along the chain direction, while a and b-axis are in the hexagonal plane. Key: C, red sphere; N, blue sphere; Cu/Ag/Au green sphere.

*The Grüneisen parameters values of unstable F and LD-point modes are not given. The modes are found to become more unstable on further compression of the lattice. Such type of modes would contribute maximum to the NTE along c-axis[1].

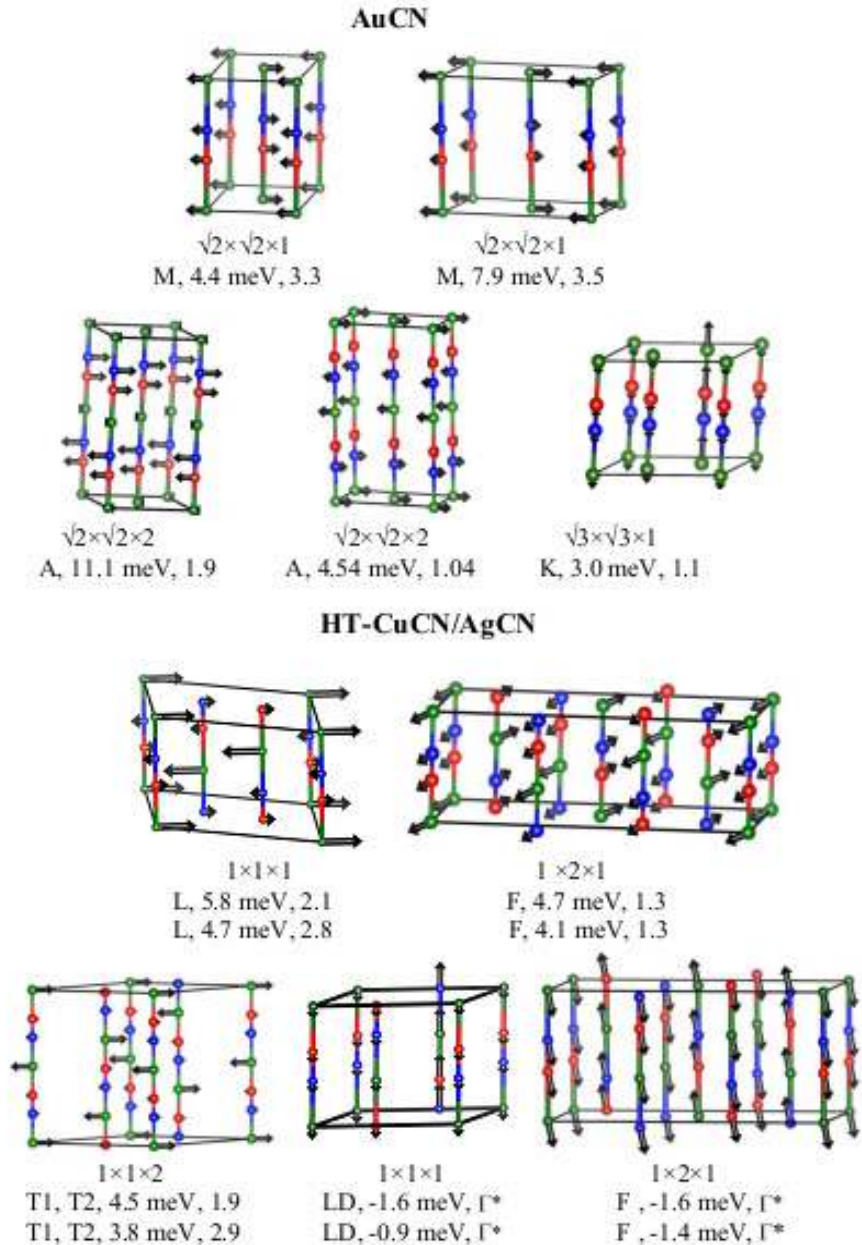